\documentclass[reqno]{amsart}
\usepackage{amsmath,amssymb,latexsym,amscd}
\usepackage{graphicx}
\newcommand{\hC}{\hat C}
\newcommand{\hd}{\hat d}
\newcommand{\h}{\hat h}
\newcommand{\F}{\hat F}

\newcommand{\tF}{\tilde F}
\newcommand{\tih}{\tilde h}
\newcommand{\tP}{\tilde \Pi}
\newcommand{\beq }{\begin{equation}}
\newcommand{\eeq }{\end{equation}}
\newtheorem{theorem}{Theorem}
\newtheorem{lemma}{Lemma}
\newtheorem{proposition}{Proposition}
\newtheorem{corollary}{Corollary}
\newtheorem{definition}{Definition}
\theoremstyle{definition}
\newtheorem{remark}{Remark}
\newtheorem{example}{Example}
\numberwithin{equation}{section}

\begin{document}
\title[Multisymplectic Theory of Balance Systems]%
{Multisymplectic Theory of Balance Systems, I}
\author{Serge Preston}\address{Department of Mathematics and Statistics, Portland State University,
Portland, OR, U.S.}\email{serge@mth.pdx.edu}

\begin{abstract}
In this paper we are presenting the theory of balance equations of the Continuum Thermodynamics (balance
systems) in a geometrical form  using Poincare-Cartan formalism of the Multisymplectic Field Theory. A
constitutive relation $\mathcal{C}$ of a balance system $\mathcal{B_{C}}$ is realized as a mapping between
a (partial) 1-jet bundle of the configurational bundle $\pi:Y\rightarrow X$ and the extended dual bundle
similar to the Legendre mapping of the Lagrangian Field Theory. Invariant (variational) form of the
balance system $\mathcal{B_{C}}$ is presented in three different forms and the space of admissible
variations is defined and studied. Action of automorphisms of the bundle $\pi$ on the constitutive
mappings $\mathcal{C}$ is studied and it is shown that the symmetry group $Sym(\mathcal{C})$ of the
constitutive relation $\mathcal{C}$ acts on the space of solutions of balance system $\mathcal{B_{C}}$.
Suitable version of Noether Theorem for an action of a symmetry group is presented with the usage of
conventional multimomentum mapping.  Finally, the geometrical (bundle) picture of the RET in terms of
Lagrange-Liu fields is developed and the entropy principle is shown to be equivalent to the holonomicy of
the current component of the constitutive section.
\end{abstract}
 \maketitle
\today
 \tableofcontents

\section{Introduction.}

This paper is the first part of a work where we are presenting the theory of balance equations of
Thermodynamics of Continuum in the framework of the variational, Multisymplectic Field Theory
(\cite{BSF,FF,EMR1,GIMMSY1,LMD,MS}). In doing so we pursue, in this, first, part of the work the following
main goals. The first is to formulate the theory of balance equations (balance systems) possibly closer to
the classical Lagrangian Field Theory in order to be able to use an extensive variety of tools developed
in this theory for the study of balance systems. The second goal is to have a united mathematical scheme
of several variants of Irreversible Thermodynamics that differs by the type of the domain of the
constitutive relations of this theory. On one side is the scheme where the state space of basic fields for
which the balance equations are present is chosen to be as small as possible and a constitutive relations
of the theory is allowed to depend on all first derivatives of these fields (in this part of the work we
consider only first order theories). See for instance \cite{ME} for the presentation of entropy principle
in such a case. On the other end there is the Rational Extended Thermodynamics where all the necessary
derivatives of the basic fields are included into the state space and the constitutive relations depend on
the fields but not on their derivatives (\cite{MR,M}).  In between these two extreme positions there is a
variety of situations where some derivatives of basic fields are included into the state space and some
are not \cite{JCL}. Quite often the choice of the derivatives (gradients of basic fields or their time
derivatives) is related to the symmetry group of the described physical situation or to the covariance
group required for the system of balance equations.
\par
The second part of this work (in preparation, see also \cite{Pr}) will be devoted to the study of the
"entropy principle" - requirement that any solution of the balance system that includes all but the
entropy balances satisfies to the entropy balance (such a requirement place a serious constraints to the
constitutive relations of the balance system).\par

In the third part of the work we study the covariance principle for the balance systems - condition that
the balance system would be covariant with respect to a (finite or infinite dimensional) Lie group. Such
study was pioneered in the Green-Naghdy-Rivlin Theorem and later on studied by J.Marsden and T.Hughes, see
\cite{MH,YMO} and M.Silhavy (\cite{Si}).\par

 This work was originated at the Conference Thermoconn 2005 in
Messina after the lecture of Professor T. Ruggeri on the Rational Extended Thermodynamics and the
discussion that I had there with Professor W. Muschik about the Entropy Principle.\par

Rational Extended Thermodynamics, (RET) which occupies an important place in this part of our work, was
initiated in the works of I.Liu and I.Muller and developed by the T.Ruggeri and I.Muller (see
\cite{MR,R1,R2,R3}). The formalism of RET "is elegant and appealing" (\cite{JCL}) and it was very tempting
to present it in a geometrical form following the framework of a Classical Field Theory
(\cite{BSF,FF,P,GMS}) and to implement the principal structures of RET in a natural geometrical way. Thus,
we present in Appendix II a sketch of the formalism of RET.\par
\par

In Section 3 we recall the basic structures of Multisymplectic Field Theory following (\cite{LMD,EMR1}).
The only new material here is the subsection 4.5 on the vertical contact structure in the space $W_{0}$ of
the united multisymplectic scheme and the characterization of Legendre mappings generated by the
Lagrangians in terms of this structure.
\par
In Section 4 we define the partial 1-jet bundles $J^{1}_{p}(\pi)$ for a configurational bundle
$\pi:Y^{(n+1)+m}\rightarrow X^{n+1}$ of m basic fields $y^{i}\in U$ over the physical or material
space-time $X^{n+1}$. We discuss two examples of such jet bundles. One, $J^{1}_{K}(\pi)$, defined by a
distribution $K\subset T(X)$ (or, with more details, by an almost product structure $T(X)=K\oplus K'$) on
the space $X$, another, $J^{1}_{S}(\pi )$, for a case where $K\oplus K'=<\partial_{t}>\oplus T(B)$, $B$
being the material or physical space and the space $U$ of the basic fields splits as the product
corresponding to the type of (first order) derivatives that enters the constitutive relations. A study of
more general types of partial jet-bundles including the jet bundles of higher order will be pursued in the
second part of this work.  In Section 6 we define and study the partial Cartan structure on the 1-jet
bundles $J^{1}_{p}(\pi)$ of these two types.
\par
In Sections 7 and 8 we study the prolongation of vector fields and connections to the partial 1-jet
bundles following the similar prolongation procedures for the conventional 1-jet bundles
(\cite{S,KMS,LMD}).\par

In Section 9 we define a general constitutive relation $\mathcal{C}$ as a smooth mapping between the
partial 1-jet bundle $J^{1}(\pi)$ and the (total) dual space ${\tilde
Z}=\Lambda^{(n+1)+(n+2)}_{2}Y/\Lambda^{(n+1)+(n+2)}_{1}Y$
\[
\mathcal{C}(x^\mu ,y^i, z^{i}_{\mu}) =(x^\mu ,y^i; F^{\mu}_{i}(x^\mu ,y^i, z^{i}_{\mu});\Pi_{i}(x^\mu
,y^i, z^{i}_{\mu})),
\]
containing the current part $F^{\mu}_{i}dy^i\wedge \eta_{\mu}$ and the source part $\Pi_{i}dy^i \wedge
\eta$.  We introduce the covering constitutive relation $ \widetilde{\mathcal{C}}$ defined by
$\mathcal{C}$, extending the Legendre transformations defined by a Lagrangian form $L\eta$.  We define the
Poincare-Cartan form $\Theta_{\mathcal{C}}=F^{\mu}_{i}dy^i \wedge \eta_{\mu}+\Pi_{i}dy^i \wedge \eta$ of a
constitutive relation $\mathcal{C}$, the Poincare-Cartan form $\Theta_{  \widetilde{\mathcal{C}} }$ of a
covering relation $ \widetilde{\mathcal{C}}$ and give several examples of types of constitutive relations:
Lagrange Type $\mathcal{C}_{L}$, mixed type with a Lagrangian current part and the source term given by a
dissipative potential ($L+D$ type), and vector-potential type.\par

In Section 10 we discuss three variational ways to get to the balance system $\mathcal{B_{C}}$
corresponding to a constitutive relation $\mathcal{C}$ (i.e. using variations $\xi \in X(J^{1}_{p}(\pi)$
and the differential of the Poincare-Cartan form $\Theta_{C}$). In doing this a traditional way, i.e.
requiring that $j^{1}(s)^{*}di_{\xi^1}\Theta_{\mathcal{C}}=0$ or $j^{1}(s)^{*}i_{\xi^1}d\Theta_{
\widetilde{\mathcal{C}}}=0$ we have, in general, to put the condition(s) $F^{\mu}_{i}D_{\mu}\xi^{i}=0 $ on
the variations $\xi$ of the Poincare-Cartan form. Locally there are always enough of such
$\mathcal{C}$-admissible variations $\xi$ to separate balance equations (Proposition 15) but globally this
may not be true.  In a case of semi-Lagrangian constitutive relations (see Sec.9) close to the
conventional Lagrangian field theory or in the case of RET constitutive relations no limitations on the
admissible variations $\xi$ are present.\par

  That is why we present the third way, using the restricted horizontal differential $\hat
d$ (see Appendix III) instead of the conventional de-Rham differential $d$ for the invariant formulation
of a balance system. In this case one does not need to restrict variations $\xi$.  In a case of Lagrangian
constitutive relation $\mathcal{C}_{L}$ the balance system coincide with the Euler-Lagrange system of
equations defined by the Lagrangian $L$ in the traditional way.

In Section 11 we discuss the properties of $\mathcal{C}$-admissible vector fields, prove that
$\mathcal{C}$-admissible vector fields form a Lie algebra with respect to the brackets of vector fields in
the $L+D$-case and study the form of $\mathcal{C}$-admissible vector fields in the case of a model
(2+2)-balance system (two fields and one space dimension) and for the five fields model of fluid
thermodynamical system with generic constitutive relations (see Sec.2).\par

In Section 12 we discuss the action of extended geometrical (lifted from $Y$) transformations on the
constitutive relations $\mathcal{C}$ and on the corresponding Poincare-Cartan form $\Theta_{C}$, define
the symmetry group $Sym(\mathcal{C})$ of a constitutive relation $\mathcal{C}$ and prove that this
symmetry group acts on the space of solutions $Sol(\mathcal{B_{C}})$ of the balance system
$\mathcal{B_{C}}$. Using a connection $\nu $ in the configurational bundle $\pi:Y\rightarrow X$ we define
the $\nu$-homogeneous constitutive relations corresponding to a case where $\mathcal{C}$ depends on the
fields and their derivatives but not on the points of space-time $X$ explicitly.
\par

In Section 13 we prove the Noether Theorem for a balance system $\mathcal{B_{C}}$ under an action of a
symmetry Lie group $G\subset Aut(\pi)$ using the multimomentum mapping of a multisymplectic field theory
\cite{LDS,MS}. The Noether Theorem leads to the family of the balance equations which reduces to the
conservation laws for special (or absent) source terms. For a semi-Lagrangian constitutive relations or
for the case where constitutive relations do not depend on the derivatives of the basic fields (RET case)
this theorem is essentially equivalent to the conventional Noether Theorems of Lagrangian field Theory,
for the general constitutive relations our version of Noether Theorem has more limited character.
\par

In Section 14 we discuss the type of the balance system $\mathcal{B_{C}}$ as a system of PDE and show it
is a combination of interacting hyperbolic, parabolic and stationary parts.\par

In Section 15 we present the dual bundle picture of the RET balance systems in terms of LL-multipliers. We
prove that the fulfillment of the entropy principle here reduces to the holonomity of the total
constitutive section of the 1-jet bundle $J^{1}(\Lambda , \Omega^{3}(X))$ of $\lambda^i$-fields  with
values in the space of semi-basic 3-forms.\par

 In Appendices I-IV we
collect the information on the properties of partial volume forms $\eta_{\nu}$, used in the text, present
the basic formalism of the Rational Extended Thermodynamics, recall the definition of the Iglesias
differential \cite{IW} and definitions and principal properties of the horizontal differential $d_{H}$ and
its restricted version $\hat d$.
\par

Results of this work were presented at the Seventh International Seminar on Geometry, Continua and
Microstructure that took place at the University of Lancaster, UK in September 2006.  Short exposition of
the the work will be published in the Proceedings of this conference.

\centerline{\textbf{Notations}.}

For a manifold $M$ we will denote by
\begin{itemize} \item $\mathcal{X}(M)$ - the Lie algebra of vector
fields on $M$, \item  $\Lambda^k M$ - the space of exterior $k$-forms on $M$, \item $\Lambda
M=\oplus^{\infty}_{k=0} \Lambda^k M$ -  the exterior algebra of the manifold $M$,
\item $J^{1}(\pi)$ - the 1-jet bundle of a bundle $\pi:Y\rightarrow X.$
\end{itemize}

\vfill \eject  \addtocontents{toc}{Chapter I. Preliminaries.}
 \centerline{\textbf{Chapter I. Preliminaries.}}

\section{Settings.}

In this section we present the bundle settings of the classical field theory in the form suited for later
presentation.  For deeper exposition we refer to the monographs \cite{BSF, FF}.

\subsection{Space-time base manifold.}

A state of material body will be described by the collection of the fields $\{ y^i ,i=1,\ldots ,m\}$
defined in a domain $X=B\times I \subset R^{n+1}$ of the physical or material space-time $R^{n+1}$ . Here
$I\subset R_{t}$ is an interval of time and $B\subset R^n$ is a domain in the n-dim physical or material
(reference) space with or without boundary.  In the first case denote by $\partial B$ the boundary of
domain $B$. That makes $\bar X$ the manifold with the boundary $\partial X=\partial B \times I \cup
B\times\partial I$.  We assume that the pseudo-Riemannian metric $G$ is defined in $X$ that can be
extended to the boundary of $X$ if such does exist. An example of such a metric is the Euclidian metric
$G=dt^2+h$, $h$ being the canonical Euclidian metric in the physical space $R^{n+1}$, but having in mind
application of our scheme to material manifold or relativistic systems we prefer to keep $G$ more
general.\par In this part of the work we will consider $B$ to be an open subset of $R^n$.\par
 We will use local coordinates $x^\mu ,\mu =1,2,\ldots ,n$ and the time variable $t=x^0$ in
$X$. We will be using Greek indices for the space-time variables and large Latin indices for space
variables only.
\par

Denote by $\eta$ the volume n-form $\eta=\sqrt{\vert G\vert} dx^0 \wedge dx^2 \ \ldots \  \wedge dx^n$
corresponding to the metric $G$.  We will be using the n-forms
\[
\eta_{\mu}=i_{\partial_{x^{\mu}}}\eta,\ \mu=0,1,2,\ldots ,n,
\]
for instance $\eta_{0}=\sqrt{\vert G\vert}dx^1\wedge dx^2\ \ldots \  \wedge dx^n$.  Necessary properties
of these forms are presented in Appendix I.
\par

For separating of space and time we employ the flat connection $\kappa$ in the bundle $X\rightarrow R_{t}:
(t=x^0, x^A )\rightarrow t$. This defines the product structure (see \cite{LR}) in the space $X$:
$T(X)=T(R_{t})\oplus T(B)$ and the corresponding decomposition in the exterior algebra $\Lambda^{*}(X)$.
In particular, we have, in the fiber over each point $x\in X$ the following decomposition

\beq \Lambda^{n+1}_{x}(X)=\mathbb{R}\eta_{0\ x}\oplus dt\wedge \Lambda^{n-1}_{x}(B^n ). \eeq

\subsection{Configurational (state) bundle.}
Basic fields of a continuum thermodynamical theory $y^i$ (except of the entropy that will be included
later) take values in the space $U\subset \mathbb{R}^m$ which we will call, a {\bf basic state space} of
the system (see \cite{Mu} for a discussion about possible choices of the basic state space and the
consequences for the structure of corresponding thermodynamical theory).\par

Following the framework of a classical field theory (see \cite{BSF,FF}) we organize these fields in the
bundle
\[
\pi_{U}: Y\rightarrow X,\ X=I\times  B, \
\]
with the base $X$.  In simple cases $Y=X\times U$ is the cylinder $\mathbb{R}\times { B}$ with the base
$X$ and the fiber $U$.
\par

Denote by $Z$ the 1-jet bundle of the bundle $\pi$: $Z=J^{1}(\pi)$. Thus, we get the double bundle
$Z\xrightarrow{\pi_{10}} Y\xrightarrow{\pi} X$ with the composition mapping $\pi^{1}=\pi \circ \pi_{10}$
defining the bundle $Z\xrightarrow{\pi^{1}} X$.\par

To formulate balance equations in terms of exterior forms we denote by $\Lambda^{k}(X)$ the bundle of
$k$-forms on $X$ and by
\[
\Lambda^{n+(n+1)}(X)=\Lambda^{n}(X)\oplus \Lambda^{n+1}(X)
\]
the bundle of exterior forms in $X$ of degree $n+(n+1)$. Space of sections of this bundle has as its basis
the forms $\eta_{\mu},\mu=0,1,2,\ldots,n ;\eta$.\par

Taking the pullback of the bundle $\Lambda^{n+(n+1)}(X)\rightarrow X$ to $Y$ (or, what is the same,
construct the fiber product of $\pi$ and $\pi_{\Lambda X}$ we get the following commutative diagram

\beq
\begin{CD}
U\times \Lambda^{n+(n+1)} @>>> Y\underset{X}{\times} \Lambda^{n+(n+1)} @>>> \Lambda^{n+(n+1)}(X)\\
@V\pi_{\Lambda U}VV   @V\pi_{\Lambda Y} VV                        @V \pi_{\Lambda X}VV\\
U @>>> Y         @>\pi >> X
\end{CD}.
\eeq

Left column of this diagram represents a typical fiber of bundle $\pi_{\Lambda Y} $ over a point $x\in X$.
Sections of the bundle $\pi \circ\pi_{\Lambda Y} : Y\underset{X}{\times} \Lambda^{n+(n+1)} \rightarrow X$
are called "semibasic" (n+(n+1)) exterior forms on the total space $Y$ of the bundle $\pi$, see \cite{LR},
Sec.4.2.\par

In the same way, taking, for arbitrary $k$ the pullback of $k$-forms on $X$ with respect to the projection
$\pi^{1}$ we get the bundle of $\pi^{1}$-semi basic $k$-forms on $Z=J^{1}(\pi)$. \par

\subsection{Balance Equations.}
Here we define the balance equations of a conventional first order field theory.  Fields $y^i$ are to be
determined as solutions of the field equations having the form of {\bf balance equations} for the currents
$F^{\mu}_{i}$, (where often $F^{0}_{i}=y^i$)
\begin{equation}
F^{\mu}_{i,\mu}=F^{0}_{i,t}+F^{A}_{i,x^{A}}  =\Pi_{i},\ i=1,\ldots, m.
\end{equation}
Here the functions $\Pi_{i}(x^\mu ,y^i ,y^{i}_{,x^\mu})$ are called the {\bf production and source} of the
components $y^i$ and $\sum_{\nu=1}^{n}F^{ A}_{i}(x^\mu ,y^i ,y^{i}_{,x^\mu})\frac{\partial }{\partial
x^A}$ - the {\bf flow } of the component $y^i$. These quantities, in general, are assumed to be function
of the fields $y^i$, of the point $x^\mu \in X $ and of (all or some of) the derivatives $y^{i}_{,x^\mu}.$
To shorten notations we will be using $z$ as the short notation of all arguments $(x^\mu ,y^i
,y^{i}_{,x^\mu}).$\par

In the Rational Extended Thermodynamics where all the derivatives entering constitutive relations are
included in between the fields $y^i$ densities, currents and sources of balance laws depend on $x^\mu,
y^i$ only (see discussion of different types of balance systems below in Sec. 14.\par

As it is customary in the classical field theory, the balance laws
could be rewritten by introducing the exterior forms:

(n+1)-form of the flows
\begin{equation}
{\mathcal F}_i=F^{\mu}_{i}(z)\eta_{\mu},\ i=1,\ldots, m,
\end{equation}
 and the

(n+2)-form of the production and source

\begin{equation}
{\Pi}_i=\Pi_{i}(z)\eta .
\end{equation}
Then the balance laws (2.4) takes the form

\begin{equation}
d{\mathcal F}_i={ \Pi }_i ,\ i=1,\ldots, m,
\end{equation}
or
\[
{\tilde d}({\mathcal F}_{i}+\Pi_i)=0
\]
if we employ the Iglesias differential $\tilde d$ introduced in \cite{IW} (see Appendix III).\par These
relations should be fulfilled for the fields $y^i =s^{i}(x)$.\par

\begin{example}\textbf{Five fields thermodynamical system - fluid (5F-fluid).}\par
As an example we consider a "five fields" thermodynamical system describing a fluid (\cite{Mus,M}.  Such a
system has 5 basic fields: mass density $\rho$, velocity vector field $v^{A}$ and the absolute temperature
$\vartheta$. Correspondingly there are five balance laws in this system - mass balance (conservation) law
for the density $\rho$, linear momentum balance law for $p_{I}=\rho h_{IJ}v^J$ ($h$ being the standard
Euclidian metric in $R^3$) and the total energy balance law for the total energy $e = \epsilon
+\frac{1}{2}\vert v\vert^2$ (sum of internal ($\epsilon$) and kinetic energy per unit of mass). To each of
these balance laws there corresponds the flux form $F$ and the source (+production) form $\Pi$:\beq
\begin{cases}
\mathcal{F}_{\rho}=\rho \eta_{0}+\rho v^{A}\eta_{A};\ \Pi_{\rho}=0\\
\mathcal{F}_{\rho v^B}=\rho v^{B}\eta_{0}+(\rho v^{B}v^{A}-t^{BA})\eta_{A};\ \Pi_{v^B}=\rho f_{B}\eta,\\
\mathcal{F}_{e }=\rho (\epsilon +\frac{1}{2}\vert v\vert^2)\eta_{0}+[\rho (\epsilon +\frac{1}{2}\vert
v\vert^2)v^{A}-t^{A}_{B}v^{B}+q^A ]\eta_{A};\ \Pi_{\vartheta }=\rho f_{A}v^A+r.
\end{cases}
\eeq Production term is zero for mass balance law, equal to the density of body forces $\rho f_{B}$ for
the linear momentum balance law and equals to the power of the body forces $f_{A}$ plus the heat source
density for the energy balance law.\par

One can introduce the internal energy $\epsilon$ as the basic variable instead of the temperature
$\vartheta$. In this case
\[
F_{\epsilon}=\rho \epsilon \eta_{0}+(\rho \epsilon v^{A}+q^{A})\eta_{A},\
\Pi_{\epsilon}=(t^{A}_{B}\frac{\partial v^{B}}{\partial x^{A}}+r)\eta,
\]
see \cite{MR}, Sec.5.3.\par Denote by $\sigma_{i} =F_{i}+\Pi_{i}$ the $(n+1)+(n+2)$-form of the
corresponding balance law.\par

Constitutive relations of this system determine, in addition to the components explicitly defined above,
the stress tensor $t^{AB}$, heat flux $q^A$, internal energy $\epsilon$, force covector $f_{A}$ and the
volume heat source density $r$ as functions of basic fields $\rho ,v^A ,\vartheta$ and \emph{some of their
derivatives}.  In addition to this, force $f_A$ and volume heat source density may explicitly depend on
the position and time $x^\mu $.  \par

A fundamental physical requirements known as "material axioms" put restrictions on the character of
dependence of density, flux and source components of the balance laws on the basic fields and their
derivatives (\cite{TN,Mus,Mus3}.  One of these  material axioms - material indifference or, more
generally, a transformation properties of a balance system under the change of observer, leads to the
independence of the heat flux $q^A$ and the stress tensor $t^{AB}$ on the velocity $v^A$ and on the
antisymmetric part of the velocity gradient.\par

Other material axioms - material symmetries, II law of thermodynamics (see \cite{Mus},Ch.6) further
restricts the form of constitutive relations.  Geometrical form of these restriction will be studied in
the continuation of this work.\par

In the simplest variant the constitutive relations of a 5F-fluid system depend on the spacial gradient of
temperature $\nabla \vartheta$ and symmetrized gradient of velocity $\nabla \mathbf{v}_{sym}$ only. Next
level of complexity is represented by the fluid with the \emph{short memory} where constitutive relations
may depend on the rate of change of temperature $\dot \vartheta$ (see \cite{TN,M}).

As a result, the domain of constitutive relations (the \textbf{state space}) of 5F-fluid system consists
of the fields $(\rho , v^A ,\vartheta ;\nabla \vartheta , \nabla \mathbf{v}_{sym}).$\par

An example of specific constitutive relations of a 5F-fluid system  is the Navier-Stokes-Fourier fluid
where

\beq
\begin{cases}
t^{A}_{B}=-p(\rho ,\vartheta )\delta^{A}_{B}+\nu(\rho ,\vartheta )Tr(\nabla
\mathbf{v}_{sym})\delta^{A}_{B}+2\mu(\rho ,\vartheta )(\nabla \mathbf{v}_{sym})^{A}_{B};\\
q^A =-\kappa(\rho ,\vartheta )\nabla^A \vartheta ;\\
\epsilon =\epsilon_{0}(\rho ,\vartheta ).
\end{cases}
\eeq Here $p$ is the pressure scalar field and $\nu ,\mu ,\kappa$ are scalar coefficients of viscosity
(the first two) and the heat conductivity respectively.\par

 Notice also that there is another, more fundamental 5-fields
thermodynamical system - $5F-solid$, where the basic fields are: mass density $\rho$, embedding
$\phi:B^3\rightarrow E^3$ of the material manifold $B$ to the physical Euclidian space $(E^3 ,h)$ and the
absolute temperature $\vartheta .$ Constitutive relations of $5F-solid$ system typically (for instance, in
thermoelasticity) depends on the spacial derivatives of embedding mapping $\phi$ (deformation gradient,
or, with the use of material indifference axiom Cauchy deformation tensor $C(\phi)=\phi^{*}h$), its time
derivative (velocity), temperature $\vartheta$ and its spacial gradient $\nabla \vartheta$. Adding of the
time derivative of Cauchy deformation tensor, or, equivalently, the symmetrized velocity gradient $\nabla
\mathbf{v}_{sym}$ (containing second derivatives of basic fields!) allows to take into account effects of
viscoelastc behavior.  The model $5F-fluid$ represents a \emph{reduction} of the $5F-solid$ system related
to the usage of the largest possible material symmetry group $SL(3,R)$ for the fluids (see \cite{TW}).
\end{example}

\begin{remark}
In the geometrical theory of differential equations (see, for instance, \cite{KV}) it is customary to
extend given system of differential equations to include all the differential equations that are
consequences of ones in a given system. It would be equally interesting to complete the system (2.3) of
the balance laws of a given thermodynamical system by all the balance laws that are their consequences. In
the second part of this work we study such "secondary balance laws" of a given balance system (of zero or
first order by the degree of derivatives of basic fields included into the constitutive relations) that
have the \emph{same domain as the initial balance laws}.  Higher order balance laws that are consequences
of a given balance system will be studied elsewhere.
\end{remark}

\subsection{Entropy condition.}

Entropy density $h^{0}$, entropy flux $h^{A},\ A=1,2,\ldots, n$ and the entropy production $\Sigma$ are
typically assumed to be a functions of the the same variables $x^\mu ,y^i, y^{i}_{,x^\mu}$ as the
coefficients of the balance laws (2.3). II law of thermodynamics requires that entropy satisfies to the
balance law
\begin{equation}
d(h^{\mu}\eta_{\mu})=\Sigma ,
\end{equation}
with the production 4-form

\beq \Sigma =\Sigma (x^\mu ,y^i, y^{i}_{,x^\mu})\eta, \sigma \geqq 0, \eeq being positive on the solutions
of the balance system (2.3).\par

{\bf Entropy principle} (\cite{M,MR}) requires that any solution of the balance equations (2.3) would also
satisfy to the equation (2.9) and that the production $\sigma$. This requirement places serious
restrictions to the form of the balance equations (2.3).
\par
To close system of equations (2.3) (or (2.3+2.9)) for $y^i$ one has to choose the {\bf constitutive
relation} $C$ of the thermodynamical system, i.e. to choose the densities,flows and production forms as
functions of $x^\mu ,y^i$ and the appropriate derivatives of fields $y^i$ . In particular, one have to
choose the domain of the constitutive relation which is typically the full or partial jet-bundle of the
configurational bundle $\pi$ of dynamical variables. By definition, the Rational Extended Thermodynamics
(see Appendix II for short exposition of the formalism of this theory) is the zero order theory in that
the domain of its constitutive relation is the space $Y$. In this article we consider the cases of
constitutive relations of zero and first order only. Constitutive relations depend also ont the background
fields (metric $G$ in $X$ or a connection $\nu$ in the bundle $\pi:Y\rightarrow X$ in this paper).  As we
will see in the part II of this work (in preparation, see also \cite{Pr}), utilizing of the entropy
condition allows to effectively reduce this process to a choice of smaller number of constitutive fields.

\vfill \eject

\section{Multisymplectic Field Theory.}
In this section we recall briefly the Poincare-Cartan formalism of the Multisymplectic Field Theory with
the modifications necessary for the formulation of the covariant theory of balance systems.  We will
follow (\cite{LMD,GMS,FF}).\par
\subsection{The 1-jet bundle}
Given a frame bundle $\pi:Y\rightarrow X$ we say that two sections $s,s':U\rightarrow Y$ defined in a
neighborhood $U$ of a point $x\in X$ define the same 1-jet $j^{1}s(x)$ if $s(x)=s'(x), \
s_{*x}=s_{*x}:T_{x}(X)\rightarrow T_{s(x)}(Y)$. This defines an equivalence relation on the set of locally
(near the point $x$) defined sections of $\pi$.  Space of equivalence classes of such local sections is
defined $J^{1}_{x}(\pi)$.\par

The total space $J^{1}(\pi)=\bigcup_{x\in X}J^{1}_{x}(\pi)$ cam be endowed with a smooth structure such
that the mappings $J^{1}(\pi)\rightarrow Y \rightarrow X$ are fibrations. The fibration $\pi_{10}:
J^{1}(\pi)\rightarrow Y$ is the affine bundle modeled in the vector bundle  $\pi^{*}(T^{*}(X))\otimes
V(\pi)$, where $V(\pi)\subset T(Y)$ is the vertical subbundle of the bundle $\pi$. \par

Let $(x^\mu ,y^i ; \mu=1,\ldots,n+1=dim(X); i=1,\ldots, m )$ be an adopted local coordinate system in $Y$.
Then the local coordinate system $(x^\mu ,y^i ,z^{i}_{\mu} ; \mu=1,\ldots,n; i=1,\ldots, m )$ can be
defined in $J^{1}(\pi)$ by the condition
\[
z^{i}_{\mu}(j^{1}_{x}s)=\frac{\partial y^i}{\partial x^\mu }(x).
\]

\subsection{Lagrangian picture: Poincar\'e-Cartan Form.}

 The volume form $\eta =\sqrt{\vert G\vert }d^{n+1}x$ permits to construct the
{\bf vertical endomorphism}
\begin{equation}
S_{\eta }=(dy^{i}-z^{i}_{\mu }dx^{\mu} )\wedge \eta_{\nu}\otimes \frac{\partial }{\partial z^{i}_{\nu}}
\end{equation}
which is a tensor field of type $(1, n+1)$ on the 1-jet bundle space $Z=J^{1}(\pi)$ of the configurational
bundle $\pi:Y\rightarrow X$. Here $\eta_{\mu}=i_{\partial_{x^{\mu}}}\eta$, see Appendix I.\par

For a Lagrangian (n+1)-form $L\eta$, $L$ being a (smooth) function on the manifold $Z=J^{1}(\pi)$ the
Poincar\'e-Cartan $(n+1)$ and $(n+2)$-forms are defined as follows:
\begin{equation}
\Theta _{L}=L\eta +S^{*}_{\eta }(dL),\ \Omega _{L}=-d\Theta _{L},
\end{equation}
where $S^{*}_{\eta }$ is the adjoint operator of $S_{\eta}$. In coordinates we have
\begin{equation}
\Theta _{L} = (L-z^{i}_{\mu} \frac{\partial L}{\partial z^{i}_{\mu}})\eta + \frac{\partial L}{\partial
z^{i}_{\mu}}dy^{i}\wedge \eta_{\mu},
\end{equation}
\begin{multline}
\Omega _{L} = -d(L-z^{i}_{\mu} \frac{\partial L}{\partial z^{i}_{\mu}}) \wedge \eta -d(\frac{\partial
L}{\partial z^{i}_{\mu}})\wedge dy^{i}\wedge \eta_{\mu}- \\
-\frac{\partial L}{\partial z^{i}_{\mu}}((-1)^{\mu}\frac{\partial ln(\sqrt{\vert G\vert })}{\partial
x^{\mu}})\wedge dy^{i}\wedge \eta = -(dy^{i}-z^{i}_{\mu}dx^{\mu})\wedge \left( \frac{\partial L}{\partial
y^{i}}\eta -d\left( \frac{\partial L}{\partial z^{i}_{\mu}}\right) \wedge \eta_{\mu} \right).
\end{multline}
\begin{remark} If the space manifold $B$ has the boundary and some boundary conditions are prescribed for the
sections $\phi :X\rightarrow Y$ of the configurational bundle in the form
\[
\phi (x)\in {\mathcal{B}c},
\]
where $\mathcal{B}c$ is a subbundle of the bundle $Y$, then, in order to correlate boundary conditions
with the variation of Poincar\'e-Cartan form it is reasonable to require that the restriction of
$\Omega_{L}$ to the boundary subbundle $\mathcal Bc$ is exact: there exists a $n$-form $\theta$ on the
subbundle ${\mathcal B}c$ such that
\begin{equation}
i_{{\mathcal B}c}\Theta_{L}=d\theta .
\end{equation}
We refer to \cite{BSF}, Chapter 7 or to the paper \cite{BS} for more details.\par
\end{remark}

Recall (\cite{LMD}) that the couple $(Z,\Omega _{L})$ is a multisymplectic manifold provided the
Lagrangian $L$ is regular, i.e. the matrix $L_{z^{i}_{\mu}z^{j}_{\nu}}$ is \textbf{nondegenerate}.
\par

An {\it extremal} of $L$ is a section of $\pi_{XY}$ such that for any vector field $\xi_{Z}$ on $Z$,
\begin{equation}
(j^{1}\phi )^{*}(i_{\xi _{Z}}d\Theta_{L})=0,
\end{equation}
where $j^{1}\phi $ is the first jet prolongation of $\phi $.

A section $\phi $ is an extremal of $L$ if and only if it satisfies the Euler-Lagrange Equation (see, for
instance, \cite{BSF,FF})

\beq (j^{1}\phi )^{*}\left( \frac{\partial (L\sqrt{\vert G\vert})}{\partial y^{i}}-\frac{d }{d
x^{\mu}}\left( \frac{\partial (L\sqrt{\vert G\vert})}{\partial z^{i}_{\mu}}\right) \right) =0,\ 1\leqq
i\leqq m. \eeq

There is an operator $\mathcal{EL}:\Gamma(\pi)\rightarrow \Gamma(V^{*}(\pi))$ (Euler-Lagrange operator)
that has the local form

\beq \mathcal{EL}(\phi)=\left(\frac{\partial (L\sqrt{\vert G\vert})}{\partial y^{i}}\circ j^{1}\phi
-\frac{d }{d x^{\mu}}\left( \frac{\partial (L\sqrt{\vert G\vert})}{\partial z^{i}_{\mu}} \circ j^{1}\phi
\right)\right) \otimes dy^i .\eeq In terms of this operator the Euler-Lagrange equations looks simply \beq
\mathcal{EL}(\phi)=0.\eeq

\subsection{Canonical multisymplectic bundles $\Lambda^{k}_{r}$.}

    Denote by $V(Y)\rightarrow Y$ the subbundle of \textbf{vertical tangent vectors} of the tangent bundle
$T(Y)$.

 Following \cite{LMD,GMS} let $\Lambda ^{k}_{r} Y$ denote the subbundle of the
vector bundle $\Lambda ^{k}Y$ of exterior $k$-forms on $Y$ consisting of those forms that vanish when $r$
of their arguments are vertical (with respect to the fibration $\pi : Y \rightarrow X$)
\[
\Lambda ^{k}_{r} Y=\{\sigma \in  \Lambda ^{k}Y\  \vert i_{\xi_{1}}\ldots i_{\xi_{r}}\sigma =0,\forall\
\xi_{i}\in V(\pi ) .\}
\]

The manifold $\Lambda ^{k}Y$ carries a canonical $k$-form $\Theta^{k} _{0}$ define as follows:
\begin{equation}
\Theta _{0}(\omega )(\xi _{1},\ldots \xi _{k})=\omega (\pi_{\Lambda^{k}} (\omega ))(\pi_{\Lambda^{k}\ *}
(\xi _{1}),\ldots \pi_{\Lambda^{k}\ *}(\xi _{k})),
\end{equation}
where $\omega \in \Lambda ^{k}Y$, $\xi _{i}\in T_{\omega}(\Lambda ^{k}Y)$, and $\pi_{\Lambda^{k}}:\Lambda
^{k}Y\rightarrow Y$ is the canonical bundle projection.\par

By restriction, this form induces an $k$-form $\Theta^{k}_{r}$ on the manifold $\Lambda^{k}_{r}Y$. We
denote $\Omega^{k}_r = - d\Theta^{k}_r$.
\subsubsection{Case k=n,n+1}
  We will
use the construction above  for $k=n+1,n+2;r=1,2 $, or, more specifically, for $k=4,5$.\par

In particular, The bundle $\Lambda^{k}_{2}(Y)$ of the exterior forms on $Y$ which are annulated if 2 of
its arguments are vertical:

\beq \omega^{k}\in  \Lambda^{k}_{2}(Y) \Leftrightarrow  i_{\xi}i_{\eta}\omega=0,\ \xi,\eta\in V(Y).\eeq

 Elements of the space $\Lambda ^{n+1}_{1}Y$ are semibasic $n$-forms locally expressed as $p(x,y) \eta
 .$ \par
 Elements of the space $\Lambda^{n+1}_{2}Y$ have, in local adapted coordinates $(x^\mu, y^i )$ the form

\[ p(x,y) \eta + p_{i}^{\mu}dy^{i}\wedge \eta_{\mu}.\]\par

 This introduces coordinates $(x^{\mu},y^{i},p)$
on the manifold $\Lambda ^{n+1}_{1}Y$ and $(x^{\mu},y^{i},p,p_{i}^{\mu})$ on the manifold
$\Lambda^{n+1}_{2}Y.$\par

Taking the case $k=n+2$ we see that the forms $dy^i\wedge \eta$ form the basis of $\Lambda^{n+2}_{2}(Y)$
while the bundle $\Lambda^{n+2}_{1}(Y)$ is \emph{zero bundle}.\par

Introduce also the notations

\[
\Lambda^{(n+1)+(n+2)}_{2}(Y)=\Lambda^{n+1}_{2}(Y)\oplus
\Lambda^{n+2}_{1}(Y),\Lambda^{(n+1)+(n+2)}_{1}(Y)=\Lambda^{n+1}_{1}(Y)\oplus \Lambda^{n+2}_{1}(Y)
\]
for the direct sum of the bundles on the right side.\par

It is clear that $\Lambda^{k}_{1}(Y)\subset \Lambda^{k}_{2}(Y)$ is the subbundle of the larger bundle.
Therefore we have the embedding of subbundles

\beq \Lambda^{(n+1)+(n+2)}_{1}(Y)\subset \Lambda^{(n+1)+(n+2)}_{2}(Y). \eeq

For the Poincare-Cartan forms  (3.10) in this case ($k=n+1,r=2$) we have local expressions
\begin{equation}
\Theta^{n+1}_{2}= p\eta+p_{i}^{\mu}dy^{i}\wedge \eta_{\mu},
\end{equation}
\beq \Omega^{n+1}_{2} =-dp\wedge \eta-dp_{i}^{\mu} \wedge dy^{i}\wedge \eta_{\mu} \nonumber
-(-1)^{\mu}p_{i}^{\mu}\frac{\partial \lambda_{G}}{\partial x^{\mu}} \wedge dy^{i}\wedge \eta, \eeq where
$\lambda_{G}=ln(\sqrt{\vert G\vert })$
\subsubsection{Dual MS-picture: Hamiltonian systems.}

Basic for the Hamiltonian form of multisymplectic field theory is the bundle: $\Lambda^{n+1}_{2}Y$ endowed
with the canonical MS-form $\Theta^{n+1}_{2}$  and its factor bundle over $Y$ (\emph{polysymlectic
bundle})
\[
Z^*=\Lambda^{n+1}_{2}Y/\Lambda^{n+1}_{1}Y,\ q :\Lambda ^{n+1}_{2}Y \rightarrow Z^{*}.
\]

Corresponding to the local adopted chart $(x^\mu ,y^i)$ the manifold $Z^*$ has the (local) coordinates
$(x^\mu ,y^i ,p^{i}_{\mu})$.\par

Pairing

\beq J^{1}(\pi)\underset{Y}{\times} Z^* \rightarrow 1_{Y}:\ (z,p)\rightarrow z^{i}_\mu p^{\mu}_{i} \eeq
identifies the bundle $Z^* \rightarrow Y$ with the linear dual to the affine bundle $\pi^{1}_{0}:J^{1}(\pi
)\rightarrow Y$.  In the similar way, bundle $\Lambda^{n+1}_{2}(Y)\rightarrow Y$ can be identified with
the affine dual to the bundle $\pi^{1}_{0}:J^{1}(\pi )\rightarrow Y$ (see \cite{LM,MS}).
\begin{remark} Another way to defined $Z^*$ is to take

\beq Z^* =\pi^{*}(T(X))\otimes V^{*}(\pi)\otimes \pi^{*}(\Lambda^{n+1}(X)), \eeq see \cite{LM,MS}.
\end{remark}

 A \textbf{Hamiltonian} is, in this approach, a section $h$ of the projection $q$. Having it
available, we define $\Theta _{h}=h^{*}\Theta^{n+1}_{2},\ \Omega_{h}=h^{*}\Omega^{n+1}_{2}$.

A section $\sigma :X\rightarrow Z^{*}$ is said to satisfy the Hamilton equation (for a given Hamiltonian
$h$) if
\[
\sigma ^{*}(i_{\xi }\Omega _{h})=0,
\]
for all vector fields $\xi$ on $Z^{*}$,

In local coordinates $(x^{\mu},y^{i},z^{i}_{\mu})$ a Hamiltonian $h$ is represented by a local function
$H$:
\[
p=-H(x^{\mu},y^{i},z^{i}_{\mu}).
\]
Then,
\begin{equation}
\Theta _{h}= -H\eta + p_{i}^{\mu}dy^{i}\wedge \eta_{\mu},
\end{equation}

\beq \Omega_{h} = -d\Theta _{h} = -dH\wedge \eta +dp_{i}^{\mu} \wedge dy^{i}\wedge \eta_{\mu} +(-1)^{\mu}
p_{i}^{\mu} \frac{\partial ln(\sqrt{\vert G\vert })} {\partial x^{\mu}} dy^{i}\wedge \eta, \eeq

 and the Hamilton equations for a section
$\sigma=(x^{\mu},\sigma^{i}(x),\sigma^{\mu}_{i}(x))$ take the form:

\beq \begin{cases}
\frac{\partial \sigma ^{i}}{\partial x^{\mu}}=\frac{\partial H}{\partial \sigma^{\mu}_{i}}, \\
div_{G}(p^{\mu}_{i}) =\sum_{\mu}[\frac{\partial \sigma ^{\mu}_{i}}{\partial x^{\mu}}+\sigma
^{\mu}_{i}\frac{\partial ln(\sqrt{\vert G\vert})}{\partial x^{\mu}}] =-\frac{\partial H}{\partial y^{i}}.
\end{cases} \eeq

In difference to the bundle $\Lambda^{n+1}(Y)$ the bundle $Z^{*}$ does not have a canonically defined form
of the Poincare-Cartan type (see for instance, Sec. below where it is shown that under the transformation
induced by an automorphism of the bundle $\pi$ the (locally defined) for).  Locally though, we can define
the form

\beq \Theta^{*}=p_{i}^{\mu}dy^{i}\wedge \eta_{\mu}\eeq

\par
Under the action of an adopted transformation $\phi \in Aut(\pi)$ considered as a change of variables and
lifted to $Z^*$ to the tensorially transformed form the term of the form $Q\eta$ is added (see below,
Section 11.1).  As a result, the form $\Theta^{*}$ is not defined canonically, \textbf{but its class $mod
(\Lambda^{n+1}_{1}(Z^*))$ is}. Taking $mod (\Lambda^{n+1}_{1}(Z^*))$ we get \textbf{canonically defined
element of the bundle} $\Lambda^{n+1}_{2}Z^{*}/\Lambda^{n+1}_{1}Z^{*}$ on $Z^*$.\par

\textbf{One may consider this class as defining the canonical $V^{*}(\pi)$-valued semi-basic n-form on
$Y$.}\par

Below we will see that it is sufficient for the separating components of a balance system to the
individual balance laws with the help of independent vertical variations.\par \vskip0.5cm

Recall (see \cite{CLM,GMS}) that given an \textbf{Ehresmann connection} $\nu :Y\rightarrow J^{1}(\pi)$ on
the bundle $\pi$ with the vertical projector
\[
P_{v}=\partial_{y^i}\otimes (dy^i+\Gamma^{i}_{\mu}dx^\mu ),
\]

defines naturally the linear section $\delta_{\nu}:  Z^*\rightarrow \Lambda^{n+1}_{2}Y$ given by \beq
\delta_{\nu}(F^{\mu}_{i}dy^i \wedge \eta_{\mu} )= (F^{\mu}_{i}\Gamma^{i}_{\mu})\eta  + F^{\mu}_{i}dy^i
\wedge \eta_{\mu}. \eeq

Section $\delta_{\nu}$  defines the pullback of the form $\Theta^{n+1}_{2}$:

\beq \delta_{\nu}^{*}\Theta^{n+1}_{2}=(F^{\mu}_{i}\Gamma^{i}_{\mu})\eta  + F^{\mu}_{i}dy^i \wedge
\eta_{\mu}.  \eeq

Form $\Theta^{n+1}_{\nu}=\delta_{\nu}^{*}\Theta^{n+1}_{2}$ is defined correctly on the manifold $Z^*$.\par

\vskip0.5cm
\subsubsection{Bundle $\tilde Z$ for the balance systems.} To present the system of balance laws in the
multisymplectic form we will need to use the vector bundles
$\Lambda^{(n+1)+(n+2)}_{i}Y=\Lambda^{n+1}_{i}Y\oplus \Lambda^{n+2}_{i}Y$, where $i=1,2$ and the vector
bundle

\beq \tilde Z=\Lambda^{(n+1)+(n+2)}_{2}Y/\Lambda^{(n+1)+(n+2)}_{1}Y=
\Lambda^{(n+1)}_{2}Y/\Lambda^{(n+1)}_{1}Y\oplus \Lambda^{(n+2)}_{2}Y/\Lambda^{(n+2)}_{1}Y, \eeq

Notice that the first term in the sum on the right is $Z^{*}$.\par

 Locally, elements of the factor bundle $\tilde Z$ can be presented in the form

\beq p^{\mu}_{i}dy^{i}\wedge \eta_{\mu}+q_{i}dy^{i}\wedge \eta . \eeq

Canonical forms $\Theta^{k}_{i}$ for $k=n+1,n+2;i=1,2$ induce on the bundle $\tilde Z$ the class of
$(n+1)+(n+2)$ form
 \beq {\tilde \Theta}=  p^{\mu}_{i}dy^{i}\wedge \eta_{\mu}+q_{i}dy^{i}\wedge \eta \   mod\  \Lambda^{*}_{1}{\tilde Z}\eeq
 where $n+1$ and
$n+2$ components of this form are lifted from the canonical forms on the components
$\Lambda^{k}_{2}Y/\Lambda^{k}_{1}Y$ for $k=n+1,n+2$. \emph{Class $mod \ \Lambda^{*}_{1}{\tilde Z}$ of this
form is defined canonically} (see above). In examples below we will be using these constructions for
$n=3$.\par

\subsection{Legendre Transformation.}

Let $L$ be a Lagrangian function. We define the fiber mapping over $Y$
\[
leg_{L}:Z\rightarrow \Lambda^{n+1}_{2}Y,
\]
as follows:
\[
leg_{L}(j^{1}_{x}\phi))(X_{1},\ldots, X_{n+1}) =(\Theta_{L})_{j^{1}_{x}\phi)}({\tilde{X}_{1},\ldots,
\tilde{X}_{n+1}}),
\]
where $j^{1}_{x}\phi \in Z, X_{i}\in T_{\phi (x)}Y$ and $\tilde{X}_{i}\in T_{j^{1}_{x}\phi (X)}Z$ are such
that $\pi_{*}(\tilde{X}_{i})=X_{i}.$

Notice that addition of a constant to the Lagrangian $L$ leads to the constant shift (in $p$) of the image
of Legendre mapping in $\Lambda^{n+1}_{2}Y$ (which is a submanifold of codimension 1 if the Lagrangian is
regular).  Thus, the space $\Lambda^{n+1}_{2}Y$ is foliated by these shifts.

In local coordinates, we have
\[
leg_{L}(x^{\mu},y^{i},z^{i}_{\mu}) = (x^{\mu},y^{i},p=L-z^{i}_{\mu}\frac{\partial L }{\partial
z^{i}_{\mu}},p^{\mu}_{i} =\frac{\partial L }{\partial z^{i}_{\mu}}).
\]

The Legendre transformation $Leg_{L}:Z\rightarrow Z^{*}$ is defined as the composition $Leg_{L}=q\circ
leg_{L}.$ Locally
\[
Leg_{L}(x^{\mu},y^{i},z^{i}_{\mu})=(x^{\mu},y^{i},p^{\mu}_{i}=\frac{\partial L }{\partial z^{i}_{\mu}}).
\]
Recall \cite{LMD} that the Legendre transformation $Leg_{L}:Z\rightarrow Z^{*}$ is a local diffeomorphism
if and only if $L$ is regular.

If, in addition, the Lagrangian $L$ is hyperregular (i.e. if $Leg_{L}$ is a global diffeomorphism), one
can define a {\bf Hamiltonian} $h:Z^*\rightarrow \Lambda^{n+1}_{2}Y$ by setting
\[
h=leg _{L}\circ Leg_{L}^{-1}.
\]
Then
\[
Leg_{L}^{*}\Theta _{h}=\Theta _{L},\ Leg_{L}^{*}\Omega _{h}=\Omega _{L}.
\]
In this case $Leg_{L}:(Z,\Omega _{L}) \rightarrow (Z^{*},\Omega_{h})$ is a multisymplectomorphism.

\subsection{Unified formalism.}

In this subsection we describe the united geometrical setting of a classical field theory developed in
(\cite{LMD},\cite{ELMMR}) as the generalization of Skinner-Rusk (Dirac) geometrical mechanics.\par

Introduce the fiber product $W_{0}=Z\times_{Y} \Lambda^{n+1}_{2}Y $ with canonical projectors $pr_{i}$ to
the $i$-th factor. We consider canonical coordinates $(x^{\mu },y^{i},z^{i}_{\mu},p,p^{\mu}_{i})$ on
$W_{0}$.

Define the $(n+1)$-forms $\Theta=pr_{1}^{*}\Theta^{n+1}_{2}$ and $\Theta ^{L}=pr_{2}^{*}\Theta_{L}$, and
the corresponding $(n+2)$-forms $\Omega =d\Theta,\ \Omega^{L}=d\Theta ^{L}$. In addition, we introduce the
differences $\Theta ^{*}=\Theta^{L}-\Theta,\ \Omega ^{*}=\Omega^{L} -\Omega$. In local coordinates we have
\begin{equation}
\Theta ^{*}=\left[ L-z_{\mu}^{i}\frac{\partial L}{\partial z^{i}_{\mu}}-p \right]\eta  +
\left[\frac{\partial L}{\partial z^{i}_{\mu}} -p^{\mu}_{i} \right] dy^{i}\wedge \eta_{\mu}.
\end{equation}

Define the submanifold $W_{1}\subset W_{0}$ as the graph of the Legendre mapping $Leg_{L}:Z\rightarrow
\Lambda^{n+1}_{2}Y$, see \cite{ELMMR} and \cite{LMD}. Locally it is given by equations
\begin{equation}
p^{\mu}_{i}-\frac{\partial L}{\partial z^{i}_{\mu}} =0.
\end{equation}
Then we have
\begin{equation}
\Theta^{*}\vert _{W_{1}}=\left[L-z_{\mu}^{i}\frac{\partial L}{\partial z^{i}_{\mu}} -p \right]\eta.
\end{equation}
Introduce the function $H_{0}$ on $W_{0}$ as follows (\cite{LMD}):
\begin{equation}
H_{0}=p+p_{i}^{\mu }z_{\mu}^{i}-pr_{2}^{*}L.
\end{equation}
This function allows to define the form
\begin{equation}
\Omega _{H_{0}}=\Omega +dH_{0}\wedge \eta = d(\Theta +H_{0}\eta ),
\end{equation}
which takes the following local expression

\beq \Omega _{H_{0}} =  -(dp+(-1)^{\mu}p_{i}^{\mu} \frac{\partial ln(\sqrt{\vert G\vert })} {\partial
x^{\mu}}dy^{i})\wedge \eta  -dp_{i}^{\mu}\wedge dy^{i}\wedge \eta_{\mu}+dH_{0} \wedge \eta. \eeq

 We can develop the above local
expression to obtain the following
\begin{eqnarray}
\Omega_{H_{0}} &=&  -dp_{i}^{\mu}\wedge dy^{i}\wedge \eta_{\mu}+p^{\mu}_{i}dz^{i}_{\mu}\wedge \eta +z^{i}_{\mu}dp^{\mu}_{i}\wedge \eta -\frac{\partial L}{\partial y^{i}}dy^{i}\wedge \eta-\\
\nonumber &&-\frac{\partial L}{\partial z^{i}_{\mu}}dz^{i}_{\mu} \wedge \eta -(-1)^{\mu}p_{i}^{\mu}
\frac{\partial ln(\sqrt{\vert G\vert })}{\partial x^{\mu}}dy^{i}\wedge \eta.
\end{eqnarray}
Along $W_{1}$ the second and fifth terms in expression for $\Omega_{H_{0}}$ cancel each other and one get
\beq
 \Omega_{H_{0}}\vert _{W_{1}}  = -dp_{i}^{\mu}\wedge dy^{i}\wedge \eta_{\mu}+
z^{i}_{\mu}dp^{\mu}_{i}\wedge \eta -[\frac{\partial L}{\partial y^{i}}+(-1)^{\mu}p_{i}^{\mu}
\frac{\partial ln(\sqrt{\vert G\vert })}{\partial x^{\mu}}]dy^{i}\wedge \eta. \eeq

 Notice that this form on $W_{1}$ does not depend on $p$.

Consider now the submanifold $W_{2}\subset W_{1}$ defined by equation $H_{0}=0$, i.e.
\begin{equation}
p=-(p_{i}^{\mu }z_{\mu}^{i}-L).
\end{equation}
which defines, for a given Lagrangian the {\it hamiltonian} section of $\mu :
\Lambda^{n+1}_{2}Y\rightarrow Z^{*}$ (a "local energy", see above).

Notice that the form $\Theta ^{*}$ (and therefore, $\Omega ^{*}$) vanish when they restrict to $W_{2}$.
This allows to identify $W_{2}$ with the graph of the Legendre mapping $Leg_{L}:Z\rightarrow Z^{*}.$

\subsection{Vertical contact structure.}
Notice that the dimensions of the fiber of the bundle $\pi^{1}_{0}:Z=J^{1}(\pi)\rightarrow Y$ and that of
the dual bundle $Z^{*}=(\Lambda^{n+1}_{2}Y/\Lambda^{n+1}_{1}Y)\rightarrow Y$ are the same. In addition to
the $(n+1)$ and $(n+2)$-forms introduced above, the bundle $\pi_{YW_{0}}:W_{0}\rightarrow Y$ has one more
geometrical structure, namely the {\it contact structure in the fibers $W_{0\ y}$ of the bundle
$\pi_{YW_{0}}:W_{0}\rightarrow Y$}. Indeed, fibers of this bundle are endowed with the canonical contact
1-form
\begin{equation}
\tilde{\theta} = dp + z^{i}_{\mu}dp_{i}^{\mu}=d(p+z^{i}_{\mu}p^{\mu}_{i})-p^{\mu}_{i}dz^{i}_{\mu}.
\end{equation}
This "vertical contact structure" allows to distinguish the Legendre mappings defined by some lagrangian
$L\in C^{\infty}(Z)$ from the general bundle mappings $\hat C: Z\rightarrow Z^*,$ namely,
\begin{proposition}
\begin{enumerate}
\item Let $L\eta$ be a Lagrangian defined on the space $Z=J^{1}(\pi)$.
Then the intersection of the graph $\Gamma_{L}$ of Legendre transformation $leg_{L}:J^{1}(\pi)\rightarrow
\Lambda^{n+1}_{2}Y$ with the fibers $W_{0\ y}$ of the bundle $W_{0}\rightarrow Y$ are the \textbf{Legendre
submanifolds} of the fibers $W_{0\ y}$.
\item Let $\hat C: J^{1}(\pi)\rightarrow \Lambda^{n+1}_{2}Y$ be any smooth bundle morphism (over $Y$) given by
\[
{\hat C}(x,y;z^{i}_{\mu})=(x,y; p(x,y;z^{i}_{\mu}),p^{\mu}_{i}(x,y;z^{i}_{\mu})).
\]
 Then the intersection of the graph $\Gamma_{\hat C}\subset W_{0}=$ of this morphism with the fibers $W_{0\
y}$ of the bundle $W_{0}\rightarrow Y$ are the \textbf{Legendre submanifolds} of the (contact) fibers
$W_{0\ y}$ if and only there exists a (locally defined) function $L\in C^{\infty}(J^{1}(\pi )) $ such that
\[
p^{\nu}_{j}=L_{z^{j}_{\nu}},\ p=L-z^{i}_{\mu}L_{z^{i}_{\mu}}.
\]
\end{enumerate}
\end{proposition}
\begin{proof} For the first statement, notice that we have in local coordinates

\[
leg_{L}(x^{\mu},y^{i},z^{i}_{\mu}) = (x^{\mu},y^{i},p=L-z^{i}_{\mu}\frac{\partial L }{\partial
z^{i}_{\mu}},p^{\mu}_{i} =\frac{\partial L }{\partial z^{i}_{\mu}}).
\]
Thus, along $\Gamma_{L}$ we have
\[
\tilde{\theta}\vert_{\Gamma_{L}}=d_{v}((L-z^{i}_{\mu}\frac{\partial L }{\partial
z^{i}_{\mu}})+z^{i}_{\mu}\frac{\partial L }{\partial z^{i}_{\mu}})-\frac{\partial L }{\partial
z^{i}_{\mu}}dz^{i}_{\mu}=0.
\]
For the second part we notice that the restriction of the 1-form $\tilde \theta $ on the fiber $W_{0\ y}$
to the graph of $\hat C$ has the form
\[
p_{,z^{j}_{\nu}}dz^{j}_{\nu}+z^{i}_{\mu}p^{\mu}_{i,z^{j}_{\nu}}dz^{j}_{\nu}=
\partial_{z^{j}_{\nu}}(p+z^{i}_{\mu}p^{\mu}_{i})-p^{\nu}_{j}.
\]
Introducing function $L=p+z^{i}_{\mu}p^{\mu}_{i}$ we immediately get the necessary expressions for
components of mapping $\hat C$.
\end{proof}

Introduce the smooth submanifold (fiberwise quadric) $W_{r}\subset W_{0}$ (reduced) defined by the
condition
\[
W_{r}=\{ w=(x,y,z^{i}_{\mu},p,p^{\mu}_{i}) \in W_{0}\vert p+z^{i}_{\mu}p^{\mu}_{i}=0  \}.
\]
\begin{remark}
Submanifold $W_{r}$ is the abstract analog of the quadric $C$ in the phase space of the homogeneous
thermodynamics which contains, due to the homogeneity requirement) all the (Legendre) constitutive
surfaces of different homogeneous TD systems with given thermodynamical phase space, see \cite{CA}.
\end{remark}

Restriction of the form $\tilde \theta$ to the fibers over $Y$ of the subbundle $W_{r}$ of the bundle
$W_{0}\rightarrow Y$ has the form
\[
{\tilde \theta}\vert_{W_{r}}=-p^{\mu}_{i}dz^{i}_{\mu}.
\]
Consider the projection $W_{0}\rightarrow Z\times Z^{*}=J^{1}(\pi)\times
\Lambda^{n+1}_{2}Y/\Lambda^{n+1}_{1}Y$ (see Sec. 3.5). Restriction of this projection to the submanifold
$W_{r}$ is the diffeomorphism of the bundles over $Y$ with the inverse $j$ given by
\[
(x,y,z^{i}_{\mu},p^{\mu}_{i})\rightarrow (x,y,z^{i}_{\mu},p=-z^{i}_{\mu}p^{\mu}_{i},p^{\mu}_{i}).
\]
Consider the fiberwise pullback $\theta^{*}=j^{*}{\tilde \theta}\vert_{W_{r}}.$  This 1-form has the same
type $\theta^{*}=-p^{\mu}_{i}dz^{i}_{\mu}.$\par  Lift it to the bundle $Z\times \tilde Z\rightarrow Y$ via
the projection to the first factor keeping the same notation for this form.\par

Let  now

\[C:J^{1}(\pi)\rightarrow {Z^*}: z\rightarrow z^{*}=F^{\mu}_{i}(x,y,z)dy^i \wedge \eta_{\mu} \]

 be
\emph{any smooth mapping of the bundles }over $Y$, let $\widehat{C}:J^{1}(\pi)\rightarrow
\Lambda^{n+1}_{2}(Y)$ be its lift to the mapping into the bundle $\Lambda^{n+1}_{2}(Y)$ defined by
$p(z)=z^{i}_{\mu}F^{\mu}_{i}(z) $ and let $C^{g}$ be the lift of the mapping $\hat C$ to the embedding
$Z\rightarrow W_{0}=Z\times \Lambda^{n+1}_{2}(Y)$. Then we define the 1-form on $Z$

\beq \theta_{C}=C^{*}(\theta^{*})=F^{\mu}_{i}(x,y,z)dz^{i}_{\mu} \eeq

on the fibers of the 1-jet bundle $\pi_{10}:J^{1}(\pi)\rightarrow Y. $ \vfill \eject
\addtocontents{toc}{Chapter II. Partial 1-jet bundles.}

\centerline{\textbf{Chapter II. Partial 1-jet bundles.}}

\section{Partial jet-bundles $Z_{p}=J^{1}_{p}(\pi)$.}

\subsection{State spaces and the partial jet-bundles.}

 State fields bundle $Y$ (with the fiber $U$ and field variables $y^i$) over the base
  $X$: $\pi_{YX}:Y\rightarrow X$, introduced in Section 2 represents the first floor
of the construction of a bundle that serves as a domain of general constitutive relations (shortly CR) of
a balance system (2.3). \par

Following types of the bundles serving as the domain for constitutive relations are most widespread:
\begin{enumerate}
\item Minimal state space (case of the total 1-jet bundle): No derivatives of physical fields are included
 into the space $U$. Constitutive relations are defined on the full 1-jet bundle $J^{1}(\pi)$ (first order theory),
 or on the full 2-jet bundle (second order theory), see \cite{TN}).  Elasticity theory is an example of such a case. Notice that the base manifold $X$ here
 can be taken ss the product of the time line $T$ and the \textbf{material manifold} $B$ (Lagrange picture)
 rather then the physical space-time. Similar bundle picture is used in astrophysics (see.
\cite{CQ}).
\item Optimal (in physical sense) state space: some fields are included into the state space $U$ with some
 of their derivatives or only these derivatives are included into the state space. For instance, it is customary to
include velocity field $v$ (which is defined by the time derivatives of the deformation embedding of the
material manifold into the physical space) in the list of basic fields and write down the balance law of
the linear momentum, corresponding to the velocity field. This is the generic case. Five field model (2.7)
is an example. In such scheme one can distinction between the fields entering the constitutive relations
alone from those entering CR with the time derivative, with the spacial gradients or with both. Such a
distinction is important if one would like to preview the type of a PDE-system that corresponds to a given
balance system - is it hyperbolic, dissipative, or some, definite mix of both, does it have a stationary,
possibly elliptic components etc.
\item Maximal state space (Rational Extended Thermodynamics - RET): In the Rational Extended Thermodynamics
 all the derivatives of physical fields entering the
constitutive relations (CR) (temperature gradients, rates of strain tensors, etc.) are included into to
the bundle $U$ of the basic fields of the theory.  It gives a tremendous technical advantage to write
constitutive relations in terms of fields $y^i$ only (without any derivatives included), to have first
derivatives only while differentiating the constitutive relations and to have a simple duality picture
(see Appendix II). On the other hand it makes the whole scheme somewhat too cumbersome: one has to include
into the system the balance laws for the derivatives of physical fields, such derivatives being listed in
between the basic fields in the space $U$.
\end{enumerate}
\par

Here we present a construction of two types of the partial jet bundles $J^{1}_{p}(\pi)$ of a fiber bundle
$\pi:Y\rightarrow X$ that will be used in the paper. One, denoted as $J^{1}_{K}(\pi)$ is related to a
subbundle $K$ of the tangent bundle $T(X)$ (or, more exact, with an almost product structure $T(X)=K\oplus
K'$ (see \cite{LR}), the other, denoted as $J^{1}_{S}(\pi)$, is defined by the decomposition $S$ of the
state space $U$ into the direct sum of subspaces with different set of derivatives entering constitutive
relations. In the second part of the work we define and study more general type of partial first and
higher higher order jet bundles.\par

\subsection{Partial 1-jet bundles $J^{1}_{K}(\pi)$.}
\begin{definition} Let $\pi_{XY}:Y\rightarrow X$ be a fiber bundle and let $K(X)\subset T(X)$ be a subbundle of the tangent
bundle of manifold $X$.
\begin{enumerate}
\item Let $\phi_{1},\phi_{2}$ be a two sections of the bundle
$\pi_{XY}$ such that $\phi_{1}(x)=\phi_{2}(x)$. We say that these two sections are \textbf{$K$-equivalent
of order 1 at a point $x\in X$} and denote this as $\phi_{1}\sim_{K_{x}} \phi_{2}  $ if the restrictions
of the tangent mappings $\phi_{i*x}:T(X)\rightarrow T_{\phi_{i}(x)}(Y)$ to the subspace $K_{x}(X)$
coincide.
\item Space of classes of $K_{x}$-equivalence of order 1 will be
called a $K$-partial 1-jet (of a section) at a point $x$. Space of $K$-partial 1-jets at a point $x\in X$
will be denoted by $J^{1}_{K\ x}(\pi_{XY})$.
\item Union $\cup_{x\in X}J^{1}_{K\
x}(\pi_{XY})$ will be denoted by $J^{1}_{K}(\pi_{XY})$ and will be called the space of $K$-partial 1-jets
of sections of the bundle $\pi_{XY}$.
\end{enumerate}
\end{definition}
\begin{example}
\begin{enumerate}
\item  For $K=\emptyset $, the bundle $J^{1}_{K}(\pi)=\{ 0\}_{Y}$ - affine bundle with zero-dimensional fiber
(this is the case of RET).
\item Let $\mathcal{F}$ be a foliation of the manifold $X$ and let $K=T(\mathcal{F})$ be the distribution
 tangent to the foliation.  Then, $J^{1}_{K}(\pi)$ is the bundle of sections of $\pi:Y\rightarrow X$ and
 their derivatives \textbf{along the foliation} $\mathcal{F}$.

\item  For $K=T(X)$ the bundle $J^{1}_{K}(\pi)=J^{1}(\pi)$ is the conventional
1-jet bundle.
\end{enumerate}
\end{example}
\begin{proposition}
\begin{enumerate}
\item   Space $J^{1}_{K}(\pi)$ of $K$-partial 1-jets of sections of the bundle
$\pi_{XY}$ has a natural structure of affine bundle $\pi_{10\ K}$ over $Y$ based on the vector bundle
$\pi^{*}(K^{*})\otimes V(\pi)\rightarrow Y$ and of the fiber bundle over $X$.  Here $K^{*}(X)$ is the
vector bundle dual to the subbundle $K(X)\subset T(X)$.
\item  There is a canonical surjection of affine bundles
\[
w_{K}: J^{1}(\pi)\rightarrow  J^{1}_{K}(\pi),
\]
associating with any class of equivalent sections of order 1 of the bundle $\pi$ containing a section
$\phi$ the class of $K_{x}$-equivalent of order 1 sections of $\pi$ containing section $\phi$.
\item Let $T(X)=K(X)\oplus K'(X)$ be a decomposition of the tangent
bundle of $X$ into the direct sub of vector subbundles (an almost product structure (AP)), then the
commutative diagram
\[
\begin{CD}  J^{1}(\pi) @> w_{K} >> J^{1}_{K}(\pi)\\
@V w_{K'} VV        @V \pi_{10\ K} VV\\
 J^{1}_{K'}(\pi) @>\pi_{10\ K'} >>  Y
\end{CD}
\]
 is the diagram determining $J^{1}(\pi)$ as the fiber product of partial affine 1-jet bundles
 with respect to $K$ and $K'$ over $Y$.
\item Let $K_{1}\subset K_{2}\subset T(X)$ be two subbundles of the tangent bundle $T(X)$.  Then there is defined
the canonical surjection $w_{21}: J^{1}_{K_{2}}(\pi)\rightarrow J^{1}_{K_{1}}(\pi)$ such that
$w_{K_{1}}=w_{21}\circ w_{K_{2}}.$
 \item (Functoriality) Let the lower square of the diagram
 \[
\begin{CD} J^{1}_{K}(\pi_{1}) @> j^{1}(f)_{K}>> J^{1}_{K}(\pi_{2})\\
@V \pi^{1}_{1}VV    @V \pi^{2}_{1}VV\\
Y_{1} @>f>> Y_{2} \\
@V\pi_{1}VV   @V\pi_{2}VV \\
X @=  X
\end{CD}
\]
represents a morphism of the bundles $f:Y_{1}\rightarrow Y_{2}$ over the manifold $B$. Then there exists
the morphism of the bundles $j^{1}(f)_{K}:J^{1}_{K}(\pi_{1})\rightarrow J^{1}_{K}(\pi_{2})$ such that the
diagram above is commutative.
\end{enumerate}
\end{proposition}
\begin{proof} Almost all statements of this proposition are simple and their proof just repeat the proof
of similar statements for the full 1-jet bundle $J^{1}(\pi)$.  To prove the last statement of the
proposition notice that for any section $s$ of the bundle $\pi_{1}$ and any  (local) section $\xi $ of the
subbundle $K\subset T(X)$ we have for the section $s'=f\circ s$ of the bundle $\pi_{2}$:
\[
\xi\cdot s'^{j}(x)=\frac{\partial f^j}{\partial y^i}(\xi \cdot s^{i})(x)
\]
and thus, the derivatives of components of a section $s'$ in the directions of the subbundle $K$ are
defined by the linear mapping of the derivatives of components of the sections $s$ in the same direction.
Therefore the mapping sending the point $(x,y,z^{i}_{\xi_k}$, $\xi_{k}$ being a local basis of the bundle
$K$ to the point $(x,y'=f(x,y),z^{'j}_{\xi_k}= \frac{\partial f^j}{\partial y^i}(x,y) z^{i}_{\xi_k}$ is
defined correctly (independent on a choice of a section $s$) and determine the mapping ${\hat f}_{K}$.
\end{proof}
\begin{remark} If a subbundle $K\subset T(X)$ is chosen, the $G$-orthogonal complement to $K$:
$K'=K^{\bot_{G}}$ can be taken as the complemental subbundle $K'$ in the AP structure $T(X)=K\oplus K'$.
\end{remark}

Let now a section $\nu$ of the bundle $(J^{1}(\pi ),\pi_{10},Y)$ over $Y$ (\textbf{a jet field or
Ehresmann connection on $\pi$}) is chosen (\cite{GMS,KMS}). The section $\nu$ of the full 1-jet bundle
determines, by composition with the surjection $w_{K}$ the section of the bundle $J^{1}_{K}(\pi ).$  Thus,
in the affine
 fibers $J^{1}_{y}(\pi )$ of $\pi^{1}_{0}$ over $Y$ (respectively $\pi^{1}_{K\ 0}:J^{1}_{K}\rightarrow Y$)
 \emph{a point $\nu (y)$ is chosen}.\par
 This defines an identification of affine space
 $J^{1}_{y}(\pi )$ with the vector space
 $V_{y}\otimes T^{*}_{\pi(y)}(X)$.
\beq I_{\nu}: J^{1}_{y}(\pi)\simeq  V_{y}\otimes T^{*}_{\pi (y)}(X), \eeq
  and similar identification of the partial 1-jet bundle
\beq
 J^{1}_{K\ y}(\pi) \simeq  V_{y}\otimes \pi^{*}(K^{*}(X)),
\eeq where $K^{*}(X)\equiv T^{*}_{x}(X)/K^{\bot}(X).$\par

 Thus, a choice of a connection $\nu$ identifies (noncanonically) 1-jet bundles
$J^{1}_{K}(\pi)$ with the \textbf{vector bundles}: $ J^{1}_{K\ \nu }(\pi)\simeq V_{\pi}\otimes
 \pi^{*}(K^{*}(X)),$ see \cite{KMS}, Sec.17.2.\par

 If an almost product structure $T(X)=K\oplus K'$ is chosen, by definition of the fiber product
  there is the bijection
between the \textbf{pairs of sections} of the bundles $J^{1}_{K\ }(\pi),J^{1}_{K'}(\pi)$ over $Y$ and the
sections of the bundle $J^{1}(\pi)$.  Thus, a choice of connection $\nu$ defines the sections in the
bundles $J^{1}_{K\ }(\pi),J^{1}_{K'}(\pi)$ over $Y$ and, therefore the commutative diagram of fiber
product of vector bundles \beq
\begin{CD}  J^{1}_{\nu}(\pi) @> w_{K} >> J^{1}_{K\ \nu}(\pi)\\
@V w_{K'} VV        @V \pi^{1\ 0}_{K} VV\\
 J^{1}_{K'\ \nu}(\pi) @> \pi^{1\ 0}_{K'} >>  Y
\end{CD}
\eeq

Splitting $T(X)=K\oplus K'$ of the almost product structure defines dual splitting of the cotangent
bundle:
\[
T^{*}(X)=K^{*}\oplus K'^{*},
\]
where $K^{*}\equiv K'^{\bot },\ K'^{*}\equiv K^{\bot }$ are annulators of the complemental subbundles.  As
a result, the isomorphism (4.2) splits \beq I_{\nu}: J^{1}_{y}(\pi)\simeq  V_{y}\otimes T^{*}_{\pi
(y)}(X)= V_{y}\otimes K^{*}_{\pi (y)}\oplus V_{y}\otimes K'^{*}_{\pi (y)}. \eeq

In particular, this defines the affine subbundle
\[ Z_{K0}=\{(x,y,z)\in Z\vert \  I_{\nu}(z)\in
V_{y}\otimes K^{*}_{\pi (y)} \} ,\]

corresponding to the vector subbundle $V(\pi)\otimes K^{*}$ in the decomposition (4.4).

\begin{example} In this example $X=T\times M\equiv R^{(1+3)}$ is the Galilean space-time,
i.e. the 4-dim space-time with the block-diagonal Euclidian metric $H=dt^2+h$ and the action of Galilean
group
\[
G=T^4\times O(3)\times V^3,
\]
where $T^4$ is the group of 4-dim translations in $X$, $O(3)$ is the orthogonal group of euclidian metric
$h$ in physical space $M$ and $V^3$ is the group of inertia frame transformations.\par
\end{example}

\begin{example}  In this example we take $\pi:R^{3}\rightarrow R^2$ with coordinates $(t,x)$ on the
base$X$ (time + one space variable), $K=<\partial_{x}>$ is the subbundle of derivative along space
direction. Consider the constitutive relation defined on the partial 1-jet bundle $J^{1}_{K}(\pi)$ leading
to the Poincare-Cartan form $\Theta_{C}=-\sigma \wedge dy= ydy\wedge dx+(\frac{y^2}{2}-\delta
z_{x})dy\wedge dt +0dy\wedge dt\wedge dx$ (conservation law).  Then the balance equation defined by this
constitutive relation is
\[
d\sigma =0\Leftrightarrow \partial_{t}y+\partial_{x}(\frac{y^2}{2}-\delta y_{,x})=y_{t}+yy_{x}-\delta
y_{xx}=0\ -
\]
Burgers equation.
\end{example}
\begin{example} For the same bundle $\pi$ as in the previous example and for the same partial 1-jet
bundle take the constitutive relation leading to the Poincare-Cartan form $-\sigma\wedge
dy=\Theta_{C}=z_{x}dy\wedge dx+cos(y)dy\wedge dt +0dy\wedge dt\wedge dx.$

Then the corresponding balance (conservation) law takes the form
\[
d\sigma =0\Leftrightarrow \partial_{t}z_{x}+\partial_{x}cos(y)=y_{,tx}-sin(y)=0\ -
\]
sin-Gordon equation.
\end{example}
\begin{remark}  If we would like to write down the KdV equation $y_{t}+6yy_{x}+y_{xxx}=0$ in the form
similar to one of the last two examples, we would need to use the Poincare-Cartan form
\[
\Theta=ydy\wedge dx+(z_{xx}-3y^2)dy\wedge dt,
\]
so that we would need to use the \emph{partial 2-jet bundle} as the domain of corresponding constitutive
relation, or to increase dimension of the state space $U$ of fields $y^i$ by adding first derivative
$z_{x}$ in their list.
\end{remark}

\begin{definition}
Let $J^{1}_{p}$ is a partial 1-jet bundle of the bundle $\pi$. Denote by $J^{2}_{p}(\pi)$ the subbundle
\[
J^{2}_{p}(\pi)=J^{2}(\pi )\cap J^{1}(J^{1}_{p}(\pi))\subset J^{2}(\pi ).
\]
\end{definition}
Sections of this bundle are 1-jets of sections of the bundle $J^{1}(J^{1}_{p}(\pi))$ modulo the mixed
derivative equality whenever one is applicable (see \cite{KMS}).\par

\subsection{Space-time splitting case - bundles $J^{1}_{S}(\pi)$.}

When the partial jet spaces $J^{1}_{K}(\pi)$ with different $K$ mixes to produce more complex partial jet
bundle we get a more complicated factor of the full 1-jet bundle. As an example consider the situation
where the fiber $U$ of the configurational bundle $Y$ splits into the subspaces of fields that enters the
constitutive relations with only time, only space and space-time derivatives (5.1):

\beq U=U_{0}\oplus U_{t}\oplus U_{x}\oplus U_{tx}, \eeq

 corresponding to the splitting $S$ of the set of
indices:

\beq S=S_{0}\cup S_{t}\cup S_{x}\cup S_{xt}:\ \ [0,m]=[0,m_{0}]\cup [m_{0}+1,m_{1}]\cup
[m_{1}+1,m_{2}]\cup [m_{2}+1,m]. \eeq

 Here $U_{0}$ includes the fields $y^i ,i\in S_{0}$ whose first
derivatives do not enter the CR, $U_{t}$ includes the fields $y^i ,i\in S_{t}$ whose time derivative
enters the CR but their spacial gradient does not, $U_{x}$ is formed by the fields $y^i ,i\in S_{x}$ whose
spacial gradient but not the time derivative enter the CR, finally, $U_{tx}$ is formed by the fields $y^i
,i\in S_{tx}$ all derivative of which enter the CR. We assume that all the fields are tensor or tensor
density fields in the space $X$. This splitting is, therefore, $Diff(T)\times Diff(B)$-invariant.\par
Using the decomposition (5.1) together with the splitting $T(X)=T(R_{t})\oplus T(B)$ (see Sec.2) we can
introduce the following

\begin{definition} Let $S$ be a diagram of a splitting the fiber of the bundle $\pi$ as the sum of subbundles (4.5).
We define the \textbf{partial 1-jet bundle} $J^{1}_{S}(\pi)$ starting with the equivalence relation for
two local sections $s_{1},s_{2}:X\rightarrow Y$ defined in a neighborhood of a point $x\in X$:

\begin{multline} s_{1}\sim_{S,x} s_{2}\Leftrightarrow s^{i}_{1,t}(x)=s^{i}_{2,t}(x),\ i\in S_{t};\
s^{i}_{1,x^A}(x)=s^{i}_{2,x^A}(x),\ i\in S_{x}, A=1,\ldots ,n;\\ s^{i}_{1,x^\mu }(x)=s^{i}_{2,x^\mu}(x),\
i\in S_{tx}, \mu=0,\ldots ,n  \end{multline} and following the steps of definition of $J^{1}_{K}(\pi)$.
\end{definition}
Bundle $J^{1}_{S}(\pi)$, as its fiber the space of the first derivatives of sections $s:X\rightarrow Y$ of
the following type
\[
u^{i}_{,t},\ i\in S_{t}=[m_{0}+1,m_{1}];\ u^{i}_{,x^{\mu}},\ i\in S_{x}=[m_{1}+1,m_{2}];\
u^{i}_{,t},u^{i}_{,x^{\mu}},\ i\in S_{tx}=[m_{2}+1,m].
\]

Statements in the next Proposition follows directly from definitions and we omit their proof.
\begin{proposition}
\begin{enumerate}
\item The bundle $J^{1}_{S}(\pi)$ is defined correctly with respect to the diffeomorphisms from
 $Diff(R_{t})\times Diff(B)$ of the base manifold $X=R_{t}\times B^3$ containing arbitrary diffeomorphisms of
 $ B^3$ and the independent time diffeomorphisms of $R_{t}$.
\item Correspondingly to the decomposition (4.5) we have the decomposition of the bundle $Y\rightarrow X$ as the
fiber product of vector bundles over $X$
 \beq Y=Y_{0}\underset{X}{\times}  Y_{t}\underset{X}{\times}
Y_{x}\underset{X}{\times}  Y_{xt}, \eeq where $\pi_{0}:Y_{0}\rightarrow X$ has $U_{0}$ as its fiber,
$\pi_{t}:Y_{t}\rightarrow X$ has $U_{t}$ as its fiber, $\pi_{x}:Y_{x}\rightarrow X$ has $U_{x}$ as its
fiber, $\pi_{tx}:Y_{tx}\rightarrow X$ has $U_{tx}$ as its fiber.
\item For the partial 1-jet bundle $J^{1}_{S}(\pi)$ we have the following decomposition into the fiber
product of \textbf{affine bundles}

\beq J^{1}_{S}(\pi)= 0(Y_{0})\underset{Y}{\times} J^{1}_{\langle \partial_t \rangle
}(Y_{t})\underset{Y}{\times} J^{1}_{\langle \partial_{x^A }\rangle }(Y_{x})\underset{Y}{\times}
J^{1}(Y_{tx}) \eeq
 over $X$.
 \item (Functoriality) Let the lower square of the diagram
 \[
\begin{CD} J^{1}_{S}(\pi_{1}) @> j^{1}(f)_{S}>> J^{1}_{S}(\pi_{2})\\
@V \pi^{1}_{1}VV    @V \pi^{2}_{1}VV\\
Y_{1} @>f>> Y_{2} \\
@V\pi_{1}VV   @V\pi_{2}VV \\
X @=  X
\end{CD}
\]
represents a morphism of the bundles $f:Y_{1}\rightarrow Y_{2}$ over the manifold $B$ such that $f(U_{1\
K_{i}})\subset U_{2\ K_{i}}$ for the subbundles $K_{i}$ be the subbundles or the splitting $S$, i.e.
$K_{i}=<0>,<\partial_{t}>,<\partial_{x^A}>, T(X)$. Then there exists the morphism of the bundles
$j^{1}(f)_{S}:J^{1}_{S}(\pi_{1})\rightarrow J^{1}_{S}(\pi_{2})$ such that the diagram above is
commutative.
\end{enumerate}
\end{proposition}

In Table 1 there are listed the derivatives of sections corresponding to a decomposition $S$  of the state
space $U$ (here $ J^{2}_{S}(\pi)=J^{2}(\pi)\cap J^{1}(J^{1}_{S}(\pi))$, see above)
\begin{table}
\begin{center}
\begin{tabular}{| c | c | c | c | c |}
\hline $Y$ & $U_{0}$ & $U_{t}$ & $U_{x}$ & $U_{tx}$ \\ \hline $J^{1}_{S}(\pi )$ & $0$ & $u^{i}_{,t}$ &
$u^{i}_{,x}$ & $u^{i}_{,t},u^{i}_{,x}$\\ \hline $J^{2}_{S}(\pi)$ & $u^{i}_{,t},u^{i}_{,x}$ &
$u^{i}_{,x},u^{i}_{,tt},u^{i}_{tx}$ & $u^{i}_{,t},u^{i}_{,tx},u^{i}_{xx}$ &
$u^{i}_{,tt},u^{i}_{,tx},u^{i}_{xx}$ \\ \hline
\end{tabular}
\caption{1- and 2-partial jet spaces $J^{i}_{S}(\pi)$.}
\end{center}
\end{table}

\begin{remark} Notice here that the case $J^{1}_{S}(\pi)$ includes, as its special cases, RET case, case
$Z _{p}=J^{1}_{K}(\pi)$ and the case of the full 1-jet bundle $J^{1}(\pi)$.  Thus, probably, this
situation is the most general case of a partial 1-jet bundle over $\pi$ important in applications (that
includes only 1-jet bundles but not higher order bundles).
\end{remark}

Introduce the notion of geometric automorphisms of the bundle $\pi$ that can be lifted to the bundle
$J^{1}_{S}(\pi)$.
\begin{definition} \begin{enumerate}
\item
An automorphism $\phi\in Aut(\pi)$ of the bundle $\pi$ is called a $S$-admissible if it is the
automorphism of the fiber product bundle decomposition (4.4),i.e. there are automorphisms: $\phi_{0}$ of
the bundle $Y_{0}\rightarrow X$, $\phi_{t}$ of the bundle $Y_{t}\rightarrow X$ etc. such that
\[
\phi=\phi_{0}\underset{Y}{\times} \phi_{t} \underset{Y}{\times} \phi_{x} \underset{Y}{\times} \phi_{tx}.
\]
Lie group of $S$-admissible automorphisms  of the bundle $\pi$ will be denoted by $Aut_{S}(\pi)$.
\item A projectable vector field $\xi \in \mathcal{X}(Y)$  is called $S$-admissible if transformations of
its local flow $\phi_{t}$ are $S$-admissible. Denote by $\mathcal{X}_{S}(\pi)$ the Lie algebra of all
$S$-admissible projectable vector fields in $Y$.
\end{enumerate}
\end{definition}
Following simple Lemma describes the structure of $S$-admissible vector fields.
\begin{lemma} A vector field $\xi\in \mathcal{X}(\pi)$ belongs to $\mathcal{X}_{S}(\pi)$ if and only if it
has the form
\begin{multline}
\xi =\xi^{\mu}(x,t)\partial_{x^\mu}+\sum_{i_{0}\in S_{0}}  \xi^{i_{0}} (x,y^{j_{0}},\ j^{0}\in
S_{0})\partial_{y^{j_{0}}}+         \sum_{i_{1}\in S_{t}}  \xi^{i_{1}} (x,y^{j_{1}},\ j^{1}\in
S_{t})\partial_{y^{j_{1}}}+     \\   + \sum_{i_{2}\in S_{x}}  \xi^{i_{2}} (x,y^{j_{2}},\ j^{2}\in
S_{x})\partial_{y^{j_{2}}}+         \sum_{i_{3}\in S_{tx}} \xi^{i_{3}} (x,y^{j_{3}},\ j^{3}\in
S_{tx})\partial_{y^{j_{3}}}.
\end{multline}
\end{lemma}
Finally, we have the following description of the 1-jet bundle of the bundle $J^{1}_{S}(\pi)$. Notice that
fibers of this bundle contains values of derivatives of first and second order of some fields $y^i$.

\begin{proposition}

Affine bundle $J^{1}(J^{1}_{S}(\pi))$ over $J^{1}_{S}(\pi)$ has its fiber modeled on the vector space
\[
(T^{*}(X)\otimes U_{0})\oplus (T^{*}_{x,xt,tt}(X)\otimes U_{t})\oplus (T^{*}_{t,xt,xx}(X)\otimes
U_{x})\oplus (T^{*}_{t,x,tx,xx}(X)\otimes U_{tx}),
\]
where the types of partial derivatives of fields $y^i$ from different components $U_{*}$ of the field
space $U$ included into the second partial jet bundle are marked.
\end{proposition}

\begin{remark}
Natural dual affine bundle to the bundle $Z_{p}=J^{1}_{p}(\pi)$ is the subbundle of the bundle $Z^*$.  To
see this we recall the natural affine projection $J^{1}(\pi)\rightarrow J^{1}_{p}(\pi)$ introduced above.
Correspondingly we get the induced monomorphism of dual affine bundles
\[
J^{1}_{p}(\pi)^{*}\rightarrow J^{1}(\pi)^{*}=Z^{*}.
\]
Yet below we will be mostly interested by the mappings from $J^{1}_{p}(\pi)$ to the whole space $Z^*$ and
$\tilde Z$.
\end{remark}

\section{Balance Equations with the domain in $J^{1}_{p}(\pi)$}
In this section we define balance equations with the domain being an open subset of a partial 1-jet bundle
and the balance systems - basic notion of this work.
\begin{definition}
\begin{enumerate}
\item A balance equation (law) with a domain $D\subset J^{1}_{p}(\pi)$ is an $n+(n+1)$ semibasic form
 $B$ defined in the domain $D$:
\[
B=F^{\mu}\eta_{\mu}+\Pi \eta,\ F^\mu ,\Pi \in C^{\infty}(D).
\]
\item A section $s\in \Gamma(\pi)$ is a solution of the balance equation $B$ if
\[
{\tilde d}j^{1}_{p}(s)^{*}(B)= dj^{1}_{p}(s)^{*}(F^\mu \eta_{\mu})-j^{1}_{p}(s)^{*}\Pi \eta =0.
\]
Here $\tilde d$ is the Iglesias differential, see Appendix II.
\item A balance law $B$ is called trivial if any section of the bundle $\pi$ is its solution.
\end{enumerate}
\end{definition}
Balance laws with a domain $D\subset J^{1}_{p}(\pi)$ form a vector space - subspace $\mathcal{BL}(D)
\subset \Lambda^{*}_{sb}(D)$ of the subalgebra of semi-basic forms $\Lambda^{*}_{sb}(D)$ of the exterior
algebra $\Lambda^{*}(D)$ in the domain $D$.\par
\begin{lemma} A balance law $B=q^{\mu}\eta_{\mu}+\Pi \eta$ is trivial if and only if $\Pi=
\sum_{\mu}d_{\mu }q^{\mu}$, where $d_{\mu}= \partial_{x^{\mu}}+z^{i}_{\mu}\partial_{y^i}$ is the total
derivative by $x^\mu$.
\end{lemma}
\begin{proof} Standard.
\end{proof}
\begin{remark} Functions $q^\mu$ should be such that the jet variables $z^{i}_{\mu}$ be admissible
variables of $J^{1}_{p}(\pi)$ provided that $\partial_{y^i}q^{\mu}\ne 0$. This places a limitations on the
type of functions $q^\mu$. Namely,
\begin{enumerate}\item For the RET case $q^{\mu}=q^{\mu}(x)$ can not depend on $y$.
\item For the full case there are no restriction to the dependence of $q^\mu(x,y)$ on $x,y$.
\item For $Z_{K}=J^{1}_{K}(\pi)$ with $K=T(B)$, one should have $\partial_{y^i }q^{0}=0$ for all $i$.
\item For $Z_{K}=J^{1}_{K}(\pi)$ with $K=T(R_{t})$ one should have  $\partial_{y^i}q^A =0$ for all
$A=1,\ldots, n.$
\end{enumerate}
\end{remark}
\begin{definition} Two balance equations (laws) $B_{k}=F^{\mu}_{k}\eta_{\mu}+\Pi_{k}\eta$ are called
Div-equivalent if $B_{2}-B_{1}=q^{\mu}\eta_{\mu}+(\sum_{\mu}d_{\mu}q^{\mu})\eta$ for some functions $q^\mu
\in C^{\infty}(Z_{p}).$
\end{definition}
It is clear that the balance equations which are Div-equivalent has the same space of solutions $s\in
\Gamma(\pi)$.\par
\begin{definition} A balance system defined in a domain $D\subset J^{1}_{p}(\pi)$ is a subspace in the
space $\mathcal{BL}(D)$ of the balance laws defined in $D$.
\end{definition}
\begin{remark} As defined, the notion of balance system is very broad.  To be more practically useful, one
has to deal with the systems large enough to specify all the components $s^{i}(x)$ of the basic fields
 $y^i$ and small enough to be determined.  Second condition is usually achieved by requiring that the number of
 equation in a balance system is equal to the number $m$ of basic fields. First condition requires the
 fulfillment of some regularity conditions (see Section 14 below) that may depend on the problem studied
 with the balance system.
\end{remark}

\section{Partial Cartan distribution.}

The 1-jet space $Z=J^{1}(\pi)$ is endowed with the canonical \emph{Cartan distribution} $Ca$ locally (in
the adapted coordinates) defined by the 1-forms
\[
\omega^i=dy^i-z^{i}_{\mu}dx^\mu .
\]
Cartan distribution is the direct sum of two distributions:
\beq Ca_{z}=D_{z}\oplus V(\pi_{10}), \eeq

where $V(\pi_{10})=<\partial_{z^{i}_{\mu }}>$ is the vertical subbundle of the tangent bundle $T(Z)$ with
respect to the projection $\pi_{10}:Z\rightarrow Y$ and
\[
D_{z} =<d_{\mu}=\partial_{x^\mu}+\sum_{i}z^{i}_{\mu}\partial_{y^i}>
\]
is the subbundle generated by the (truncated) total derivatives by $x^\mu$. Distribution $D$ is defined
correctly with respect to the automorphisms of the bundle $\pi$ but it is not integrable.\par Distribution
$D$ allows to define the contact lift of vector fields from $X$ to $Z$:
\[
\xi=\sum_{\mu}\xi^{\mu}(x)\partial_{x^\mu}\ \rightarrow {\hat \xi}=\sum_{\mu}\xi^{\mu}(x)D_{\mu}.
\]
In a case where the subbundle $K$ is not integrable we will have to use non-holonomic frames in $X$ and
the corresponding coframes.\par

Let $(x^\mu ,y^i )$ is a local adopted coordinate chart in the bundle $\pi$. Let
$\xi_{\mu}=\xi_{\mu}^\lambda \partial_{\lambda}$ be a (local) nonholonomic frame of the tangent bundle
$T(X)$. Denote by $\psi^{\mu}$ its dual coframe: $<\psi^{\mu},\xi_{\nu}>=\delta_{\nu}^{\mu}$. Using this
definition it is easy to see that
\[
{\hat \psi}^{\sigma}= (\xi^{\alpha}_{\beta} )^{-1\ \sigma}_{\mu}dx^{\mu}.
\]

  Let ${\hat
\psi}^\mu$ be the pullback of the 1-form $\psi^\mu$ to $Z=J^{1}(\pi)$ by the projection $\pi^1
:Z=J^{1}(\pi)\rightarrow X$.\par

Following simple Lemma gives the representation of the Cartan distribution in $Z$ in terms of such a
non-holonomic frame.
\begin{lemma} Let $(x^\mu ,y^i )$ is a local adopted coordinate chart in the bundle $\pi$. Let $\xi_{\mu}$
be a (local) nonholonomic frame of the tangent bundle $T(X)$. Denote by $\psi^{\mu}$ its dual coframe
($<\psi^{\mu},\xi_{\nu}>=\delta_{\nu}^{\mu}$).  Let ${\hat \psi}^\mu$ be the pullback of the 1-form
$\psi^\mu$ to $Z=J^{1}(\pi)$ by the projection $\pi^1$. Introduce the (local) coordinates in the fibers of
the bundle $Z=J^{1}(\pi)\rightarrow Y$  by
\[
\widetilde{z}^{i}_{\mu}(j^{1}(s)(x))=(\xi_{\mu}\cdot s^{i})(x)
\]
for all sections $s:X\rightarrow Y$.  Then
\begin{enumerate}
\item We have ${\tilde z}^{i}_{\mu}=\xi^{\lambda}_{\mu}z^{i}_{\mu}.$
\item Cartan distribution $Ca$ in $Z$ is defined by the forms
\[
\widetilde{\omega}^{i}=dy^i-\sum_{\mu}\tilde{z}^{i}_{\mu}{\hat \psi}^{\mu},
\]
\item Cartan distribution is generated by the vector fields
\[
Ca(z)=<\widehat{\xi}_{\mu}=\xi_{\mu}+\sum_{i}\tilde{z}^{i}_{\mu}\partial_{y^i},\ \partial_{z^{i}_{\mu}},\
\mu=1,\ldots ,n+1>.
\]
\end{enumerate}
\end{lemma}
\begin{proof}
We know that Cartan distribution is generated by the vertical vector fields $\partial_{z^{i}_{\mu}}$
trivially  annulated by the forms $\widetilde{\omega}^{i}$ and by the linearly independent vector fields
${\widehat{\xi}}_{\mu}=\xi^{\lambda }_{\mu}\widehat{\partial_{\mu}}$.  Thus it is sufficient to check that
$\widetilde{\omega}^{i}(\widehat{\xi}_{\mu})=0$ for all $i,\mu.$  We have
\begin{multline}
\widetilde{\omega}^{i}(\widehat{\xi}_{\mu})=\langle dy^i-\sum_{\mu}\tilde{z}^{i}_{\mu}{\hat \psi}^{\mu},
\xi^{\lambda}_{\mu}d_{\lambda}\rangle =\langle dy^i-\sum_{\sigma }\tilde{z}^{i}_{\sigma}{\hat
\psi}^{\sigma},
\xi^{\lambda}_{\mu}(\partial_{x^\lambda }+\sum_{i}z^{i}_{\lambda}\partial_{y^i})\rangle =\\
=\xi^{\lambda}_{\mu} \left( z^{i}_{\lambda} - \tilde{z}^{i}_{\sigma}{\hat
\psi}^{\sigma}(\partial_{\lambda}) \right)=\xi^{\lambda}_{\mu} \left( z^{i}_{\lambda} -
\tilde{z}^{i}_{\sigma}(\xi^{\alpha}_{\beta} )^{-1\ \sigma}_{\lambda }
\right)=\xi^{\lambda}_{\mu}z^{i}_{\lambda}-\tilde{z}^{i}_{\sigma}
\delta^{\sigma}_{\lambda}=\xi^{\lambda}_{\mu}z^{i}_{\lambda}-\tilde{z}^{i}_{\lambda}=0.
\end{multline}
To prove the third statement we notice that
\[ \widehat{\xi}_{\mu}=\xi^{\lambda}_{\mu}(\partial_{\lambda}+z^{i}_{\lambda}\partial_{y^i})= \xi_{\mu}+
\xi^{\lambda}_{\mu}z^{i}_{\lambda}\partial_{y^i}=\xi_{\mu}+\tilde{z}^{i}_{\mu}\partial_{y^i}.\]
\end{proof}
Recall the following
\begin{definition} An exterior form $\nu^k$ on the 1-jet space $Z=J^{1}(\pi)$ is called \textbf{contact} if for all
sections $s\in \Gamma(\pi)$, $j^{1}(s)^{*}\nu=0$.
\end{definition}
Contact forms on $Z$ form the ideal $C\Lambda^{*}(Z)$ of the exterior algebra $\Lambda^{*}(Z)$.  Forms
$\omega^i$ defined above in the case of a holonomic frame or forms ${\tilde \omega}^{i}$ in a case of a
non-holonomic frame generate the ideal $C\Lambda^{*}(Z)$.\par

In the 2-jet bundle $J^{2}(\pi)$ with local coordinates $x^\mu, y^i ,z^{i}_{\mu}, z^{i}_{\mu \nu}$ similar
Cartan distribution is defined generated by the ideal of contact form with the generators

\beq \omega^i = dy^i -z^{i}_{\mu}dx^mu , \omega^{i}_{\mu} =dz^{i}_{\mu}-z^{i}_{\mu \nu}dx^\nu .\eeq

In a contrast to the full 1-jet bundle in the maximal (RET) case the fiber of $J^{1}_{p}(\pi)$ is one
point and has no local geometrical structure.  We will show that in the intermediate case a partial 1-jet
bundles $J^{1}_{K}(\pi), J^{1}_{S}(\pi)$ have the "partial Cartan distribution", corresponding to the
structure of the fibers of $J^{1}_{p}(\pi)\rightarrow Y.$  This distribution although depending not just
on the subbundle $K$ but on the complemental distribution $K'$ as well (i.e. on the whole almost product
structure $T(X)=K\oplus K'$, \cite{LR}) plays an important role for the partial jet bundles similar to
that of the conventional Cartan distribution.
\par
\subsection{Case of $J^{1}_{K}(\pi)$, $K$ - integrable.} We start with the case of a decomposition
$T(X)=K\oplus K'$ of the tangent bundle of the base $X$ and the corresponding decomposition
$T^{*}(X)=K^{*}\oplus K'^{*}$ of the cotangent bundle into the direct sum of two \emph{integrable}
subbundles. Locally, one can choose a coordinate chart $x^{\mu}=<x^{\nu};x^{\sigma}>$ such that (with
respect to the index splitting $\mu =<\nu,\sigma >$)
\[
K=<\partial_{x^\nu} >;\ K'=<\partial_{x^\sigma} >.
\]
Almost product structure allows to split both subdistributions of the decomposition (5.1) as the sums of
$K$- and $K'$-subdistributions

\beq \begin{cases} V(\pi_{1})=V(\pi_{1})_{K}\oplus V(\pi_{1})_{K'},\
V(\pi_{1})_{K}(z)=<\partial_{z^{i}_{\nu}}>,
V(\pi_{1})_{K'}(z)=<\partial_{z^{i}_{\sigma}}>,\\
D(z)=D_{K}(z)\oplus D_{K'}(z),\ D_{K}(z)=<D_{\nu}(z)>,\ D_{K'}(z)=<D_{\sigma}(z)>.
\end{cases} \eeq
This decomposition is invariant under the automorphisms $\phi$ of the bundle $\pi$ whose projection $\bar
\phi$ to $X$ \emph{preserves the almost product structure} $T(X)=K\oplus K'$.\par

 Now  we define the 1-forms on the partial 1-jet bundle $J^{1}_{K}(\pi )$:
\[
\omega^{i}_{K}=dy^i-\sum_{\nu}z^{i}_{\nu}dx^\nu
\]
(summation by $\nu$ only!) in the domain of the chart $x^\mu$.  These 1-forms are defined correctly
\emph{with respect to the diffeomorphisms of $X$ preserving the decomposition } $T(X)=K\oplus K'$ (i.e.
leaving both distributions of this decomposition invariant). It is easy to check that a section $q$ of the
bundle $J^{1}_{K}(\pi )$ is the (partial) 1-jet of a section $s:X\rightarrow Y$ if and only if
$q^{*}(\omega^j )\vert_{K}=0$ for all $i=1,\ldots ,m$.\par

Thus, we have
\begin{proposition} Let $T(X)=K\oplus K'$ be a decomposition of the tangent bundle of the base $X$ into
the direct sum of integrable subbundles. \begin{enumerate} \item The one forms $\omega^i$ defined in the
local coordinate chart $x^{\mu}=<x^{\nu};x^{\sigma}>$ integrating the subbundles $K,K'$  by

\beq \omega^{i}=dy^i-\sum_{\nu}z^{i}_{\nu}dx^\nu \eeq

generate a distribution on the partial jet space $J^{1}_{K}(\pi )$ of codimension $m$ invariant under the
diffeomorphisms of $X$ preserving the decomposition $T(X)=K\oplus K'$.
\item A section $q$ of the bundle
$J^{1}_{K}(\pi )$ is the (partial) 1-jet of a section $s:X\rightarrow Y$ if and only if $q^{*}(\omega^j
)\vert_{K}=0$ for all $i=1,\ldots ,m$.
\item Partial Cartan distribution is the linear span of the vector fields
\[
\partial_{x^\nu}+z^{i}_{\nu}\partial_{y^i},\ \partial_{x^\sigma},\ \partial_{z^{i}_{\nu}}.
\]
\item Let $\chi$ be a connection in the bundle $\pi$. Define the affine subbundle $Z_{K0}$, (depending on
 the connection $\nu$ and the integrable almost product structure $T(X)=K\oplus K'$) by the equations
 $z^{i}_{\sigma}=0$. Then the intersection of
Cartan distribution $Ca$ of $T(Z)$ with the tangent to the subbundle $Z_{K0}$ is the linear span of the
tangent vectors
\[
Ca \cap T(Z_{K0})=<D_{\nu}, \partial_{x^{\sigma}},\partial_{z^{i}_{\nu}}.>
\]
\item Restriction to $Z_{K0}$ of the projection $Z\rightarrow J^{1}_{K}(\pi)$ defined the
isomorphism of affine bundles $Z_{K0}\rightarrow J^{1}_{K}(\pi)$ mapping the distribution $Ca \cap
T(Z_{K0})$ isomorphically onto the partial Cartan distribution $CA_{K}$ in $J^{1}_{K}(\pi)$.
\end{enumerate}
\end{proposition}

\subsection{Case $J^{1}_{K}(\pi)$, $K$ - general.}

 Let now $K$ be a \textbf{general} vector subbundle of $T(X)$ and let $T(X)=K\oplus K'$ is the almost
product structure containing $K$ as one of the subbundles. Choose a local basis of distribution $K$
(respectively of $K'$) consisting of the vector fields $\xi_{\nu},\nu=1,\ldots ,k$ (respectively
$\xi_{\sigma},\sigma=1,\ldots ,n+1-k$). Vector fields $\xi_{\mu}, \mu=1,\ldots ,n+1$ form a local frame.
Introduce the dual coframe $\psi^{\mu }$ of this frame by requiring that
($<\psi^{\mu},\xi_{\alpha}>=\delta^{\mu}_{\alpha}$).\par

Frame $\xi_{\mu}$ defines in a domain $W\subset X$ determines the zero curvature connection (absolute
parallelism) $\tau$ in $W$. This connection has, in general, a non-zero torsion $T^{\mu}_{\alpha \beta}$
that can also be defined in terms of the commutators of vector field of the frame or, what is equivalent,
in terms of the differentials of the coframe 1-forms:

\beq [\xi_{\alpha},\xi_{\beta}]=T^{\mu}_{\alpha \beta }\xi_{\mu};\ d\psi^{\mu}=T^{\mu}_{\alpha
\beta}\psi^{\alpha} \wedge \psi^\beta, \eeq

 with the tensor
$T^{\mu}_{\alpha \beta }$ being the torsion $T$of the connection $\tau$.\par

 Take the pullbacks ${\hat \psi}_{\nu}$ of the 1-forms
$\psi_{\nu}$ to the bundle $J^{1}_{K}(\pi )$. A fiber $J^{1}_{K\ x,y}$ of the bundle $J^{1}_{K}(\pi)$ will
be endowed with the defined above (local) coordinates $\tilde{z}^{i}_{\nu}=(\xi_{\nu}\cdot s^{i})(x)$
where $s(x)$ is a (local) section of $\pi$ such that $s(x)=y$ (Notice that these coordinates are defined
by the distribution $K$ \textbf{only} but not on the complemental distribution $K'$).\par

 Consider the set of 1-forms on $J^{1}_{K}(\pi )$

\beq \widetilde{\omega}^{j}=dy^j-\sum_{\nu}z^{j}_{\nu}{\hat \psi}_{\nu}. \eeq

Let $\sigma=(x^{\mu};s^{i}(x),s^{i}_{\nu}(x)) :X\rightarrow J^{1}_{K}(\pi )$ be a local section of
$J^{1}_{K}(\pi )$ such that $\sigma^{*}\omega^{j}=0$ for all $j$. This condition is equivalent to the
fulfillment of the conditions
\[
s^{i}_{\nu}(x)=\xi_{\nu}s^{i}(x), j=1,\ldots ,k,
\]
in other words to the integrability of the section $\sigma$. Action of a diffeomorphism of $X$ preserving
the almost product structure $T(X)=K\oplus K'$ transforms vector fields of the frame $\xi_{\nu}$ of $K$,
these of the subbundle $K'$, dual coframe and the vertical coordinates $\tilde{z}^{i}_{\nu}(x)$ of the
section by the action of Jacoby matrix in a coherent way ensuring the correctness of the following
definition

\begin{definition} Let $K$ be a general (locally trivial) subbundle of $T(X)$ and let $T(X)=K\oplus K'$ is the almost
product structure containing $K$ as one of the subbundles. Let $\xi_{\nu},\nu=1,\ldots ,k$ be a local
frame of distribution $K$, $\xi_{\sigma}, \nu=k+1,\ldots ,n+1$ is the local frame of $K'$, let
$\psi^{\mu}$ be a dual coframe of the local frame $\xi_{\mu}$. Let ${\hat \psi}_{\nu}$ be the pullback of
the 1-forms $\psi_{\nu}$ to the bundle $J^{1}_{K}(Y)$. Then we define the (partial) Cartan distribution
$Ca_{K\oplus K'}$ on the bundle $J^{1}_{K}(\pi )$  as the one determined by the 1-forms  \beq
\widetilde{\omega}^{j}=dy^j-\sum_{\nu =1}^{k}\tilde{z}^{j}_{\nu}\widehat{\psi}_{\nu}. \eeq
\end{definition}
\begin{proposition}
\begin{enumerate}
\item Distribution $Ca_{K}$ has the property
that a section $\sigma: X\rightarrow Z_{K}=J^{1}_{K}(\pi )$ is the K-partial 1-jet of a section
$s:X\rightarrow Y$ iff $\sigma^{*}\widetilde{\omega}^{j}\vert_{K}=0$ for all $j$.
\item Distribution $Ca_{K}$ is invariant under the flow lifts of diffeomorphisms of $X$ preserving the AP
structure $T(X)=K\oplus K'$.
\item Cartan distribution is generated by the (locally defined) vector fields
\[
\widehat{\xi}_{\nu}=\xi_{\nu}+\sum_{i}\tilde{z}^{i}_{\nu}\partial_{y^i},\ \xi_{\sigma},
\partial_{z^{i}_{\nu}}.
\]
\item Let $\nu$ be a connection in the bundle $\pi$. Affine subbundle $Z_{K0}$, defined by the connection
$\nu$ and the almost product structure $T(X)=K\oplus K'$ (see ()) by the equations
 $\tilde{z}^{i}_{\sigma}=0$ and the intersection of
Cartan distribution $Ca$ of $T(Z)$ with the tangent to the subbundle $Z_{K0}$ is the linear span of the
tangent vectors
\[
Ca \cap T(Z_{K0})=<\widehat{\xi}_{\nu}, \xi_{\sigma},\partial_{\widetilde{z}^{i}_{\nu}}.>
\]
\item Restriction to $Z_{K0}$ (see Prop. 5 above) of the projection $Z\rightarrow J^{1}_{K}(\pi)$ defined the
isomorphism of affine bundles $Z_{K0}\rightarrow J^{1}_{K}(\pi)$ mapping the distribution $Ca \cap
T(Z_{K0})$ isomorphically onto the partial Cartan distribution $Ca_{K}$ in $J^{1}_{K}(\pi)$.
\end{enumerate}
\end{proposition}
\begin{proof}
To prove the second statement of the proposition we write the basic contact forms on the full 1-jet
bundles as
\[
\widetilde{\omega}^{i}=\partial_{y^i}-\sum_{\mu}\tilde{z}^{i}_{\mu}{\hat \psi}^\mu.
\]
  The Cartan distribution $Ca$ is generated, in this basis, by the vector fields
$\xi_{\nu}+\tilde{z}^{i}_{\nu}\partial_{y^i}, \xi_{\sigma}+\tilde{z}^{i}_{\sigma}\partial_{y^i},\
\partial_{\tilde{z}^{i}_{\nu}},
\partial_{\tilde{z}^{i}_{\sigma}}$. Using the connection $\nu$ for identifying the affine 1-jet bundles with the corresponding
vector bundles (see (5.5)) we see that the projection $J^{1}(\pi)\rightarrow J^{1}_{K}(\pi )$ has, in the
chosen adopted coordinates nonholonomic in the jet fibers, the form
$(x,y,\tilde{z}^{i}_{\nu},\tilde{z}^{i}_{\sigma})\rightarrow (x,y,\tilde{z}^{i}_{\nu},0).$  Under this
projection the vector fields $\partial_{z^{i}_{\sigma}}$ go to zero while others projects to the vector
fields $\xi_{\nu}+\tilde{z}^{i}_{\nu}\partial_{y^i},\ \xi_{\sigma},\
\partial_{\tilde{z}^{i}_{\nu}},\ 0$ respectively (if we assume that $\tilde{z}^{i}_{\sigma}$ goes to zero.  These vector
fields are horizontal with respect to the partial contact structure on the partial 1-jet space
$J^{1}_{K}(\pi)$.
\end{proof}
\vskip0.5cm  Let only subbundle $K\subset T(X)$ is given and we complete it to the AP-structure in two
different ways:
\[
T(X)=K\oplus K_{1}=K\oplus K_{2}.
\]
Let $\eta_{\nu}$ (respectively $\eta_{\sigma\ 1},\eta_{\sigma\ 2}$ be a local basis of distribution $K$
(respectively, of $K_{1},\ K_{2}$).  Mapping $\beta:T(X)\rightarrow T(X)$ given by
\[
\eta_{\nu}\rightarrow \eta_{\nu},\ \eta_{\sigma\ 1}\rightarrow \eta_{\sigma\ 2}
\]
defines the pure gauge automorphism of the tangent bundle $T(X)$ preserving subbundle $K$ and exchanging
subbundles $K_{1}$ and $K_{2}$. Dual mapping $\beta^{*}$ defines the automorphism of $T^{*}(X)$ sending
$\psi^{\sigma\ 1}\rightarrow \psi^{\sigma\ 2 }$ but sending $\psi^{\nu\ 1}$ into other 1-forms $\psi^{\nu\
2}$. Coordinates $\tilde{z}^{i}_{\nu}$ in the fiber of the partial 1-jet bundle $J^{1}_{K }(\pi)$ defined
by the condition $\tilde{z}^{i}_{\nu\ 1}(j^{1}_{K}(s))=\eta_{\nu}\cdot s^{i}=<\eta_{\nu},ds^{i}>$ are
mapped under the isomorphism $\beta $ to the \emph{same} coordinates $\tilde{z}^{i}_{\nu\
2}(j^{1}_{K}(s))=\eta_{\nu}\cdot s^{i}=<\eta_{\nu},ds^{i}>$ and the pullbacked formes ${\hat \psi}^{\nu\
1}$ in $J^{1}_{K}(\pi)$ are mapped to the forms  ${\hat \psi}^{\nu\ 2}$ - pullbacks of the forms
${\psi}^{\nu\ 2}$.  Thus, the generating forms $\widetilde{\omega}^{i}_{K1}=dy^i -\tilde{z}^{i}_{\nu}{\hat
\psi}^{\nu\ 1}$ of the first partial Cartan distribution are mapped to the generating forms
$\widetilde{\omega}^{i}_{K2}=dy^i -\tilde{z}^{i}_{\nu}{\hat \psi}^{\nu\ 2}$ of the second Cartan
distribution. In terms of vector fields we have the mapping $(\hat{\xi}_{\nu},\xi_{\sigma\
1},\partial_{z^{i}_{\nu}})\rightarrow N_{\nu},\xi_{\sigma\ 2},\partial_{z^{i}_{\nu}})$ of the first PCS to
the second PCS.  This mapping is the (pure) gauge isomorphism of the first structure to the second.  It is
clear that if the AP-structures $K\oplus K_{j}$, $J=1,2$ are integrable with a local integrating charts
$(x^{nu},x^{\sigma\  j})$ then the geometrical mapping of the change of coordinates between these charts
$(x^{nu},x^{\sigma\  1})\rightarrow (x^{nu},x^{\sigma\  2})$ determines the isomorphism of partial Cartan
structures defined above for general case.  Thus we have proved the following

\begin{proposition} Let a subbundle $K\subset T(X)$ is given. Let
\[
T(X)=K\oplus K_{1}=K\oplus K_{2}
\]
are two ways to complete $K$ to an AP-structure.  Then the partial Cartan structures $CA_{K\ j},\ j=1,2$
in $J^{1}_{K}(\pi)$ defined by these two AP-structures are isomorphic.  Isomorphism between these
structures leaves invariant the sub-distribution of the (partial) Cartan distribution
$<\widehat{\xi}_{\nu},
\partial_{z^{i}_{\nu}}>.$  If the AP-structures $K\oplus K_{j}$ are integrable, isomorphism between
corresponding partial Cartan distributions is generated by the (geometrical) coordinate change of adopted
(local ) charts.
\end{proposition}

\subsection{Case of $J^{1}_{S}(\pi)$.}

consider now the partial 1-jet bundle $J^{1}_{S}(\pi)$ corresponding to the decomposition () of the basic
field space $U$ and to the fiber product decompositions for $Y$ and for $J^{1}_{S}(\pi)$ given in
Proposition (). Integrable product structure $T(X)=\langle \partial_{t}\rangle \oplus \langle
\partial_{x^A}\rangle $ allows to define conventional Cartan distribution $Ca_{tx}$ in the 1-jet bundle
 $J^{1}(\pi_{tx})$ and
partial Cartan structures $Ca_{t},Ca_{x}$ in the partial 1-jet bundles $J^{1}_{\partial_{t}}(\pi_{t}),
J^{1}_{\partial_{x^A}}(\pi_{x})$ respectively.  Define now the distribution \beq Ca_{S}=Ca_{t}\oplus
Ca_{x}\oplus Ca_{xt}. \eeq

Combining the results for all partial bundles (three of them since the bundle $Y_{0}$ has zero fibers) we
come to the following

\begin{proposition} Let $(S,U_{i})$ is the splitting of the fields space $U$ of the form (). Define in the
partial 1-jet bundle $J^{1}_{S}(\pi)$ the distribution $Ca_{S}$ as the direct (fibered) sum of the partial
Cartan distribution for all four partial 1-jet bundles $J^{1}_{i}(\pi)$
\[
Ca_{S}=Ca_{t}\oplus Ca_{x}\oplus Ca_{xt}.
\]
Then,
\begin{enumerate}
\item Distribution $Ca_{S}$ is generated by the following 1-forms
\[
\omega^{i}=dy^i -z^{i}_{t}dt,\ i\in S_{t},\ \omega^{i}=dy^i -z^{i}_{A}dx^A,\ i\in S_{x}; \omega^{i}=dy^i
-z^{i}_{\mu}dx^\mu,\ i\in S_{tx}.
\]
\item A section $\sigma:X\rightarrow J^{1}_{S}(\pi)$ is integrable: $\sigma =i^{1}_{p}(s)$ for some
section $s:X\rightarrow Y$ if and only if
\[
\sigma^{*}\omega^i =0,\ \text{for\ all}\ i.
\]
\end{enumerate}
\end{proposition}
\subsection{Contact ideal on $Z_{p}$}

\begin{definition} An exterior form $\nu^k$ on the 1-jet space $Z=J^{1}_{p}(\pi)$ is called
 \textbf{contact} if for all sections $s\in \Gamma(\pi)$, $j^{1}_{p}(s)^{*}\nu=0$.
\end{definition}

Contact forms on $Z_{p}$ form the ideal $I_{p}(Ca)$ of the exterior algebra $\Lambda^{*}(Z_{p})$. In local
coordinates $(x^\mu ,y^i )$ denote by $P$ the set of pairs of indices $(\mu ,i)$ such that coordinate
$z^{i}_{\mu}$ is defined in $Z_{p}=J^{1}_{p}(\pi)$.\par
 Forms $\omega^i =dy^i-\sum_{(\mu,i)\in P}z^{i}_{\mu}dx^mu $ defined above generate the ideal
$I(Ca)$.\par

In the 2-jet bundle $J^{1}(J^{1}_{p}(\pi))$ with local coordinates $x^\mu, y^i ;z^{i}_{\mu},(\mu,i)\in P;
z^{i}_{\mu nu},(\mu,i)\in P$ similar partial Cartan distribution is defined, generated by the ideal of
contact form with the generators

\beq \omega^i = dy^i -\sum_{(\mu,i)\in P}z^{i}_{\mu}dx^mu ; \omega^{i}_{\mu} =dz^{i}_{\mu}-z^{i}_{\mu
nu}dx^\nu ,\sum_{(\mu,i)\in P}.\eeq

 \vfill \eject

\section{Lift of vector fields to $Z_{p},{\tilde Z}, W.$}

\subsection{Transformations of $X$.}

In the space-time manifolds $X=T\times \bar B$ of a concrete physical systems or in some natural bundles
over $X$ (tangent bundle, frame bundle, etc.) that are the place for the field variables there are usually
defined and undergone the study different groups of transformations reflecting the covariance and
invariance properties of the geometrical structures of this theory or even their dynamical behavior.
Examples of such groups are:
\begin{enumerate}
\item Diffeomorphism group of $X$,
\item Automorphism group of the space-time bundle $\pi_{BX}: X\rightarrow B,$
\item Group of diffeomorhpismes of the manifold with the boundary $\bar B$,
\item A Lie group $G$ of material symmetries of material manifold $B$ (in a case where $B$
 is a material manifold) acting on the frame bundle $F(B)$,
\item Group of Galilean Transformations acting in the Newtonian space-time $X=T\times E^3$,
\item Subgroup $V$ of the last group of the transition to the frame moving with constant velocity,
\item Poincare group acting in the space-time of special relativity $(R^{1+3},\eta)$,
\item A gauge group $G^X $ corresponding to a Lie subgroup $G\subset GL(n,R)$ and acting on the tangent or
frame bundle of $X$,
\item Affine group acting on Euclidian space $E^3=(R^3,h)$.
\end{enumerate}

\subsection{Transformation groups in $Y$.}

If a bundle $Y$ is a natural bundle (\cite{FF}) or if $Y=Y_{1}\oplus Y_{2}$, where bundle $Y_{1}$ is
natural, then the action of a group $G$ on $X$ is naturally lifted to the action in $Y$ (respectively in
$Y_{1}$ and then in $Y$ by trivial extension) in such a way that the projection $\pi:Y\rightarrow X$
becomes a $G$-morphism.

If $Y$ is not a natural bundle, one can use an Ehresmann connection $\nu:Y\rightarrow J^{1}(Y)$ in the
bundle $\pi:Y\rightarrow X$ to lift vector fields of infinitesimal action of $G$ (and, possibly, the
action of the group $G$ itself) to the $\nu$-horizontal vector fields in $Y$: $\xi \rightarrow {\hat
\xi}$.  Vector fields $\hat \xi $ are projectable vector fields in $Y$. Such a lift will be the morphism
of Lie algebras (i.e. $[{\hat \xi},{\hat \eta}]={\hat {[\xi ,\eta] }}$) \emph{provided the curvature of
connection $\nu$ vanishes}. Denote by $Aut(\pi)$ the group of automorphisms of the bundle $\pi$ -
diffeomorphisms $\phi \in Diff(Y)$  projecting to the diffeomorphisms $\bar \phi$ of $X$ and by
$\mathcal{X}(\pi)$ the Lie algebra of $Aut(\pi )$ formed by the projectable vector fields in $Y$.\par
  Consider the
situation where $G$ is a subgroup of the group $\mathcal{X}(\pi)$ of automorphisms of the bundle $\pi$ - a
group of diffeomorphisms of $Y$ preserving fibers of the bundle $\pi$. Transformations $g\in G$ project to
the diffeomorphisms $g_{0}$ of $X$ forming the subgroup $G_{0}\subset Diff(X)$. Epimorphism $G\rightarrow
G_{0},\ g\rightarrow g_{0}$ of groups has a normal subgroup $N$ of $G$ as its kernel:
\[
1\rightarrow N\rightarrow G\rightarrow G_{0}\rightarrow 1
\]
is the corresponding exact sequence. $N$ is the intersection of $G$ with the group $G\mathcal{X}(\pi)$ of
pure gauge automorphisms - automorphisms of bundle $\pi$ acting in fibers and, therefore, generating
identity diffeomorphism of the base (see \cite{KMS}).\par
\begin{remark}
It is possible that the transformations from $G_{0}$ can be naturally lifted to the automorphisms of
$\pi$: $g\rightarrow {\hat g}.$ This happens for instance if the fields $y^{i}$ are {\bf tensor fields or
tensor densities fields} on $X$.  Lifts of elements $h\in G_{0}$ form a subgroup ${\hat G}_{0}\subset
\mathcal{X}(\pi).$  Nothing guarantees that ${\hat G}_{0}\subset G$ but if this happens, then one get the
semidirect product decomposition $G={\hat G}_{0}\times N$ with $G$ acting by automorphisms of $N$.\par
\end{remark}
One may consider projection $g\rightarrow g_{0}$ as the action of $G$ on $X$. This action naturally lifts
to the action of $G$ by automorphisms of the bundles of exterior forms $\Lambda^{k}(X)\rightarrow X.$ Then
action of $G$ on $Y$ and $\Lambda^{n+(n+1)}X$ \emph{leaves invariant its subbundle}
$\Lambda^{n+(n+1)}_{r}Y$ over $Y$ because the action of $G$ by automorphisms of the bundle $\pi$ send
fibers of $\pi:Y\rightarrow X$ into fibers and therefore leaves the vertical subbundle $V(Y)\subset T(Y)$
invariant.  We formulate this result as the following
\begin{lemma} Let $G\subset Aut(\pi)$ be a Lie group of automorphisms of the bundle $\pi$.
\begin{enumerate}
\item
The projection $G_{0}$ of the group $G$ to $X$ lifts to the natural bundle of exterior algebras
$\Lambda^{*}(X)$ such that the pullback of the forms
\[
\pi^{*}:\Lambda^{*}(X)\rightarrow \Lambda^{*}(Y)
\]
is equivariant with respect to the projection $g\rightarrow g_{0}.$
\item Subbundles $\Lambda^{k}_{r}(Y)$ are invariant under the lifted action of $G$.
  \end{enumerate}
\end{lemma}
\par

Let now action of $G$ on $X$ by $g\rightarrow g_{0}$ \emph{preserves the subbundle} $K\subset T(X)$.  Then
one can naturally define the action of $G$ on the partial 1-jet bundle $J^{1}_{K}(\pi)$ in such a way that
the projections $J^{1}_{K}(\pi )\rightarrow Y\rightarrow X$ become $G$-morphisms (see below). \par

Similarly, if an action of the group $G$ leaves the splitting (5.1) invariant and its projection $G_{0}$
leaves invariant the space-time decomposition $T(X)=T(\mathbb{R}_t )\oplus T(B)$ of the tangent bundle,
one may lift its action to $J^{1}_{S}(\pi)$.

If an action of $G$ can be lifted to the bundle $J^{1}_{p}(\pi )$ and to $\Lambda^{n+(n+1)}_{2}Y$, by
taking the fiber product of these actions we may lift the action of $G$ to the space
$W_{0p}=J^{1}_{p}\times \Lambda^{n+(n+1)}_{2}Y $.  As a result we may pose a question to study the lifted
action of the group $\hat G$ on the constitutive relations $\mathcal{C}$, lifted CR $\mathcal{\hat C}$,
Cartan-Poincare forms $\Theta_{C}$ and the balance system generated by $\hat C$ (see below, Sec.).

Let us look in more details at these prolongations of transformations (in global as well as in
infinitesimal variants).

\subsection{Lift of vector fields and transformations to $J^{1}_{p}(\pi )$.}

\begin{definition}
Denote by $Aut(\pi^{1}_{p})$  the automorphism group of the double bundle $\pi^{1}_{p}:
J^{1}_{p}(\pi)\rightarrow Y\rightarrow X$ i.e. diffeomorphisms of $J^{1}_{p}(\pi)$ projecting to $Y$ and
$X$. Introduce the corresponding Lie algebra $\mathcal{X}(\pi^{1}_{p})$ of vector fields $\xi
$.\end{definition}

Recall that for $K=T(X)$, $J^{1}_{p}(\pi)=J^{1}(\pi)$.\par
  Vector
fields $\xi \in \mathcal{X}(\pi^1 )$ have, in adapted local coordinates $(x^\mu ,u^i, z^{i}_{\mu})$, the
form

\beq \xi =\xi^{\mu}(x)\partial_{x^\mu}+\xi^{i}(x^\nu,y^j )\partial_{y^i}+\xi^{i}_{\mu}(x^\nu ,y^j ,
z^{j}_{\nu})\partial_{z^{i}_{\mu}}. \eeq \par
\subsubsection{Case $Z_{p}=Z=J^{1}(\pi)$.}
Recall (\cite{KMS,S}) that there exists the natural lift $\xi \rightarrow \xi^1$ of an arbitrary vector
field $\xi \in \mathcal{X}(\pi )$ to the vector field $\xi^{1}$ in $Z=J^{1}(\pi)$  defined by the
conditions described in the following

\begin{proposition} (see \cite{KMS,S}).
\begin{enumerate}
\item
For any vector field $\xi \in \mathcal{X}(Y)$ there is a unique vector field $\xi^{1}\in \mathcal{X}(Z)$
(\textbf{1-jet prolongation of $\xi$}) defined by the conditions:
\begin{enumerate}
\item Vector field $\xi^{1} \in \mathcal{X}(Z)$ is projectable to $Y$ and
\[
\pi_{10*}(\xi^{1})=\xi,
\]
\item Local flow of the vector field $\xi^{1}$ preserves the Cartan distribution $Co$ (such a vector
 field is called an infinitesimal contact transformation).
\end{enumerate}
\item The lift $\xi^1$ of a vector field $\xi =\xi^{\mu}(x,y)\partial_{x^{\mu}}+\xi^{i}(x,y)\partial_{y^{i}}$
 has in local adapted coordinates the form
\beq \xi^{1}=\xi^{\mu}(x)\partial_{x^{\mu}}+\xi^{i}(x,y)\partial_{y^{i}} +\left(\frac{d \xi^{i}}{d
x^{\mu}}-z^{i}_{\nu}\frac{d \xi^{\nu}}{d x^{\mu}} \right)\partial_{z^{i}_{\mu}}, \eeq
 where
$D_{\mu}\xi^{i}=\frac{d \xi^{i}}{d x^\mu }=\frac{\partial \xi^i }{\partial x^\mu
}+z^{j}_{\mu}\frac{\partial \xi^{i}}{\partial y^j}$ is the total derivative of the function $\xi^i$ and
similarly for $\xi^\mu$.
\item The mapping $\xi \rightarrow \xi^{1}$ is the homomorphism of Lie algebras:
\[
[\xi,\eta]^{1}=[\xi^{1},\eta^{1}]
\]
for all $\xi,\eta \in \mathcal{X}(Y).$
\item For a projectable vector field
$\xi\in \mathcal{X}(\pi),\ \xi =\xi^{\mu}(x)\partial_{x^{\mu}}+\xi^{i}(x,y)\partial_{y^{i}}$ \textbf{1-jet
prolongation} $\xi^1$ coincide with the \textbf{flow prolongation} (see below).
\end{enumerate}
\end{proposition}

The \textbf{flow prolongation (lift)} is defined by the local flow $\phi_{t}(x,y)$ of the vector field
$\xi \in \mathcal{X}(\pi)$.
 Let ${\bar \phi}_{t}(x)$ be the flow induced by $\phi_{t}$ in $X$ (having the vector field
$\xi^{\mu}(x)\partial_{x^\mu}$ as the generator.  Flow $\phi_{t}$ acts on sections $y=s(x)$ of the bundle
$\pi$ by the rule: $s\rightarrow (\phi s)(x)=\phi_{t}s({\bar \phi_{t}}^{-1}(x))$.  Differentiating by $t$
at $t=0$ we get the generator of action on the 1-jet part $s^{i}_{,x^{\mu}}$ in the form (7.2) (see
(\cite{KMS,S})).\par
 The flow lift of automorphisms from $Aut (\pi)$ and of
corresponding vector fields is the homomorphism of groups (Lie algebras)

  \beq Aut(\pi)\rightarrow Aut(\pi^{2}_{0}),\ \mathcal{X}(\pi)\rightarrow
\mathcal{X}(\pi^{2}_{0}) \eeq

 that locally, with respect to the adopted chart have the (7.2)

\subsubsection{Case of $Z_{p}=J^{1}_{K}(\pi)$.}

Let now $K$ be a subbundle of the tangent bundle $T(X)$ and let $T(X)=K\oplus K'$ be an AP-structure
containing $K$. Let $\eta_{\nu } (=\partial_{x^\nu}\ \text{in\ an\ integrable\ case})$ be a local basis of
$K$ and denote by $\tilde{z}^{i}_{\nu }$ the corresponding local coordinates in the fiber of the bundle
$\pi_{10}:J^{1}_{K}(\pi )\rightarrow Y$ (see above).

\begin{definition} \begin{enumerate}
\item Denote by $\mathcal{X}_{K}(\pi )$ the Lie algebra of  $\pi$-projectable vector fields
 $\xi $ in $Y$ such that the field $\bar \xi$ generated by $\xi $ in $X$ \emph{preserves the distribution}
 $K\subset T(M)$: ${\bar \phi}_{t*}K =K$ for the local flow ${\bar \phi}_{t}$ of the vector field $\bar \xi .$
\item
 Denote by $\mathcal{X}_{K\oplus K'}(\pi )$ the Lie algebra of  $\pi$-projectable vector fields
 $\xi $ in $Y$ such that the field $\bar \xi$ generated by $\xi $ in $X$ \emph{preserves the distributions}
 $K,K'\subset T(M)$: ${\bar \phi}_{t*}K =K$ for the local flow ${\bar \phi}_{t}$ of the vector field $\bar \xi$
 (and the same for $K'$).
 \end{enumerate}
\end{definition}

\begin{lemma} Let the AP-structure $T(X)=K\oplus K'$ is integrable and let $(x^\nu ,x^\sigma )$ be  a
(local) integrating chart. Then
\begin{enumerate}
\item A $\pi$-projectable vector field $\xi=\xi^{\mu}(x)\partial_{x^\mu}+\xi^{i}(x,y)\partial_{y^i}$ belongs
to $\mathcal{X}_{K}(\pi )$ if and only if
\[
{\bar \xi}=\xi^{\mu}(x^\nu ,x^\sigma )\partial_{x^\mu}=\xi^{\nu}(x)\partial_{x^\nu}+\xi^{\sigma}(x^\sigma
)\partial_{x^\sigma},
\]
i.e. if the components $\xi^{\sigma}(x)$ do not depend on the variables $x^\nu$.
\item A $\pi$-projectable vector field $\xi=\xi^{\mu}(x)\partial_{x^\mu}+\xi^{i}(x,y)\partial_{y^i}$
belongs to $\mathcal{X}_{K\oplus K'}(\pi )$ (preserves the almost product structure $T(X)=K\oplus K'$) if
and only if
\[
{\bar \xi}=\xi^{\mu}(x )\partial_{x^\mu}=\xi^{\nu}(x^{\nu_{1}})\partial_{x^\nu}+\xi^{\sigma}(x^\sigma
)\partial_{x^\sigma},
\]
\end{enumerate}
\end{lemma}
\begin{proof}
\[
[{\bar \xi},\partial_{x^\nu}]=-(\partial_{x^\nu}\cdot
\xi^{\nu_{1}})\partial_{x^{\nu_{1}}}-(\partial_{x^\nu}\cdot \xi^{\sigma})\partial_{x^\sigma}.
\]
This vector field belongs to $K$ if and only if $\partial_{x^\nu}\cdot \xi^{\sigma}=0$ for all $\nu$ and
$\sigma$.  The second statement is proved in the same way.
\end{proof}

\begin{proposition} Let the AP-structure $T(X)=K\oplus K'$ is integrable and let $(x^\nu ,x^\sigma )$ be a
(local) integrating chart. \begin{enumerate}\item For a vector field
$\xi=\xi^{\mu}(x)\partial_{x^\mu}+\xi^{i}(x,y)\partial_{y^i} \in \mathcal{X}_{K}(\pi)$ the following
properties are equivalent
\begin{enumerate}
\item There exist a vector field $\xi^1 \in \mathcal{X}(J^{1}_{K}(\pi))$ such that
\begin{enumerate}
\item Local flow of the vector field $\xi^1$ preserves the partial Cartan distribution $Ca_{K}$.
\item $\pi_{10\ *}\xi^1 =\xi$.
\end{enumerate}
\item Vector field $\xi$ has, in a local integrating chart $(x^\nu ,x^\sigma )$ the form
\[
\xi
=\xi^{\nu}(x^{\nu_{1}})\partial_{x^{\nu}}+\xi^{\sigma}(x^{\sigma_{1}})\partial_{x^{\sigma}}+\xi^{i}(x^\nu
,y)\partial_{y^i}.
\]
In particular the projection $\bar \xi$ of the vector field $\xi $ in $X$ preserved the almost product
structure $K\oplus K'$.
\end{enumerate}
\item In the case where these conditions are fulfilled the vector field $\xi^1$ is unique and is given by the formula

\beq
\xi^1=\xi^{\nu}(x^{\nu_{1}})\partial_{x^{\nu}}+\xi^{\sigma}(x^{\sigma_{1}})\partial_{x^{\sigma}}+\xi^{i}(x^\nu
,y)\partial_{y^i}+\left(d_{\nu}\xi^i -z^{i}_{\nu_{1}}\frac{\partial \xi^{\nu_{1}}}{\partial x^\nu} \right)
\partial_{z^{i}_{\nu}}\eeq
\item Mapping $\xi \rightarrow \xi^{1}$ is the homomorphism of Lie algebras:
\[
[\xi,\eta]^{1}=[\xi^{1},\eta^{1}]
\]
for all $\xi,\eta \in \mathcal{X}_{K,K'}(\pi).$
\end{enumerate}
\end{proposition}
\begin{proof} Let
${\hat
\xi}=\xi^{\mu}(x)\partial_{x^\mu}+\xi^{i}(x,y)\partial_{y^i}+\lambda^{i}_{\nu}\partial_{z^{i}_{\nu}}$ be a
prolongation to the partial jet bundle $Z_{K}$ of the vector field $\xi$.  Then, condition of the
preservation of the partial Cartan structure is equivalent to the condition that for all the generators
$\omega^i_{K}=dy^i-\sum_{\nu}z^{i}_{\nu}dx^\nu$ of the contact ideal of exterior forms,
\[
\mathcal{L}_{\hat \xi}\omega^{i}_{K}=\sum_{j}q^{i}_{j}\omega^{j}_{K},
\]
for some functions $q^{i}_{j}\in C^{\infty}(Z_{p})$. We calculate
\begin{multline}
\mathcal{L}_{\hat \xi}\omega^{i}_{K}=(di_{\hat \xi}+i_{\hat \xi}d)(dy^i-\sum_{\nu}z^{i}_{\nu}dx^\nu)=
d[\xi^i -z^{i}_{\nu}\xi^{\nu}]+i_{\hat \xi }(-dz^{i}_{\nu}\wedge dx^\nu )=\\
=d\xi^{i}-\xi^{\nu}dz^{i}_{\nu}-z^{i}_{\nu}d\xi^{\nu}-\lambda^{i}_{\nu}dx^\nu+\xi^{\nu}dz^{i}_{\nu} =
\xi^{i}_{,x^\mu}dx^\mu+\xi^{i}_{,y^j}dy^j-z^{i}_{\nu}[\xi^{\nu}_{,x^{\nu_{1}}}dx^{\nu_{1}}
+\xi^{\nu}_{,x^\sigma}dx^\sigma]-
\lambda^{i}_{\nu}dx^\nu =\\
=\sum_{j}q^{i}_{j}(dy^j-\sum_{\nu}z^{j}_{\nu}dx^\nu),
\end{multline}
or
\[
(\xi^{i}_{,x^\sigma}-z^{i}_{\nu}\xi^{\nu}_{,x^\sigma})dx^\sigma +\xi^{i}_{,y^j}dy^j+[\xi^{i}_{,x^\nu }
-\lambda^{i}_{\nu}-z^{i}_{\nu_{1}}\xi^{\nu_{1}}_{,x^{\nu}}]dx^\nu =
\sum_{j}q^{i}_{j}(dy^j-\sum_{\nu}z^{j}_{\nu}dx^\nu).
\]
This equality is fulfilled if and only if we have
\[
\begin{cases}
\xi^{i}_{,x^\sigma}-z^{i}_{\nu}\xi^{\nu}_{,x^\sigma}=0,\\
q^{i}_{j}=\xi^{i}_{,y^j},\\
\xi^{i}_{,x^\nu } -\lambda^{i}_{\nu}-z^{i}_{\nu_{1}}\xi^{\nu_{1}}_{,x^{\nu}}=-q^{i}_{j}z^{j}_{\nu}.
\end{cases}
\]
Since neither $\xi^i$ nor $\xi^\nu$ depend on $z^{i}_{\mu}$ first system is equivalent to the requirement
that both $\xi^i$ and $\xi^\nu$ are independent on $x^\sigma$. Then the second condition determines
$q^{i}_{j}$ and third - $\lambda^{i}_{\nu}=\xi^{i}_{,x^\nu
}+\xi^{i}_{,y^j}z^{j}_{\nu}-z^{i}_{\nu_{1}}\xi^{\nu_{1}}_{,x^{\nu}}$ and the prolongation $\hat \xi$ takes
the form described in the Proposition.
\end{proof}

\subsubsection{Case of $Z_{p}=Z_{S}=J^{1}_{S}(\pi)$.}
Consider now the case of partial 1-jet bundle $J^{1}_{S}(\pi)$ generated by the (x,t)-decomposition $S$
(see section ??).  By Proposition 3 the bundle $Y$ is the fiber product of the bundles
$Y=Y_{0}\underset{X}{\times}  Y_{t}\underset{X}{\times} Y_{x}\underset{X}{\times}  Y_{xt}$ and the partial
1-jet has the form of the the fiber product $J^{1}_{S}(\pi)= 0(Y_{0})\underset{Y}{\times}
J^{1}_{t}(Y_{t})\underset{Y}{\times} J^{1}_{x}(Y_{x})\underset{Y}{\times} J^{1}(Y_{tx}).$ \par A natural
class of automorphisms of the bundle $\pi$ is the class of $S$-automorphisms of $\pi$ (see Definition 4)
and corresponding class of $S$-admissible vector fields in $Y$ $\xi\in \mathcal{X}_{S}(\pi)$. In simple
words these are geometrical or infinitesimal automorphisms of the bundle $\pi$ that preserve the $S$-type
of fields under transformation.\par
    In this case we have the canonical integrable AP-structure $T(X)=T(B)\oplus <\partial_{t}>$ with a local
 chart $(x,t)$. Applying
the arguments used for the study of prolongation $\xi \rightarrow \xi^1$ to the partial 1-jet bundles
$J_{K}(\pi)$ we get the following analog of previous Proposition:
\begin{proposition}
\begin{enumerate}
\item A vector field $\xi\in \mathcal{X}_{S}(\pi)$ preserves the AP-structure $T(B)\oplus <\partial_{t}>$
if and only if
\[
{\bar \xi}=\xi^{\mu}(x,t)\partial_{x^\mu}=\xi^A (x)\partial_{x^A}+\xi^{t}(t )\partial_{t},
\]
\item
For any $S$-admissible $\pi$-projectable vector field $\xi \in \mathcal{X}_{S}(\pi)$ following statements
are equivalent
\begin{enumerate}
 \item There is a vector field $\xi^{1}\in \mathcal{X}(Z_{S})$  such that
\begin{enumerate}
\item Vector field $\xi^{1} \in \mathcal{X}(Z_{S})$ is $\pi_{10}$-projectable and
\[
\pi_{10*}(\xi^1 )=\xi,
\]
\item Local flow of the vector field $\xi^{1}$ preserves the partial
Cartan distribution $Co_{S}$ at $Z_{S}(\pi)$.
\end{enumerate}
\item Vector field $\xi$ has the following form
\begin{multline}
\xi =\xi^A (x)\partial_{x^A}+\xi^{t}(t )\partial_{t}+     \sum_{i_{0}\in S_{0}}  \xi^{i_{0}}
(x,t;y^{j_{0}},\ j^{0}\in S_{0})\partial_{y^{j_{0}}}+      \sum_{i_{1}\in S_{t}}  \xi^{i_{1}}
(t;y^{j_{1}},\ j^{1}\in S_{t})\partial_{y^{j_{1}}}+ \\     + \sum_{i_{2}\in S_{x}}  \xi^{i_{2}}
(x;y^{j_{2}},\ j^{2}\in S_{x})\partial_{y^{j_{2}}}+       \sum_{i_{3}\in S_{tx}} \xi^{i_{3}}
(x,t;y^{j_{3}},\ j^{3}\in S_{tx})\partial_{y^{j_{3}}}.
\end{multline}
\end{enumerate}
where dependence of vertical components $\xi^i$ of the vector field $\xi$ on the variables $x^A ,t$ is
specified by the subset $S_{*}$ containing index $i$.

\item In the case where these conditions are fulfilled the vector field $\xi^1$ is unique and is given by
the formula (recall that $x^0 =t$)

\begin{multline} \xi^1=\xi+\sum_{i_{1}\in S_{t}} \left( \frac{d\xi^{i_{1}}}{dx^0 }-z^{i_{1}}_{x^0}\frac{\partial
{\bar \xi}^{0}}{\partial x^0} \right) \partial_{z^{i_{1}}_{x^0}}+ \sum_{i_{2}\in S_{x}} \left(
\frac{d\xi^{i_{2}}}{dx^A }-z^{i_{2}}_{x^B}\frac{\partial {\bar \xi}^{B}}{\partial x^A} \right)
\partial_{z^{i_{2}}_{x^A}}+\\ + \sum_{i_{3}\in S_{tx}} \left(
\frac{d\xi^{i_{3}}}{dx^\mu }-z^{i_{3}}_{x^\nu}\frac{\partial {\bar \xi}^{\nu}}{\partial x^\mu} \right)
\partial_{z^{i_{3}}_{x^\mu}}.
\end{multline}
\item Mapping $\xi \rightarrow \xi^{1}$ is the homomorphism of Lie algebras:
\[
[\xi,\eta]^{1}=[\xi^{1},\eta^{1}]
\]
for all $\xi,\eta \in \mathcal{X}_{K}(Y).$
\end{enumerate}
\end{proposition}
\vskip0.5cm
\subsection{Case of a general AP-structure $T(X)=K\oplus K'$}
Consider now a case where AP-structure $T(X)=K\oplus K'$ is not integrable.  Denote by $\xi_{\nu}$
(respectively by $\xi_{\sigma}$ local frames of distributions $K, K'$ respectively, by
$\psi^{\nu},\psi^{\sigma}$ - dual coframe. Introduce the structural equation

\beq d\psi^{\mu}=T^{\mu}_{\beta \gamma}(x)\psi^{\beta} \wedge \psi^{\gamma} \eeq
 of the coframe $\psi^\mu$, where $T$ is the tensor defined above.\par

 Denote by $\tilde{z}^{i}_{\nu}$ the vertical coordinates in the partial
frame bundle $Z_{K}=J^{1}_{K}(\pi)$ defined by the condition
$\tilde{z}^{i}_{\nu}(j^{1}_{p}(s))=\xi_{\nu}s^i$ for all sections $s$ of the bundle $\pi$.\par

Recall that the partial Cartan distribution in $J^{1}_{K}(\pi)$ is generated by the 1-forms
$\widetilde{\omega}^{i}=dy^i-\sum_{\nu \in K}\tilde{z}^{i}_{\nu}{\hat \psi}^{\nu}$, where ${\hat
\psi}^{\nu}=\pi^{1\
*}\psi^{\nu}$ is the pullback of a coframe 1-form to the partial jet bundle $Z_{K}$.

Let now $\xi ={\bar \xi}^{\mu}\xi_{\mu} +\xi^{i}\partial_{y^i}$ be a projectable vector field in $Y$ with
the projection $\bar \xi$ in $X$ and let \[\xi^1 ={\bar \xi}^{\mu}\xi_{\mu} +\xi^{i}\partial_{y^i}
+\lambda^{i}_{\nu}\partial_{\tilde{z}^{i}_{\nu}}\]
 be some prolongation of vector field $\xi $ to the bundle $Z_{K}$.\par

We would like to find conditions on the field $\xi$ under which there exists its prolongation to
$J^{1}_{K}(\pi)$ preserving the partial Cartan structure $Ca_{K}$.

We calculate:
\begin{multline}
\mathcal{L}_{\xi^1 }\widetilde{\omega}^{i}=\mathcal{L}_{\xi^1}(dy^i-\sum_{\nu \in
K}\tilde{z}^{i}_{\nu}{\hat \psi}^{\nu})=
(di_{\xi^1}+i_{\xi^{1}}d)(dy^i-\sum_{\nu \in K}\tilde{z}^{i}_{\nu}{\hat \psi}^{\nu})=\\
-i_{\xi^1}(d\tilde{z}^{i}_{\nu}\wedge {\hat \psi}^{\nu}+\tilde{z}^{i}_{\nu}d{\hat \psi}^{\nu})+d(\xi^i
-\tilde{z}^{i}_{\nu}<{\hat \psi}^{\nu},{\bar \xi})= -(\xi^{1}\cdot \tilde{z}^{i}_{\nu}){\hat
\psi}^{\nu}+<{\hat \xi},\psi^{\nu}>d\tilde{z}^{i}_{\nu}-\tilde{z}^{i}_{\nu}i_{\xi^1}(d{\hat \psi}^{\nu})+\\
+d\xi^i -\tilde{z}^{i}_{\nu}d<\psi^\nu ,{\bar \xi}>-<\psi^\nu ,{\bar
\xi}>d\tilde{z}^{i}_{\nu}=d\xi^i-\tilde{z}^{i}_{\nu}d<\psi^\nu ,{\bar \xi}> -(\xi^{1}\cdot
\tilde{z}^{i}_{\nu}){\hat \psi}^{\nu}-\tilde{z}^{i}_{\nu}i_{\xi^1}(d{\hat \psi}^{\nu}).
\end{multline}
In the last term $i_{\xi^1}(d{\hat \psi}^{\nu})=i_{\bar \xi}(d{\psi}^{\nu})$ since $d{\hat
\psi}^{\nu}={\hat {d\psi^{\nu}}}$.\par

Flow of the vector field $\xi^1$ preserves the partial Cartan distribution if and only if
\[
\mathcal{L}_{\xi^1 }\widetilde{\omega}^{i}=q^{i}_{j}\widetilde{\omega}^{j}=q^{i}_{j}(dy^i-\sum_{\nu \in
K}\tilde{z}^{i}_{\nu}{\hat \psi}^{\nu})
\]
for some functions $q^{i}_{j}$ on $J^{1}_{K}(\pi) $.  Using the calculation above we get to the condition

\beq
 {\widehat {d\xi^i}}-\tilde{z}^{i}_{\nu}{\widehat {d<\psi^\nu ,{\bar \xi}>}} -\lambda^{i}_{\nu}{\hat \psi}^{\nu}-
 \tilde{z}^{i}_{\nu}{\widehat {i_{\bar \xi}(d\psi^{\nu})}}=q^{i}_{j}(dy^i-
 \sum_{\nu \in K}\widetilde{z}^{i}_{\nu}{\hat \psi}^{\nu})
\eeq

 for all $i$ (we have used $\xi^{1}\cdot \tilde{z}^{i}_{\nu}=\lambda^{i}_{\nu}$).\par We have in the last
formula $<\psi^\nu ,{\bar \xi}>=\xi^{\nu}$ and we will use the relations
\[
d\xi^i =(\xi_{\mu}\cdot \xi^{i})\psi^\mu +(\partial_{y^j}\xi^i )dy^j,\ d\xi^{\nu}=(\xi_{\mu}\cdot
\xi^{\nu})\psi^\mu .
\]
Using these two formulas together with (7.8) in (7.10) we write it in the form

\beq (\xi_{\mu}\cdot \xi^{i}){\widehat {\psi^\mu}} +(\partial_{y^j}\xi^i ){\widehat
{dy^j}}-\tilde{z}^{i}_{\nu}(\xi_{\mu}\cdot \xi^{\nu}){\widehat {\psi^\mu}} -\lambda^{i}_{\nu}{\widehat
{\psi^\nu}}-\tilde{z}^{i}_{\nu} i_{\xi^1}(T^{\nu}_{\beta \gamma}(x){\widehat {\psi^{\beta}\wedge
\psi^{\gamma})}}=q^{i}_{j}{\widehat {dy^i}}-q^{i}_{j}\tilde{z}^{i}_{\nu}{\hat \psi}^{\nu}. \eeq

Comparing coefficients of $dy^j$ we get

\beq q^{i}_{j}=\partial_{y^j}\xi^i  \eeq

and rewrite the rest of (6.12) as follows

\beq (\xi_{\mu}\cdot \xi^{i}){\widehat {\psi^\mu}} -\tilde{z}^{i}_{\nu}(\xi_{\mu}\cdot \xi^{\nu}){\widehat
{\psi^\mu}} -\lambda^{i}_{\nu}{\widehat {\psi^\nu}}-\tilde{z}^{i}_{\nu} T^{\nu}_{\beta \gamma}(x){\widehat
{(\xi^{\beta} \psi^{\gamma} - \xi^{\gamma} \psi^{\beta })}}=-(\partial_{y^j}\xi^i )
\tilde{z}^{i}_{\nu_1}{\hat \psi}^{\nu_1}. \eeq

We remind that in this formula $\nu$ runs through indices in $K$ while $\mu,\alpha ,\gamma ,\beta $
through all indices from $0$ to $n$.\par

Present $\lambda^{i}_{\nu}$ in the form
\[
\lambda^{i}_{\nu}={\bar \lambda}^{i}_{\nu}(x,y)+\tilde{z}^{k}_{\nu_{1}}\lambda^{i \nu_{1}}_{\nu
k}(x,y,\tilde{z}),
\]
where first term does not depend on the jet coordinates.  Such a representation can always done locally.
Substitutive this decomposition into (7.13) and extract the terms that does not contain variables
$\tilde{z}^{i}_{\nu}$ as a factor

\[
(\xi_{\mu}\cdot \xi^i )\psi^{\mu} -{\bar \lambda}^{i}_{\nu}\psi^{\nu}=0.
\]
This equality is equivalent to two statements

\beq
\begin{cases}
\xi_{\sigma}\cdot \xi^{i}=0 \Leftrightarrow \xi^{i}=\xi^{i}(x^\nu ,y),\\
{\bar \lambda}^{i}_{\nu}=\xi_{\nu}\cdot \xi^{i}.
\end{cases}
\eeq After excluding terms without $z$-variables and using the equality
\[T^{\nu}_{\beta
\gamma}{\widehat {(\xi^{\beta} \psi^{\gamma} - \xi^{\gamma} \psi^{\beta })}}=(T^{\nu}_{\beta
\mu}(x(\xi^{\beta}-T^{\nu}_{\mu \beta}(x)\xi^{\beta}){\widehat {\psi^\mu}}=2T^{\nu}_{\beta
\mu}(x)\xi^{\beta}{\widehat {\psi^\mu}}\]

valid due to the antisymmetry of $T^{\nu}_{\alpha \beta}(x) $ by lower indices, the equality (6.14) will
take the form

\beq  -\tilde{z}^{i}_{\nu}(\xi_{\mu}\cdot \xi^{\nu}){\widehat {\psi^\mu}} -\tilde{z}^{k}_{\nu}\lambda^{i
\nu}_{\nu_{1} k}{\widehat {\psi^{\nu_{1}}}}-2\tilde{z}^{i}_{\nu}T^{\nu}_{\beta \mu}(x)\xi^{\beta}{\widehat
{\psi^\mu}}=-(\partial_{y^j}\xi^i ) \tilde{z}^{j}_{\nu_1}{\hat \psi}^{\nu_1}. \eeq

Equating here coefficients of the 1-forms $\psi^{\mu}$ with $\mu=\sigma $ we get
\[
-\tilde{z}^{i}_{\nu}(\xi_{\sigma}\cdot \xi^{\nu})-2\tilde{z}^{i}_{\nu}T^{\nu}_{\beta
\sigma}(x)\xi^{\beta}=0,
\]
or \beq \xi_{\sigma}\cdot \xi^{\nu}+2T^{\nu}_{\beta \sigma}(x)\xi^{\beta}=0. \eeq These are structural
equations for the $X$-components of the vector field $\xi$.\par

Equating coefficients of the form $\psi^{\nu_{1}}$ in (7.15) we finally get

\beq
 -\tilde{z}^{i}_{\nu}(\xi_{\nu_{1}}\cdot \xi^{\nu}) -\tilde{z}^{k}_{\nu}\lambda^{i \nu}_{\nu_{1}
k}(x,y,\tilde{z})-2\tilde{z}^{i}_{\nu}T^{\nu}_{\beta \nu_{1}}(x)\xi^{\beta}=-(\partial_{y^k}\xi^i )
\tilde{z}^{k}_{\nu_1}. \eeq

All terms in this formula except the second one on the left side are linear by $\tilde{z}$-variables.
Therefore this second term on the left side is also linear by $\tilde{z}$ and it follows from this that
the functions $\lambda^{i \nu}_{\nu_{1} k}$ depend on $x^\mu, y^i$ but not on $\tilde{z}$. \par

Equating coefficients at $\tilde{z}^{k}_{\nu}$ we get
\[
\begin{cases}
\lambda^{i \nu}_{\nu_{1} k}(x,y)=(\partial_{y^k}\xi^i )\delta^{\nu}_{ \nu_{1}},\ \text{for}\ k\ne i,\\
(\xi_{\nu_{1}}\cdot \xi^{\nu})+\lambda^{i \nu}_{\nu_{1} i}(x,y)+2T^{\nu}_{\beta
\nu_{1}}(x)\xi^{\beta}=(\partial_{y^k}\xi^i )\delta^{\nu}_{ \nu_{1}},\ \text{for}\ k = i.
\end{cases}
\]

From this we get

\beq
\begin{cases}
\lambda^{i \nu}_{\nu_{1} k}(x,y)=(\partial_{y^k}\xi^i )\delta^{\nu}_{ \nu_{1}},\ \text{for}\ k\ne i,\\
\lambda^{i \nu}_{\nu_{1} i}(x,y)=(\partial_{y^k}\xi^i )\delta^{\nu}_{ \nu_{1}}-(\xi_{\nu_{1}}\cdot
\xi^{\nu})-2T^{\nu}_{\beta \nu_{1}}(x)\xi^{\beta},\ \text{for}\ k = i.
\end{cases}
\eeq

Substituting these expressions and (7.14) into the formulas for $\lambda^{i}_{\nu}$ (and reversing places
of $\nu$ and $\nu_{1}$) we find $\lambda^{i}_{\nu}$ in the form

\begin{multline}
\lambda^{i}_{\nu}= \xi_{\nu}\cdot \xi^{i}+\tilde{z}^{i}_{\nu_{1}}\lambda^{i\nu_{1}}_{\nu i}=
\xi_{\nu}\cdot \xi^{i}+\sum_{k}\tilde{z}^{i}_{\nu_{1}}(\partial_{y^k}\xi^i )
\delta^{\nu_{1}}_{\nu}-\tilde{z}^{i}_{\nu_{1}}[(\xi_{\nu}\cdot \xi^{\nu_{1}})-
2T^{\nu_{1}}_{\beta \nu}(x)\xi^{\beta}]=\\
=d_{\nu}\xi^{i}-\tilde{z}^{i}_{\nu_{1}}[(\xi_{\nu}\cdot \xi^{\nu_{1}})-2T^{\nu_{1}}_{\beta
\nu}(x)\xi^{\beta}].
 \end{multline}

Thus, we have proved the following
\begin{theorem} Let $T(X)=K\oplus K'$ be an almost product structure on $X$.  Let
$\{\xi_{\mu} \} =(\xi_{\nu},\xi_{\sigma})$ be a (local) frame adopted to the AP-structure and let
$\\psi^{\nu},\psi^{\sigma}$ be the dual coframe. Let the structural equations of this coframe be
\[
d\psi^{\mu}=T^{\mu}_{\beta \gamma}(x)\psi^{\beta} \wedge \psi^{\gamma}.
\]
A vector field $\xi =\xi^{\mu}\xi_{\mu}+\xi^{i}\partial_{y^i}$ in $Y$ have a prolongation to a vector
field ${\hat \xi}=\xi +\sum_{\nu \in K}\lambda^{i}_{\nu}\partial_{\tilde{z}^{i}_{\nu}}$ in the partial
1-jet bundle $Z_{p}=J^{1}_{K}(\pi)$ if and only if the condition
\[
\xi_{\sigma}\cdot \xi^{\nu}+2T^{\nu}_{\beta \sigma}(x)\xi^{\beta}=0,\ \text{for\ all}\ \sigma\in K',\nu
\in K
\]
is fulfilled.  In such a case, this prolongation is unique and is given by

\beq \xi^1 = \xi+ \left( d_{\nu}\xi^{i}-\tilde{z}^{i}_{\nu_{1}}[(\xi_{\nu}\cdot
\xi^{\nu_{1}})-2T^{\nu_{1}}_{\beta \nu}(x)\xi^{\beta}\right)\partial_{\tilde{z}^{i}_{\nu}}. \eeq
\end{theorem}
\begin{remark} Theorems on the prolongation proved before for $J^{1}_{K}(\pi)$ (Proposition 13) for an integrable
AP-structure $T(X)=K\oplus K'$ and that for the full 1-jet bundle (Proposition 12) are special cases of
the last result.
\end{remark}
\vskip1cm

 Under the action of an automorphism $\phi \in Aut_{p}(\pi)$ a vector field $\xi
 =\xi^{\mu}\partial_{\mu}+\ xi^{i}\partial_{i}$ is transformed as follows
 \[
\phi_{*}(\xi )=({\bar
\phi}^{\mu}_{,nu}\xi^{\nu})\partial_{\mu}+(\phi^{i}_{,\nu}\xi^{\nu}+\phi^{j}_{,i}\xi^{i})\partial_{j}.
 \]
Automorphism $\phi$ can be flow lifted to the bundle space $J^{1}_{p}(\pi)$ as follows: Let $s(x)$ be a
local section of the bundle $\pi$, then for the action of $\phi$ to the section $s$: $\phi^{*}s(x)=\phi
(s({\bar \phi}^{-1}(x)))$ we find
\begin{multline}
z^{i}_{\mu}(j^{1}_{p}(\phi^{*}s)(x))=\partial_{\mu}\phi^i (s({\bar \phi}^{-1}(x)))=(\phi^{i}_{,y^j}\circ
s)(x) \partial_{\lambda}s^{j} (x)\partial_{\mu}({\bar \phi}^{-1\ \lambda})(x)+
((\partial_{\lambda}\phi^{i})\circ s)(x)\partial_{\mu}({\bar \phi}^{-1\ \lambda})(x)=\\=
J(\phi)^{i}_{j}\frac{\partial s^j}{\partial x^\lambda}J({\bar
\phi}^{-1})^{\lambda}_{\mu}+J(\phi)^{i}_{\lambda}J({\bar \phi}^{-1})^{\lambda}_{\mu} ,
\end{multline}
in other words $z^{i}_{\mu}$ transforms by affine transformation

 \beq z^{i}(\phi^1
(x,y,z))=J(\phi)^{i}_{j}J({\bar \phi}^{-1})^{\lambda}_{\mu}z^{j}_{\lambda}+J(\phi)^{i}_{\lambda}J({\bar
\phi}^{-1})^{\lambda}_{\mu}. \eeq

For the vertical vector fields we have

\[
\phi^{i}_{*}(\partial_{z^{i}_{\mu}})f=\partial_{z^{i}_{\mu}}(f\circ \phi^{i}(x,y,z))=\frac{\partial
f}{\partial z^{j}_{\nu}}J(\phi)^{i}_{j}J({\bar \phi}^{-1})^{\lambda}_{\mu},
\]
therefore, for a projectable vector field
$\xi=\xi^{\mu}(x)\partial_{\mu}+\xi^{i}(x,y)\partial_{i}+\xi^{i}_{\mu}\partial_{z^{i}_{\mu}}$ we have

\beq (\phi^{1}_{*}\xi )=\left(\xi^{i}_{\mu}J(\phi)^{i}_{j}J({\bar \phi}^{-1})^{\lambda}_{\mu}
\right)\partial_{z^{j}_{\lambda}}+\ldots   \eeq

\subsection{Prolongation of $\pi$-automorphisms to the dual bundles $\Lambda^{k}_{r}Y$ and $\tilde Z$.}

    Automorphisms of the bundle $\pi$ (and, correspondingly, projectable vector fields $\xi \in \
\mathcal{X}(\pi)$) have a  natural (flow) prolongation to the projectable diffeomorphisms (and projectable
vector fields) of the double bundle $ \Lambda^{k}_{r}Y\rightarrow Y\rightarrow X$ of exterior forms on $Y$
(see \cite{KMS} or \cite{LMD}). This lift $\xi \rightarrow \xi^{1*}$ is defined by the pullback
$\phi_{t}^{*}$ of the exterior forms on $Y$ by the local flow $\phi_{t}$ of a vector field $\xi \in
\mathcal{X}(\pi )$. Since the commutator of a projectable vector field
$\xi^{\mu}(x)\partial_{x^\mu}+\xi^{i}(x^\nu,y^j )\partial_{y^i}$ and an arbitrary vertical vector field
$\eta =\eta^{i}(x^\nu,y^j )\partial_{y^i}$ \emph{is vertical}, lifted local automorphisms of $\pi$ (and
the corresponding infinitesimal transformations - Lie derivatives with respect to the lifted vector
fields) preserve the subbundles $\Lambda^{k}_{r}Y\subset \Lambda^{k}Y$ and, therefore, define the lifts of
automorphism transformations (global, local or infinitesimal) to the corresponding automorphisms of the
(double) bundles $\Lambda^{k}_{r}Y\rightarrow Y\rightarrow X.$\par
\par

Another way to lift a general vector field $\xi \in \mathcal{X}(Y)$ is defined by the following
construction that was studied in \cite{LMD} for the case where the metric $G$ is Euclidian (put
$\lambda=0$ in the formulas of following definition).

\begin{definition}(Definition-Proposition, \cite{LMD}.) Let $\alpha$ be a pullback to
$\Lambda^{n+1}_{2}Y$ of a $\pi$-semibasic form $\alpha =\alpha^{\nu}(x,y)\eta_\nu$ on $Y$. Let $\xi \in
\mathcal{X}(Y)$.
\begin{enumerate}
\item
 Then there exist and is unique a vector field
$\xi^{*\alpha}$ on $\Lambda^{n+1}_{2}$ satisfying to the following conditions
\begin{enumerate}
\item Vector field $\xi^{*\alpha}$ is $\pi_{\Lambda^{n+1}_{2}Y  \ Y}$-projectable and
\[
\pi_{\Lambda^{n+1}_{2}Y  \ Y\ *} \xi^{*\alpha}=\xi,
\]
\item
\[
\mathcal{L}_{\xi^{*\alpha}}\Theta^{n+1}_{2}=-d\alpha.
\]
\end{enumerate}
\item Vector field $\xi^{*\alpha}$ has the local form
\begin{multline}
\xi^{*\alpha}=\xi +\xi^{*\alpha \ p}\partial_{p}+\xi^{*\alpha \ p^{\mu}_{i}}\partial_{p^{\mu}_{i}},\ \text{where} \\
\xi^{*\alpha \ p}= -p\left( \frac{\partial \xi^\mu }{\partial x^\mu }+\xi^{\mu}\frac{\partial \lambda
}{\partial x^\mu}\right) -p^{\mu}_{i}\left(\frac{\partial \xi^{i}}{\partial x^\mu }-\xi^{i}\frac{\partial
\lambda }{\partial x^\mu}\right)-\frac{\partial \alpha^\mu }{\partial x^\mu }-\alpha^{\nu}\lambda_{,x^\nu};\\
\xi^{*\alpha \ p^{\mu}_{i}}= -p^{\nu}_{i}\left(\frac{\partial \xi^\mu }{\partial x^\nu
}-\xi^{\mu}\lambda_{,x^\nu }\right)-p^{\mu}_{j}\frac{\partial \xi^j }{\partial y^i}-p^{\mu}_{i}\left(
\frac{\partial \xi^\nu }{\partial x^\nu
}-\xi^{\nu}\lambda_{,x^\nu}\right)-p_{i}^{\nu}\xi^{\mu}\lambda_{,x^\nu}-\frac{\partial \alpha^\mu
}{\partial y^i}.
\end{multline}
\item Let a vector field $\xi \in \mathcal{X}(Y)$ be $\pi$-projectable.  Then the 0-lift $\xi^{*0}$  of $\xi $
coincide with the flow prolongation $\xi^{ 1*}$ defined above.
\end{enumerate}
\end{definition}
\begin{remark} In the work \cite{MS} there was defined the class of "covariant canonical transformations of
$\Lambda^{n+1}_{2}(Y)$" as $\pi_{}$-projectable transformations of $\Lambda^{n+1}_{2}(Y)$ preserving the
multisymplectic form $\Theta^{n+1}_{2}.$  Later on we will use transformations from this class to discuss
the transformations of constitutive relations.
\end{remark}
Let now $\phi\in Aut(\pi)$ be an automorphism of the bundle $\pi$.  Arguments in the beginning of this
subsection shows that the flow lift $\phi^{1*}$ of $\phi$ to the bundle $\Lambda^{k}Y$  leaves its
subbundles $\Lambda^{k}_{r}Y$ invariant.  In particular, $\phi^{1*}$ acts on the subbundles
$\Lambda^{n+1}_{2}Y,\Lambda^{n+2}_{2}Y$ leaving their subbundles $\Lambda^{n+1}_{1}Y,\Lambda^{n+2}_{1}Y$
invariant and leaving canonical forms $\Theta^{n+1}_{2}Y$ and $\Theta^{n+2}_{2}Y$ invariant.  Therefore,
$\phi^{1*}$ generates the automorphism $\widetilde{\phi}^{*}$ of the bundle ${\tilde
Z}=\tilde{Z}^{n+1}\oplus \tilde{Z}^{n+2}$ leaving both terms invariant.\par

 Let  $H:
\tilde{Z}^{n+1}\rightarrow \Lambda^{n+1}_{2}Y$ to be a section (see Sec.) of the bundle
$\Lambda^{n+1}_{2}Y\rightarrow \tilde{Z}^{n+1}$.  Then for the induced form
$\Theta_{H}=H^{*}\Theta^{n+2}_{2}=H(x,y,p^{\mu}_{i})\eta+p^{\mu}_{i}dy^i \wedge \eta_{\mu}$ we have

\beq {\tilde \phi}^{*}\Theta_{H}(x,y,p^{\mu}_{i})= {\tilde \phi}^{*}H^{*}\Theta^{n+2}_{2}= (H\circ {\tilde
\phi})^{*}\Theta^{n+2}_{2}= H\circ {\tilde \phi}(x,y,p^{\mu}_{i})\eta+p^{\mu}_{i}dy^i \wedge \eta_{\mu}.
\eeq

Thus, though the $(0,n+1)$-term of the form $\Theta_{H}$ is changed, its $(1,n)$-term is invariant. For
the infinitesimal action of vector field ${\tilde \xi}^{*}$ - generator of the 1-parametrical group of
difeomorphisms $\phi^{1*}_{t}$ we get from the previous formula
\[
\mathcal{L}_{{\tilde \xi}^{*}}\Theta_{H}=({\tilde \xi}^{*}\cdot H)\eta.
\]
In particular, we will be using this formula for the sections defined by a connection $\nu$ in the bundle
$\pi$ with $H=p^{\mu}_{i}\Gamma^{i}_{\mu}(x)$.

\par

For our study we need to lift a projectable vector field $\xi$ to the bundle ${\tilde Z}={\tilde
Z}^{n+1}\oplus {\tilde
Z}^{n+2}=\Lambda^{(n+1)+(n+2)}_{2}Y/\Lambda^{(n+1)+(n+2)}_{1}Y=\Lambda^{n+1}_{2}Y/\Lambda^{n+1}_{1}Y\oplus
\Lambda^{(n+2)}_{2}Y/\Lambda^{(n+2)}_{1}Y$.  Next result allows to lift $\xi$ to the bundle ${\tilde
Z}^{n+2}=\Lambda^{(n+2)}_{2}Y.$

\begin{proposition} For any projectable vector field  $\xi \in X(\pi)$ there exists unique projectable
vector field $\xi^{*(n+2)}$ on the bundle $\Lambda^{(n+2)}_{2}Y$ that leaves the canonical form
$p_{i}dy^{i}\wedge \eta$ invariant.  That vector is given by the relation

\beq \xi^{*(n+2)}=\xi + \xi_{k}\partial_{p_{k}},\ \xi_{k}= p_{k}(\xi^{\mu}\frac{\partial \lambda
}{\partial x^\mu }-\frac{\partial \xi^\mu}{\partial x^\mu})-p_{j}\frac{\partial \xi^j}{\partial y^k}, \eeq

where $\lambda =ln(\vert G\vert)$.
\end{proposition}
\begin{proof} We have, for a vector field of the form  $\xi^{*(n+2)}=\xi + \xi_{k}\partial_{p_{k}}$
\begin{multline}
\mathcal{L}_{\xi^{*(n+2)}}(p_{i}dy^{i}\wedge \eta)=
(di_{\xi^{*(n+2)}}+i_{\xi^{*(n+2)}}d)(p_{i}dy^{i}\wedge \eta)=d[p_{i}\xi^{i}\eta-p_{i}\xi^{\mu}dy^i \wedge
\eta_{\mu}] +i_{\xi^{*(n+1)}}(dp_{i}\wedge dy^{i}\wedge
\eta_{\mu})=\\
=[ d(p_{i}\xi^{i})\eta-d(p_{i}\xi^{\mu})\wedge \eta_{\mu})-p_{i}\xi^{\mu}dy^{i}\wedge d\eta_{\mu}]
+[\xi^{i}dy^{i}\wedge \eta -\xi^{i}dp_{i}\wedge \eta +\xi^{\mu}dp_{i}\wedge dy^{i}\wedge \eta_{\mu}]=\\
=[\xi^{i}dy^{i}\wedge \eta-\xi^{i}dp_{i}\wedge \eta +\xi^{\mu}dp_{i}\wedge dy^{i}\wedge \eta_{\mu}]+
[\xi^{i}dp_{i}\wedge \eta +p_{i}\frac{\partial \xi^{i}}{\partial y^{j}}dy^{j}\wedge \eta
-\xi^{\mu}dp_{i}\wedge dy^{i}\wedge \eta_{\mu}+\\ +\frac{\partial \xi^\mu}{\partial
x^\mu}p_{i}dy^{i}\wedge \eta -(p_{i}\xi^{\mu}\frac{\partial \lambda}{\partial x^\mu})dy^{i}\wedge \eta ]=
(\frac{\partial \xi^\mu}{\partial x^\mu}p_{i}+p_{i}\frac{\partial \xi^{i}}{\partial
y^{j}}-(p_{i}\xi^{\mu}\frac{\partial \lambda}{\partial x^\mu})+\xi_{i})dy^{i}\wedge \eta .
\end{multline}
Here we have used the relation $d\eta_{\mu}=\lambda_{,x^\mu} \eta$.\par Equating the obtained expression
 to zero we get the expression for $\xi_{i}$ as in the Proposition.
\end{proof}

 Combining the last result with the prolongation $\xi^{*0}$
from the Definition-Proposition 10 and with the prolongation $\xi^{*(n+2)}$ from the previous Proposition
and using factorization by $\Lambda^{(n+1)+(n+2)}_{1}Y$ we get the following
\begin{corollary} For any projectable vector field  $\xi \in X(\pi)$ there exists unique projectable
vector field $\xi^{1*}$ in the space ${\tilde Z}=\Lambda^{(n+1)+(n+2)}_{2}Y/\Lambda^{(n+1)+(n+2)}_{1}Y$ -
prolongation of $\xi$, \emph{preserving the $(n+1)+(n+2)$ form} $p^{\mu}_{i}dy^{i}\wedge
\eta_{\mu}+p_{k}dy^{k}\wedge \eta$.
\end{corollary}

\subsection{Transformations of $ W_{0}$ and $W_{1}$.}

Taking the fiber product of the action of $A=\mathcal{X}(\pi^1)$ (global, local or infinitesimal) in
$J^{1}_{p}(\pi)$ and its action by the (global, local or infinitesimal)  automorphisms of the the bundle
$\Lambda^{k}_{r}Y\rightarrow Y\rightarrow X$ induced first by the projection to $Y$ and then by the lift
to $\Lambda^{k}_{r}Y$, described in the last subsection, we define the (global, local or infinitesimal)
action of the group $Aut(\pi^1)$ on the bundles $W^{co}= J^{1}_{p}(\pi)\underset{Y}{\times}
\Lambda^{(n+1)+(n+2)}_{2}Y$ and
$W_{p}=J^{1}_{p}(\pi)\underset{Y}{\times}(\Lambda^{(n+1)+(n+2)}_{2}Y/\Lambda^{(n+1)+(n+2)}_{1}Y)=Z_{p}\times
{\tilde Z}.$\par

Combining this action with the homomorphism $\mathcal{X}(\pi)\rightarrow \mathcal{X}(\pi^1 )$ induced by
the lift prolongation we get the action of $\mathcal{X}(\pi)$ by the projected diffeomorphisms of $W^{co}$
and $W_{p}$.

The action of the group $A$ in $W^{co}$ and $W_{p}$ will allow to define its action on the vector spaces
$\tilde{\mathcal{CR}}$ and $\mathcal{CR}$ of covering and usual constitutive relations  respectively (see
Section 9 below).\par

\section{Prolongation of the connections to $Z_{p}$ and $\tilde Z$.}
A (pseudo-Riemannian) metric $G$ in $X$ determines the linear Levi-Civita connection $\Gamma^{G}$ in the
tangent bundle $T(X)$. By duality an, by tensor and exterior  product it defines connection in the
cotangent bundle $T^{*}(X)$, in the bundles of tensors, in the exterior forms bundles $\Lambda^{k}(X)$ and
similarly in other natural bundles over $X$, see \cite{FF}. On the frame bundle $F(X)$ connection
$\Gamma^{G}$ is defined by the $so(n,R)$-valued 1-form
\[
\omega^{G}(x,f)=\Gamma^{i}_{kj}(x)dx^k+\tau^{i}_{j}(x),
\]
where $\tau (x)$ is the $so(n,R)$-valued 1-form of Maurer-Cartan, while on the tangent bundle $T(X)$ it is
defined by the equations
\[
d\xi^{\mu}- \Gamma^{\mu}_{\nu \sigma}(x)\xi^{\nu}dx^\sigma=0,
\]
where $(x^\mu,\xi^{\nu})$ are the adopted coordinates in the tangent bundle $T(X)$. Thus, the lift to the
tangent bundle of the vector field $\partial_{x^\mu }$ is
\[
{\hat \partial_{x^\mu}}(x,\xi)=\partial_{x^\mu}+ \Gamma^{\sigma}_{\nu \mu}\xi^{\nu }\partial_{\xi^ \sigma
}.
\]
\vskip 1cm
\par

A (nonlinear, Ehresmann) connection $\nu $ (\cite{KMS,MaS}) in the bundle $\pi:Y\rightarrow X$ determines
and is determined by the section $q_\nu :(x^\mu ,y^i )\rightarrow (x^\mu ,y^i  ,\Gamma^{i}_{\mu}(x,y))$ of
the bundle $J^{1}(\pi)\rightarrow Y$ (\cite{KMS}). We will always assume this connection to be complete.
In a local adapted chart $(x^\mu ,y^i)$ this connection is defined by the equations \beq
dy^i-L^{i}_{\mu}(x,y)dx^\mu =0, \eeq
 so that the horizontal lift of a basic vector field
$\partial_{x^\mu }$ is the (projectable) vector field in $Y$ of the form \beq
\partial_{x^\mu }+L^{i}_{\mu}(x,y)\partial_{y^i}.
\eeq

 Connection $\nu$ defines the connection on the bundle $VY\rightarrow Y$ \emph{linear over }$Y$.
Namely, applying the functor of vertical tangent bundle to the section $q_{\nu}:Y\rightarrow J^{1}(\pi)$
we get a mapping $Vq_{\nu}:VY\rightarrow VJ^{1}(\pi)$.  Let

\beq \begin{CD} VJ^{1}(\pi) @> i_{Y}>> J^{1}(VY)\\
@V \pi VV    @V \pi_{10}VV \\
VY @>=>> VY
 \end{CD}\eeq

be the canonical involution (\cite{S}), then the composition
\[
\mathcal{V}_{Y}\nu :=i_{Y}\circ Vq_{\nu}:VY\rightarrow  J^{1}(VY)
\]
determines the connection on $VY \rightarrow X$ called the \emph{vertical prolongation of the connection}
$\nu$.  If $Y^i=dy^i$ are coordinates in $VY$ complemental to the adopted coordinates $(x^\mu ,y^i )$ then
the connection $\mathcal{V}_{Y}\nu $ is defined by the equations (7.1) and

\beq dY^i -\frac{\partial L^{i}_{\mu}}{\partial y^j}Y^j dx^\mu =0. \eeq

More then this, connection $\nu$ defined canonically (by the flow prolongation of the flows of horizontal
vector fields, see \cite{KMS}, Ch.X) the connections on the bundles $T(Y)\rightarrow X,\
T^{*}(Y)\rightarrow X,\ \Lambda^{k}Y\rightarrow X$ satisfying to the proper forms of Leibniz relations
with respect to the pairing, tensor and exterior products (see \cite{KMS}) which we denote by the same
letter $\nu$.
\par
If we would like this extension to preserve the subbundles $\Lambda^{k}_{r}$ we would have to modify it
using the connection $\Gamma^G$ on the base manifold $X$ in order to extend the horizontal translation to
the bundle $Z^* \simeq T(X)\otimes_{Y}V^{*}(\pi)\otimes_{Y}\wedge \Lambda^{n}(X)$. We denote by $\hat
\Gamma$ the obtained connection.  We have

\par
\begin{proposition}
\begin{enumerate}
\item Vertical subbundle $V(Y)\rightarrow X$ of the bundle $T(Y)\rightarrow X$ is invariant under the
$\nu$-parallel translation along any curve $\gamma : (a,b)\rightarrow X .$
\item Ehresmann connection $\nu$ in the bundle $\pi$ and linear connection $\Gamma$ on $X$ (i.e. on the
tangent bundle $T(X)\rightarrow X$) define canonically the connection $\hat \Gamma$ in the bundle
$Z^{*}\rightarrow X$ whose horizontal lift is defined as follows \beq
\partial_{x^\mu}\rightarrow \partial_{x^\mu}+\Gamma^{i}_{\mu}(\nu ) \partial_{y^i}
+\left(-\frac{\partial \Gamma^{i}_{\mu}(\nu )}{\partial y^j}p^{\sigma }_{i}+\Gamma^{\gamma}_{\mu
\gamma}p^{\sigma}_{j} \right)\partial_{p^{\sigma}_{j}}. \eeq
\item Lift of vector fields $\xi \in \mathcal{X}(X)$ with the help of $\nu$ to $Y$ and then - by flow
lift to $Z^*$ coincide with the $\hat \Gamma$-horizontal lift of $\xi.$
\item Subbundles $\Lambda^{k}_{r}Y$ of the bundles $\Lambda^{k}(Y)$ are invariant under the $\hat \Gamma $-parallel
translation along any curve $\gamma : (a,b)\rightarrow X .$
\item Canonical multisymplectic forms $\Omega^{k}_{r}$ and $\tilde \Omega$ are invariant under the (local)
flows of $\hat \Gamma$-horizontal vector fields ${\hat \xi},\ \xi \in \mathcal{X}(X)$:
\[
L_{\hat \xi}\Omega^{k}_{r}=0.
\]
\item Pullback $\pi^{*}:\Lambda^{k}(X)\rightarrow \Lambda^{k}(Y)$ embeds the exterior k-form bundle of $X$ into
the exterior k-form bundle of $Y$. Subbundles $\pi^{*}(\Lambda^s X)\subset \Lambda^{s}Y$ are invariant
under the parallel translation.
\end{enumerate}
\end{proposition}
\begin{proof} First statement follows from the fact that parallel translation in $Y$ maps fibers of $\pi$
into fibers.  For the proof of second statement see \cite{CLM}. Third statement follows from comparison of
the comparison of expressions of vector fields lifted in two ways.  Third statement follows from the first
one. Forth and the fifth statements follows from the fact that (local) flows of $\nu$-horizontal vector
fields in $Y$ are projectable to $X$ and, therefore, the pullback by these transformations of the forms
from $\pi^{*}(\Lambda^s X)$ is, by duality, realized by the projected flows.
\end{proof}

Using the connection $\Gamma^{G}$ one can lift the connection $\nu$ in the bundle $\pi$ to the connection
$\hat \nu$ in the bundle $\pi^{1}:J^{1}(\pi)\rightarrow X.$  This lift can be achieved by different ways,
(see the proof proved by I.Kolar and others (\cite{KMS}) that "all the natural operators transforming a
general connection on $\pi:Y\rightarrow X$ and a linear connection on $M$ (here $\Gamma^G$) into a general
connection on $J^{1}(\pi)\rightarrow X$ form the one-parametrical family $tP+(1-t)\mathcal{T}^{1},\ t\in
R$" where $P, \mathcal{T}^{1}$ are two distinguished connections.\par
  Here we will use only one of these
connections, namely the version of connection $P$ for the partial 1-jet bundles defined as follows (see
\cite{KMS}, Sec.45.7). Section $q_{\nu}$ determines the identification of the 1-jet bundle
$J^{1}_{p}(\pi)$ with the associated vector bundle $\mathcal{V}Y\otimes T^{*}_{p}(X)$
\[I:J^{1}(\pi)\simeq VY\otimes T^{*}_{p}(X)\]
 similar to one for
conventional 1-jet bundle. Here $T^{*}_{p}(X)$ is the subbundle of $T^* (X)$ dual to the subbundle
$K\subset T(X)$ defined by the AP structure $K\oplus K'$. Vertical prolongation ${\mathcal V}\nu $ of the
connection $\nu$ was defined above. On the other hand, connection $\nu$ determines the horizontal lift of
vector fields in $X$ preserving the AP structure $K\oplus K'$ to the vector fields in the cotangent bundle
$T^{*}(X)$ whose (local) flow leaves the dual decomposition $T^{*}(X)=K^{\bot}\oplus K'^{\bot}$ invariant.
Therefore, vector fields from $X_{K\oplus K'}(X)$ are lifted to the vector fields in $T^{*}_{K}(X)\equiv
K'^{\bot}.$ As a result we get the lift of the vector fields

\beq T_{K\oplus K'}(X) \rightarrow \mathcal{X}(VY\otimes T^{*}_{K\oplus K'}(X))\rightarrow
\mathcal{X}(J^{1}_{K}(\pi)).\eeq

This lift determines the subbundle $H_{K\oplus K'}(\nu )$ of the bundle $T(J^{1}_{K}(\pi))$ complemental
to the vertical tangent subbundle of the bundle $\pi^{1}_{K}: J^{1}_{K}(\pi)\rightarrow X$ and, therefore,
the connection in the bundle $\pi^{1}_{K}$. More then this, horizontal distribution $H_{K\oplus K'}(\nu )$
of this connection splits naturally as the sum of two subbundles - horizontal lifts of bundles $K$ and
$K'$ respectively:

\beq H_{K\oplus K'}(\nu )=K(\nu )\oplus K'(\nu). \eeq

Denote obtained connection by $\nu^{1}_{K}$.  We get\par
\begin{proposition} Let $T(X)=K\oplus K'$ be an AP-structure on $X$. Let $\nu$ be a connection on the
bundle $\pi:Y\rightarrow X$ and $\Gamma^G$ is the Levi-Civita connection in $X$. There is the canonically
defined connection $\nu^{1}_{K}$  in the bundle $\pi^{1}_{K}: J^{1}_{K}(\pi)\rightarrow X$ whose
horizontal distribution splits
\[
H_{K\oplus K'}(\nu )=K(\nu )\oplus K'(\nu)
\]
as the direct sum of subbundles - horizontal lifts of distributions $K,K'$.
\end{proposition}

In the case of partial 1-jet space $J^{1}_{S}(\pi)$ where we have the natural product structure
$<\partial_{t}>\oplus T(B)$ similar result is valid for a connection $\nu$ on the bundle $\pi$
\emph{provided the parallel translations by the lifts of vector fields from
$\mathcal{X}_{<\partial_{t}>\oplus T(B)}(X)$ leaves the fiber product structure (5.6) of the bundle $\pi:
Y\rightarrow X$ invariant}.  It is easy to prove the following
\begin{proposition} Let $\nu$ be a connection on the
bundle $\pi:Y\rightarrow X$ such that the fiber product structure $Y=Y_{0}\underset{X}{\times}
Y_{t}\underset{X}{\times} Y_{x}\underset{X}{\times}  Y_{xt}$  is invariant under the $\nu$-parallel
translation.  Then there is the prolongation of $\nu$ to the connection $\nu^{1}_{S}$ in the bundle
$\pi^{1}_{S}:J^{1}_{S}(\pi)\rightarrow X$ such that the parallel translation with respect to $\nu^{1}_{S}$
preserves the fiber product structure $J^{1}_{S}(\pi)= 0(Y_{0})\underset{Y}{\times}
J^{1}_{t}(Y_{t})\underset{Y}{\times} J^{1}_{x}(Y_{x})\underset{Y}{\times} J^{1}(Y_{tx}).$ Horizontal
distribution of this connection splits
\[
H_{S}(\nu^{1}_{S} )=<\partial_{t}>^1 \oplus T(B)^1
\]
as the sum of two distributions - horizontal lifts of sub-distributions $<\partial_{t}>, \  T(B)$
respectively.
\end{proposition}

\begin{remark} Connection $\nu^{1}_{S}$ is the fiber product of the connections in the four
 bundles in the decomposition $J^{1}_{S}(\pi)= 0(Y_{0})\underset{Y}{\times}
J^{1}_{t}(Y_{t})\underset{Y}{\times} J^{1}_{x}(Y_{x})\underset{Y}{\times} J^{1}(Y_{tx})$ over $X$. First
component of this product is the component of connection $\nu$ in the bundle $Y_{0}\rightarrow X$ itself -
no prolongation is necessary for this bundle.
\end{remark}

In the future will need the following result allowing to lift a connection $\nu$ $q_\nu : Y\rightarrow
J^{1}(\pi)$ in the bundle $\pi$ to the vertical part ${\hat \nu}:J^{1}(\pi)\rightarrow J^{1}(J^{1}(\pi))$
of the second order connection on the bundle $\pi^{1}$ that defines the connection on the bundle
$\pi_{10}: J^{1}(\pi)\rightarrow Y$.

\begin{proposition}(\cite{GMS}, Proposition 2.6.1)  Let $\Gamma $ be a symmetric linear connection on $X$ and
$\nu : J\rightarrow J^{1}(\pi)$ be a connection in the bundle $\pi$. Then there is an involution
$s_{\Gamma }:J^{1}(J^{1}(\pi))\rightarrow J^{1}(J^{1}(\pi))$ over $J^{1}(\pi)$  such that composition

\beq {\hat \nu}_{\Gamma}=s_{\Gamma}\circ \Gamma\nu :J^{1}(\pi)\rightarrow J^{1}(J^{1}(\pi)) ,\eeq given in
local coordinates by

\beq {\hat \nu}_{\Gamma}=dx^{\lambda }\otimes (\partial_{x^\lambda}+L^{i}_{\lambda}\partial_{y^i}+
[\partial_{x^\lambda }L^{i}_{\mu}+y^{j}_{\lambda}\partial_{y^j}L^{i}_{\mu}+\Gamma^{G\ \nu}_{\lambda \mu
}(y^{i}_{\nu}-L^{i}_{\nu})]\partial_{z^{i}_{\mu}}) \eeq

\end{proposition}

Connection $\hat \nu$ allows to extend the vertical 1-form $\theta_{C}\in \Gamma(V(\pi_{10})^{*})$ to the
1-form on the whole space $J^{1}_{p}(\pi )$ assuming that it annulate the horizontal subspaces $Hor({\hat
\nu})\subset T(J^{1}_{p}(\pi )).$

\vfill \eject

\addtocontents{toc}{Chapter III. Constitutive Relations and Balance Systems.}

\centerline{\textbf{Chapter III. Constitutive Relations and Balance Systems.}} \vskip1cm

 Below we will use the notation $Z_{p}=J^{1}_{p}(\pi)$ that unite the following special cases:
RET, where $J^{1}_{p}(\pi)=\{ 0\}$ is the trivial (with the one point fiber) bundle over $Y$;
$Z_{K}=J^{1}_{K}(\pi)$ for a subbundle $K$ in a AP-structure $T(X)=K\oplus K'$, case
$Z_{S}=J^{1}_{S}(\pi)$ for a splitting $S$ of the field bundle (see (4.5-6)),\r, finally $Z=J^{1}(\pi)$ -
full 1-jet bundle. We will be using both longer and shorter notations whichever is more convenient at a
point. If we consider a prolongation $\xi \rightarrow \xi^1$ of a vector field $\xi \in \mathcal{X}(Y)$ to
the partial 1-jet bundle, \emph{we will always presume that the vector field $\xi $ satisfies to the
conditions for the existence of the prolongation $\xi^1$ preserving the partial Cartan structure (see
Section 6)}.\par \vskip0.5cm

\section{General Constitutive Relations (CR).}
In this section we define general constitutive relations and the Poincare-Cartan forms defined by these
relations.  We also give examples of several types of constitutive relations.\par

 Consider the following (constitutive) commutative diagram

 \beq
\begin{CD}
Z_{p}=J^{1}_{p}(\pi ) @>{\hat C} >>\Lambda^{(n+1)+(n+2)}_{2}(Y) \\
@V= VV   @V\chi VV\\
Z_{p}=J^{1}_{p}(\pi ) @>C >> {\tilde Z}=\Lambda^{(n+1)+(n+2)}_{2}(Y)/\Lambda_{1}^{(n+1)+(n+2)}(Y) \\
@V \pi^{1}_{0}VV  @V \pi^{(n+1)+(n+2)} VV\\
Y @=  Y\\
@V \pi VV  @V\pi VV \\
 X @ =  X
\end{CD}.
\eeq
\begin{definition}
\begin{enumerate}
\item A (general) \textbf{constitutive relation} (CR) $C$  of a field theory with the configurational bundle $\pi:
Y\rightarrow X$ and the partial 1-jet space $Z_{p}=J^{1}_{p}(\pi )$ is a smooth morphism of bundles over
$Y$
\[
C: J^{1}_{p}(\pi )\rightarrow  \tilde Z =\Lambda_{2}^{(n+1)+(n+2)}Y/\Lambda_{1}^{(n+1)+(n+2)}Y.
\]
 \par

In local coordinates $(x^\mu ,y^i, z^{i}_{\mu})$ on $J^{1}_{p}(\pi )$ and $(p^{\mu}_{i},q_{i})$ on $\tilde
Z$ a CR-mapping $\mathcal{C}$ has the form

\beq C(x^\mu ,y^i, z^{i}_{\mu}) =(x^\mu ,y^i; F^{\mu}_{i}(x^\mu ,y^i, z^{i}_{\mu});\Pi_{i}(x^\mu ,y^i,
z^{i}_{\mu}))\eeq
\par
\item A general constitutive relation $\mathcal{C}$ is called \textbf{regular }if the mapping $\mathcal{C}$ is the diffeomorphism
 of $Z_{p}$ onto the submanifold of $\tilde Z$.
\item
A covering constitutive relation $\hat C$ of the field theory with the configurational bundle $\pi:
Y\rightarrow X$ and the partial 1-jet space $Z_{p}=J^{1}_{p}(\pi )$ is a smooth mapping of bundles
\[
{\hat C}: J^{1}_{p}(\pi )\rightarrow  \Lambda_{2}^{(n+1)+(n+2)}Y.
\]
 In
local coordinates $(x^\mu ,y^i, z^{i}_{\mu})$ on $J^{1}_{p}(\pi )$ and $(p, p^{\mu}_{i},q_{i})$ on
$\Lambda^{n+1}_{2}(Y) \oplus \Lambda^{n+2}_{2}(Y)$ a CCR-mapping $\widehat{\mathcal{C}}$ has the form

\beq  \widehat{\mathcal{C}}(x^\mu ,y^i, z^{i}_{\mu}) =(x^\mu ,y^i;p(x^\mu ,y^i, z^{i}_{\mu});
F^{\mu}_{i}(x^\mu ,y^i, z^{i}_{\mu});\Pi_{i}(x^\mu ,y^i, z^{i}_{\mu}))\eeq
\par

\item A constitutive relation $C$ (respectively a covering CR $\hat C$) is called a conservative relation
 (CR)  (resp. a covering CCR) if $\Pi_{i}=0, i=1,\ldots ,m$.
 \item For a given constitutive relation $\mathcal C$ denote by ${\mathcal C}_{-}$ the constitutive
 relation obtained from $C$ by the changing sign of the production ($(n+2)$) part:

\beq C_{-}(x^\mu ,y^i, z^{i}_{\mu}) =(x^\mu ,y^i; F^{\mu}_{i}(x^\mu ,y^i, z^{i}_{\mu});-\Pi_{i}(x^\mu
,y^i, z^{i}_{\mu}))\eeq

\end{enumerate}
\end{definition}
\par
\begin{remark}
Physical case corresponds to the choice $n=3$.
\end{remark}
\begin{remark}
Definition given here is very broad, including, in particular, a zero mapping.  Thus, to get a useful
class of constitutive relations one has to put some nondegeneracy conditions to this mapping including but
not reducing to the regularity of a CR defined above.
\end{remark}
\begin{remark} We can also define constitutive relations defined in a domain $U\subset Z_{p}$ instead of
the whole space $Z_{p}$. This may be necessary in a situation where some constraints in the from of
inequalities on the derivatives of the fields $y^i$ are present.
\end{remark}
\par
\begin{example}
In the maximal (RET) case $U=U_{0}$, the partial 1-jet space $J^{1}_{p}(\pi )$ coincide with $Y$ (its
fiber is a point $R^0$) and the constitutive relation is just the section of the bundle ${\tilde
Z}\rightarrow Y$. As we will see in the next section it is convenient and natural to consider the CCR
$\widetilde{\mathcal{C}}$ for the RET constitutive relation $\mathcal{C}$ defined on the full 1-jet bundle
$Z$ of the bundle $\pi$ (formally we could take it have $\eta$-component zero, but it would be a less
convenient choice).
\end{example}
We have the following simple
\begin{proposition}
\begin{enumerate}
\item Constitutive relations (and the covering constitutive relations) form the
$C^{\infty}(J^{1}_{p}(\pi))$-module $\mathcal{CR}$ and $ \mathcal{\widehat{CR}}$ respectively.
\item Let ${\hat C}\in   \widehat{\mathcal{CR}}$ be a CCR, then combining the defining mapping ${\hat C}:
J^{1}_{p}(\pi)\rightarrow \Lambda^{(n+1)+(n+2)}_{2}(Y)$ with the projection by
$\Lambda^{(n+1)+(n+2)}_{1}(Y)$ we associate with a CCR $\hat C$ the constitutive relation $C\in
\mathcal{CR}.$
\end{enumerate}
\end{proposition}

Using the canonical forms on the bundle $\Lambda^{(n+1)+(n+2)}_{2}Y$ we \textbf{define} the
\textbf{Poincare-Cartan form of the covering constitutive relation} $\hat C$

\beq \Theta_{\hat C}=\mathcal{C}^{*}(\Theta^{n+1}_{2}+\Theta^{n+2}_{2})=p\eta+ F^{\mu}_{i}dy^i \wedge
\eta_{\mu}+\Pi_{i}dy^i \wedge \eta. \eeq

\begin{definition} Canonical linear mapping $~: \mathcal{CR}\rightarrow \widehat{\mathcal{CR}}$ (section
 of the projection above) is defined by the formula
\[
(\tilde {\mathcal{C}})(z)=(x^\mu ,y^i;-z^{j}_{\nu}F^{\nu}_{j}(x^\mu ,y^i, z^{i}_{\mu}); F^{\mu}_{i}(x^\mu
,y^i, z^{i}_{\mu});\Pi_{i}(x^\mu ,y^i, z^{i}_{\mu})).
\]
CCR $\tilde C$ will be called the \textbf{lifted CCR of the constitutive relation} $C$.
\end{definition}

For the lifted CCR $\tilde C$ of a CR $C$ defined in the full 1-jet bundle $Z=J^{1}(\pi)$ we have

\beq \Theta_{\tilde {\mathcal{C}}}=S_{\eta}( F^{\mu}_{i}dz^{i}_{\mu} )   =-z^{i}_{\mu}F^{\mu}_{i}\eta+
F^{\mu}_{i}dy^i \wedge \eta_{\mu}+\Pi_{i}dy^i \wedge \eta. \eeq

Here $S_{\eta}$ is the vertical endomorphism (3.1). Notice that we get the same result by applying the
vertical endomorphism to any 1-form $\lambda \in \Lambda^{1}(Z)$ of the form $\lambda =
F^{\mu}_{i}dz^{i}_{\mu}+F_{i}dy^i +F_{\mu}dx^{\mu}.$
\par
\vskip0.5cm
\begin{definition}
Let $\nu$ be an Ehresmann connection in the bundle $\pi:Y\rightarrow X$.  Let $C$ be a general
constitutive relation. Determine the $\nu$-lift ${\tilde C}_{\nu}$ of constitutive relation $C$ by section
$\delta_{\nu}: {\tilde Z} \rightarrow \Lambda^{(n+1)+(n+2)}$ (see (3.20), Sec.3)
\[
{\tilde C}_{\nu}(z)=(x^\mu ,y^i;\Gamma^{j}_{\mu}F^{\mu}_{j}(x^\mu ,y^i, z^{i}_{\mu}); F^{\mu}_{i}(x^\mu
,y^i, z^{i}_{\mu});\Pi_{i}(x^\mu ,y^i, z^{i}_{\mu})).
\]
\end{definition}

 Taking the pullback of the canonical form ${\tilde \Theta}_{\nu}$ on the bundle $\tilde Z$ we
get the $\nu$-induced \textbf{Poincare-Cartan form of the constitutive relation} ${\tilde C}_{\nu}$

\beq \Theta_{{\tilde C}_{\nu}}={\tilde
C}_{\nu}^{*}(\delta^{*}_{\nu}\Theta^{n+1}_{2}+\Theta^{n+2}_{2})=(F^{\mu}_{i}\Gamma^{i}_{\mu})\eta +
F^{\mu}_{i}dy^i \wedge \eta_{\mu}+\Pi_{i}dy^i \wedge \eta. \eeq It is the special case of the following
construction.\par

\begin{remark} Below (Sec.12) we show that an action of a transformations on the Poincare-Cartan form $\Theta_{C}$
corresponding to a CR $C$ produce an additional term of the type $A\eta$. As a result for the
compatibility with the action of transformations one have to consider the classes of Poincare-Catran forms
$\Theta_{C}$ - their images in the factor bundle
$\Lambda^{(n+1)+(n+2)}_{2}(J^{1}_{p}(\pi))/\Lambda^{(n+1)+(n+2)}_{1}(J^{1}_{p}(\pi))$ rather then simply
the forms on $\tilde Z$.
\end{remark}
\begin{remark} Equivalent definition of the general constitutive relations can be given in terms of
section of the corresponding bundles: \beq
\begin{CD}
\pi^{*}_{1}(\Lambda^{(n+1)+(n+2)}_{2}Y/\Lambda_{1}^{(n+1)+(n+2)}Y)
 @>\pi^{*}_{(n+1)+(n+2)} >> J^{1}_{p}(Y)\\
@V\pi_{1\ (n+1)+(n+2)} VV   @V\pi_{1}VV\\
\tilde Z =\Lambda_{2}^{(n+1)+(n+2)}Y/\Lambda_{1}^{(n+1)+(n+2)}Y @>\pi_{(n+1)+(n+2)}>> Y\\
@.  @V\pi VV \\
 @. X
\end{CD}
\eeq Then we can use the following
\begin{definition}
A (general) constitutive relation $C$ of the field theory with the configurational bundle $\pi:
Y\rightarrow X$ and the partial 1-jet space $Z_{p}=J^{1}_{p}(\pi )$ is a smooth section $\mathcal C$ of
the vector bundle $\pi^{*}_{(n+1)+(n+2)}$ in the diagram above.\par

In the local fiber coordinates $(x^{\mu},y^{i},z^{i}_{\mu})$ a section of the bundle
$\pi^{*}_{(n+1)+(n+2)}$ has the form

\beq \mathcal{C }(x^{\mu},y^{i},z^{i}_{\mu})=
F^{\mu}_{i}dy^{i}\wedge\eta_{\mu}+\Pi_{i}dy^{i}\wedge\eta_{\mu}, \eeq

where $F^{\mu}_{i},\Pi_{i}$ are functions on the space $J^{1}_{p}(Y)$.\par
\end{definition}
\end{remark}

Below we list several types of the constitutive relations that are widely used in physics and continuum
mechanics.

\begin{example} A \textbf{Lagrange constitutive relation} defined by a smooth (Lagrangian) function
$L\in C^{\infty}(Z_{p})$ is given by the mapping:
\beq C_{L}(x^\mu ,y^i,
z^{i}_{\mu})=(p^{\mu}_{i}=\frac{\partial L}{\partial z^{i}_{\mu}};\Pi_{i}=\frac{\partial L}{\partial
y^{i}}). \eeq

Correspondingly, a \textbf{covering Lagrange constitutive relation} is defined by a smooth function $L\in
C^{\infty}(Z_{p})$ giving the mapping $Z_{p} \rightarrow \Lambda^{n+(n+1)}_{2}Y$

\beq {\hat C}_{L}(x^\mu ,y^i, z^{i}_{\mu})=(p=-z^{i}_{\mu}\frac{\partial L}{\partial z^{i}_{\mu}}
,p^{\mu}_{i}=\frac{\partial L}{\partial z^{i}_{\mu}};\Pi_{i}=\frac{\partial L}{\partial y^{i}}). \eeq

Notice that the covering Lagrange relation defined here does not coincide with the covering CR defined by
the Legendre transformation of the Lagrangian $L\eta$. Relation between these two covering CR will be
studied elsewhere.
\end{example}

\begin{example} A \textbf{semi-lagrangian CR} is defined by a smooth function $L\in
C^{\infty}(Z_{p})$ and an arbitrary functions $Q_{i}\in C^{\infty}(Z_{p}), i=1,\ldots ,m$:
\beq
C_{L,Q_{i}}(x^\mu ,y^i, z^{i}_{\mu})=(p^{\mu}_{i}=\frac{\partial L}{\partial
z^{i}_{\mu}};\Pi_{i}=Q_{i}(x^\mu ,y^i, z^{i}_{\mu})).
\eeq
\end{example}
\begin{remark} In a case when the domain of $C$ is a partial 1-jet bundle $Z_{p}=J^{1}_{p}(\pi)$ the component
$F^{\mu}_{i}$ of CR $C$ is equal zero if the derivative $z^{i}_{\mu}$ is absent from the fibers of the
partial jet bundle $J^{1}_{p}(\pi)$. In the case of RET semi-Lagrangian CR is trivially zero.
\end{remark}
\begin{remark} For a semi-Lagrangian CR there exists natural - Lagrangian lift to the CCR:
\beq {\hat C}_{L,Q}=(L-z^{i}_{\mu}L_{,z^{i}_{,mu}})\eta +\frac{\partial L}{\partial z^{i}_{\mu}}dy^i\wedge
\eta_{\mu}+Q_{i}dy^i \wedge \eta.  \eeq It will be used below for formulating corresponding Noether
Theorem.
\end{remark}
A very important example of a semi-Lagrangian CR is the following

\begin{example}\textbf{$L+D$-system}.  Let $L$ be a smooth function $L\in
C^{\infty}(Z_{p}).$ Let the time derivatives $z^{i}_{0}$ of all basic fields belong to $Z_{p}$ and let
$D\in C^{\infty}(Z_{p})$ be one more function (dissipative potential). Define the constitutive relation
$\mathcal{C}_{L,D}$ that differs from Lagrangian CR $C_L$ by the condition
$\Pi_{i}=L_{,y^i}+D_{,z^{i}_{0}}$. Thus, the corresponding Poincare-Cartan form is

\beq \Theta_{L,D}=\Theta_{L}+D_{z^{i}_{,0}}dy^i \wedge \eta. \eeq

\end{example}
\begin{example}\textbf{Vector-potential CR}. Consider a RET case. Let $h=h^\mu(x,y)\eta_{\mu}$ be a
semi-basic n-form on $Y$.  Define a constitutive relation by the formula

\beq C_{h}(x^\mu ,y^i)=(p^{\mu}_{i}=\frac{\partial h^\mu}{\partial y^i};\Pi_{i}=\Pi_{i}(x,y)). \eeq

 This is the case of the \emph{dual formulation in terms of
Lagrange-Liu variables} ($y^i$ replaces the $\lambda^i$ here), see Sections 15 and 17 below.
\end{example}
\begin{example} One can combine Semi-Lagrangian and vector-potential examples into the following one.  Let
$L, Pi_{i},\ i=1,\ldots ,m\in C^{\infty}(J^{1}_{p}(\pi)$ and let Let $h=h^\mu(x,y)\eta_{\mu}$ be a
semi-basic n-form on $Y$.  Define a constitutive relation by the mapping \beq C_{h}(x^\mu ,y^i,
z^{i}_{\mu})=(p^{\mu}_{i}=L_{z^{i}_{\mu}}+\frac{\partial h^\mu}{\partial y^i};\Pi_{i}=\Pi_{i}(x,y)). \eeq
\end{example}
\begin{example}  For the 5F-fluid system (see Sec.2 above) with the trivial bundle
$Y=X\times R^5 \rightarrow X$ and the basic fields
 $(\rho , v^A, \vartheta )$ the constitutive relations define and are defined by its
Poincare-Cartan form
\begin{multline} \Theta_{C_5}=[\rho d\rho \wedge \eta_{0}+\rho v^B d\rho \wedge \eta_{B}]+
   [(\rho v^A )dv^A \wedge
\eta_{0}+(\rho v^A v^B -t^{AB})dv^A \wedge \eta_{B}]+\\ + [\rho \epsilon d\vartheta \wedge \eta_{0} +(\rho
\epsilon V^A +q^A )d\vartheta \wedge \eta_{B}] ] + [ f_{A} dv^A \wedge \eta +t^{A}_{B}\frac{\partial
V^B}{\partial x^A} d\vartheta \wedge \eta +r].
\end{multline}
In the simplest case of $5F-fluid$ system considered in Sec.2, the \emph{state space} represents the fiber
of the bundle $\pi^{1}:J^{1}_{S}(\pi)\rightarrow X$ where $S_{0}=\{\rho \}, S_{x}=\{v^A ,\vartheta \},
S_{t}=S_{tx}=\emptyset .$ Thus, only derivatives of velocity components and of temperature by spacial
coordinates $x^A$ are present in the 1-jet fiber of the state space. State space $S$ itself is the fiber
bundle over the basic fields space $U$: $\varrho:S\rightarrow U$.\par
\end{example}

\vfill \eject

\section{Balance System $\mathcal{B_{C}}$ defined by a Constitutive Relation $\mathcal{C}$.}

In the Lagrangian Field Theory the Poincare-Cartan form $\Theta_{L}$ appears in the second term of a local
variation in the direction of a vector field $\xi \in \mathcal{X}(J^{1}(\pi))$ obtained by the variational
version of the Cartan formula $L_{\xi}(L\eta)=(i_{\xi}d+di_{\xi})(L\eta)$

\beq \delta_{\xi}(L\eta )(j^{1}(s)) =j^{1\ *}(s)[  i_{\xi}e(L)(s)+di_{\xi}\Theta_{L}(z)] \eeq
see
\cite{FF}. Here $s\in \Gamma (\pi)$ is a section of bundle $\pi$. \par The semi-basic $n+1$-form $e(L)\in
\Lambda^{n+1}(J^{1}(\pi))$ is the Euler-Lagrange form and the Euler-Lagrange system of the field theory
with the Lagrangian $L$ for a section $s$ has the form:

\beq j^{1\ *}(s)i_{\xi}e(L)=0,\ \forall \xi. \eeq

In its turn, the form $\Theta_{L}$ is the Poincare-Cartan form (4.3) and the same Euler-lagrange system of
equations is obtained by (4.6) \beq j^{1\ *}(s)di_{\xi}\Theta_{L}(z)=0. \eeq

For a general constitutive relation (9.2) and the corresponding Poincare-Cartan form $\Theta_{C}$ we have
an analog of variational formula (10.1) given by Cartan formula

\beq L_{\xi}\Theta_{C}=i_{\xi}d\Theta_{C}+di_{\xi}\Theta_{C}. \eeq

Thus, we can try to formulate the balance laws corresponding to the CR mapping $C$ by following one of two
ways suggested above. We can take the pullback via $j^{1\ *}_{p}(s)$ of the first or second term in (10.4)
and request it to be zero for a set of variation vector fields $\xi$, large enough to separate the balance
equations corresponding to the CR $\mathcal{C}$.  Yet, as we see below, both of these cases meet some
interesting restrictions. In order to get the balance equations the variations $\xi$ should satisfy some
conditions defined by the $F^{\mu}_{i}$-part of the constitutive relation $C$.  More specifically we
should have $m$ linearly independent vector fields $\xi$ in order to extract all $m$ balance equations
from the invariant formulation of the type (10.2) or (10.3). Locally this is always possible but still
leads to some restrictions to the type of variations. We will see that there is a way around this
difficulty if one uses in CR version of the formula (10.2) the  \emph{reduced horizontal  differential}
$\hat d$ (comp. \cite{KV} or Appendix IV) instead of the conventional De-Rham differential $d$.  On the
other hand studying these restrictions we will determine the special place of the \emph{semi-Lagrangian
constitutive relations} in between the general CR - these are CR defined on the full 1-jet bundle
$J^{1}(\pi)$ for which there are no limitations for them on the nature of variations vector fields.\par

\subsection{Poincare-Cartan formulation of a balance system.} We assume that a connection
$\nu:Y\rightarrow J^{1}(\pi)$ is fixed.  We start with the Poincare-Cartan way of obtaining the balance
equations and for this we take an arbitrary vector field $\xi \in \mathcal{X}(Z_{p})$ locally having form

\beq \xi =\xi^{\mu}\partial_{x^{\mu}}+\xi^{i}\partial_{y^{i}}+\xi^{i}_{\mu}\partial_{z^{i}_{\mu}}, \eeq
and plug it into the ($\nu$-dependent) Poincare-Cartan form $\Theta_{C_{\nu}}$

\beq i_{\xi}\Theta_{ C_{\nu}}=(F^{\mu}_{i}\Gamma^{i}_{\mu})\xi^{\lambda}\eta_{\lambda}+
F^{\mu}_{i}\xi^{i}\eta_{\mu}-F^{\mu}_{i}\xi^{\nu}dy^{i}\wedge\eta_{\mu
\nu}-\Pi_{i}\xi^{\mu}dy^{i}\wedge\eta_{\mu}+(\Pi_{i}\xi^{i})\eta. \eeq

Take the vector field $\xi$ vertical, i.e. assume that $\xi^\mu=0, \mu=1,\ldots, n+1$.  Then \emph{any}
addition of a term of the form $h(z)\eta$ choused by an adopted change of coordinates or by another choice
of a connection $\nu$ will be eliminated and the result, $i_\xi \Theta_{ C} $ is defined canonically.
\par Now we apply the I-differential ${\tilde d}$

\begin{multline} {\tilde d}i_{\xi}\Theta_{ C}= d(F^{\mu}_{i}\xi^{i})\wedge\eta_{\mu}+
(F^{\mu}_{i}\xi^{i}(\partial_{x^{\mu}}\lambda_{G} )\eta- d(F^{\mu}_{i}\xi^{\nu})dy^{i}\wedge
\eta_{\mu\nu}+\\ \sum_{\mu <\nu} F^{\mu}_{i}\xi^{\nu}dy^{i}\wedge
((\partial_{x^{\nu}}\lambda_{G})\eta_{\mu}-(\partial_{x^{\mu}}\lambda_{G})\eta_{\nu})
-(\Pi_{i}\xi^{i})\eta +\Pi_{i}\xi^{\mu}dy^i\wedge \eta_{\mu}.
 \end{multline}
Thus, requesting the vector field $\xi $ to be \textbf{vertical} (i.e. putting $\xi^{\mu}=0$) we get

\begin{multline} {\tilde d}i_{\xi}\Theta_{C}=
d(F^{\mu}_{i}\xi^{i})\wedge\eta_{\mu}+(F^{\mu}_{i}\xi^{i}\partial_{x^{\mu}}\lambda_{G} )\eta
-(\Pi_{i}\xi^{i})\eta = \\ \xi^{i}(F^{\mu}_{i,x^{\mu}}\eta+F^{\mu}_{i,y^{j}}dy^{j}\wedge
\eta_{\mu}+F^{\mu}_{i,z^{j}_{\nu}}dz^{j}_{\nu}\wedge
\eta_{\mu})+(F^{\mu}_{i}\xi^{i}\partial_{x^{\mu}}\lambda_{G} )\eta -\Pi_{i}\xi^{i}\eta
+F^{\mu}_{i}d\xi^{i}\wedge \eta_{\mu}.
 \end{multline}
\par

Applying now the pullback by the 1-jet $j^{1}_{p}(s)$ of a section $s\in \Gamma(\pi)$ and using
\[
j^{1*}_{p}(s) [F^{\mu}_{i,y^{j}}dy^{j}\wedge \eta_{\mu}+F^{\mu}_{i,z^{j}_{\nu}}dz^{j}_{\nu}\wedge
\eta_{\mu}]=F^{\mu}_{i,y^{j}}s^{j}_{,\mu}\eta+F^{\mu}_{i,z^{j}_{\nu}}s^{j}_{\nu \mu}\eta,
\]

we get

\begin{multline} j^{1\ *}_{p}(s){\tilde d}i_{\xi}\Theta_{C}= \xi^{i}[(F^{\mu}_{i}\circ j^{1}_{p}(s))_{,x^{\mu}}+
F^{\mu}_{i}(\partial_{x^{\mu}}\lambda_{G}) -\Pi_{i}\circ j^{1}_{p}(s)]\eta + \\ +F^{\mu}_{i}\circ
j^{1}_{p}(s) \left(
\xi^{i}_{,x^{\mu}}+\xi^{i}_{,y^j}s^{j}_{,x^{\mu}}+\xi^{i}_{,z^{j}_{\nu}}s^{j}_{,x^{\nu}x^{\mu}}\right)
\eta
 \end{multline}
In the right side in parentheses stays the pullback by $j^{1}(s)$ of the total derivative $
d_{\mu}\xi^{i}= \xi^{i}_{,x^{\mu}}+z^{i}_{\mu}\xi^{i}_{,y^j} +z^{i}_{\mu \nu }\xi^{i}_{,z^{j}_{\nu}}$
 of a component $\xi^i$ of vector field $\xi^{i}$ along section $s$ by $x^\mu$ contracted with the
 form $F^{\mu}_{i}$: $j^{1*}(s)(F^{\mu}_{i}d_{\mu}\xi^{i})\eta $.

 \par In order to
extract the balance equations from the invariant form of variational principle obtained by equating to
zero the obtained expression we have to require the second term in (10.9) to be zero.  Thus, vector field
$\xi$ should satisfy to the additional condition that is explicitly formulated in the next definition.

\begin{definition}
\begin{enumerate}
\item
 For a constitutive relation $C$ (or, more precise, for current form
 $F=F^{\mu}_{i}dy^{i}\wedge \eta_{\mu}\in \Lambda^{n}_{2}Z/\Lambda^{2}_{1}Z$)
 denote by $\mathcal{X}(\mathcal{C})$ the sheaf associated to the \emph{pre-sheaf of vector fields} over $Y$
 that for an open set $U\subset Y$ consists of vertical vector fields $\xi\in \Gamma (U,V(\pi ))$
 whose flow prolongation $\xi^1=\xi^{i}\partial_{y^i}+(d_{\mu}\xi^{i})\partial_{z^{i}_{\mu}}$ satisfies
  to the condition ( $\xi^{1\ i}=\xi^{i}$)

\beq FDiv (\xi )=F^{\mu}_{i}d_{\mu}\xi^{i}=0 \eeq in $U$.
 Vector fields - sections of the sheaf $\mathcal{X}(\mathcal{C})$  will be called
  $\mathcal{C}$-\textbf{admissible}. \par
\item  A constitutive relation $\mathcal{C}$ with the current form $F$ is called \textbf{locally separable} if
 each point $y\in Y$ has a neighborhood $U_{y}$ such that  there are $m$ vector fields $\xi_{k}$ in the
space of sections $\Gamma (U_{y},\mathcal{X(C)})$ linearly independent at each point $y_{1}\in U_{y}.$
\item A constitutive relation $C$ is called \textbf{separable in an open subset} $W\subset Y$ if there are $m$
vector fields $\Gamma (W,\mathcal{X(C)})$ linearly independent at each point of $W$.
\end{enumerate}
\end{definition}

Having introduced these notions we can now formulate the Poincare-Cartan version of the variational
principle for the balance system with the domain $J^{1}_{p}(\pi)$ and the constitutive relation $C$.

\begin{definition} Let $\mathcal{C}$ be a constitutive relation with the domain $J^{1}_{p}(\pi)$.
 We say that a section
$s: U_{s}\rightarrow Y$ of the bundle $\pi:Y\rightarrow X$, $U_{s}$ being an open subset in $X$,
  satisfies to the \textbf{balance system (system of balance laws)  defined by CR $\mathcal{C}$}
 if \textbf{for \emph{all $\mathcal{C}$-admissible} vector fields}
 $\xi \in \mathcal{X}(C)\vert_{\pi_{1}^{-1}(U_{s})}$ (i.e. over $U_{s}$)

 \beq
{\tilde d}(j^{1}_{p}(s)^{*}(i_{\xi}\Theta_{C}))=j^{1}_{p}(s)^{*}({\tilde d}(i_{\xi}\Theta_{C}))=0.
 \eeq
\end{definition}
Here $\tilde d $ is the Iglesias differential of a (n+(n+1))-form on $X$ (see Appendix II).  In simple
terms
\[
{\tilde d}(\omega^{n}+\omega^{n+1})=d\omega^{n}-\omega^{n+1}.
\]

 Let $z\in Z_{p}$ and let $V\subset X$ be a neighborhood of the projection
$x=\pi_{1}(z)\in X$ over which the bundle $Y$ is trivial $Y\vert_{V}\approx V\times U$ and $W\subset U$ be
an open set such that $V\times W$ is the neighborhood of the point $y=\pi_{10}(z)$. We may assume that $V$
and $W$ are domains of adopted chart $(x^\mu ,y^i )$. Vector fields $\xi_{j}=\partial_{y^{j}}\in
V(\pi)\vert_{V\times W}$ have the property that in this local chart $\frac{d\xi^{j}}{dx^{\mu}}=0$ in
$\pi^{-1}_{10}(V\times W).$  Therefore these $m$ vector fields in $X(V\times W)$ are $C$-admissible for
any constitutive relation $C$ with the domain in $J^{1}_{p}(\pi)$ for all four choices of partial 1-jet
bundles (see Proposition 10b and Theorem 1 for the case of $J^{1}_{K}(\pi)$, Proposition 11b for the case
of $J^{1}_{S}(\pi)$). As a result we get
\begin{proposition} Any constitutive relation $C$ is locally separable.
\end{proposition}
Globally defined $C$-admissible vector fields have an important meaning for the balance system
$\mathcal{B_{C}}$ (see next section). In the next section the case of a special situation will be
described, for some type of the bundles $\pi$ a natural class of $m$ linearly independent globally defined
vertical vector fields $\xi \in \mathcal{X}(Y)$ that are admissible for all CR $\mathcal{C}$. Now we
formulate the main result of this section in the Poincare-Cartan formulation.

\begin{theorem}
If a constitutive relation $\mathcal{C}$ is locally separable, then the following  statements for a
section $s\in \Gamma(\pi)(D_{s}),\ D_{s}\subset X$ are equivalent:
\begin{enumerate}
\item
\beq
 {\tilde d}j^{1\ *}_{p}(s)i_{\xi}\Theta_{\mathcal{C}}\equiv j^{1\ *}_{p}(s){\tilde d}i_{\xi}\Theta_{\mathcal{C}}=0,\
 \text{for\ all}\  \xi \in \mathcal{X}(\mathcal{C})\vert_{D_s },
\eeq
\item Section $s$ is the solution of the following system of balance laws - \textbf{balance system}:
\beq
 (F^{\mu}_{i}\circ j^{1}_{p}(s))_{,x^{\mu}}+F^{\mu}_{i}(\partial_{x^{\mu}}\lambda_{G})=
 \Pi_{i}(j^{1}_{p}(s)),\ i=1,\ldots, m.
\eeq
\end{enumerate}
\end{theorem}
\begin{proof}
It is clear that (1) follows from (2).  If (1) is true, choose $m$ (local) linearly independent vector
fields $\xi_{k}\in \mathcal{X}(C)$ in a neighborhood of a point $x\in X$. Then, for such (locally defined)
vector fields $\xi$ the last term in (10.9) is zero and we get the system of linear equations

\[
\xi^{i}_{k}(s(x))((F^{\mu}_{i}\circ
j^{1}_{p}(s))_{,x^{\mu}}+F^{\mu}_{i}(\partial_{x^{\mu}}\lambda_{G})-\Pi_{i}\circ j^{1}_{p}(s))=0
\]
at each point $y$ in a neighborhood of the point $x$ for $i$ unknowns $(F^{\mu}_{i}\circ
j^{1}_{p}(s))_{,x^{\mu}}+F^{\mu}_{i}(\partial_{x^{\mu}}\lambda_{G})-\Pi_{i}\circ j^{1}_{p}(s)$ in the
parenthesis.  By the condition, matrix $\xi^{i}_{k}(s(x))$ of this system is nondegenerate, so system has
only zero solution. Being true in a neighborhood of any point $x\in X$, (2) is true in all $X$.
\end{proof}
Below the balance system system (10.13) will be refereed to as the $\bigstar$.
\begin{example}
For a Lagrangian constitutive relation (See Example 3) the balance system (10.13) takes the form \beq
(\frac{\partial L}{\partial z^{i}_{\mu}}\circ j^{1}_{p}(s))_{,x^{\mu}}+\frac{\partial L}{\partial
z^{i}_{\mu}}\circ j^{1}_{p}(s)(\partial_{x^{\mu}}\lambda_{G})=\frac{\partial L}{\partial
y^{i}}(j^{1}_{p}(s)), \eeq or
\[
\frac{\partial}{\partial x^{\mu}}(\frac{\partial L}{\partial z^{i}_{\mu}}\circ
j^{1}_{p}(s))+\frac{\partial L}{\partial z^{i}_{\mu}}\circ
j^{1}_{p}(s)(\partial_{x^{\mu}}\lambda_{G})-\frac{\partial L}{\partial y^{i}}(j^{1}_{p}(s))=0,
\]
i.e. is the system of Euler-Lagrange equations for the Lagrangian form $L\eta$.  Here
$\lambda_{G}=ln(\vert G\vert ).$
\end{example}
\begin{example}\textbf{$L+D$-system}.  For a $L+D$-system where the Poincare-Cartan form is
\[
\Theta_{L,D}=\Theta_{L}+D_{z^{i}_{0}}dy^i \wedge \eta,
\]
the corresponding balance system has the form

\beq \mathcal{E}_{L}(s)_{i}=\frac{\partial}{\partial x^{\mu}}(\frac{\partial L}{\partial z^{i}_{\mu}}\circ
j^{1}_{p}(s))+\frac{\partial L}{\partial z^{i}_{\mu}}\circ
j^{1}_{p}(s)(\partial_{x^{\mu}}\lambda_{G})-\frac{\partial L}{\partial y^{i}}(j^{1}_{p}(s))=\frac{\partial
D}{\partial z^{i}_{0}}\circ j^{1}(s). \eeq

This system has the form of Euler-Lagrange equations with the dissipative Rayleigh potential $D$ (see
(\cite{Ma})).
\end{example}
\begin{example} \textbf{System of conservation laws.} If we take $\Pi_{i}=0$ in the constitutive relation
$C$ then the balance system $\bigstar$ takes the form of the system of conservation laws \beq
(F^{\mu}_{i}\circ j^{1}_{p}(s))_{,\mu }=0. \eeq

\end{example}

\subsection{Euler-Lagrange formulation of the balance system.}

Now we will see what happens if we apply the standard order $i_{\xi} d$ of operations that is used in the
Lagrangian Theory (10.2) to the  \emph{lifted Poincare-Cartan form}  $\Theta_{{\tilde C}_{-}}$ of the
constitutive relation $\mathcal{C}$ (see ()), or, more generally, to an arbitrary covering constitutive
relation of the form

\[ \Theta_{\widehat{ \mathcal{C_{-}}}}=
p\eta +F^{\mu}_{i}dy^i \wedge \eta_{\mu}-\Pi_{i}dy^i\wedge \eta .\]

We reversed the sign of the source term in the Poincare-Cartan form to compensate for the different order
of operation of contraction and applying the differential.\par Doing these calculations we will be
repeatedly using the relation between the contact forms of partial contact structures on $Z_{p}$ and the
differentials of basic variables (see Sec.4):
\[
dy^i =\omega^{i}+\sum_{(\mu ,i)\in P}z^{i}_{\mu}dx^\mu ,\ dz^{i}_{\mu}=\omega^{i}_{\mu }+\sum_{(\mu ,i)\in
P} z^{i}_{\mu \sigma} dx^\sigma,
\]
and the total derivative $d_{\nu}$ on $Z_{p}$ (see Appendix IV), or, more exactly, on the
$J^{1}(Z_{p})=J^{2}_{p}(\pi)$
\[
d_{\nu}f=\partial_{x^\nu}f+\sum_{(\nu ,i)\in P} z^{i}_{\nu}\partial_{y^i}f+ \sum_{(\sigma,i)\in
P}z^{i}_{\nu \sigma}\partial_{z^{i}_{\sigma}}f.
\]

For a given section $s\in \Gamma_{V}(\pi ),\ V\subset X$ we request the fulfilment of the equation
 \beq j^{1}(s)^{*}(i_{\xi}{\tilde d}\Theta_{\hat C_{-}})=0 \eeq
for large enough family of (locally defined) vector fields $\xi \in \mathcal{X}(Z_{p})$ (not necessary
vertical with respect to the projection $\pi^{1}:Z_{p}\rightarrow X$) guarantying the sections $s$ to be a
solution of the \emph{balance system} of $m$ independent balance equations.  Remark that vector fields
$\xi^1$ for $\xi \in \mathcal{V}(\pi)$ in the first subsection above are a special case of considered
vector fields (for $\xi \in \mathcal{V}(\pi)$,
$\xi^{1}=\xi^{j}\partial_{y^{j}}+d_{\mu}\xi^{j}\partial_{z^{j}_{\mu}}$).\par

 Thus, we take the $(n+1)+(n+2)$-form $\Theta_{{\hat C}_{-}}$ of the form (9.5) and apply
first the Iglesias differential ${\tilde d}$ and then $i_{\xi}$ for a vector field $\xi
=\xi^{\nu}\partial_{\nu}+ \xi^{j}\partial_{y^{j}}+\xi^{i}_{\mu}\partial _{z^{i}_{\mu}}$.\par We will
denote by $Con$ an arbitrary contact forms that appears in calculations.  Assuming the summation by
repeated indices agreement we recall that only $z^{i}_{\mu}$ or derivatives by these variables with $(\mu
,i)\in P$ are present on the formulas.\par

  We get, using that $d\eta_{\mu}=\lambda_{G,\mu}\eta$,

\begin{multline}
i_{\xi}{\tilde d}\Theta_{{\hat C}_{-}}=i_{\xi}[d(p\eta +F^{\mu}_{i} dy^i \wedge \eta_{\mu})+\Pi_{i}dy^i
\wedge \eta] = \\
i_{\xi}\left[ dp \wedge \eta +dF^{\mu}_i \wedge dy^i \wedge \eta_\mu -F^{\mu}_{i} dy^i \wedge d\eta_\mu
+\Pi_i
dy^i \wedge \eta \right]=\\
(i_\xi dp)\wedge \eta -dp\wedge i_{\bar \xi}\eta +(\xi \cdot F^{\mu}_{i})dy^i \wedge \eta_\mu -\xi^i
dF^{\mu}_i \wedge \eta_\mu +dF^{\mu}_i \wedge dy^i \wedge i_{\bar \xi}\eta_{\mu}-\\
-F^{\mu}_{i}\lambda_{G,\mu}(\xi^i \eta -dy^i \wedge i_{\bar \xi}\eta )+\Pi_{i}\xi^i \eta -\Pi_{i}dy^i
\wedge i_{\bar \xi}\eta =\\
= (\xi \cdot p)\eta +(\xi\cdot F^{\mu}_{i})(\omega^i +z^{i}_{\nu}dx^\nu )\wedge \eta_{\mu}-\xi^i
dF^{\mu}_i \wedge \eta_\mu  -\xi^i F^{\mu}_{i}\lambda_{G,\mu}\eta +\xi^i \Pi_{i}\eta -\\
-dp\wedge i_{\bar \xi}\eta +dF^{\mu}_{i}\wedge (\omega^i +z^{i}_{\nu}dx^\nu ) \wedge i_{\bar \xi}\eta_\mu
+\lambda_{G,\mu}F^{\mu}_{i}dy^i \wedge i_{\bar \xi}\eta - \Pi_{i}(\omega^i +z^{i}_{\nu}dx^\nu ) \wedge
i_{\bar \xi}\eta ] =
\end{multline}
Now we are using the fact that $dF^{\mu}_{i}=(d_\nu F^{\mu}_{i})dx^\nu +F^{\mu}_{i,y^j}\omega^j
+F^{\mu}_{i,z^{j}_{\nu}}\omega^{j}_{\nu}$ and, similarly, $dp=d_{\nu}dx^\nu +Con$ and continue
\begin{multline}
=(\xi \cdot p)\eta +(\xi\cdot F^{\mu}_{i})z^{i}_{\nu}dx^\nu \wedge \eta_{\mu}-\xi^i d_{\nu}F^{\mu}_i
dx^\nu \wedge \eta_\mu -\xi^i F^{\mu}_{i}\lambda_{G,\mu}\eta+\xi^i \Pi_{i}\eta +\\
+[-d_{\nu}pdx^\nu -\Pi_{i}z^{i}_{\nu}dx^\nu +\lambda_{G,\mu}F^{\mu}_{i}z^{i}_{\nu}dx^\nu ]\wedge i_{\bar
\xi}\eta +(d_{\nu}F^{\mu}_{i})dx^{\nu}\wedge z^{i}_{\lambda}dx^\lambda \wedge i_{\bar \xi}\eta_\mu +Con =\\
[ \xi \cdot p +(\xi \cdot F^{\mu}_{i})z^{i}_{\mu}-\xi^i d_{\mu}F^{\mu}_i -\xi^i
F^{\mu}_{i}\lambda_{G,\mu}+\xi^i \Pi_{i}]\eta + [-d_{\nu}p -\Pi_{i}z^{i}_{\nu}
+\lambda_{G,\mu}F^{\mu}_{i}z^{i}_{\nu}]dx^\nu \wedge \xi^{\sigma}\eta_{\sigma}+\\
+(d_{\nu}F^{\mu}_{i})dx^{\nu}\wedge z^{i}_{\lambda} dx^\lambda \wedge \xi^{\sigma}\eta_{\mu \sigma}+Con=.
\end{multline}

Now we will use the formula (16.5, Appendix I) for $dx^\lambda \wedge \eta_{\mu \sigma}$ from which it
will follow that if $\nu =\mu ,\lambda =\sigma$, then $dx^\nu \wedge dx^\lambda \wedge \eta_{\mu
\sigma}=\eta$ and when $\nu =\sigma ,\lambda =\mu$, then $dx^\nu \wedge dx^\lambda \wedge \eta_{\mu
\sigma}=-\eta.$ Using this in the last term in the previous formula we get

\beq  (d_{\nu}F^{\mu}_{i})dx^{\nu}\wedge z^{i}_{\lambda} dx^\lambda \wedge \xi^{\sigma}\eta_{\mu \sigma}=
[z^{i}_{\sigma}d_{\mu}F^{\mu}_{i}\xi^\sigma -z^{i}_{\mu}\xi^{\sigma}d_{\sigma}F^{\mu}_{i}]\eta.
 \eeq
Using this result we continue

\begin{multline}
=[ \xi \cdot p +(\xi\cdot F^{\mu}_{i})z^{i}_{\mu}-\xi^i d_{\mu}F^{\mu}_i -\xi^i
F^{\mu}_{i}\lambda_{G,\mu}+\xi^i \Pi_{i}+[-d_{\nu}p -\Pi_{i}z^{i}_{\nu}
+\lambda_{G,\mu}F^{\mu}_{i}z^{i}_{\nu}] \xi^{\nu} + \\ +z^{i}_{\sigma}d_{\mu}F^{\mu}_{i}\xi^\sigma
-z^{i}_{\mu}\xi^{\sigma}d_{\sigma}F^{\mu}_{i} ]\eta +Con=\\
= \{ \xi \cdot (p+z^{i}_{\mu}F^{\mu}_{i})-\xi^{i}_{\mu}F^{\mu}_{i}
-\xi^{\nu}d_{\nu}(p+z^{i}_{\mu}F^{\mu}_{i})+\xi^{\sigma}z^{i}_{\mu\sigma}F^{\mu}_{i}+
\xi^{\sigma}z^{i}_{\sigma}[d_{\mu}F^{\mu}_{i}+\lambda_{G,\mu}F^{\mu}_{i}-\Pi_{i}]-\\-
\xi^{i}[d_{\mu}F^{\mu}_{i}+\lambda_{G,\mu}F^{\mu}_{i}-\Pi_{i}]\}\eta+Con=\\=\{(\xi
-\xi^{\nu}d_{\nu})(p+z^{i}_{\mu}F^{\mu}_{i})-\xi^{i}_{\mu}F^{\mu}_{i}+\xi^{\sigma}z^{i}_{\mu\sigma}F^{\mu}_{i}+
(\xi^{\sigma}z^{i}_{\sigma} -\xi^{i})[d_{\mu}F^{\mu}_{i}+\lambda_{G,\mu}F^{\mu}_{i}-\Pi_{i}]\}\eta +Con
\end{multline}
We notice now that for a function $f\in C^{\infty}(Z_{p}),$ lifted to the partial 2-jet bundle
$J^{2}_{p}(\pi)$, $\xi \cdot f
-\xi^{\nu}d_{\nu}f=(\xi^{i}-\xi^{\nu}z^{i}_{\nu})\partial_{y^i}f+(\xi^{i}_{\sigma}- \xi^{\nu}z^{i}_{\sigma
\nu})\partial_{z^{i}_{\sigma}}f.$  Starting from this moment we assume that all the forms are lifted to
$J^{2}_{p}(\pi)$, vector fields are flow prolonged there (we will see that our considerations do not
depend on this prolongation).  Applying this for $f=p+z^{i}_{\mu}F^{\mu}_{i}$ we see that the expression
in figure brackets is equal to

\begin{multline} (\xi^{\sigma}z^{i}_{\sigma} -\xi^{i})[d_{\mu}F^{\mu}_{i}+\lambda_{G,\mu}F^{\mu}_{i}-\Pi_{i}]
+(\xi^{\sigma}z^{i}_{\mu\sigma}-\xi^{i}_{\mu}) F^{\mu}_{i}+
(\xi^{i}-\xi^{\nu}z^{i}_{\nu})\partial_{y^i}(p+z^{j}_{\mu}F^{\mu}_{j})+(\xi^{i}_{\sigma}-
\xi^{\nu}z^{i}_{\sigma \nu})\partial_{z^{i}_{\sigma}}(p+z^{j}_{\nu}F^{\nu }_{j})=\\=
(\xi^{\sigma}z^{i}_{\sigma}
-\xi^{i})[d_{\mu}F^{\mu}_{i}+\lambda_{G,\mu}F^{\mu}_{i}-\Pi_{i}-\partial_{y^i}(p+z^{j}_{\nu}F^{\nu}_{j})]+
(\xi^{\sigma}z^{i}_{\mu\sigma}-\xi^{i}_{\mu})[
F^{\mu}_{i}-\partial_{z^{i}_{\mu}}(p+z^{j}_{\nu}F^{\nu}_{j})].  \end{multline}

Thus, we get, finally

\begin{multline} i_{\xi}{\tilde d}\Theta_{{\hat C}_{-}}=\\ =\{(\xi^{\sigma}z^{i}_{\sigma}
-\xi^{i})[d_{\mu}F^{\mu}_{i}+\lambda_{G,\mu}F^{\mu}_{i}-\Pi_{i}-\partial_{y^i}(p+z^{j}_{\nu}F^{\nu}_{j})]+
(\xi^{\sigma}z^{i}_{\mu\sigma}-\xi^{i}_{\mu})[
F^{\mu}_{i}-\partial_{z^{i}_{\sigma}}(p+z^{j}_{\nu}F^{\nu}_{j})]\}\eta +Con =\\
\{-\omega^{i}(\xi)[d_{\mu}F^{\mu}_{i}+\lambda_{G,\mu}F^{\mu}_{i}-\Pi_{i}-\partial_{y^i}(p+z^{j}_{\nu}F^{\nu}_{j})]-
\omega^{i}_{\mu}(\xi)[ F^{\mu}_{i}-\partial_{z^{i}_{\mu}}(p+z^{j}_{\nu}F^{\nu}_{j})]\}\eta +Con
 .\end{multline}
Here
\[ \begin{cases} \omega^{i}=dy^i-\sum_{(\mu ,i)\in P}z^{i}_{\mu}dx^\mu , \\ \omega^{i}_{\mu}=dz^{i}_{\mu}-\sum_{(\mu ,i)\in P}z^{i}_{\mu
\sigma}dx^\sigma \end{cases}\]

are generating (partial) contact forms on the bundle $J^{1}(J^{}_{p}(\pi))$.
\begin{remark}
 Notice that the quantities
$Q_{i}=\omega^{i}(\xi)= \xi^{i}-z^{i}_{\mu}\xi^{\mu}$ form the \emph{characteristic of the vector field}
$\xi =\xi^{\mu} \partial_{\mu}+ \xi^{i}\partial_{y^i}$, see \cite{O1}, Ch.2.\end{remark}
\par

These arguments proves the following
\begin{proposition} Let $\hat C$ be a CCR defined in a domain of the partial 1-jet bundle $Z_{p}$.  Then,
for any $\xi \in \mathcal{X}(Z_{p})$,

\beq i_{\xi}{\tilde d}\Theta_{{\hat C}_{-}}=- \omega^{1}_{\widehat{C}}(\xi)-\omega^{2}_{\widehat{C}}(\xi),
\eeq
 where
 \beq \begin{cases}
\omega^{1}_{\hat C}(\xi)=\omega^{i}(\xi)[d_{\mu}F^{\mu}_{i}+\lambda_{G,\mu}F^{\mu}_{i}-\Pi_{i}-\partial_{y^i}(p+z^{j}_{\nu}F^{\nu}_{j})],\\
\omega^{2}_{\hat C}(\xi) =\sum_{(\mu ,i)\in
P}(F^{\mu}_{i}-\partial_{z^{i}_{\mu}}(p+z^{j}_{\nu}F^{\nu}_{j}))\omega^{i}_{\mu}(\xi)
\end{cases}\eeq
will be called respectively as the \textbf{first (\emph{Euler-Lagrange}) and the second contact forms of
the covering constitutive relation} $C$.
\end{proposition}
\begin{remark} Calculation leading to the last Proposition is valid in a case of a covering constitutive relation
depending on the derivatives of higher order.  Such a CCR, which we write, for (formal) simplicity as
defined on the infinite jet bundle of the bundle $\pi$ (see Appendix IV or \cite{FF,KV,O1}:
\[
{\hat C}:J^{\infty}(\pi) \rightarrow \Lambda^{(n+1)+(n+2)}(Y)
\]
but depending on derivatives of order $\leqq N$ defines in the same way the Poincare-Cartan form
$\Theta_{\hat C}$ and we can formulate the balance system in the same way, postulating the fulfillment of
the equation $j^{\infty}(s)^{*}(i_{\xi}{\tilde d}\Theta_{{\hat C}_{-}})=0$ for variations $\xi \in
\mathcal{X}(J^{infty}(\pi))$ in the number sufficient for separating the balance laws.  Using the
corresponding properties of higher order contact form
$\omega^{i}_{\Lambda}=dz^{i}_{\Lambda}-\sum_{\lambda}z^{i}_{\Lambda +\lambda}dx^\lambda$, $\Lambda$ being
a multi-index and total derivatives (see Appendix IV), for instance,
\[
dz^{i}_{\Lambda}=\omega^{i}_{\Lambda}+Con,\ (\xi-\xi^{\sigma}d_{\sigma})\cdot f
=\omega^{i}(\xi)f_{,y^i}+\sum_{\Lambda}\omega^{i}_{\Lambda}(\xi)f_{,z^{i}_{\Lambda}},
\]
we get the result similar to (10.24):

\beq i_{\xi}{\tilde d}\Theta_{{\hat C}_{-}}=-
\omega^{1}_{\widehat{C}}(\xi)-\omega^{2}_{\widehat{C}}(\xi)+\sum_{\Lambda \vert \vert\Lambda \vert
>1}\omega^{i}_{\Lambda}(\xi)\partial_{z^{i}_{\Lambda}}(p+z^{i}_{\mu}F^{\mu}_{i}).
\eeq It follows from this that for the CCR $\tilde C$ of a constitutive relation $C$ of higher order where
$p+z^{i}_{\mu}F^{\mu}_{i}=0$ no new limitations for the admissible variations $\xi$ beyond those
considered here will appear.  More detailed study of constitutive relations of higher order will be done
in the continuation of this work.
\end{remark}
\begin{remark} Equality (10.23) contains the jet variables of the second order $z^{i}_{\mu \sigma}$.  Yet,
\textbf{only} $\omega^{i}_{\mu}$ with $(\mu ,i)\in P$ \textbf{are present in the formula} (10.23)!  For
instance in the RET case \emph{all these terms are absent} from (10.23).\par

In order that the equation resulting from taking the pullback by $j^{1}(s)$ would not depend on the
variables not in $J^{1}_{p}(\pi)$ we require that all the coefficients of these variables would be zero.
This leads to the strong conditions to the admissible variations $\xi$.
\end{remark}

\begin{proposition} Let $\hat C$ be a covering constitutive relation.  The following properties 1),2) are equivalent
\begin{enumerate}
\item Second contact form  of CCR $\hat C$ is identically zero:
\beq \omega_{\hat C} =F^{\mu}_{i}-\partial_{z^{i}_{\mu}}(p+z^{j}_{\nu}F^{\nu}_{j})\omega^{i}_{\mu}=0, \eeq
\item Locally
\[
F^{\mu}_{i}=\partial_{z^{i}_{\mu}}L,\ L\in C^{\infty}(Z_{p}),
\]
i.e. CR $\mathcal{C}$ is (locally) semi-Lagrangian.
\item If properties 1),2) are fulfilled, then
\[
p=L-z^{i}_{\mu}\partial_{z^{i}_{\mu}}L +l(x,y)
\]
with an arbitrary function $l(x,y).$

\end{enumerate}
\end{proposition}
\begin{proof}Rewrite the first equality as
\beq
\partial_{z^{i}_{\sigma}}p=-z^{i}_{\mu}\partial_{z^{i}_{\sigma}}F^{\mu}_{i}.
\eeq

 Right sides of these equalities satisfy to the mixed derivative test
\[
\partial_{z^{j}_{\lambda}}(-z^{k}_{\mu}\partial_{z^{i}_{\sigma}}F^{\mu}_{k})=
\partial_{z^{i}_{\sigma}}(-z^{k}_{\mu}\partial_{z^{j}_{\lambda}}F^{\mu}_{k}),
\]
or $\partial_{z^{j}_{\lambda}}F^{\sigma}_{i}=\partial_{z^{i}_{\sigma}}F^{\lambda}_{j}.$  From this
equality valid for all couples of indices  $(j,\lambda ),(i,\sigma )$ the second statement follows.  \par
To prove the opposite - reverse the arguments.\par

\end{proof}

\begin{theorem} Let $\mathcal{C}_{L,\Pi}$ be a semi-Lagrangian CR: $F^{\mu}_{i}=L_{,z^{i}_{\mu}}$. Let $\hat C_{L,\Pi}$ be a CCR covering $\mathcal{C}$ and such
that $p=L-z^{i}_{\mu}\partial_{z^{i}_{\mu}}L +l(x,y)$ with some function $l(x,y)$.  Then,
\begin{enumerate}
\item Equality 10.23 takes the form
\[
i_{\xi}{\tilde d}\Theta_{{\hat C_{L,\Pi,-
}}}=-\omega^{i}(\xi)[d_{\mu}L_{,z^{i}_{\mu}}+\lambda_{G,\mu}L_{,z^{i}_{\mu}}-\Pi_{i}-\partial_{y^i}(L+l(x,y))]+Con.
\]
\item Following statements are equivalent
\begin{enumerate}
\item For  a section $s\in \gamma(\pi)$ and for all  $\xi \in \mathcal{X}(Z_{p})$
\[ j^{1}_{p}(s)^{*}i_{\xi}{\tilde d}\Theta_{{\hat
C_{L,\Pi\ -}}}=0.  \]
\item Section $s$ is the solution of the system of balance equations

\beq (L_{,z^{i}_{\mu}}\circ j^{1}_{p}(s))_{,\mu} +(\lambda_{G,\mu}\circ s) L_{,z^{i}_{\mu}}\circ
j^{1}_{p}(s)=\Pi_{i}\circ j^{1}_{p}(s)+\partial_{y^i}(L+l(x,y)).
\eeq
\end{enumerate}
\end{enumerate}
\end{theorem}
\begin{proof}
Proof of the first statement follows from the equation (10.23) if we substitute
$F^{\mu}_{i}=\partial_{z^{i}_{\mu}}L$.\par Equivalence of the statements in second part follows from the
fact that when we will apply the pullback by the 1-jet section $j^{1}_{p}(s)$ of the bundle
$\pi^{1}:Z_{p}\rightarrow X$ to the equality in the first statement of Theorem, contact form vanishes and
from the fact that \emph{locally} there always exist $m$ linearly independent vector fields $\xi_{k}$ such
that the $m\times m$-matrix $\omega^{i}(\i_{k})$ is invertible.
\end{proof}

\begin{corollary} Let $C_{L,\Pi}$ be a semi-Lagrangian  CR: $F^{\mu}_{i}=L_{z^{i}_{\mu}}$. Then,
\begin{enumerate}
\item For $\Pi_{i}=l(x,y)=0$ the balance system (10.29) takes the form of the system of Euler-Lagrange
Equations (10.14).
\item For $\Pi_{i}=\Pi^{1}_{i}-\partial_{y^i}(L+l(x,y))$, the balance system (10.29) takes the form of the
balance system $\bigstar$ with the source term $\Pi^{1}_{i}dy^i \wedge \eta$.
\end{enumerate}
\end{corollary}
\vskip0.5cm

Consider now the case of a RET constitutive relation $C:Y\rightarrow {\tilde Z}$ and a CCR ${\hat C}:$
covering $C$ (i.e. having the same (1,n)- and (n+2)- components but an arbitrary (0,n+1)-component
$p(x,y)$, with the Poincare-Cartan form

\[
\Theta_{\hat C}=p\eta +F^{\mu}_{i}dy^i \wedge \eta_{\mu}+\Pi_{i}dy^i \wedge \eta.
\]

Following the arguments leading to the basic relation (10.23), namely continuing with the first equality
in (10.21) we get, due to the independence of $\hat C$ on the jet variables
\begin{multline}
i_{\xi}{\tilde d}\Theta_{{\hat C}_{-}}=(\xi^{i}-z^{i}_{\nu}\xi^\nu )\Pi_{i}+(-\xi^{i}+z^{i}_{\nu}\xi^\nu
)(d_{\mu}F^{\mu}_{i}+F^{\mu}_{i}\lambda_{G,\mu})+(\xi\cdot p -\xi^{\nu}d_{\nu}p )+z^{i}_{\mu}(\xi\cdot
F^{\mu}_{i}-\xi^{\sigma}d_{\sigma}F^{\mu}_{i})+Con=\\
=(z^{i}_{\nu}\xi^\nu-\xi^{i}
)(d_{\mu}F^{\mu}_{i}+F^{\mu}_{i}\lambda_{G,\mu}-\Pi_{i})+\omega^{i}(\xi)\partial_{y^i}(p+z^{j}_{\mu}F^{\mu}_{j})+Con=\\
-\omega^{i}(\xi)\left[
 d_{\mu}F^{\mu}_{i}+F^{\mu}_{i}\lambda_{G,\mu}-\Pi_{i}-\partial_{y^i}(p+z^{i}_{\mu}F^{\mu}_{i})
 \right]+Con.
\end{multline}
Here we have used the fact $\xi\cdot p -\xi^{\nu}d_{\nu}p
=\xi^{i}p_{,y^i}+\xi^{\sigma}p_{,x^\sigma}-\xi^{\nu}(p_{,x^\nu}+z^{i}_{\nu}p_{,y^i})=(\xi^{i}-z^{i}_{\nu}\xi^\nu
)\partial_{y^i}p=\omega^{i}(\xi )\partial_{y^i}p$ and, similarly, $z^{i}_{\mu}(\xi\cdot
F^{\mu}_{i}-\xi^{\sigma}d_{\sigma}F^{\mu}_{i})=z^{i}_{\mu}\omega^{j}(\xi)\partial_{y^j}F^{\mu}_{i}$ (we
remind that in the RET case all functions depend on $(x,y)$ only).
\par
If we take ${\hat C}={\tilde C}$ to be lifted CCR then the term $\partial_{y^i}(p+z^{i}_{\mu}F^{\mu}_{i})$
vanishes and, using the Proposition 18 above we finish the proof of the following

\begin{theorem} Let $C$ be a constitutive relation of the RET type and $\tilde C$ - corresponding lifted CCR. Then the
following statements are equivalent
\begin{enumerate}
\item For a given section $s\in \Gamma(\pi)$ and for all $\xi \in \mathcal{X}(Y)$
\beq s^{*}(i_{\xi}{\tilde d}\Theta_{{\hat C}_{-}})= 0\eeq
\item Section $s$ is the solution of the balance system $\bigstar$.
\end{enumerate}
\end{theorem}

Consider now the \emph{general case} but take $p+z^{j}_{\nu}F^{\nu}_{j}=0$ i.e. consider ${\hat C}=\tilde
C$ to be the \emph{lifted CCR }of a CR $\mathcal{C}$ (see previous section). Then equality (10.23-24)
takes the form

\beq i_{\xi}{\tilde d}\Theta_{{\tilde C}_{-}}=\{(\xi^{\sigma}z^{i}_{\sigma}
-\xi^{i})[d_{\mu}F^{\mu}_{i}+\lambda_{G,\mu}F^{\mu}_{i}-\Pi_{i}]+
(\xi^{\sigma}z^{i}_{\mu\sigma}-\xi^{i}_{\mu}) F^{\mu}_{i}\}\eta +Con \eeq

 Taking the pullback of
equality (10.23) by the 1-jet $j^{1}(s)$ of a section $s\in \Gamma(\pi)$ we get
\begin{multline}
j^{1}(s)^{*}i_{\xi}{\tilde d}\Theta_{{\tilde C}_{-}}=j^{1}(s)^{*}((\xi^{\sigma}z^{i}_{\mu
\sigma}-\xi^{i}_{\mu})F^{\mu}_{i})\eta +
j^{1}(s)^{*}(\xi^{\sigma}z^{i}_{\sigma}-\xi^{i})[d_{\mu}F^{\mu}_{i}+\lambda_{G,x^\mu}F^{\mu}_{i}-\Pi_{i}]\eta.
\end{multline}
Equating this to zero we see that if we want the $\xi$-weighted balance equation
\[
j^{1}(s)^{*}(\xi^{\sigma}z^{i}_{\sigma}-\xi^{i})[d_{\mu}F^{\mu}_{i}+\lambda_{G,x^\mu}F^{\mu}_{i}-\Pi_{i}]=0
\]
to be true for a section $s$ \emph{under some conditions independent on section $s$} (i.e. independent on
the $z^{i}_{\mu \sigma}$!) we have to require that for the vector field $\xi$ used for the "variation"
\[
\omega^{2}_{\hat C}(\xi) =(\xi^{\sigma}z^{i}_{\mu \sigma}-\xi^{i}_{\mu})F^{\mu}_{i}=0.
\]
Since $F^{\mu}_{i}$ and $\xi$ do not depend on  is $z^{i}_{\mu \sigma}$ vector field $\xi$ should be
 such that $\xi^\sigma =0$ for $\sigma$ such that \emph{for some} $i$ $(\sigma ,i)\in P$
(this is trivially true if $\xi$ is $\pi^{1}$-\emph{vertical}).\par  Then the $\xi$-weighted balance
equation
 for such vector field $\xi$ will be fulfilled if and only if
 $j^{1}(s)^{*}(\xi^{i}_{\mu}F^{\mu}_{i})=0$ and, for $s$-independent condition we have to require
 $\xi^{i}_{\mu}F^{\mu}_{i}=0$.  This brings us to the extension of the condition $FDiv(\xi)=0$ to the
space of all $\pi^1$-vertical vector fields (see above)
\begin{definition}\begin{enumerate}
\item A vector field $\xi \in X(Z_{p})$ is called $P$-\textbf{vertical} if $\xi^\sigma =0$ for $\sigma$ such
 that \emph{for some} $i$ $(\sigma ,i)\in P$.
\item
 A $\pi_{10}$-projectable $P$-vertical vector field $\xi \in X(U),U\subset Z_{p}$ is called
 $\mathcal{C}$-\textbf{admissible} if $\sum_{(\mu ,i)\in P}\xi^{i}_{\mu}F^{\mu}_{i}=0$
in $U$.\par

Denote by $\mathcal{X}(\mathcal{C})$ the sheaf generated by the pre-sheaf of the $C$-admissible vector
fields in $Z_{p}$.
\item A constitutive relation ${\mathcal{C}}$ is called \textbf{separable in} $U$ if the space of
projections to $Y$ of the space of $C$-admissible vector fields in $U$ has, at each point $\pi_{10}(U)$
dimension $m$.
\end{enumerate}
\end{definition}
\begin{remark} Flow prolongation $\xi^1$ of the $C$-admissible $\pi$-vertical vector fields
 $\xi \in \mathcal{X}(Y)$ is a special case of $C$ admissible vector field $\xi \in \mathcal{X}(Z_{p}).$
 Notice, though, that while the condition  $F^{\mu}_{i}\xi^{i}_{\mu}$ is \emph{linear algebraic} for a general vector
 field $\xi\in \mathcal{X}(Z_{p})$,  it is differential for the lifts $\xi^1$ of vector fields $\xi \in
 \mathcal{X}(Y)$ (see next section for examples of specific forms of these relations).
\end{remark}
\begin{example} Let $Z_{p}=J^{1}_{<\partial_{x}>}(\pi)$ be a partial 1-jet bundle defined by the
distribution $<\partial_{x}>$. In other words we assume that a constitutive relation $C$ depends on the
spacial but not on the time variables of the fields $y^i$. Then a vector field $\xi$ is $P$-vertical if
$\xi^{x^A}=0,\ A=1,2,\ldots ,n$ while component $\xi^{0}$ corresponding time derivative $\partial_{t}$ may
be arbitrary.
\end{example}

Thus, previous arguments proves the following

\begin{theorem}
If a constitutive relation $\mathcal{C}$ is locally separable, then the following  statements for a
section $s\in \Gamma(\pi)(U),\ U\subset X$ are equivalent:
\begin{enumerate}
\item
\beq j^{1}(s)^{*}(i_{\xi}{\tilde d}\Theta_{{\tilde C}_{-}})=0 \ \text{for\ all}\ \xi \in
\mathcal{X(C}\vert_{U}).\eeq
\item Section $s$ is the solution of the following system of balance laws - \textbf{balance system}:
\[
 (F^{\mu}_{i}\circ j^{1}_{p}(s))_{,x^{\mu}}+F^{\mu}_{i}\circ j^{1}_{p}(s)(\partial_{x^{\mu}}\lambda_{G})=
 \Pi_{i}(j^{1}_{p}(s)),\ i=1,\ldots, m.\hskip3cm (\bigstar)
\]
\end{enumerate}
\end{theorem}

\vskip0.5cm

\subsection{Reduced horizontal differential formulation of the balance system.}

Recall (see \cite{KV,GMS} or Appendix IV) that the reduced horizontal differential $\hat d$ acts from
$J^{k}(\pi)$ to $J^{k+1}(\pi)$ for all $k$ by the formulas (20.9-10).\par

Now, let us postulate the balance system corresponding to the CR $\mathcal{C}_{-}$ in the form

\beq  j^{1\ *}_{p}(s) i_{\xi^1}{\tilde {\hat d}} \Theta_{\mathcal{C}_{-}}=0, \eeq for \emph{all}
variations $\xi \in
 \mathcal{X}(Z_{p})$.\par
Notice that the additional term in $\Theta^{n+1}_{C_{-}}$ of the form $h(z)\eta$ produced by an arbitrary
transformation by an adopted transformation $\phi \in Aut(\pi)$ \emph{will be eliminated} by applying the
reduced horizontal differential $\hat d$ (see formula (20.9-10), Appendix IV) from which it follows that
$\hd (q\eta)=0$), so that this equation is independent on a choice of representation of the
Poincare-Cartan form $\widetilde{\Theta}_{C}$.\par We have, \emph{for any vector field} $\xi \in
\mathcal{X}(Z_{p})$

\begin{multline}
i_{\xi}{\tilde {\hat d}}\Theta_{\mathcal{C}_{-}}=i_{\xi}[{\hat d}(F^{\mu}_{i}dy^i\wedge \eta_{\mu})
+\Pi_{i}dy^i \wedge \eta ]=i_{\xi}[-(d_\mu F^{\mu}_{i})dy^i\wedge \eta -F^{\mu}_{i}dy^i\wedge {\hat
d}\eta_{\mu})
+\Pi_{i}dy^i \wedge \eta ]=\\
=i_{\xi}[ (-d_\mu F^{\mu}_{i}-F^{\mu}_{i}\lambda_{G,x^\mu}+\Pi_{i} )dy^i \wedge \eta ] =\xi^{i}[-d_\mu
F^{\mu}_{i}-F^{\mu}_{i}\lambda_{G,x^\mu }+\Pi_{i}]\eta -\\ - \xi^{\mu}[-d_\mu
F^{\mu}_{i}-F^{\mu}_{i}\lambda_{G,x^\mu } +\Pi_{i}]dy^i \wedge \eta_{\mu},
\end{multline}
since ${\hat d}\eta_{\mu}=\lambda_{G,x^\mu }\eta.$ Now we take the pullback by the section $j^{1}_{p}(s)$
and get \beq i_{\xi}{\tilde {\hat d}}\Theta_{\mathcal{C}_{-}}= [-d_\mu
F^{\mu}_{i}-F^{\mu}_{i}\lambda_{G,x^\mu }+\Pi_{i}](\xi^{i}-\xi^{\mu}s^{i}_{,\mu})\eta \eeq

 Requiring (10.35) to be fulfilled \textbf{for all} vector fields $\xi \in
\mathcal{X}(Z_{p})$ we guarantee the possibility to have, for a given section $s$ an $m$ vector fields
$\xi$ such that for their component $\xi^{i}\partial_{y^i}+\xi^{\mu}\partial_{\mu}$ the differences
$(\xi^{i}-\xi^{\mu}s^{i}_{,\mu})$ are linearly independent in a neighborhood of any point $x\in X$ (we
actually can choose these vector fields to be 1-jet prolongations of vector fields $\xi \in
\mathcal{X}_{\pi}(Y)$). Therefore, the condition (10.35) will be fulfilled for all vector fields $\xi \in
\mathcal{X}(Z_{p})$ if and only if the balance system of equations $\bigstar $
\[
j^{1\ *}_{p}(s)[d_\mu F^{\mu}_{i}+F^{\mu}_{i}\lambda_{G,x^\mu }]=\Pi_{i}(j^{1}_{p}(s)),\ i=1.\ldots ,m
\]
is satisfied by the section $s$. Thus, we get

\begin{theorem} Let $\mathcal C$ be a constitutive relation.  For a section $s\in \Gamma (\pi)$ the following
 statements are equivalent
\begin{enumerate}
\item For all vector fields $\xi \in
\mathcal{X}(Z_{p}),$
\[
j^{1\ *}_{p}(s) i_{\xi^1}{\tilde {\hat d}} \Theta_{\mathcal{C}_{-}}=0.
\]

\item Section $s$ is the solution of the balance system $\bigstar $

\[
j^{1\ *}_{p}(s)[d_\mu F^{\mu}_{i}-F^{\mu}_{i}\lambda_{G,x^\mu }]=\Pi_{i}(j^{1}_{p}(s)),\ i=1.\ldots
,m.\hskip3cm (\star)
\]
\end{enumerate}
\end{theorem}

\begin{remark} If we would like to use the conventional horizontal differential $d_H$ instead of reduced
one in the formulated above we would still remove the term of the form $q(z)\eta$ of the Poincare-Cartan
form of CR $\mathcal{C}$, but in the calculation above we would get, for a \emph{vertical vector field}
$\xi$ on $Y$ an extra term
 \[i_{\xi^1}[F^{\mu}_{i}d_{H} dy^i \wedge
\eta_{\mu}]=i_{\xi^1}[F^{\mu}_{i}dx^\lambda \wedge dz^{i}_{\lambda} \wedge
\eta_{\mu}]=-i_{\xi^1}[F^{\mu}_{i}dz^{i}_{\mu}\wedge \eta]=F^{\mu}_{i}\xi^{1\ i}_{\mu}\eta \]
 and, after
taking the pullback by a section $s:X\rightarrow Y$ we would get an extra term
$(F^{\mu}_{i}d_{\mu}\xi^{i})\eta.$  This brings us back to the requirement that vector field $\xi$ of
variation is $C$-admissible.
\end{remark}

 \vfill \eject

\section{$\mathcal{C}$-admissible vector fields.}
In this section we will start studying the vector space $\mathcal{X(C)}$ of $C$-admissible vertical vector
fields $\xi=\xi^{i}\partial_{y^i} \in V(\pi),$ i.e. vector fields satisfying to the condition

\[
\omega^{2}_{C}(\widehat{\xi} )=FDiv(\xi)=F^{\mu}_{i}d_{\mu}\xi^{i}=0.
\]
Here $\widehat{\xi}$ is the (arbitrary) lift of vector field $\xi\in \mathcal{X}(\pi)$.  Let $\phi \in
Aut_{p}(\pi )$ be an automorphism of the bundle $\pi$ and let
$\xi=\xi^{\mu}\partial_{\mu}+\xi^{i}\partial_{y^i} \in \mathcal{X}(\pi)$
 be any projectable vector field in $Y$. Then we have

 \beq \phi_{*}(\xi)=({\bar \phi}^{\mu}_{,\nu }\xi^\nu )\partial_{\mu}
 +(\phi^{i}_{,\nu}\xi^{\nu}+\phi^{j}_{,y^i} \xi^{i})\partial_{y^j}. \eeq

Let $\xi=\xi^{i}\partial_{y^i}\in \mathcal{X(C)}$ and let $\phi \in Aut_{p}(\pi )$ be as above.
\begin{lemma} For a transformed constitutive relation  $C^{\phi}=\widetilde{\phi}^{*}\circ C\circ \phi^{1\ -1}$
we have $Fdiv(C^\phi)=\phi^{*}Fdiv(C)$.
\end{lemma}
\begin{proof} By the (12.5-6), (where we substitute $\phi^{-1}$ for $\phi$!) for the transformed CR $C^{\phi}$ the (1,n)-component of its Poincare-Cartan
form  $\Theta_{C,\nu}$ is transformed as follows
\[
F^{\mu}_{i}dy^{i}\wedge \eta_{\mu}\rightarrow  detJ({\bar \phi}^{-1})J({\bar
\phi})^{\mu}_{\nu}(F^{\nu}_{j}\circ (\phi^{-1})^{1})(\phi^{-1})^{j}_{,y^i}dy^i \wedge \eta_{\nu}.
\]
On the other hand, by () the $\pi_{10}$-vertical component of a $\pi^1$ vertical vector field $\xi$
transforms under the flow lifted transformation of $Z_{p}$ as
$\xi^{i}_{\mu}\partial_{z^{i}_{\mu}}\rightarrow \left(\xi^{i}_{\mu}J(\phi)^{i}_{j}J({\bar
\phi}^{-1})^{\lambda}_{\mu} \right)\partial_{z^{j}_{\lambda}}.$ Combining these two laws of transformation
we see that
\[
F(C^\phi )^{\mu}_{i}(\phi_{*}\xi)^{i}_{\mu}=detJ({\bar
\phi}^{-1})\left(F(C)^{\mu}_{i}\xi^{i}_{\mu}\right),
\]
and, therefore, these two quantities equals zero simultaneously.
\end{proof}
Considering change of local admissible variables and corresponding local automorphism in the intersection
of the domains of local charts we see that

\begin{corollary}  Condition $F^{\mu}_{i}\xi^{i}_{\mu}=0$ defining the class of
$\pi^{1}$-vertical vector fields $\mathcal{X(C)}$ is independent on the local adopted chart $(x^\mu
,y^i)$.\end{corollary}
\par

A natural question that leads directly to the "entropy condition" for a balance system (9.13) (see Part II
of this work for more details)
\[
(F^{\mu}_{i}\circ j^{1}_{p}(s))_{,x^{\mu}}=\Pi_{i}(j^{1}_{p}(s)),\ i=1,\ldots, m,\hskip 4cm (\bigstar )
\]
defined by a CR $\mathfrak{C}$ is - are there, except of the linear combinations of balance equations in
the system  $\star$, balance laws for the bundle $Y^{1}_{p}(\pi)$ that follows from  the balance system
$\star$ in the following sense:

\begin{definition} Fix a CR $\mathcal{C}$ and consider the corresponding balance system ($\star$).
  We call a balance law
  \beq (K^{\mu}\circ
j^{1}_{p}(s))_{,\mu}=Q\circ j^{1}_{p}(s). \eeq of the same type (i.e. with the coefficients defining on
$J^{1}_{p}(\pi)$) given by a (n+1)+(n+2)-form $K^{\mu}\eta_{\mu}+Q\eta$ on the space $Y^{1}_{p}(\pi)$
\textbf{generated by the CR} $\mathcal{C}$ (or the \textbf{secondary balance laws} for the system
($\bigstar$)) if any solution $s:X\rightarrow Y$ of the balance system ($\bigstar$) is at the same time
solution of the balance law (11.2).
\end{definition}

All the balance laws that follows from the balance system (including the balance laws in the system
($\bigstar$) themselves and their linear combinations) form the vector space $\mathcal{BL}_{\mathcal{C}}.$
The simple class of secondary balance laws beyond the linear combinations of the balance laws of the
system ($\bigstar$) is determined by the following

\begin{proposition} Let a vertical vector field $\xi =\xi^i \partial_{y^i}\in V(\pi)$
 belongs to the $\mathcal{X(C)}$, i.e. the condition $FDiv (\xi)=0$ is fulfilled. Then the balance law
\[
j^{1\ *}_{p}(s)d(\xi^{i}F^{\mu}_{i}\eta_{\mu})=\xi^{i}\Pi_{i}\eta \Leftrightarrow
((\xi^{i}F^{\mu}_{i})\circ j^{1}(s))_{,x^\mu}=(\xi^{i}\Pi_{i})\circ j^{1}(s)
\]
belongs to the space $\mathcal{BL}_{\mathcal{C}}.$
\end{proposition}
\begin{proof} Follows from $d(j^{1\ *}_{p}(s)\xi^{i}F^{\mu}_{i}\eta_{\mu})=
\xi^{i}d(j^{1\ *}_{p}(s)F^{\mu}_{i}\eta_{\mu})+j^{1\ *}_{p}(s)FDiv(\xi)\eta .$
\end{proof}

\begin{remark} Vector fields $\xi =\xi^{i}\partial_{y^i}$ with \emph{constant} components  $\xi^i$ in
a local coordinate system  are obviously $C$-admissible. To such a vector field $\xi$ there correspond, by
the Proposition the (secondary) balance law that is, of course, the linear combination of the original
balance laws with constant coefficients.  More geometrically, one may consider the abelian m-dim
subalgebras of the Lie algebra $V(\pi)$ of vertical vector fields on the bundle $\pi$. Vector fields of
such a subalgebra (generating, by Frobenius theorem the local charts) gives the necessary number of
vertical vector fields satisfying to the $C$-admissibility condition.\par On the contrary, variable
$C$-admissible vector fields $\xi$ generate some nontrivial secondary balance laws. In the part II of this
work we will study such secondary laws more detailed.
\end{remark}
\par
As it is well known, the vector space of divergent-free vector field (divergence being defined by a
pseudo-Riemannian metric or, more fundamental, by a volume form - exterior form of maximal degree nonzero
at every point of the manifold $M$) is closed under the bracket of vector fields and, therefore form the
Lie subalgebra of Lie algebra $\mathcal{X}(Y)$. Asking the same question about the vector space
$\mathcal{X}(C)$ we get, in general, the negative answer. \par

To illustrate the notion of an $\mathcal{C}$-admissible vector field we consider two simple examples of CR
with 1 space variable
\begin{example} Consider a case of the full 1-jet bundle $J^{1}(\pi)$ and of one field $y(t,x)$ being
the function of time $t$ and one space variable $x$. There is only one balance law
\[
\partial_{t}F^{0}+\partial_{x}F^1 =\Pi,
\]
and the corresponding balance relation has the Poincare-Cartan form $\Theta_{C}=F^{0}dy \wedge
dx+F^{1}dy\wedge dt +\Pi dy\wedge dt\wedge dx$. Here $F^\mu$ are, in general, functions of all variables
$(t,x;y; z_{t},z_{x})$. As a result, the condition for a vertical vector field $\xi
=\xi(t,x;y)\partial_{y}$ to be $C$-admissible takes the form
\[
F^{0}d_{t}\xi+F^{1}d_{x}\xi=0,
\]
or
\[
(F^{0}\partial_{t}  +F^{1}\partial_{x})\xi +(F^{0}z_{t}+F^{1}z_{x})\partial_{y}\xi =0.
\]
\par
 Let $F^\mu$ are independent of the jet variables - $F^\mu =F^\mu (t,x,y)$.  Then from the equation
above it follows that $\partial_{y}\xi =0$, i.e. $\xi =\xi (t,x)$ and $(F^{0}\partial_{t}
+F^{1}\partial_{x})\xi=0.$  This last equation tells that the function $\xi (t,x)$ is constant along the
trajectories of the vector field $F^{0}\partial_{t} +F^{1}\partial_{x}$ on the plane.  Locally, in a
neighborhood of points where this vector field is nonsingular, it tells that the function $\xi$ is an
arbitrary function of a transverse variable.  Consider an example where $F^{0}=y,F^{1}=c\ - \ const$. Then
the condition reduces to the equation
\[
(y\partial_{t}+c\partial_{x})\xi=0,
\]
so that $\xi =f(x-\frac{c}{y}t)$ with an arbitrary differentiable function of one variable. \par If
$\xi_{i}=f_{i}(x-\frac{c}{y}t)\partial_{y}$ are two such $C$-admissible vector fields, then their
commutator
\[
[\xi_{1},\xi_{2}]=(-\frac{ct}{y^2})(f_{1}f'_{2}-f_{2}f'_{1})\partial_{y}
\]
\emph{is not a $C$-admissible vector field} because the coefficient of $\partial_{y}$ does not have the
required form.
\end{example}

\vskip0.5cm
\begin{example} As a second example we consider a balance system with  2-dim space-time $X$ (with coordinates $x^0=t,x^1=x$) and
two fields $y^1,y^2$. For such a system the condition $FDiv(\xi)=0$ takes the form
\[
F^{0}_{i}d_{x^0}\xi^i +F^{1}_{i}d_{x^1}\xi^i = F^{0}_{i}\partial_{t}\xi^i +F^{1}_{i}\partial_{x}\xi^i +
[z^{k}_{0}F^{i}_{0}+z^{k}_{1}F^{1}_{i}]\partial_{y^k}\xi^i=0.
\]
Restrict to the RET case where $F^{\mu}_{i}=F^{\mu}_{i}(x,y)$ do not depend on the jet variables
$z^{i}_{\mu}$. Then the condition above splits into three conditions

\beq
\begin{cases}
F^{0}_{i}\partial_{y^k}\xi^{i}=0,\\
F^{1}_{i}\partial_{y^k}\xi^{i}=0,\\
(F^{0}_{i}\partial_{t}+F^{1}_{i}\partial_{x})\xi^i=0.
\end{cases}
\eeq

 First two equations show that the (nonzero) covectors $F^{\mu}_{i}dy^i,\ \mu=0,1$ are annulated by
the linear transformation with the matrix $J(\xi)=\partial_{y^k}\xi^{i}.$ This is (generically) possible
in two cases:
\begin{enumerate}
\item $J(\xi)\equiv 0$.  This means that $\xi^i =\xi^i (x)$ do not depend on the fiber coordinates $y^i$.
  This being true, the
third condition takes the form similar to that in the previous example
\[
(F^{0}_{i}(x,y)\partial_{t}+F^{1}_{i}(x,y)\partial_{x})\xi^{i}(x)=0, \forall y.
\]
Let, for instance, $F^{0}_{i}=y^i$ be the density of the field $y^i$.  Then the last condition takes the
form
\[
(y^i \partial_{t}+F^{1}_{i}(x,y)\partial_{x})\xi^{i}(x)=0, \forall y.
\]
Decomposing the flux term $F^{1}_{i}$ by $y^i$ into Taylor series (or differentiating it by $y^i$) we get
the evolutional equation

\beq \partial_{t}\begin{pmatrix} \xi^1\\  \xi^2\end{pmatrix}  +\begin{pmatrix}
F^{1}_{i,y^j}(t,x,y=0)\end{pmatrix}
\partial_{x}\begin{pmatrix} \xi^1\\  \xi^2\end{pmatrix} =0\eeq

 and the family of
ordinary differential equations $Q^{\alpha}_{i}\partial_{x}\xi^{i}=0,\ \alpha =(\alpha_{1},\alpha_{2})$
for the components $\xi^i$ of vector field $\xi$ number and type of which is determined by the character
of dependence of $F^{1}_{i}(x,y)$ on $y$.\par If, for instance $F^{1}_{i}=F^{1}_{ij}(x)y^j$ are linear by
$y$, no additional conditions are present and the evolutional system (11.4) is locally solvable for a good
enough initial condition $\xi^{i}(t,x)\vert_{t=0}=\xi^{i}(x).$  This gives us a family of vector fields
$\xi \in \mathcal{X(C)}$ depending on two functions of one variable as in the previous example.
\item Case $rk(J(\xi))=1.$ In this case $F^{\mu}_{i},\mu=0,1$ belongs to the kernel of the $2\times 2$ matrix $J(\xi)$
and are, therefore, proportional:

\beq F^{1}_{i}=\lambda (x,y)F^{0}_{i}, \eeq

with some function $\lambda(x,y)$. Third equation takes the form
\[
F^{0}_{i}(\partial_{t} +\lambda \partial_{x})\xi^{i}=0.
\]
If, for instance $F^{0}_{i}=y^i$, last equation takes the form
\[
(\partial_{t} +\lambda \partial_{x})(y^{i}\xi^{i})=0.
\]
Starting with a CR $\mathcal{C}$ of the form considered in this example that satisfies the relation (11.5)
we find the function $\lambda$, then
 the system of equations (11.3) for the components $\xi^i$ of the vector field $\xi(x,y)$ takes the form

\beq
\begin{cases}
y^{i}\partial_{y^k}\xi^{i}=0,\ i=1,2,\\
(\partial_{t} +\lambda \partial_{x})(y^{i}\xi^{i})=0.
\end{cases}
\eeq
\end{enumerate}
First two equations can be rewritten in the form
\[
\partial_{y^k}(y^1\xi^i +y^2 \xi^2 )-\xi^k =0.
\]
Introduce the function $q(x,y)=y^1 \xi^i +y^2 \xi^2$. Then first two equations give us $\xi^i
=\partial_{y^i}q$.  substituting $\xi^i$ in such a form into the first two equations we get for the
function $q(x,y)$ two equations
\[
\begin{cases}
y^1 q_{,11}+y^2 q_{,12}= (y^1 \partial_{y^1})q_{,y^1}=0,\\
y^1 q_{,21}+y^2 q_{,22}=(y^1 \partial_{y^1})q_{,y^2}=0.
\end{cases}
\]
In terms of polar coordinates $\rho ,\theta$ in the plane $(y^1 ,y^2 )$ these two conditions means that
$\xi^{i}=\xi^{i}(\theta)$ are independent on the radial variable $\rho$  and depend on the angular
variables $\theta =tan^{-1}(\frac{y^2}{y^1})$ and $t,x$. \par
  This is equivalent to the statement that $q=y^1 {\tilde q}(x,\frac{y^2}{y^1}).$\par
Substituting this to the third equation we get it in the form
\[ (\partial_{t}+\lambda(x,y)\partial_{x}){\tilde q}=0.\]
  Any solution of this first order wave type equation with the
parameter $\frac{y^2}{y^1}$ determines the vector field $\xi \in \mathcal{X(C)}$.
\end{example}
\vskip0.6cm

 There exists a geometrical situation where there is a natural class of $m$
linearly independent at each point globally defined vector fields $\xi$ admissible for all the
constitutive relations.
\begin{proposition}
Let $\pi:Y\rightarrow X$ be a trivial principal bundle of a connected abelian n-dimensional Lie group $A$.
Then the (globally defined) fundamental vector fields on $Y$ (generated by the right action of $A$ on $Y$)
satisfy to the condition $\omega^{2}_{\mathcal{C}}(\widehat{\xi})=0$.
\end{proposition}
\begin{proof} Trivial.
\end{proof}
\begin{corollary} Let $\pi:Y\rightarrow X$ be a trivial principal bundle of a connected abelian
 m-dimensional Lie group $A$ (in particular, a \emph{trivial vector bundle over} $X$). Let $\mathcal{C}$ be
 an arbitrary constitutive relation.  Then the equivalence statement of
 Theorem 2 is valid for any constitutive relation $\mathcal{C}$.
\end{corollary}
\begin{remark} One might expect that in a case of a non-trivial bundle $\pi:Y\rightarrow X$ there may be
less then $n$ linearly independent \emph{global} vector fields in $\mathcal{X}(F)$. It would be
interesting to study a topological meaning of such a phenomena.
\end{remark}

 For a \textbf{projectable vertical vector field} we can formulate this condition (and more generally,
  condition (7.7)) in a covariant way.  Recall that
$F^{\mu}_{i}dy^{i}\wedge \eta_{\mu}$ is the section of the bundle $\pi^{1*}_{0}(V^{*}(\pi)\otimes
\Lambda^{n-1}(X))$ over $Z_{p}$.  Last bundle have, as its $\Lambda^{n}(X)$-dual, the bundle
$\pi^{1*}_{0}(V(\pi)\otimes \Lambda^{1}(X))$ over $Z_{p}$ and we may consider the associating of
$d_{\mu}\xi^{i}$ to the vector field $\xi $ as the mapping of\textbf{ total differential}

\beq D: \xi=\xi^{i}(x,y)\frac{\partial}{\partial y^{i}} \rightarrow  d_{\mu}\xi^{i}dx^{\mu}\otimes
\frac{\partial}{\partial y^i}: V(\pi)\rightarrow \pi^{1*}_{0}(V(\pi)\otimes \Lambda^{1}(X)).\eeq

Then condition (9.10) takes the form
\[
FDiv(\xi )=\langle F^{\mu}_{i}dy^{i}\wedge \eta_{\mu}, D\xi \rangle =0,
\]

where notation $FDiv(\xi )$ for the expression on the right was introduced in the previous section.\par

\begin{example}
 Consider the case of a
vector fields of the type

\[
\xi =\xi^{i}(x^\mu ,y^i ,z^{i}_{\mu} ) \frac{\partial}{\partial y^i }.
\]
For such a vector field and an arbitrary section $s\in \Gamma (\pi)$

\beq j^{1*}(s)\langle F^{\mu}_{i}dy^{i}\wedge \eta_{\mu}, D\xi \rangle = [F^{\mu}_{i}\circ j^{1}(s)
(\xi^{i}_{,x^\mu}+\xi^{i}_{y^j}s^{j}_{,x^\mu})+F^{\mu}_{i}\circ
j^{1}(s)\xi^{i}_{,z^{j}_{\nu}}s^{j}_{x^{\nu}x^{\mu}} ]\eta .\eeq

Since second derivative components of 2-jets of sections $s$ at a point $(x,y,z)$ \emph{can be arbitrary},
condition of $C$-admissibility for these vector fields $\xi$ splits into two conditions:

\beq
\begin{cases}
F^{\mu}_{i} (\xi^{i}_{,x^\mu}+\xi^{i}_{,y^j}z^{j}_{\mu})=
F^{\mu}_{i} (\frac{\partial}{\partial x^\mu}+z^{j}_{\mu}\frac{\partial}{\partial y^j})\xi^{i} &=0,\\
F^{(\mu}_{i}\xi^{i}_{,z^{j}_{\nu )}}&=0,\ \text{for\ all}\ \mu,\nu,j,
\end{cases}
\eeq where in the second condition the symmetrization by $\mu \nu$ is done.\par

One can rewrite second condition in the form
\[
F^{\mu}_{i}\xi^{i}_{,z^{j}_{\nu }}=\omega^{\mu \nu}_{j},
\]
for an arbitrary family of skew-symmetrical tensors $\omega^{\mu \nu}_{j}$ on $Z_{p}$.  For a projectable
vector field $\xi $ the second equation is trivially satisfied.\par
\end{example}
\subsection{$\mathcal{C}$-admissible vertical vector fields: 5F-fluid system}

Here we consider the condition of $C$-admissibility for a  vector field $\xi \in \mathcal{X}(Y)$ in the
case of the 5F-fluid balance system $C_{5}$, see (2.7) in the Newtonian space-time with Euclidian metric
$(E^h,h=dt^2 +\sum_{A}dx^{A\ 2})$ and the global coordinates $(t,x^A)$. In order to simplify calculations
we choose basic fields to be $(\rho ;v^A ,A=1,2,3; \vartheta)$ and will use the internal energy $\epsilon$
balance instead of the full energy density $e$ balance. We assume that the constitutive relations for
5-fields system, i.e. functions $\epsilon, t^{A}_{B},q^A$ may depend on the spacial gradients of dynamical
variables $ \nabla v , \nabla \vartheta$.\par

We have for $C_{5}$:

\begin{multline} \Theta_{C_5}=[\rho d\rho \wedge \eta_{0}+\rho v^B d\rho \wedge \eta_{B}]+
   [(\rho v^A )dv^A \wedge
\eta_{0}+(\rho v^A v^B -t^{AB})dv^A \wedge \eta_{B}]+\\ + [\rho \epsilon d\vartheta \wedge \eta_{0} +(\rho
\epsilon V^A +q^A )d\vartheta \wedge \eta_{B}] ] + [ f_{A} dv^A \wedge \eta
+(f_{A}v^A+t^{A}_{B}\frac{\partial V^B}{\partial x^A}+r) d\vartheta \wedge \eta].
\end{multline}

Since we assume that the constitutive relations $C_{5}$ are independent on the derivatives of $\rho$, for
a general vector field ${\hat \xi}\in \mathcal{X}(Z_{p})$, condition of $C$-admissibility
$F^{\mu}_{i}\xi^{i}_{\mu}=0$ has the form
\[
(z^{i}_{A \mu}\xi^{\mu}-\xi^{i}_{A})F^{A }_{i}=0; A=1,2,3,\mu=0,1,2,3\ i=v^B ,\vartheta.
\]
Due to the independence of the 2-jet variables $z^{i}_{A\mu}$ and to the fact that none of the flow
components $F^{\mu}_{i}$ is zero, we get: $\xi^{\mu}=0,\ \mu=0,1,2,3$.  Thus, $C_{5}$-admissible vector
field $\xi$ is $\pi$-vertical and condition of $C_{5}$-admissibility takes the form

\beq \sum_{(A,i)\in P}\xi^{i}_{A}F^{A}_{i}= (\rho v^A v^B-t^{AB}) \xi^{v^B}_{A}+(\rho \epsilon v^A +q^A
)\xi^{\vartheta}_{A}=0. \eeq

Any $\pi^1$-vertical vector field $\xi \in \mathcal{X}(Z_{p})$ that satisfy to this linear algebraic
condition is $C_{5}$-admissible.

\par
Consider now the case where the vector field above is the flow prolongation $\xi^1$ of a $\pi$-vertical
vector field  $\xi
=\xi^{\rho}\partial_{\rho}+\xi^{v^A}\partial_{v^A}+\xi^{\vartheta}\partial_{\vartheta}$, i.e. use
\[
\xi^1 =\xi +d_{\mu}\xi^i \partial_{z^{i}_{\mu}}=\xi +d_{\mu}\xi^{\rho}\partial_{z^{\rho}_{\mu}}+
d_{\mu}\xi^{v^A}\partial_{z^{v^A}_{\mu}}+d_{\mu}\xi^{\vartheta}\partial_{z^{\vartheta}_{\mu}}.
\]

Condition of admissibility $\sum_{(A,i)\in P}F^{A}_{i}d_{A}\xi^{i}=0$ (summation goes over the space
derivatives since no time derivatives of dynamical fields enters the constitutive relations) is obtained
by substituting $\xi^{i}_{A}=d_{A}\xi^i $ into the algebraic equation (11.11).
\beq
 \rho v^A v^B d_{A}\xi^{v^B}+\rho \epsilon v^A d_{A}\xi^{\vartheta}=t^{AB} d_{A}\xi^{v^B}-q^A d_{A}\xi^{\vartheta}.
\eeq

This equation splits corresponding to the order of the terms in the total derivatives
$d_{A}=\partial_{x^A}+z^{i}_{A}\partial_{y^i}$ and we get:

\beq
\begin{cases}
\text{Coefficient\  of}\ z^{i}_{A}:\ \rho v^A v^B \xi^{v^B}_{,y^i}+\rho \epsilon v^A
\xi^{\vartheta}_{,y^i}=t^{AB}\xi^{v^B}_{,y^i}-q^A \xi^{\vartheta}_{,y^i} ,
\ i=\rho ,v^C ,\vartheta; A=1,2,3,\\
\text{No}\ z^{i}_{A}: \ \ \rho v^A v^B \xi^{v^B}_{,A}+\rho \epsilon v^A
\xi^{\vartheta}_{,A}=t^{AB}\xi^{v^B}_{,A}-q^A \xi^{\vartheta}_{,A}.
\end{cases}
\eeq

or, in more details,

\beq \begin{cases} \rho v^A v^B \xi^{v^B}_{,\rho }+\rho \epsilon v^A
\xi^{\vartheta}_{,\rho }=t^{AB}\xi^{v^B}_{,\rho }-q^A \xi^{\vartheta}_{,\rho} ,\\
\rho v^A v^B \xi^{v^B}_{,v^C}+\rho \epsilon v^A \xi^{\vartheta}_{,v^C}=t^{AB}\xi^{v^B}_{,v^C}-q^A
\xi^{\vartheta}_{,v^C},\\
\rho v^A v^B \xi^{v^B}_{,\vartheta }+\rho \epsilon v^A \xi^{\vartheta}_{,\vartheta
}=t^{AB}\xi^{v^B}_{,\vartheta}-q^A \xi^{\vartheta}_{,\vartheta},\\
\rho v^A v^B \xi^{v^B}_{,A}+\rho \epsilon v^A \xi^{\vartheta}_{,A}=t^{AB}\xi^{v^B}_{,A}-q^A
\xi^{\vartheta}_{,A}.
\end{cases}\eeq

Looking at the first three (families of ) equations in this system we see that the left sides of these
equations do not depend on the jet variables.  Therefore right sides of these equations do not depend on
jet- variables $\nabla v ,\nabla \vartheta $ too.\par

For the second equation this gives, since $t^{A}_{B},q^A$ do not depend on $v^C$ that

\[
t^{AB}\xi^{v^B}_{,v^C}-q^A \xi^{\vartheta}_{,v^C}=(t^{AB}\xi^{v^B}-q^A
\xi^{\vartheta})_{,v^C}=\lambda^{A}_{v^C}(\rho, v,\vartheta)
\]
with some functions $\lambda^{A}_{v^C}(\rho, v,\vartheta)$ of named variables.\par
 Since function $(t^{AB}\xi^{v^B}-q^A \xi^{\vartheta})$ in
the left side does not depend on $C$ and due to the mixed derivative test
$\partial_{v^C_{1}}\lambda^{A}_{v^C_{2}}=\partial_{v^C_{2}}\lambda^{A}_{v^C_{1}},$ we have
$\lambda^{A}_{v^C}=\partial_{v^C}h^{A}(\rho, v,\vartheta)$ with some functions $h^{A}(\rho, v,\vartheta)$.
From this it follows that

\beq (t^{AB}\xi^{v^B}-q^A \xi^{\vartheta}-h^A )_{,v^C}=0\Rightarrow t^{AB}\xi^{v^B}-q^A
\xi^{\vartheta}=h^{A}(\rho, v,\vartheta)+h^{A}_{1}((\rho, \vartheta ,\nabla v ,\nabla \vartheta)). \eeq

 Thus, second equation takes the form

\beq \rho v^A v^B \xi^{v^B}_{,v^C}+\rho \epsilon v^A \xi^{\vartheta}_{,v^C}=h^{A}_{,v^C}(\rho ,v ,\theta
). \eeq

 Take here derivative by $\nabla \theta$.  We get
\[
\rho \frac{\partial \epsilon}{\partial \nabla \theta} v^A \xi^{\vartheta}_{,v^C}=0.
\]
Thus, either $\frac{\partial \epsilon}{\partial \nabla \theta}=0$ or  $\xi^{\vartheta}_{,v^C}=0.$  Taking
derivative by $\nabla v$ we get that either $\frac{\partial \epsilon}{\partial \nabla
\theta}=\frac{\partial \epsilon}{\partial \nabla v}=0$ or $\xi^{\vartheta}_{,v^C}=0.$
\par Consider first the second alternative - $\xi^{\vartheta}_{,v^C}=0.$  substituting this into the
equation (11.16) we get

\beq t^{A}_{B}\xi^{v^B}_{,v^C}=h^{A}_{,v^C} \eeq for all $A,C$. Thus, either $t^{A}_{B}$ does not depend
on the gradient variables or $\xi^{v^B}_{,v^C}$ is degenerate and the part of $t^{A}_{B}$ that depend on
gradients belongs to the kernel of $\xi^{v^B}_{,v^C}$.

Consider, \emph{generically}, a case where $\xi^{v^B}_{,v^C}=0\rightarrow h^{A}_{,v^C}=0$ as well.\par
Thus, $\xi^{\vartheta},\xi^{v^B}$ do not depend on $v^C$. Taking derivatives by $v^A,v^B$ in the first
equation we get $\xi^{\vartheta}_{,\rho}=0$ ($\epsilon$ does not depend on $v^C$), then, using this and
differentiating by $v^A$ we get $\xi^{v^B}_{,\rho}=0$ as well.  Repeating this procedure with the third
equation we will see that $\xi^{\vartheta},\xi^{v^B}$ do not depend on $\vartheta$ as well.  Thus,
\[
\xi^{v^B}_{,v^C}=0\rightarrow \xi^{\vartheta},\xi^{v^B}-const.
\]

Another choice would be to have $\epsilon =\epsilon (\rho ,\theta )$ (which is realized, for instance in
the case of Navier-Stokes fluid (\cite{M}).\par

In such a case we return to the relation (11.16)
\[
t^{A}_{B}\xi^{v^B}_{,v^C}-q^A \xi^{\vartheta}_{,v^C}=h^{A}_{,v^C}.
\]
Taking here derivative by $\nabla v$ and \emph{assuming generically that}  $\xi^{v^B}_{,v^C}$
\emph{nondegenerate}  we get
\[
\frac{t^{A}_{B,\nabla v}}{q^{A}_{,\nabla v}}=(\xi^{v^B}_{,v^C})^{-1}\cdot
\xi^{\vartheta}_{,v^C}=\lambda_{B}(\rho ,\theta )
\]
does not depend on $v^C$ or on any gradients.  Similarly we get
\[
\frac{t^{A}_{B,\nabla \vartheta }}{q^{A}_{,\nabla \vartheta }}=(\xi^{v^B}_{,v^C})^{-1}\cdot
\xi^{\vartheta}_{,v^C}=\lambda_{B}(\rho ,\theta ).
\]

  Then $q^{A}_{,\nabla
v}\lambda_{B}=t^{A}_{B,\nabla v}, q^{A}_{,\nabla \vartheta}\lambda_{B}=t^{A}_{B,\nabla \vartheta }$ and,
integrating by $\nabla v, \nabla \vartheta$ we get:
\[
\begin{cases}
q^A \lambda_{B}(\rho ,\vartheta )=t^{A}_{B}+\mu^{A}_{B}(\rho ,\vartheta ),\\
\xi^{\theta}_{,v^C}=\xi^{v^B}_{,v^C}\lambda_{B}(\rho ,\vartheta ).
\end{cases}
\]
Here $\lambda_{B}(\rho ,\vartheta ),\mu^{A}_{B}(\rho ,\vartheta )$ are \emph{constitutional functions},
depending on the constitutive relations $C_{5}$ but not on the vector field $\xi$.\par

 From the second equation we get
\[
\xi^{\vartheta}=\xi^{v^B}\lambda_{B}(\rho ,\vartheta)+\zeta (\rho ,\vartheta).
\]
Using expressions for  $t^{A}_{B}$ and $\xi^\vartheta$ from these two equations  we get

\[
t^{A}_{B}\xi^{v^B}-q^A \xi^{\vartheta}=-q^A \zeta(\rho ,\vartheta)-\mu^{A}_{B}\xi^{v^B}_{,v^C}.
\]
Taking here derivative by $v^C$ and substituting into the second equation of system (11.14) we get
\[
\rho v^A (v^B \xi^{v^B}_{,v^C}+ \epsilon \xi^{\vartheta}_{,v^C})=-\mu^{A}_{B}\xi^{v^B}_{,v^C}.
\]
substituting here $\xi^{\vartheta}_{,v^C}=\xi^{v^B}_{,v^C}\lambda_{B}$ we get

\beq [\rho v^A v^B +\rho v^A \epsilon \lambda_{B}(\rho ,\vartheta)+\mu^{A}_{B}(\rho
,\vartheta)]\xi^{v^B}_{,v^C}=0. \eeq Generically, to be true in an open set of basic fields space $U$ we
have to have $\xi^{v^B}_{,v^C}=0$.  Then $\xi^{\vartheta}_{,v^C}=0$ and from this it follows as above that
$\xi^{v^B},\xi^{\vartheta}-const$.\par

\begin{proposition} Assume that in the 5F-fluid system constitutive fields (i.e. $t^{A}_{B},q^A ,\epsilon)$
 do not depend on the velocity $v^A$ and on the $\nabla \rho$.  Then, \textbf{generically}, i.e. without
 special conditions for the constitutive relations, components
 $\xi^{v^C},\xi^{\vartheta} $ of a vector field $\xi \in
 V(Y)\cap \mathcal{X}(C)$ are constant and an admissible $\pi$-vertical vector field $\xi \in V(Y)$ has
 the form

 \[ \xi =\xi^{\rho}(t,x,\rho ,v,\vartheta)\partial_\rho + \xi^{v^B}(t)\partial_{v^B}+\xi^{\vartheta}\partial_{\vartheta}. \].
\end{proposition}

More detailed study of the geometrical properties of the 5F-fluid balance system including the
consideration of special, non-generic cases of the constitutive relations will be done in other paper.\par

\vfill \eject
\section{Action of geometrical transformations on the constitutive relations.}
In this section we study the action on the constitutive relations of the natural prolongations  of the
projectable transformations $\phi\in Aut_{p}(\pi)$ of $Y$ studied in the Section 5.\par

\subsection{Action of $\phi\in Aut_{\pi}(Y)$ on the covering constitutive relations}
Let ${\hat C}:J^{1}_{p}(\pi)\rightarrow \Lambda^{n+1}Y\oplus \Lambda^{n+2}Y$ be a covering constitutive
relation. Decompose corresponding Poincare-Cartan form as follows:

\beq \Theta_{\hat C}=\Theta^{0,n+1}_{\hat C}+\Theta^{1,n}_{\hat C}+\Theta^{n+2}_{\hat
C}=p\eta+F^{\mu}_{i}dy^i \wedge \eta_{\mu}+\Pi_{i}dy^i \wedge \eta. \eeq

\par

An automorphism $\phi \in Aut_{p}(\pi)$ of the bundle $\pi$ can be lifted to the contact automorphism
$\phi^1$ of the partial 1-jet bundle $Z_{p}=J^{1}_{p}(\pi)$ over $Y$ and $X$ (see Sec. 5).  It can also be
lifted to the bundle automorphism $\phi^{1*}=(\phi^{*(n+1)},\phi^{*(n+2)})$ of the bundle
$\Lambda^{n+1}_{2}Y\oplus \Lambda^{n+2}_{2}Y$ preserving its subbundle $\Lambda^{n+1}_{2}Y\oplus
\Lambda^{n+2}_{2}Y$. Automorphism $\phi^{1*}$ leaves canonical form(s) $\Theta^{n+1}_{2}\oplus
\Theta^{n+2}_{2}$ invariant (Sec.5). \par

More generally, let $\psi \in Aut(\pi_{1})$ belongs to the group of automorphisms of the double bundle
$\pi^{1}:J^{1}_{p}(\pi) \rightarrow Y\rightarrow X$. Transformation $\psi \in Aut(\pi^{1})$ generate the
automorphism $\phi \in Aut(\pi)$. In its turn, transformation $\phi$ extends to the automorphism $\phi^1
\in Aut_{p}(\pi_{1})$ (see Section 6). This allows to present
\[
\psi = \phi^1 \circ \psi_{gau},
\]
where the automorphism $\psi_{gau}$ projects to the identity diffeomorphism of $Y$ and, thus, represents
pure gauge transformation of $J^{1}_{p}(\pi )$. Correspondingly, the action of $\psi $ on the
Poincare-Cartan form of a CCR $\hat C$ is the composition
\[
\psi^{*}\Theta_{\hat C}= \psi^{*}_{gau}\phi^{1\ *}\Theta_{\hat C}.
\]
Calculate this action explicitly.\par

  Lifted automorphism $(\phi^{1},\phi^{1*})$ transforms the constitutive relation $\hat C$
into the constitutive relation

\beq \hat{C}^{\phi}=\phi^{1*}\circ \hat{C}\circ \phi^{1\ -1}.\eeq

 For the Poincare-Cartan form of $\hat{C}^\phi $ we have
(using for the transformed CCR $\hat{C}^\phi $ the fact that $\phi^{1*}$ preserves the multisymplectic
forms $\Theta^{n+1}_{2}$ and $\Theta^{n+2}_{2}$)

\begin{multline}
\Theta_{\hat{C}^\phi}=(\hat{C}^\phi)(\Theta^{n+1}_{2}\oplus \Theta^{n+2}_{2})= (\phi^{1*}\circ {\hat
C}\circ \phi^{1\ -1})^{*}(\Theta^{n+1}_{2}\oplus \Theta^{n+2}_{2})=\\= \phi^{1\ -1\  *}\circ
\hat{C}^{*}\circ (\phi^{1*})^{*}(\Theta^{n+1}_{2}\oplus \Theta^{n+2}_{2})= \phi^{1\ -1\  *}\circ
\hat{C}^{*}(\Theta^{n+1}_{2}\oplus \Theta^{n+2}_{2})=\phi^{1\ -1\ *}\Theta_{\hat{C}}.
\end{multline}
On the other hand

\begin{multline}
\phi^{1*-1}\Theta_{\hat{C}}=\phi^{1*-1}[p\eta+F^{\mu}_{i}dy^i \wedge \eta_{\mu}+\Pi_{i}dy^i \wedge
\eta]=(p\circ \phi^{1\ -1})\cdot {\bar \phi}^{*-1}\eta +\\ + (F^{\mu}_{i}\circ \phi^{1\ -1})d(\phi^{-1\
i})(x,y)\wedge {\bar \phi}^{*-1}\eta_{\mu}+\Pi_{i}\circ \phi^{1\ -1} d(\phi^{-1\ i}) \wedge {\bar
\phi}^{-1\ *}\eta.
\end{multline}
here we have used the fact that $dy^i$ and $\eta ,\eta_{\mu}$ are pullbacked form the spaces $Y$ and $X$
respectively and that pullback by $\pi_{10}$ or by $\pi^1$ commutes with the pullback by $\phi^{1}$ and
its projections to $Y$ and $X$ respectively.\par

To shorten the notations we will make the next calculation for $\widehat{\phi} \in Aut_{p}(\pi^{1})$ -
automorphsim of the double bundle $Z_{p}\rightarrow Y\rightarrow X$ rather then $\phi^{-1}$ (lift
$\widehat{\phi}=\psi^1$ of a automorphism $\psi \in Aut_{p}(\pi)$ is a special case of this more general
case). Automorphism $\widehat{\phi}$ induces automorphism  $\phi \in Aut_{p}(\pi)$ and automorphism of
splitting structure (if $J^{1}_{P}(\pi)$ is the proper partial 1-jet bundle) in $X$. We notice that ${\bar
\phi}^{*}\eta =detJ({\bar \phi})\eta$ where $detJ({\bar \phi})$ is the Jacobian of the (local)
diffeomorphism $\bar \phi$ defined by the volume form $\eta$.  On the other hand
\[
{\bar \phi}^{*}\eta_{\mu}={\bar \phi}^{*}i_{\partial_{x^\mu}}\eta=i_{{{\bar
\phi}^{-1}}_{*}\partial_{x^\mu}}{\bar \phi}^{*}\eta=detJ({\bar \phi})J({\bar
\phi}^{-1})^{\nu}_{\mu}\eta_{\nu}
\]
since ${{\bar \phi}^{-1}}_{*}\partial_{x^\mu}= J({\bar \phi}^{-1})^{\nu}_{\mu}\partial_{x^\nu}.$\par

Altogether
\begin{multline}
\widehat{\phi}^{*}\Theta_{\hat{C}}=p\circ \widehat{\phi}\cdot detJ({\bar \phi})\eta +(F^{\mu}_{i}\circ
\widehat{\phi})(\phi^{i}_{,x^\sigma}dx^\sigma +\phi^{i}_{,y^j}dy^j)\wedge [detJ({\bar \phi})J({\bar
\phi}^{-1})^{\nu}_{\mu}\eta_{\nu}]+\\ +\Pi_{i}\circ \widehat{\phi}(\phi^{i}_{,x^\sigma}dx^\sigma
+\phi^{i}_{,y^j}dy^j)\wedge detJ({\bar \phi})\eta = [detJ({\bar \phi})J({\bar
\phi}^{-1})^{\nu}_{\mu}(F^{\mu}_{i}\circ \widehat{\phi})\phi^{i}_{,y^j}] dy^j \wedge \eta_{\nu}+\\ +
[detJ({\bar \phi})(\Pi_{i}\circ \widehat{\phi}) \phi^{i}_{,y^j}]dy^j \wedge \eta +[p\circ
\widehat{\phi}\cdot detJ({\bar \phi})+ (F^{\mu}_{i}\circ \widehat{\phi})(\phi^{i}_{,x^\nu}[detJ({\bar
\phi})J({\bar \phi}^{-1})^{\nu}_{\mu}]\eta.
\end{multline}
Splitting the terms we can write last result as follows.
 \beq
\begin{cases}
[\widehat{\phi}^{*}\Theta_{\hat{C}}]=^{\widehat{\phi}}p \eta + ^{\widehat{\phi}}F^{\mu}_{i}dy^i \wedge
\eta_{\mu}+ ^{\widehat{\phi}}\Pi_{i}dy^i \wedge \eta, \
\text{where} \\
^{\widehat{\phi}}p= [p\circ \widehat{\phi}+ (F^{\mu}_{i}\circ \widehat{\phi})(\phi^{i}_{,x^\nu}J({\bar
\phi}^{-1})^{\nu}_{\mu}]\cdot detJ({\bar \phi}),
 \\  ^{\widehat{\phi}}F^{\mu}_{i}= detJ({\bar \phi})J({\bar
\phi}^{-1})^{\mu}_{\nu}(F^{\nu}_{j}\circ
\widehat{\phi})\phi^{j}_{,y^i},\ \\
^{\widehat{\phi}}\Pi_{i}=detJ({\bar \phi})(\Pi_{j}\circ \widehat{\phi}) \phi^{j}_{,y^i}.
\end{cases}
\eeq To use these formulas for calculating $C^{\widehat{\phi}}$ and $\Theta_{C^{\widehat{\phi}}}$ one
should replace $\widehat{\phi}$ in (12.3) by inverse mapping $\widehat{\phi}^{-1}$.\par

From the last result it follows that $\Theta^{1,n}_{\hat C}$ and $\Theta^{n+2}_{\hat C}$ transforms
\emph{tensorially} under the action of $\widehat{\phi}$ while the component $\Theta^{0,n+1}_{\hat C}$
transforms affine.\par

\vskip0.5cm \par

\begin{remark}
If we take $\widehat{\mathcal{C}}$ to be $\nu$-lifted CCR $\hat{C}_{\nu}=q_{\nu}\circ \mathcal{C}$ (see
Sec.  ) for an arbitrary constitutive relation $\mathcal{C}:Z_{p}\rightarrow {\tilde Z}$ and
$\Theta_{\mathcal{C},\nu}$ - corresponding $\nu$-lifted Poincare-Cartan form of $\mathcal{C}$. for a
one-parameter group of automorphisms $\widehat{\phi}_{t}$ of $Z_{p}$ we define one-parameter group of
automorphisms $\psi_{t}$ of $\Lambda_{2}^{(n+1)+(n+2)}$ by projecting ${\hat \phi}_{t}$ to the
one-parameter group of automorphisms $\phi_{t}$ of $Y$ and then lifting it to $\Lambda_{2}^{(n+1)+(n+2)}$
using results of section 5. We have

\begin{multline} \widehat{\phi}^{*}_{-t}\Theta_{\hat{C},\nu}=
\widehat{\phi}^{*}_{-t}(\hat{C}^{*}{\tilde \Theta}_{\nu})=
 \widehat{\phi}^{*}_{-t}C^{*}\psi_{t}^{*}\psi_{-t}^{*}{\tilde \Theta}_{\nu}=(\phi^{1*}_{-t}C^{*}\psi_{t}^{*})[{\tilde
\Theta}_{\nu}+(F^{\mu}_{Pi}\Gamma^{i}_{\mu})\circ \psi_{t}\eta]=\\= (\psi_{t} \circ C\circ
\widehat{\phi}_{-t})^{*}[{\tilde \Theta}_{\nu}+(F^{\mu}_{Pi}\Gamma^{i}_{\mu})\circ \psi_{t}\eta]=
C^{\phi_{t}\
*}[{\tilde \Theta}_{\nu}+((F^{\mu}_{Pi}\Gamma^{i}_{\mu})\circ \psi_{t})\eta]=
\Theta_{C^{\widehat{\phi}_t},\nu}+C^{\widehat{\phi}_t \
*}((F^{\mu}_{Pi}\Gamma^{i}_{\mu})\circ \psi_{t})\eta,
\end{multline}
 where we have used the fact that $\psi_{t}$ acts on the
form ${\Theta}_{C,\nu}$ leaving its $dy^i \wedge \eta_{\mu}$ part invariant.  Here
$C^{\widehat{\phi}_t}=\psi_{t} \circ C\circ \widehat{\phi}_{-t}.$\par

If we take the expression obtained above mod $\Lambda^{n+1}_{1}$ the last term vanished and we get the
formula for transformation of the $(1,*)$-part of the Poincare-Cartan form $\Theta_{\mathcal{C},\nu}$ of a
CR $\mathcal{C}$ \emph{independent on a choice of connection} $\nu$! (another argument would be that the
last term vanishes if we contract it with a vertical vector field $\xi$):

\beq \phi^{1*}\Theta^{(1,*)}_{C,\nu}=[detJ({\bar \phi})J({\bar \phi}^{-1})^{\nu}_{\mu}(F^{\mu}_{i}\circ
\phi^{1})\phi^{i}_{,y^j}] dy^j \wedge \eta_{\nu}+ [detJ({\bar \phi})(\Pi_{i}\circ \phi^{1})
\phi^{i}_{,y^j}]dy^j \wedge \eta. \eeq
\end{remark}
In terms of separate balance laws

\beq \sigma_{i}=F^{\mu}_{i}\eta_{\mu}+\Pi_{i}\eta, \eeq

 using the
pullback of the basic forms $\eta ,\eta_{\nu}$ by $\bar \phi$
\[
\begin{cases}
\eta^{\bar \phi}=detJ({\bar \phi})\eta;\\
 \eta^{\bar \phi}_{\mu}=detJ({\bar \phi})J({\bar \phi}^{-1})^{\nu}_{\mu}\eta_{\nu}
\end{cases}
\]
we have for the transformed balance laws

 \beq \sigma^{\phi}_{i}=((F^{\nu}_{j}\circ \phi^{1} )\phi^{j}_{,y^i})\eta^{\bar \phi}_{\nu}
 +((\Pi_{j}\circ \phi^{1})\phi^{j}_{,y^i})\eta^{\bar \phi },\eeq

Transformation $C\rightarrow \mathcal{C}^\phi $ acts, in a natural way, on the sheaf of solutions
$Sol(\mathcal{C})$ transforming it to the sheaf of solutions of the balance system $\mathcal{B_{C}^\phi}:$

\beq  Sol(C) \rightleftarrows  Sol(C^\phi ). \eeq
\par

A pure gauge automorphisms $\psi_{gau}$ in the decomposition $\psi = \phi^1 \circ \psi_{gau}$ acts simply
by

\beq \psi^{*}_{gau}\Theta_{C}=F^{\mu}_{i}\circ \psi_{gau} dy^i \wedge \eta_{\mu}+\Pi_{i}\circ \psi_{gau}
dy^i\wedge \eta. \eeq

and the individual balance laws $\sigma_{i}$ after transformation take the form

\beq \sigma_{i}^{\psi_{gau}}=F^{\mu}_{i}\circ \psi_{gau}\eta_{\mu}+ \Pi_{i}\circ \psi_{gau}\eta. \eeq

\vskip1cm

Let now $\xi\in \mathcal{X}_{p}(\pi)$  be an \emph{infinitesimal automorphism} (vector field) of the
bundle $\pi$, i.e. a projectable vector field in $Y$ satisfying to the conditions of Section 7 for lifting
to the partial 1-jet bundle $J^{1}_{p}(\pi)$:
\[
  \xi =\xi^{\mu}(x)\partial_{x^\mu}+\xi^{i}(x,y)\partial_{y^i}.
\]
Let $\xi^{1}$ be its prolongation to the projectable contact vector field in $J^{1}_{p}(\pi)$ (see Sec.7).
Thus, we have
\[
\xi^{1}=\xi^{\mu}(x)\partial_{x^{\mu}}+\xi^{i}(x,y)\partial_{y^{i}} +\left(d_{\mu}\xi^{i}
-z^{i}_{\nu}\frac{\partial \xi^{\nu}}{\partial x^{\mu}} \right)\partial_{z^{i}_{\mu}},
\]
where summation in the last term is taken over the $z^{i}_{\mu}$ that are present in the partial 1-jet
bundle. In the RET case we do not need to introduce any prolongation.
\par

 Let $\widetilde{\xi}^{*}$ be the prolongation of
$\xi$ to the projectable vector field in ${\tilde Z}=\Lambda^{(n+1)+(n+2)}_{2}/\Lambda^{(n+1)+(n+2)}_{1}$
preserving canonical multisymplectic forms $\Theta^{n+1}_{2},\Theta^{n+2}_{2}$ (See Sec.7):
\begin{multline}
\widetilde{\xi}^{*}=\xi^{\mu}(x)\partial_{x^{\mu}}+\xi^{i}(x,y)\partial_{y^{i}}+
\left(-p^{\nu}_{i}\left(\frac{\partial \xi^\mu }{\partial x^\nu }-\xi^{\mu}\lambda_{,x^\nu
}\right)-p^{\mu}_{j}\frac{\partial \xi^j }{\partial u^i}-p^{\mu}_{i}\left( \frac{\partial \xi^\nu
}{\partial x^\nu }-\xi^{\nu}\lambda_{G,x^\nu}\right)-p_{i}^{\nu}\xi^{\mu}\lambda_{G,x^\nu} \right)
\partial_{p^{\mu}_{i}}+\\ +\left( p_{k}\left( \xi^{\mu}\frac{\partial \lambda_{G} }{\partial x^\mu }-\frac{\partial
\xi^\mu}{\partial x^\mu}\right)-p_{j}\frac{\partial \xi^j}{\partial u^k}\right)
\partial_{p_{k}}.
\end{multline}
Let now $\phi^{1}_{t}$ be a local flow in $J^{1}_{p}(\pi)$ of the vector field $\xi^1$ and $\psi_{t}$ be a
local flow in ${\tilde Z}$ of the vector field $\widetilde{\xi}^{*}$.
\par

 Taking in the expression for the transformed mapping $\psi_{t} \circ C\circ \phi^{1}_{-t}$ derivative by $t$ at
$t=0$ we get the generalized Lie derivative of mapping $C$ with respect to the vector fields
$(\xi^{1},\widetilde{\xi}^{*})$ (see \cite{KMS}, Chapter 11) - the vector field over the mapping
$C:Z_{p}\rightarrow {\tilde Z}$:

\beq \mathcal{L}_{(\xi^{1},\widetilde{\xi}^{*})}\mathcal{C} = C_{*}(\xi^{1})- \widetilde{\xi}^{*}\circ
\mathcal{C}.\eeq

In local adapted coordinates we have

\begin{multline} L_{(\xi^{1},\xi^{1*})}\mathcal{C}=
[ \left(\xi^{\nu}\partial_{x^{\nu}}+\xi^{j}\partial_{u^{j}} +\left(d_{\sigma}\xi^{j}
-z^{j}_{\nu}\frac{\partial \xi^{\nu}}{\partial x^{\sigma}} \right)\partial_{z^{j}_{\sigma}}
\right)F^{\mu}_{i}+\\+\left(\left( F^{\nu}_{i}\left(\frac{\partial \xi^\mu }{\partial x^\nu
}-\xi^{\mu}\lambda_{G,x^\nu }\right)+F^{\mu}_{j}\frac{\partial \xi^j }{\partial u^i}+F^{\mu}_{i}\left(
\frac{\partial \xi^\nu }{\partial x^\nu
}-\xi^{\nu}\lambda_{G,x^\nu}\right)+F_{i}^{\nu}\xi^{\mu}\lambda_{G,x^\nu} \right) \right) ]
\partial_{p^{\mu}_{i}}+\\ + \left[\left(\xi^{\nu}(x)\partial_{x^{\nu}}+\xi^{i}(x,y)\partial_{u^{i}}
+\left(d_{\mu}\xi^{i}-z^{i}_{\nu}\frac{\partial \xi^{\nu}}{\partial x^{\mu}} \right)\partial_{z^{i}_{\mu}}
\right)\Pi_{k}- \left( \Pi_{k}(\xi^{\mu}\frac{\partial \lambda_{G} }{\partial x^\mu }-\frac{\partial
\xi^\mu}{\partial x^\mu})-\Pi_{j}\frac{\partial \xi^j}{\partial u^k}\right)\right]\partial_{p_{k}}
\end{multline}

In the case of partial 1-jet bundles we assume restrictions to the automorphisms and vector fields that
were introduced in Section 7. For instance in a case of $J^{1}_{S}(\pi)$ we assume that the automorphisms
of $\pi$ preserve the structure of fiber product (4.8).\par

\begin{definition}
\begin{enumerate}

\item  A
diffeomorphism $\Phi$  of $W_{p}=Z_{p}\times {\tilde Z}$ is called a \textbf{generalized symmetry
transformation} of constitutive relation $\mathcal{C}$ if
$\Phi(\Gamma_{\mathcal{C}})=\Gamma_{\mathcal{C}}$ for the graph $\Gamma_{\mathcal{C}}$ of the mapping
$\mathcal{C}$.   A generalized symmetry $\Phi$ of $\mathcal{C}$ is called a \textbf{trivial symmetry} of
$\mathcal{C}$ if restriction of $\Phi $ to $\Gamma_{\mathcal{C}}$ is identity.

\item A couple of diffeomorphisms $\phi^{1}\in Diff (Z_{p})$, $\psi\in Diff(\tilde Z )$ is said to generate
the \textbf{symmetry transformation} of $\mathcal{C}$ if the diffeomorphism $\Psi=\psi \times \phi^{-1}$
of $W_{p}$ is the generalized symmetry of $\mathcal{C}$. This is equivalent to the condition
\[
\psi \circ \mathcal{C}(z)=\mathcal{C}\circ \phi (z)\  \text{for\  all}\  z\in Z_{p}.
\]
A symmetry is, of course, a special case of a generalized symmetry.
\item An automorphism $\phi \in Aut_{p}(\pi)$ is called a \textbf{ geometrical symmetry transformation} of a constitutive relation
$C$ if the diffeomorphism $\Psi ={\tilde \phi}^{*}\times \phi^{1\ -1}$ of $W_{p}$ is the symmetry of $C$,
i.e. if $\mathcal{C}^{\phi_{t}}=\mathcal{C}$.
\item An automorphism $\phi \in Aut_{p}(\pi)$ is called a \textbf{ geometrical symmetry transformation} of
a covering constitutive relation $\widehat{\mathcal{C}}$ if the diffeomorphism $\Psi ={\tilde
\phi}^{*}\times \phi^{1\ -1}$ of $W_{p}$ is the symmetry of $ \widehat{\mathcal{C}}$, i.e. if
$\widehat{\mathcal{C}}^{\phi_{t}}=\widehat{\mathcal{C}}$.
\item
Let $\xi\in \mathcal{X}_{p}(\pi)$ be a projectable vector field.  We say that $\xi $ is a
\textbf{geometrical infinitesimal symmetry} of the constitutive relation $\mathcal{C}$ if
$L_{(\xi^{1},\widetilde{\xi}^{*})}\mathcal{C}=0.$
\end{enumerate}
\end{definition}
Properties presented in the next Proposition follows directly from the given definitions. Last statement
follows from (12.2-3)
\begin{proposition}
\begin{enumerate}
\item A vector field $\xi \in \mathcal{X}_{p}(\pi)$ is an infinitesimal symmetry of $\mathcal{C}$ if
 (and only if) the
(local) phase flow diffeomorphisms $\Phi^{\xi}_{t}=\psi_{t}\times \phi^{1}_{-t}$ of $W_{p}$ defined by the
prolongation of $\xi$ map $\Gamma_{\mathcal{C}}$ into itself:
\[
\Phi^{\xi}_{t}(\Gamma_{\mathcal{C}})=\Gamma_{\mathcal{C}},
\]
i.e. if the (local) phase flow $\phi_{t}$ of vector field $\xi$ is the geometrical symmetry of
$\mathcal{C}$.
\item Generalized symmetries of $\mathcal{C}$ form the group $GSym(\mathcal{C})\subset Diff (W_{p}).$
\item Trivial symmetries of $\mathcal{C}$ form the normal subgroup  $TSym(\mathcal{C})$ of $GSym(\mathcal{C})$.
\item Geometrical symmetries $\Phi$ form the subgroup $Sym(\mathcal{C})\subset Aut_{p}(\pi)$.
\item Infinitesimal symmetries of $\mathcal{C}$ form Lie algebra $\mathfrak{g}(\mathcal{C})\subset \mathcal{X}_{p}(\pi)$
with the bracket of vector  fields in $Y$ as the Lie algebra operation.
\item A vector field $X\in \mathcal{X}(W_{p})$ is the generator of the 1-parametrical group of generalized
 symmetries of $\mathcal{C}$ if and only if it is tangent to the graph $\Gamma_{\mathcal{C}}$.
 \item If $\xi\in \mathcal{X}_{p}(Y)$ is the infinitesimal geometrical symmetry, then for the local phase
 flow $\phi_{t}$ of $\xi$
 \[
\Theta_{{\hat C}^{\phi_{t}}}=\phi^{1\ -1\ *}\Theta_{\hat C}=\Theta_{\hat C}.
 \]
\end{enumerate}
\end{proposition}

Condition that generalized Lie bracket (12.17) is zero has the form of a system of differential equation
of the first order for the components of the constitutive relation $\mathcal{C}$:

\beq
\begin{cases}
 \xi^{1}\cdot F^{\mu}_{i}+\left(\left( F^{\nu}_{i}\left(\frac{\partial \xi^\mu }{\partial x^\nu
}-\xi^{\mu}\lambda_{G,x^\nu }\right)+F^{\mu}_{j}\frac{\partial \xi^j }{\partial y^i}+F^{\mu}_{i}\left(
\frac{\partial \xi^\nu }{\partial x^\nu
}-\xi^{\nu}\lambda_{G,x^\nu}\right)+F_{i}^{\nu}\xi^{\mu}\lambda_{G,x^\nu} \right) \right)=0,\\
\xi^{1}\cdot \Pi_{k}- \left( \Pi_{k}(\xi^{\mu}\frac{\partial \lambda_{G} }{\partial x^\mu }-\frac{\partial
\xi^\mu}{\partial x^\mu})-\Pi_{j}\frac{\partial \xi^j}{\partial y^k}\right)=0
\end{cases}
\eeq

Vector field $\xi^1$ in these equations for a fixed $\mu$ represents the acts on the components of the
vector function (with values in the space dual to the vertical tangent vector of the bundle $\pi$, i.e. in
$V(\pi^{*})$ lifted to the space $Z_{p}$.
\par
For the vertical vector fields $\xi =\xi^{i}\partial_{y^{i}}$ the system (11.5) takes the form

\beq
\begin{cases}
\left(\xi^{j}\partial_{y^{j}}+d_{\nu}\xi^{j}\partial_{z^{j}_{\nu}}\right)
F^{\mu}_{i}+F^{\mu}_{j}\frac{\partial \xi^j }{\partial y^i} =\left(\delta^{j}_{i}\xi^{1}+
\frac{\partial \xi^j}{\partial y^i} \right)F^{\mu}_{j}=0,\\
\left(\xi^{j}\partial_{y^{j}}+d_{\nu}\xi^{j}\partial_{z^{j}_{\nu}}\right)\Pi_{k}+ \Pi_{j}\frac{\partial
\xi^j}{\partial y^k}=\left(\delta^{j}_{i}\xi^{1}+ \frac{\partial \xi^j}{\partial y^i} \right)\Pi_{j}=0
\end{cases}
\eeq We can rewrite last system as the system of conditions to the vertical vector field $\xi
=\xi^{i}\partial_{y^{i}}$:

\beq
\begin{cases}
\left( F^{\mu}_{i,z^{j}_{\sigma}}d_{\sigma}+F^{\mu}_{j}\frac{\partial }{\partial
y^i}\right)\xi^{j} +F^{\mu}_{i,y^j}\xi^{j}=0,\ \mu=1,\ldots ,n;i=1,\ldots,m,\\
\left( \Pi_{k,z^{j}_{\sigma}}d_{\sigma}+\Pi_{j}\frac{\partial }{\partial
y^k}\right)\xi^j+\Pi_{k,y^j}\xi^{j}=0,\ k=1,\ldots ,m.
\end{cases}
\eeq Recall that here $D_{\sigma} \xi^i=\frac{\partial \xi^i}{\partial
x^\sigma}+z_{\sigma}^{j}\frac{\partial \xi^i}{\partial y^j}.$\par

\vskip0.5cm

Let $\phi \in Aut_{p}(\pi)$ be a geometrical symmetry of a CR $\mathcal{C}$.  For a solution $s\in \Gamma
(\pi)$ of the balance system (10.12-13), i.e. for a section $s\in \Gamma (\pi)$ such that
\[
j^{1\ *}_{p}(s)i_{\xi}{\tilde d}\Theta_{\widehat{\mathcal{C}_{-}}}=0
\]
for all $\xi\in \mathcal{X}(\mathcal{C})$  we have
\begin{multline}
(j^{1}_{p}(\phi^{*}s))^{*}i_{\xi}{\tilde d}\Theta_{\widehat{\mathcal{C}_{-}}}=[\phi^{1\
*}(j^{1}_{p}(s))]^{*}i_{\xi}{\tilde d}\Theta_{\widehat{\mathcal{C}_{-}}}=
(j^{1}_{p}(s))^{*}\circ \phi^{1\  *}i_{\xi}{\tilde d}\Theta_{\widehat{\mathcal{C}_{-}}}=\\
=(j^{1}_{p}(s))^{*} i_{\phi^{1}_{*}\xi}\phi^{1\  *}{\tilde d}\Theta_{\widehat{\mathcal{C}_{-}}}=
(j^{1}_{p}(s))^{*} i_{\phi^{1}_{*}\xi}{\tilde d}\phi^{1\
*}\Theta_{\widehat{\mathcal{C}_{-}}}=(j^{1}_{p}(s))^{*}i_{\phi^{1}_{*}\xi}{\tilde d}\Theta_{\widehat{\mathcal{C}_{-}}}.
\end{multline}
Here we have used the symmetry condition in the form presented in Proposition 24, 7).  Last expression is
equal zero if (vertical) vector field $\phi^{1}_{*}\xi \in \mathcal{X(C)}.$ But, by Lemma 6, Sec.10
$\phi^{1}_{*}\mathcal{X(C)}=\mathcal{X(C}^{\phi}).$  Since for a geometrical symmetry transformation
$\mathcal{C}^{\phi}=\mathcal{C}$  we have proved the following

\begin{theorem} Let $\phi \in Aut_{p}(\pi)$ be a symmetry of the CR
 $\mathcal{C}$. Then the mapping $s\rightarrow
 \phi^{*}s$ maps the set $Sol(C)$ of solutions of the balance system (8.12-13) into itself.
\end{theorem}
\vskip1cm

In the second part of the work we will study action of transformations on the balance systems in more
details, including covariance transformations, equivalence relations etc.

 \vskip0.5cm
\subsection{Homogeneous constitutive relations.}
If the state space of a theory contains enough fields to make the constitutive relations free from the
explicit dependence on $(t,x)\in X$ (general relativity or theory of uniform materials are two examples),
then the corresponding balance system simplifies and while studying it one does not need to introduce
assumptions on the character of the space-time dependence of the balance system.  Definition given below
is an invariant way to distinguish a class of such CR.

Any local chart $x^\mu$ in $X$ defined the local (translational) action of $R^n$ in $X$ associating with
the basic vectors $e_{\mu}$ the vector field $\partial_{x^\mu}.$  Vice versa, any n-dimensional
commutative subalgebra $\mathfrak{h}$ of the Lie algebra of vector fields $\mathcal{X}(U)$, $U$ being an
open connected subset of $X$, defines the locally transitive action of $R^n$ in $U$ and, therefore, a
local chart in a neighborhood of any point in $U$.

\begin{definition}
\begin{enumerate}
\item
Let $\nu $ be a connection in the bundle $\pi$ satisfying to the conditions of Propositions 14 or 15 with
"partial" meaning $K\oplus K'$ or $S$ respectively.   We will call a constitutive relation $\mathcal{C}$
\textbf{$\nu $-homogeneous} if any point $z\in Z_{p}$ there exists a local chart in a neighborhood
$U_{x},\ x=\pi^{1}(z)$ such that the Poincare-Cartan form $\Omega_{\mathcal{C}}$ of the CR $\mathcal{C}$
is invariant under the local flows $\phi^{\xi^1}_{t}$ of the lifts ${\hat \xi}^1$ of $\nu$-horizontal
vector fields ${\hat \xi}, \xi \in \mathfrak{h}$ in the neighborhood of $y=\pi_{10}(z)$:
\[
\mathcal{L}_{{\hat \xi}^1}\Omega_{\mathcal{C}}=0\ mod\ q\eta.
\]
\item A constitutive relation $C$ is called a homogeneous if there is a connection $\nu$ on the bundle $\pi$ such
that $C$ is $\nu$-homogeneous.
\end{enumerate}
\end{definition}
\vskip1cm

\begin{proposition} Let $\nu$ be a connection in the bundle $\pi$. Then the following properties of a
constitutive relation $C$ are equivalent:
\begin{enumerate}
\item $C$ is $\nu$-homogeneous,
\item For all $\xi \in \mathfrak{h}$, the $\nu$-horizontal lift $\hat \xi$ is the infinitesimal symmetry
 of the constitutive mapping $\mathcal{C}$ in sense of Definition 22.
 \item The graph $\Gamma_{C}\subset Z_{p}\times \tilde Z$ of mapping $\mathcal{C}$ is invariant
under the flow generated by (flow) lifts of $\nu$-horizontal vector fields $\hat \xi $, $\xi \in
\mathfrak{h}$.
\end{enumerate}
\end{proposition}
\begin{proof} Trivially follows from the Definition 23 and Proposition 24.
\end{proof}
\begin{remark} In a case where connection $\nu$ is flat, the association $\xi \rightarrow {\hat \xi}^{1}$
is the Lie algebra endomorphism $\mathfrak{h} \rightarrow Aut(\pi^{1})\subset \mathcal{X}(Z_{p})$.
\end{remark}

\begin{remark}
It would be interesting to study the influence of the curvature of connection $\nu$ on the properties of
$\nu$-homogeneous constitutive relations.
\end{remark} \vskip1cm

\vfill \eject

\section{Noether Theorem.}

Noether Theorem of the Lagrangian Field Theory associates the conservation law with one-parameter groups
of symmetries (or with the corresponding vector fields) the \emph{conservation laws} that is valid for any
solution of the Euler-Lagrange equations (on shell).  Conserved currents are defined in terms of the
(multi)-momentum mapping that in the case of a multisymplectic field theory was constructed in \cite{MS}.
\par

In the situation considered in this work we might expect a similar result to be true at least for the
semi-Lagrangian constitutive relation $C_{L,\Pi}$ or RET case (see Sec.9). On the other hand, in the
general case, with serious restrictions to the admissible variations $\xi$ one can hardly expect the
Noether Theorem type results. In this section we study possible formulations of the (first) Noether
Theorem in for semi-Lagrangian, RET and general constitutive relation $C$. We follow the works
\cite{LDS,MS,CGM} in the presentation of Noether Theorem of multisymplectic field theory. \par

We consider separately cases of semi-Lagrangian constitutive relation $C$ and corresponding Lagrangian
lift to the CCR $\hat C$ (see Sec.9) and the general case. In the first case results are parallel to the
Lagrangian case, in the second one they are much more limited.\par

Let a Lie group $G\subset Sym(C)\subset Aut_{p}(\pi)$ be a subgroup of the geometrical symmetry group of a
constitutive relation $\mathcal{C}$. Let $\mathfrak{g}$ be the Lie algebra of the group $G$ and
$\mathfrak{g}^{*}$ be its dual space.  Lie algebra $g$ acts on $Y$ by projectable infinitesimal
transformations, i.e there exists homomorphism of Lie algebras $\mathfrak{g}\rightarrow X_{p}(\pi).$   For
an element $\xi \in \mathfrak{g}$ we denote by the same letter the corresponding vector field in $Y$, by
$\xi^1$ - the lifted vector field in $J^{1}_{p}(\pi)$ preserving Cartan distribution, by $\xi^{1*}$ - the
vector field in $\Lambda^{(n+1)+(n+2)}_{2}Y$ leaving invariant the canonical multisymplectic form:
\[
L_{\xi^{1*}}(\Theta^{n+1}_{2}+\Theta^{n+2}_{2})=0.
\]
In a more general fashion consider the Lie subalgebra $\mathfrak{g}\subset \mathcal{X}_{\pi^1}(Z_{p})$ of
the Lie algebra of projectable (to $Y$ and to $X$) vector fields $\hat \xi$ in $Z_{p}$ which consists of
the \emph{infinitesimal symmetries} of the CCR $\hat C$.  In other words we assume that the projection
$\xi \in \mathcal{X}(\pi)$ of $\hat \xi$ in $Y$ is defined and being lifted to the vector field $\xi^{*}$
in the bundle $\Lambda^{(n+1)+(n+2)}$ preserving canonical form(s) $\Theta^{n+1}_{2}+\Theta^{n+2}_{2}$ is
such that $\mathcal{L}_{({\hat \xi},\xi^{*})}C=0$ (see Sec.12)). Then as is proved in Sec.12 in terms of
Poincare-Cartan form $\Theta_{\hat C}$ this condition takes the form

 \beq L_{\xi^{1}}(\Theta^{n+1}_{\hat
C}+\Theta^{n+2}_{\hat C})=0, \eeq
 obtained by differentiating condition 7) in the Proposition 23.  This
splits into two conditions - independent preservation of forms $\Theta^{n+1}_{\hat C}$ and
$\Theta^{n+2}_{\hat C}$. First condition is the natural generalization of the invariance condition of the
Lagrangian field Theory (\cite{LDS}).\par Former situation ($\xi \in \mathcal{X}_{p}(\pi)$) is the special
case of the later one where ${\hat \xi}=\xi^1 .$

\begin{definition} Let $\hat C$ be a covering constitutive relation in $Z_{p}$.
\begin{enumerate}
\item A vector field $\xi \in X(Y)$ is called a \textbf{variational symmetry} if the Lie derivative
\[
\mathcal{L}_{\xi^1}\Theta^{n+1}_{\hat C}\in \mathcal{I}(Ca)
\]
(belongs to the differential ideal of (partial) contact structure of $Z_{p}$, see Sec.6) and also
$\xi^{1}$ is tangent to the boundary subbundle $B$ and verifies $\mathcal{L}_{\xi^1}\Pi_{C}=0.$
\item A vector field $\xi \in X(Y)$ is called a \textbf{Noether (divergence) symmetry} if there is a $n$-form
$\alpha\in \Lambda^{n}(Y)$ whose pullback $\alpha$ to $Z_{p}$ is exact on $B$: $\alpha_{B}=d\beta$ and
such that
\[
\mathcal{L}_{\xi^1}\Theta^{n+1}_{\hat C}-d\alpha \in \mathcal{I}(Ca),
\]
and vector field $\xi^{1}$ is tangent to $B$ and verifies $\mathcal{L}_{\xi^1}\Pi_{C}=0.$
\item A vector field $\xi_{Z}\in \mathcal{X}(Z_{p})$ is called a \textbf{Cartan symmetry} of $C$ if
\begin{enumerate}
\item Flow of $\xi_{Z}$ preserves the differential ideal $\mathcal{I}(Ca)$: $\mathcal{L}_{\xi_Z}\theta \in \mathcal{I}(Ca)$
for all $\theta \in \mathcal{I}(Ca)$,
\item There exists a n-form $\alpha$ on $Z_{p}$ that is exact on $B$: $\alpha_{B}=d\beta$ and such that
\[
\mathcal{L}_{\xi_Z}\Theta^{n+1}_{\hat C}-d\alpha \in \mathcal{I}(Ca),
\]
\item Vector field $\xi_Z$ is tangent to $B$ and verifies $\mathcal{L}_{\xi_Z}\Pi_{C}=0.$
\end{enumerate}
\end{enumerate}
\end{definition}
Every variational symmetry is Noether symmetry as well. If $\xi$ is a Noether symmetry, then its flow
prolongation is a Cartan symmetry. Vice versa, a $\pi_{10}$-projectable Cartan symmetry is the flow
prolongation of its projection which is the Noether symmetry. In the next proposition proof of which is
the same as in \cite{LDS} some properties of symmetries of these three types are collected.
\begin{proposition} Let $\hat C$ be a covering constitutive relation defined at $Z_{p}$.
\begin{enumerate}
\item Variational symmetries form the Lie subalgebra $\mathfrak{vg}_{\widehat{C}}$ of $\mathcal{X}(Y)$.
\item Noether symmetries form the Lie subalgebra $\mathfrak{ng}_{\widehat{C}}$ of $\mathcal{X}(Y)$.
\item Cartan symmetries form the Lie subalgebra $\mathfrak{cg}_{\widehat{C}}$ of $\mathcal{X}(Z_{p})$.
\item For the prolongations of the first two types of vector fields we have the following sequence of
embeddings of Lie subalgebras of $\mathcal{X}(Z_{p})$:
\[
 \mathfrak{vg}^{1}_{\widehat C}\hookrightarrow
 \mathfrak{ng}^{1}_{\widehat C}\hookrightarrow  \mathfrak{cg}_{\widehat C}
\]
\item A geometrical infinitesimal symmetry $\xi \in X_{p}(\pi)$ of CCR $\hat C$ is the variational symmetry of the CCR
$\hat C$.
\item An infinitesimal symmetry ${\hat \xi}\in \mathcal{X}_{\pi^1}(Z_{p})$ is the Cartan symmetry of the
CCR $\hat C$.
\end{enumerate}
\end{proposition}
Now we define the canonical multimomentum mapping (MM) following \cite{MS}.
\begin{definition} The multimomentum mapping $J:\Lambda^{n+1}_{2}Y \rightarrow \Lambda^{n}(X)\otimes \mathfrak{g}^{*}$
is defined as
\[
J(z^{*})(\xi)=i_{\xi^{1*}}\Theta^{n+1}_{2}(z^{*}),\ \text{for\ all}\ z^{*}\in \Lambda^{n+1}_{2},\ \xi \in
\mathfrak{g}.
\]
\end{definition}
\begin{lemma}(\cite{MS,LDS}) For the MM-mapping $J$ and an arbitrary ${\hat \xi}\in \mathcal{X}(Z_{p})$
\[
i_{\widehat{\xi}^{*}}d\Theta^{n+1}_{2}=-dJ(z^{*})(\widehat{\xi}).
\]
\end{lemma}
\begin{proof} We have $0=L_{\widehat{\xi}^{*}}\Theta^{n+1}_{2}=i_{\widehat{\xi}^{*}}d\Theta^{n+1}_{2}
+di_{\widehat{\xi}^{*}}\Theta^{n+1}_{2}=i_{\widehat{\xi}^{*}}d\Theta^{n+1}_{2}+dJ(z^{*})(\widehat{\xi}).$
\end{proof}
The MM mapping $J^{\hat C}$ for an arbitrary covering constitutive relation ${\hat C}$ is defined here in
the same way as it was defined in \cite{MS} for the Legendre transformation corresponding to a Lagrangian
$L\eta$.\par

\begin{definition} A multimomentum mapping of a covering constitutive relation ${\hat C}$ is the mapping
$J^{\hat C}:Z_{p}=J^{1}_{p}(\pi)\rightarrow \Lambda^{n}(X)\otimes \mathfrak{g}^{*}$
\[
J^{\hat C}(z)(\hat{\xi})={\hat C}_{z}^{*}J([{\hat C}^{n+1}(z))({\hat \xi} )=i_{\hat{\xi} }\Theta_{\hat
C}^{n+1}(z),
\]
where $\mathcal{\hat C}^{n+1}$ is the (n+1)-component of the constitutive mapping $\mathcal{\hat C}$.
\end{definition}
\begin{remark} Notice that $J^{\hat C}(z)$ depends only on the current component of the constitutive mapping
and, therefore, $J^{\hat C}(z)=J^{{\hat C}_{-}}(z).$
\end{remark}
\begin{lemma} If the mapping $\mathcal{\hat C}$ is regular (i.e. if it is the diffeomorphism onto its image), then
\[
i_{\hat{\xi}}d\Theta^{n+1}_{\mathcal{\hat C}}=-dJ^{\mathcal{\hat C}}(z)(\hat{\xi}),\ \ \text{for\ all}\
z\in J^{1}_{p}(\pi),\ \xi \in \mathfrak{g}.
\]
\end{lemma}
\begin{proof} Follows from the previous Lemma by using the $\mathcal{G}$-equivariance of the constitutive relation
$C$ giving $\mathcal{\hat C}_{*}\hat{\xi }=\xi^{*}\circ \mathcal{\hC}$ (recall that $\xi$ is the
projection of $\hat \xi$ to $Y$). More specifically, we have
\begin{multline}
(dJ^{\hC}(z)(\hat{\xi})=(d\hC^{*}J)(z)(\hat{\xi})=\hC^{*}(dJ)(\hat{\xi})=dJ(\hC(z))(\hC_{*}(\hat{\xi}))=
dJ(\hC(z))(\xi^{*}(\hC(z)))=\\
=\hC^{*}i_{\xi^{*}(\hC(z))}d\Theta^{n+1}_{2}=\hC^{*}i_{\hC_{z*}(\hat{\xi}(z))}d\Theta^{n+1}_{2}=
i_{\hat{\xi}}\hC^{*}_{z}d\Theta^{n+1}_{2}=i_{\hat{\xi}}d\hC^{*}_{z}\Theta^{n+1}_{2}=i_{\hat{\xi}}d\Theta^{n+1}_{\hC}
\end{multline}
\end{proof}

\begin{theorem}(Noether Theorem)  Let $\hat C$ be a semi-Lagrangian covering constitutive relation with \[
\Theta_{\hat C}=(L-z^{i}_{\mu}L_{z^{i}_{\mu}})\eta+L_{z^{i}_{\mu}}dy^i \wedge
\eta_{\mu}+(\Pi_{i}-L_{,y^i})dy^i \wedge \eta,
\]
and let $\xi \in X_{p}(Y)$ be a variational symmetry of $\hat C$. Then for all solutions $s\in
\Gamma(\pi)$ of the balance system $\bigstar$ the following balance equation is true

\beq d [(j^{1}(s))^{*}J^{\mathcal{\hat C}}(z)(\xi)]=(\omega^{i}(\xi)\Pi_{i})\circ j^{1}(s))^{*}\eta.
 \eeq
\end{theorem}
\begin{proof} We have, by the Cartan formula for the Lie derivative and using Theorem 3
\[
0=(j^{1}(s))^{*}\mathcal{L}_{\xi^1}\Theta^{n+1}_{\hat C}=(j^{1}(s))^{*}\left( di_{\xi^1}\Theta^{n+1}_{\hat
C}+ i_{\xi^1}d\Theta^{n+1}_{\hat C}\right)=d (j^{1}(s))^{*}i_{\xi^1}\Theta^{n+1}_{\hat
C}+(\omega^{i}\Pi_{i})\circ j^{1}(s))^{*}\eta,
\]
since $s$ is a solution of the balance system (10.29) with $\Pi_{i}$ replaced by $\Pi_{i}-L_{,y^i}.$
\end{proof}
In the same way the following statement is proved
\begin{theorem}(Noether Theorem) Let $\hat C$ be a semi-Lagrangian covering constitutive relation with \[
\Theta_{\hat C}=(L-z^{i}_{\mu}L_{z^{i}_{\mu}})\eta+L_{z^{i}_{\mu}}dy^i \wedge
\eta_{\mu}+(\Pi_{i}-L_{,y^i})dy^i \wedge \eta,
\]
and let $\xi_{Z}\in X(Z_{p})$ be a Cartan symmetry of $\hat C$ (in particularly, $\xi_{Z}=\xi^1$ for $\xi
\in X_{p}(Y)$ to be a Noether symmetry of $\hat C$). Then for all solutions $s\in \Gamma(\pi)$ of the
balance system $\bigstar$ the following balance equation is true

\beq d [(j^{1}(s))^{*}i_{\xi_Z}\Theta^{n+1}_{\hat C}-\alpha ]=(\omega^{i}(\xi_{Z})\Pi_{i})\circ
j^{1}(s))^{*}\eta.
 \eeq
\end{theorem}

\begin{corollary} If, in addition to the conditions of the Theorem 8 the balance system $\mathcal{B_{C}}$ is the
conservation system (i.e. if $\Pi_{i}=0,\ i=1,\ldots ,m$), then for all $\xi \in \mathfrak{g}$ and for all
solutions $s\in \Gamma(\pi)$ of the balance system $\mathcal{B_{C}}$ the \emph{Noether conservation law}
holds:
 \beq d [(j^{1}(s))^{*}i_{\xi_Z}\Theta^{n+1}_{\hat C}-\alpha ]=0 ,\eeq
\end{corollary}
\begin{remark} Associating  to each $\xi \in \mathfrak{g}$  the corresponding balance law (13.3) defined
the linear mapping
\[
\mathfrak{g}\rightarrow \mathfrak{BL}_{\mathcal{\widehat{C}}}
\]
to the space of secondary balance laws of the system $B_{C}$ (see Sec.11).
\end{remark}

Let now condition (13.1) is fulfilled i.e. $G$ is symmetry of \emph{both flux \textbf{and} source terms}
of the constitutive relation $C$. Then
\begin{multline} di_{\hat{\xi}}\Theta_{\hat{C}}^{n+2}=-i_{\hat{\xi}}d\Theta_{\hat{C}}^{n+2}= -i_{\hat{\xi}}(d\Pi_{i}\wedge dy^i
\wedge \eta )=\\ =[-(\hat{\xi}\cdot \Pi_{i})dy^i \wedge \eta -
\xi^{i}d\Pi_{i}\wedge \eta +\xi^{\mu}d\Pi_{i} \wedge dy^i \wedge \eta_{\mu}]=\\
=-(\hat{\xi}\cdot \Pi_{i})(\omega^i +z^{i}_{\nu}dx^\nu )\wedge \eta -\xi^{i}(\Pi_{i,x^\sigma}dx^\sigma
+\Pi_{i,y^j}(\omega^j +z^{j}_{\sigma}dx^\sigma )+\Pi_{i,z^{j}_{\sigma}}(\omega^{j}_{\sigma }-z^{j}_{\sigma
\lambda} dx^\lambda  )\wedge \eta +\\ + \xi^{\mu}d\Pi_{i} \wedge (\omega^i+z^{i}_{\nu} dx^\nu ) \wedge
\eta_{\mu}] = \text{Con}+\xi^{\mu}d\Pi_{i} \wedge z^{i}_{\nu} dx^\nu \wedge \eta_{\mu} =\\ =
\text{Con}+\xi^{\mu}[\Pi_{i,x^\sigma}dx^\sigma +\Pi_{i,y^j}(\omega^j +z^{j}_{\sigma}dx^\sigma
)+\Pi_{i,z^{j}_{\sigma}}(\omega^{j}_{\sigma }-z^{j}_{\sigma \lambda} dx^\lambda )]\wedge z^{i}_{\mu}\eta
=\\ = \text{Con}.
\end{multline}
Here $Con$ means a contact form.  During this calculation we repeatedly used the equality $dx^\nu \wedge
\eta =0$. Notice that the same statement follows directly from the fact that the pullback by $j^{1}(s)$ of
the $n+1$-form is necessary closed.  Applying now the pullback by $j^{1}(s)$ we get the following

\begin{proposition} Let, in addition to the conditions of Theorem 7,  $G$ is the symmetry of the source part
of the constitutive relation, i.e. (13.1) is true. Then
\[
di_{\hat{\xi}}\Theta^{n+2}_{\hat{C}}=\text{Cont}\Rightarrow
dj^{1}(s)^{*}i_{\hat{\xi}}\Theta^{n+2}_{\hat{C}}=0
\]
 for all sections $s$. Therefore, locally (and in a top. trivial domain, globally)
\[
j^{1}(s)^{*}i_{\hat{\xi}}\Theta^{n+2}_{\hat{C}}=d\Phi_{\hat{C}}(s,\hat{\xi} ,z)
\]
for some (n+1) form $\Phi_{\hat{C}}$ ($\mathfrak{g}$-potential of the source $\hat{C}^{n+2}$) linearly
depending on the vector field $\xi$.
\end{proposition}
\begin{remark} notice that in the last Proposition, as in the Noether Theorems above
$i_{\hat{\xi}}\Theta^{n+2}_{\hat{C}}=(\Pi_{i}\xi^{i})\eta.$
\end{remark}
\begin{corollary} If $G$ is the Lie group of symmetries of a regular constitutive relation $C$ then
(locally) in the conditions of Theorem 9,
\[
d[J^{\widehat{\mathcal{C}}}(j^{1}(s)(x)(\xi)-\Phi_{\mathcal{C}}(s,\xi ,z) ]=0
\]
for all solutions $s\in \Gamma(\pi)$ of the balance system $\mathcal{B_{C}}.$
\end{corollary}
For the RET constitutive relations we get the results similar to those for semi-Lagrangian case valid for
the \emph{lifted} covering constitutive relations (comp. Sec.10, Thm.4):

\begin{theorem}(Noether Theorem)  Let $\tilde C$ be a lifted covering constitutive relation of the RET type
with \[ \Theta_{\tilde C}=-z^{i}_{\mu}F^{\mu}_{i}\eta+F^{\mu}_{i}dy^i \wedge \eta_{\mu}+\Pi_{i}dy^i \wedge
\eta,
\]
and let $\xi \in X(Y)$ be a variational symmetry of $\tilde C$. Then for all solutions $s\in \Gamma(\pi)$
of the balance system $\bigstar$ the following balance equation is true

\beq d [(j^{1}(s))^{*}J^{\mathcal{\tilde C}}(z)(\xi)]=(\omega^{i}(\xi)\Pi_{i})\circ j^{1}(s))^{*}\eta.
 \eeq
\end{theorem}

Now we formulate the Noether Theorem for the balance system $\mathcal{B_{C}}$ corresponding to a general
regular constitutive relation. This result is limited since the infinitesimal symmetry vector fields $\xi$
should be $C$-admissible.

\begin{theorem} Let $\mathcal{C}$ be a \emph{regular} constitutive relation defined on a partial 1-jet bundle
$J^{1}_{p}(\pi)$ and $\tilde C$ - its lifted (covering) constitutive relation. Let a Lie group $G\subset
Sym(C)\subset Aut(\pi)$ be a symmetry group of the flux part $\Theta^{n+1}_{C}$ of the constitutive
relation $\mathcal{C}$ such that its Lie algebra $\mathfrak{g}$ consists of $C$-admissible vector fields
$\mathfrak{g} \subset \mathcal{X(C)}$ on $Z_{p}$. Then for all $\hat{\xi} \in \mathfrak{g}$ and for all
solutions $s\in \Gamma(\pi)$ of the balance system $\mathcal{B_{C}},$

\beq d[J^{\tilde{C}}(j^{1}(s)(x)(\widehat{\xi})]=j^{1}(s)^{*}i_{\xi}\Theta^{n+2}_{\tilde{C}}=
j^{1}(s)^{*}[\omega^{i}(\widehat{\xi})\Pi_{i}]\eta ,\eeq where $\omega^{i}=dy^i-\sum_{\mu,(\mu ,i)\in
P}z^{i}_{\mu}dx^\mu $ are the basic Cartan forms in $J^{1}_{p}(\pi)$.
\end{theorem}
\begin{proof} We have
\begin{multline}
dJ^{\tilde C}(j^{1}(s))(\hat{\xi})=j^{1}(s)^{*}dJ^{{\tilde
C}_{-}}(\hat{\xi})=-j^{1}(s)^{*}i_{\hat{\xi}}d\Theta^{n+1}_{\tilde{C}_{-}} =\\
=-j^{1}(s)^{*}i_{\hat{\xi}} \Theta^{n+2}_{\tilde{C}_{-}}=j^{1}(s)^{*}[\omega^{i}(\xi^{1})\Pi_{i}]\eta.
\end{multline}
where $\omega$ is some contact form (we have used the result of Lemma 12 Appendix IV). Here we have used
formulation (10.34) (see Theorem 5) of the balance system. In the last equality we have used the relation
$\omega^{i}(\hat{\xi})=\xi^{i}$.  As a result we get
\[
dJ^{C}(j^{1}(s))(\xi)=j^{1}(s)^{*}i_{\xi^{1}}
\Theta^{n+2}_{C}=j^{1}(s)^{*}[\omega^{i}(\xi^{1})\Pi_{i}]\eta.
\]
\end{proof}
\begin{corollary} If, in addition to the conditions of the last Theorem the balance system $B_{C}$ is the
conservation system (i.e. if $\Pi_{i}=0,\ i=1,\ldots ,m$), then for all $\xi \in \mathfrak{g}$ and for all
solutions $s\in \Gamma(\pi)$ of the balance system $B_{C}$ the Noether conservation law holds:
 \beq d[J^{C}(j^{1}(s)(x)(\xi)]=0 ,\eeq
\end{corollary}

\subsection{Energy-Momentum Balance Law.}
Let $\nu$ be a connection in the bundle $\pi:Y\rightarrow X$ with the form
$dy^{i}-\Gamma^{i}_{\mu}dx^\mu.$  Consider a $\nu$-homogeneous constitutive law $\mathcal{C}$ and the
corresponding balance system $\mathcal{B_{C}}$. Let $(x^\mu ,y^i)$ be a local adopted chart in the bundle
$\pi$.\par Let $\partial_{x^\mu }$ be a basic vector field in $X$ and let $\xi_{\mu}=\partial_{x^\mu
}+\Gamma^{i}_{\mu}\partial_{y^i}$ be its horizontal lift in $Y$. Flow lift of the vector field $\xi_{\mu}$
is
\[
{\hat \xi}_{\mu}=\partial_{x^\mu
}+\Gamma^{i}_{\mu}\partial_{y^i}+d_{\nu}\Gamma^{i}_{\mu}\partial_{z^{i}_{\nu}}.
\]
\textbf{Now we assume that} ${\hat \xi}_{\mu}\in \mathcal{X(C)}.$ Remind that this is true for all $\mu$
and all connections $\nu$ in semi-Lagrangian case and in the RET case. In the general case that requires
fulfillment of two conditions: ${\hat \xi}_{\mu}$ is $P$-vertical and

\beq \omega_{C}({\hat \xi}_{\mu})=F^{\nu}_{i}d_{\nu}\Gamma^{i}_{\mu}=0.\eeq
\par

 Calculate now
\[
i_{{\hat \xi}_{\mu}}\Theta_{C}^{n+1}=F^{\nu}_{i}\Gamma^{i}_{\mu}\eta_{\nu}-F^{\nu}_{i}dy^i \wedge
\eta_{\mu \nu},
\]
and $\omega^{i}({\hat \xi}_{\mu})=\Gamma^{i}_{\mu}-z^{i}_{\mu}$.  Therefore, the balance law (13.3,13.7)
takes, for the vector field ${\hat \xi}_{\mu}$, the form

\beq dj^{1}_{p}(s)^{*} [F^{\nu}_{i}\Gamma^{i}_{\mu}\eta_{\nu}-F^{\nu}_{i}dy^i \wedge \eta_{\mu \nu}]=
j^{1}_{p}(s)^{*}(\Pi_{i}(\Gamma^{i}_{\mu}-z^{i}_{\mu}))\eta.\eeq

Introducing 1-jet of section $s$ into the form in brackets and omitting the form $\eta$ we get the
energy-momentum balance law in the form (comp. \cite{GMS}, Chapter 3)

\beq d[(F^{\nu}_{i}\Gamma^{i}_{\mu})\circ j^{1}_{p}(s)-\delta^{\nu}_{\mu}F^{\sigma}_{i}s^{i}_{\sigma}
+F^{\nu}_{i}s^{i}_{,\mu} ]_{,x^{\nu}}= \Pi_{i}(s)(\Gamma^{i}_{\mu}(s)-s^{i}_{,\mu})). \eeq

Energy-momentum Tensor for the constitutive relation $C$ has, thus, the form

\beq
T^{\nu}_{\mu}=F^{\nu}_{i}\Gamma^{i}_{\mu}-\delta^{\nu}_{\mu}F^{\sigma}_{i}z^{i}_{\sigma}+F^{\nu}_{i}z^{i}_{\mu}.
\eeq
\subsection{Case of pure gauge symmetry transformation.}

Let $\xi=\xi^{i}\partial_{y^i}$ be a vertical (pure gauge) symmetry transformation of a constitutive
relation $\mathcal{C}$. Then, the flow lift of vector field $\xi$ to $Z_{p}$ is  $\xi^{1}=\xi
+d_{\mu}\xi^{i}\partial_{z^{i}_{\mu}}$.  \par We calculate
\[
i_{\xi^{1}}\Theta_{C}^{n+1}=\xi^{i}F^{\mu}_{i}\eta_{\mu}, \omega^{i}(\xi^{1})=\xi^{i}.
\]
Therefore, the Noether balance equation corresponding to the vector field  $\xi$ has the form

\beq dj^{1}_{p}(s)^{*}(\xi^{i}F^{\mu}_{i}\eta_{\mu})= j^{1}_{p}(s)^{*}(\xi^{i}\Pi_{i}\eta ). \eeq
Substituting $s$ explicitly we get this equation in the form

\beq [(\xi^{i}F^{\mu}_{i})( j^{1}_{p}(s))]_{,x^\mu}= (\xi^{i}\Pi_{i})( j^{1}_{p}(s)) \ -\eeq - the
secondary balance law defined by the $\mathcal{C}$-admissible vector field $\xi$ (see Sec. 11) and part II
of this work.

\vfill \eject
\section{Evolutional balance systems.}

Here we consider the case where $Z_{p}=Z_{S}=J^{1}_{S}(\pi)$.  A general balance system (9.13) does not
necessary produce an evolutional dynamical system for all the state fields $y^i$, the extreme case of such
a situation be when no time derivatives enters the constitutive relation and $F^{0}_{i}=0$ for all $i$. To
specify type of balance systems that produce a dynamical system for all the fields $y^i$ we have to put
some restrictions on the CR $\mathcal{C}$. \par

We simplify our consideration here by assuming here that the CR $C$ satisfies to the condition:
\vskip0.4cm \textbf{Time derivatives $y^{i}_{t}$ of the fields $u^i$ may enter the constitutive relation
$C$ only through the term $F^{0}_{i}$ of } $\mathcal{C}$. \vskip0.4cm

Then, the balance system $\bigstar $ can be schematically written in the form

\beq \sum_{j\in S}\frac{\partial F^{0}_{j}}{\partial y^k}y^{k}_{,t}+\sum_{S_{t}\cup S_{tx}}\frac{\partial
F^{0}_{j}}{\partial y^{k}_{t}}y^{k}_{,tt}=G_{j}(y,y_{x},y_{xx}), \eeq

 where terms containing time derivatives are gathered on the left.\par

\begin{example} Consider a $\nu$-horizontal constitutive relation $\mathcal{C}$  such that $U_{t}=U_{tx}=0$ (so that no time derivatives of the fields $y^i$
 enters $\mathcal{C}$), If in such a case the condition of \emph{regularity} is fulfilled

\beq det\left( \frac{\partial F^{0}_{i}}{\partial y^{j} }\right) \ne 0,\eeq guarantees that the system can
be written in the normal form
\[
y^{i}_{t}=H^{i}(y,y_{x}),
\]
and, in an analytical case, the Cauchy problem for this system is is locally solvable.
\end{example}
Restrict to the case where the bundle $Y\rightarrow X$ is the \emph{vector bundle}. Introduce the pullback
of the bundle $\pi:Y\rightarrow X$ to $Z_{p}$: \beq
\begin{CD} \pi^{1\ *}(Y) @>>> Y\\
@VVV  @V \pi VV \\
Z_{p} @> \pi^{1} >> X
\end{CD}.
\eeq

 Let the $m\times m$ matrix $A_{2}(z)=\frac{\partial F^{0}_{j}}{\partial
y^{k}_{t}}(z)$ has, in a neighborhood of a point $z$ a constant rank $s_{2} \leqq \vert S_{t}\cup
S_{tx}\vert $. Then, the kernel $K_{2}(z)$ of the matrix $\frac{\partial F^{0}_{j}}{\partial y^{k}_{t}}$
defines, at each point $z$ the vector subspace $K_{2}(z)\subset \pi_{1\ z}^{*}(Y_{x}).$  If the rank of
matrix $A_{2}(z)$ is constant, we get (locally) the subbundle $K_{2}$  of the pullback bundle $\pi^{1\
*}(Y)$ of the fields $y^i$ whose \emph{second time derivatives} $y^{i}_{tt}$ do not enter the balance
system (14.1).\par
 In the same way, at each point there is defined the rank $s_{1}$ of the matrix
 $A_{1}(z)=\frac{\partial F^{0}_{j}}{\partial
 y^k}(z)$ and the subspace $K_{1}(z)$ - kernel of the linear mapping defined by the matrix $A_{1}(z)$.  If
 that rank is locally constant, then the vector subbundle $K_{1}$ of the bundle $\pi_{1\ z}^{*}(Y)$ is
 defined in the same way as $K_{2}$ - subbundle of vector fields whose \emph{first time derivatives} $y^{j}_{t}$
  do not enter the balance system at the point $z$.
\par
Define the intersection $K(z)=K_{1}(z)\cap K_{2}(z)$. Generically, if the rank of this intersection
$k_{e}(z)$ is (locally) constant one get the subbundle $K\subset \pi_{1\ z}^{*}(Y)$ of the fields
\emph{whose time derivative do not enter the balance system} (9.13).\par

We have for the defined sub-bundles the inclusion $K\subset K_{1}$. Choose a compliment $K^1$ to the
subbundle $K$ of the bundle $K_1$ (if a Riemannian metric is defined on the fibers $U_{x}$ of the vector
bundle $\pi:Y\rightarrow X$ then it is natural to take $K^{1}=K^{\bot}$).\par

In the same way, choose a vector subbundle $K^{2}$ complemental to $K_{2}$ in the bundle $\pi^{1\ *}(Y)$
(orthogonal if a Riemannian metric is defined on the fibers $U_{x}$ of the bundle $\pi:Y\rightarrow X$
).\par

Thus, we get the decomposition

\beq \pi_{z}^{1\ *}(Y)=K_{z}\oplus K^{1}_{z}\oplus K^{2}_{z} \eeq

of the pullback of the state bundle $Y$ into the sum of subbundles with the corresponding fields that

\begin{enumerate}
\item $y^i \in K^{2}$ enters the $\mathcal{C}^{n+1}_{0}$ (i.e. $dy^i\wedge \eta_{0}$-term of CR) with their time
derivative $y^i_{,t}$,
\item $y^i \in K^{2}$ enters the $\mathcal{C}^{n+1}_{0}$ (i.e. $dy^i\wedge \eta_{0}$-term of CR) but their time
derivative $y^i_{,t}$ does not enter $\mathcal{C}$,
\item Neither $y^i \in K^{2}$ no its time derivative $y^i_{,t}$ enter the $\mathcal{C}$.
\end{enumerate}

\par
\begin{remark} If the bundle $Y\rightarrow X$ is not a vector bundle, similar decomposition exists for the
vertical tangent $V(\pi)$ of the bundle $Y$ and can be used instead.
\end{remark}

Decomposition (14.4) allows to split the system of balance laws (locally, if the ranks of matrices
$A_{i},i=1,2$ and dimension of intersection $K_{1}\cap K_{2}$ are locally constant) into the three
subsystems - hyperbolic for the fields in $K^{2}$, parabolic - for the fields in $K^{1}$ and stationary -
for the fields in $K$.

\begin{theorem}
Let $\pi:Y\rightarrow X$ is the vector bundle and let the CR $\mathcal{C}$ satisfies to the condition:
\textbf{time derivatives $y^{i}_{t}$ of the fields $y^i$ may enter the constitutive relation $C$ only
through the terms $F^{0}_{i}$ of } the CR $\mathcal{C}$. Assume that the the ranks of matrices
$A_{i},i=1,2$ and dimension of intersection $K_{1}\cap K_{2}$ of the subbundles $K_{1},K_{2}\subset
\pi^{1\ *}(Y)$ defined above are constant throughout the $Z_{S}$. Then the pullback of the bundle $Y$ to
the partial 1-jet bundle $Z_{S}$ splits into the sum of three vector subbundles \beq \pi_{1\
z}^{*}(Y)=K\oplus K^{1}\oplus K^{2} \eeq and the balance system (14.1) splits into the hyperbolic,
parabolic and stationary subsystems
\begin{multline}
\begin{cases}
y^{i}_{tt}&=P_{i}(x,y^j,z^{j}_{\mu}),\ y^{i}\in K^{2},\\
y^{i}_{t}&=P_{i}(x,y^j ,z^{j}_{\mu},\mu \ne 0;y^{k}_{t}, y^{k}\in K^{2}),\ y^{i}\in K^{1},\\
0&= P_{i}(x,y,z^{i}_{\mu},\mu \ne 0,y^{k}_{t}, y^{k}\in K^{2}\cap K^{1}.
\end{cases}
\end{multline}
\end{theorem}

At a point $z\in Z_{p}$ let $(h(z),p(z),e(z))$ be corresponding dimensions of subbundles $K^2,\ K^1,\ K$.
we call the triple of numbers $(h(z),p(z),e(z))$ the index of a system (14.1) at a point $z$ and numbers
in this index - hyperbolic, parabolic and stationary dimensions at  the point $z$.  It is clear that sum
of these dimensions is equal to $m$: $h(z)+p(z)+e(z)=m$.

\section{RET balance systems. Lagrange-Liu dual formulation.}
In this section we suggest a bundle picture of the Rational Extended Thermodynamics in terms of dual
variables.  We will be using terminology from Sec.3. Recall that for the conventional RET case where
$F^{0}_{i}=y^i$ (\cite{MR} or Sec.3 above) whenever the entropy density $h^{0}(x,y)$ is convex by vertical
variables $y^i$, the change of variables $\{ y^i\} \rightarrow \lambda^{i}=\frac{\partial h^0}{\partial
y^i}$ is globally defined diffeomorphsim $\wp_{x}: U_{x}\rightarrow \Lambda_{x}$ of the fibers $U_{x}$
onto the space $\Lambda $ of variables $\lambda^i$. This allows to introduce the dual bundle
$\pi_{Y^{*}X}:Y^{*}\rightarrow X$ with the fiber $\Lambda $ with the corresponding
isomorphism of bundles \beq \begin{CD}Y @>\wp >>Y^{*}\\
@V\pi_{Y^{*}X} VV    @V\pi_{YX} VV\\
X @= X \end{CD} \eeq

Since in this section we are repeatedly using notation $\Lambda$ for the space of dual variables, it will
be convenient to change the notation for for the space (or bundle) of exterior k-forms from the
$\Lambda^k$ to $\Omega^k$.\par

 Taking the pullback of the bundle of $n+(n+1)$-forms on $X$ via the projection $\pi_{X Y^*}$ or, what
 is the same, forming the fiber product of the bundle $\pi_{Y^{*}X}$ with the (n+(n+1))-bundle
$\pi_{n+(n+1)}$ (see Sec.2) we get the following commutative square

\beq
\begin{CD}
\Lambda \times \Omega^{n+(n+1)} @>>> Y^{*}\underset{X}{\times} \Omega^{n+(n+1)} @>>> \Omega^{n+(n+1)}(X)\\
@V\pi_{\Omega \Lambda }VV   @V\pi_{\Omega Y^{*}} VV                        @V \pi_{\Omega X}VV\\
\Lambda @>>> Y^{*}         @>\pi_{Y^{*}X} >> X
\end{CD},
\eeq where the left column represent a typical fiber of the middle
column bundle over a point $x\in X$.\par

A  point of a fiber of the bundle $\pi_{\Omega \Lambda}$ can be
presented as
\[
(\lambda, \sum_{\mu} q^{\mu}\eta_{\mu}+p\eta ),
\]
where $q^\mu (x,\lambda ),p(x,\lambda )$ are functions defined on the space $Y^*$.\par

Introduce the 1-jet bundle $J^{1}(\pi_{Y^{*}\Omega})$ of the bundle $\pi_{\Omega Y^{*} }$. A point of the
fiber of this 1-jet bundle $\pi_{Y^{*}\Omega }: J^{1}(\pi_{Y^{*}\Omega})\rightarrow \Lambda (=Y^{*}_{x})
\times (\Omega^{n+(n+1)}_{x})$ (over a fixed base point $x\in X$) can be presented as \beq
 q^{\mu}_{i}\eta_{\mu
}\wedge d\lambda^{i}+p_{i}\eta\wedge d\lambda^{i}, \eeq
 were we
have used the standard isomorphism $J^{1}(E\rightarrow U)\simeq  E\bigotimes T^{*}(U)$ of bundles over $U$
induced by a connection in the bundle $E\rightarrow U$.  In this case we are using a connection induced in
the central column of the bundle (15.2) by the connection $\Gamma^G$ in the bundle of n+(n+1)-forms over
$X$. \vskip0.5cm

Organize the spaces introduced above into the following bundle
picture, where on the right are the local coordinates in the fibers
of the bundles

\begin{equation}
\begin{CD}
J^{1}(\Lambda \times \Omega^{n+(n+1)})@> \hookrightarrow  >> J^{1}(Y^{*}\underset{X}{\times}
\Omega^{n+(n+1)}) @.\hskip0.3cm
  (\lambda^{i};q^{\mu}, q^{\mu}_{i};p ,p_{i})\\
@V\pi^{*}_{1} VV  @V\pi^{*}_{1}VV\\
(\Lambda \times \Omega^{n+(n+1)})@> \hookrightarrow  >>
 Y^{*}\underset{X}{\times} \Omega^{n+(n+1)}  @. (\lambda^{i};q^{\mu},p )\\
@V \pi_{\Lambda \Omega} VV  @V\pi_{\Lambda \Omega }VV \\
\Lambda @>\hookrightarrow >> Y^{*} @. (\lambda^{i})\\
@V\pi VV  @V\pi_{Y^{*}X} VV\\
. @>\hookrightarrow >> X
\end{CD}
\end{equation}

A choice of a section $\h $ of bundle $\pi_{\Lambda \Omega}$
determines the dual entropy density $\h^{0}(\lambda)$, its flow
$\h^{\nu}(\lambda )$ and the entropy production $\Sigma (\lambda )$
as the function of dual variables $\lambda_i$. \par

A choice of a section $\mathfrak{c}=(q^{\mu}_{i}(\lambda ),p_{i}(\lambda ))$ of the 1-jet bundle
$\pi^{*}_{2}=\pi_{\Lambda \Omega }\circ \pi^{*}_{1}$ determines, in addition to the previous quantities,
the quantities $q^{\mu}_{i}$ and $p_{i}$ as functions of dual variables $\lambda^{i}$.\par

 If we identify
\begin{equation}
\tF^{\mu}_{i} \equiv q^{\mu}_{i}(\lambda ),\ {\tilde {\Pi}}_{i}\equiv p_{i}(\lambda ),
\end{equation}
we see that a choice of a section $\mathfrak{c}$ of the jet bundle $\pi^{*}_{2}$ {\bf is equivalent to the
choice of \emph{all the constitutive relations} of the theory simultaneously}.\par

Recall that a section $\mathfrak{c}$ of the bundle $\pi^{*}_2$ is called {\bf holonomic} if it is a 1-jet
of a section $\h $ of the bundle $\pi_{Y^{*}\times\Omega^{n}}$:
\[
\mathfrak{c}(\lambda)=j^{1}(\h )(\lambda ).
\]

Now we notice that {\bf if the $\Omega^n$-component $\mathfrak{c}^n$ of the section $\mathfrak{c}$ is
holonomic, fields $\tF^{\mu}_{i}(\lambda ),\h^{\nu}(\lambda )$ satisfy to the relations
\begin{equation}
d_{y}\h^\mu =\tF^{\mu}_{i}d\lambda^{i}\Leftrightarrow \tF^{ \mu}_{i}=\frac{\partial \h^{\mu}}{\partial
\lambda^{i}}
\end{equation}
and vice versa}.

To see this we recall (see, for instance \cite{KV,KMS}) that the 1-jet space $J^{1}(\Lambda
\underset{X}{\times} \Omega^n) $ is endowed with the canonical contact structure defined by the forms
\[
\theta^{\mu}= dq^\mu -q^{\mu}_{i }d\lambda^i .
\]

Necessary and sufficient conditions for a section $\mathfrak{c}=(\h^{\mu}(\lambda ),q^{\mu}_{i
}(\lambda))$ to be holonomic is the fulfillment of relations
\[
\mathfrak{c}^{*}(\theta^{\mu})=d\h^{\mu}-q^{\mu}_{i }d\lambda^{i}= 0
\]
for all $\mu =0,1,\ldots, n$ which is the other form of relations (15.6) with the identification (15.5)
above.\par

Assume now that the dual space $\Lambda$ of variables $\lambda^i$ is the vector space and consider now the
Liouville vector field $\zeta$ in the (vector) space
 $\Lambda $
\[
\zeta =\lambda^{i}\frac{\partial}{\partial \lambda^{i}}.
\]
We \textbf{require additionally that the section $\mathfrak{c}$ satisfies to the (residual entropy)}
condition
\begin{equation}
i_{\zeta }(\Pi_{i}(\lambda )\eta\wedge d\lambda^{i})=\Pi_{i}\lambda^{i}=\Sigma (\lambda )\geqq 0
\end{equation}
In such a way we ensure the fulfilment of condition (2.9) \emph{including the positivity of entropy
production} $\Sigma$ (see 2.10).

As a result we have proved the following

\begin{proposition}
The following statements are equivalent
\begin{enumerate}
\item Constitutive relations defined by the section $\mathfrak{c}$ of the
bundle $\pi^{*}_{2}$ satisfy to the entropy principle. \item $\Omega^n$-component of section
$\mathfrak{c}$ \textbf{is holonomic} and $\Omega^(n+1)$-component $\Pi_{i}d\lambda^i \wedge \eta$ of
section $\mathfrak{c}$ satisfies to the \textbf{positivity condition}
\begin{equation}
i_{\zeta }[\Pi_{i}(\lambda )d\lambda^i \wedge \eta]=\Pi_{i}\lambda^{i}\eta=\Sigma (\lambda )\eta \geqq 0.
\end{equation}
In the last inequality we use the nonnegativity defined by the mass
form $dM=\rho\eta $.
\end{enumerate}
\end{proposition}

\begin{example}
Let a function $\Psi (\lambda )$ be given such that the radial monotonicity condition \beq \zeta \cdot
\Psi \geqq 0 \eeq is fulfilled. This condition is equivalent to the geometrical requirement that the
sublevel domains $\Psi^{-1}(-\infty ,c)$ of the function $\Psi $ are "star-shaped" domains with respect to
the origin.\par
 Consider a production vector $\Pi_i$ of the form
\[
\Pi_{i}=\frac{\partial \Psi }{\partial \lambda^{i}} \Leftrightarrow \Pi_{i}d\lambda^{i}\wedge \eta = d\Psi
\wedge \eta
\]
with the function $\Psi (\lambda )$.  Then the positivity condition $\lambda^{i}\Pi_{i}\geqq 0$ is
fulfilled due to the condition (15.9).
\end{example}

Now we would like to present the balance system in terms of dual fields $\lambda^i(x)$ instead of the
original fields $y^{i}(x)$ in the way similar to the Euler-Lagrange Equations in the multisymplectic
Poincare-Cartan formalism (see above):
\begin{equation}
\partial_{x^{\mu}}[(j^{1\ *}(\lambda)\F^{\mu}_{i})(\lambda (x))]=(j^{1\ *}(\lambda
)\Pi_{i}))(\lambda (x)).
\end{equation}
\par

To do this we start with a section
\beq
 \mathfrak{c}=j^{1}(\h )-\Pi
=(\lambda^{i};\h^{\mu}(\lambda)\eta_{\mu}+\Sigma(\lambda)\eta ; \frac{\partial \h^\mu }{\partial
\lambda^i} d\lambda^i \wedge \eta_{\mu}-\Pi_{i}(\lambda)d\lambda^i \wedge \eta ) \eeq
 of the 1-jet bundle $\pi^{*}_2$ satisfying to the conditions of
the Proposition 28 above.\par

Taking the differential $\tilde d$ of the vertical part of section $\mathfrak{c}$ - the (n+1)+(n+2) form
$\mathfrak{c}_{v}=\frac{\partial \h^\mu }{\partial \lambda^i} d\lambda^i \wedge
\eta_{\mu}-\Pi_{i}(\lambda)d\lambda^i \wedge \eta$ we get

\[
d(\frac{\partial \h^\mu }{\partial \lambda^i})\wedge d\lambda^i \wedge
\eta_{\mu}+\Pi_{i}(\lambda)d\lambda^i \wedge \eta = -\partial_{x^\mu }\left( \frac{\partial \h^\mu
}{\partial \lambda^i}\right)d\lambda^i \wedge \eta -\frac{\partial \h^\mu }{\partial
\lambda^i}\lambda_{G,x^\mu }d\lambda^i \wedge \eta +\Pi_{i}(\lambda)d\lambda^i \wedge \eta.
\]

Now we take the interior derivative of this form in the direction of an arbitrary vertical vector field
$\xi \in T(\Lambda )$ (corresponding, in Poincare-Cartan formalism, to the vertical variation of a section
$(\h ,\Sigma )$ in the direction of $\xi $) and get
\[
i_{\xi}\mathfrak{c}_{v}=-\partial_{x^\mu }\left( \frac{\partial \h^\mu }{\partial \lambda^i}\right)\xi^i
\wedge \eta -\frac{\partial \h^\mu }{\partial \lambda^i}\lambda_{G,x^\mu }\xi^i \wedge \eta
+\Pi_{i}(\lambda)\xi^i \wedge \eta
\]
Taking now the pullback of this n+(n+1) form with respect to a section $\lambda =\lambda(x)$ of the bundle
$\pi_{Y^{*}X}:Y^{*}=X\times \Lambda \rightarrow X$ we get
\begin{multline}
\lambda^{*}(i_{\xi}\mathfrak{c}_{v})=i_{\xi \circ \lambda (x)}\lambda^{*}d \mathfrak{c}_{v}= i_{\xi \circ
\lambda (x)}\lambda^{*}d \lambda^{*}\mathfrak{c}_{v}=\\ =\xi^{i}(x,\lambda(x))[-(D_{\mu}\left(
\frac{\partial \h^\mu }{\partial \lambda^i}\right) (x,\lambda(x))-\frac{\partial \h^\mu }{\partial
\lambda^i}\lambda_{G,x^\mu }+\Pi_{i}(x,\lambda(x))]\eta .
\end{multline}

Equating this expression to zero and requiring that the last equation would be fulfilled for a section
$\lambda (x) $ {\bf for arbitrary (vertical) vector field $\xi $ in the space $\Lambda $} we see that the
condition  $\lambda^{*}(i_{\xi}\mathfrak{c}_{v})=0$ is equivalent to the fulfillment of the balance system
of equations (15.10)
\[
\partial_{x^{\mu}}\left(\frac{\partial \h^\mu }{\partial
\lambda^i}(\lambda (x))\right) = \Pi_{i}(\lambda(x))
\]
which is, with the identification $F^{\mu}_{i}=\frac{\partial \h^\mu }{\partial \lambda^i}$ equivalent to
the dual system of balance equations (15.10).  Thus we have proved the following statement

\begin{theorem}
Let
\[
S=j^{1}(\h )-\Pi =(\lambda^{i};\h^{\mu}(\lambda)\eta_{\mu}+\Sigma(\lambda)\eta ; \frac{\partial \h^\mu
}{\partial \lambda^i} d\lambda^i \wedge \eta_{\mu}-\Pi_{i}(\lambda)d\lambda^i \wedge \eta )
\]
 be a (constitutive) section of the 1-jet bundle $\pi^{*}_2$ satisfying to the conditions of
the Proposition 28 above.  Then the following statements about a section $\lambda (x)$ of the bundle
$\pi_{Y^{*}X}:Y^{*}=X\times \Lambda \rightarrow X$ are equivalent:
\begin{enumerate}
\item For any vertical vector field $\xi $ in the space $\Lambda$
\[
\lambda^{*}(i_{\xi}\mathfrak{c}_{v})=0
\]
\item With the identification $\F^{\mu}_{i}(\lambda )=\frac{\partial \h^\mu
}{\partial \lambda^i}$, the system of dual fields $\lambda =\lambda (t,x^\nu )$ satisfy to the balance
system (15.10), to the entropy principle and to the second law of thermodynamics.
\end{enumerate}
\end{theorem}
\vskip 0.6cm

\section{Conclusion.}
Basic structures of a multisymplectic theory of systems of balance laws (balance systems) was developed in
this paper. Constitutive relations of balance systems appears in this scheme as a generalized Legendre
transformations $C$ between the (partial) 1-jet bundles of the configurational bundle $\pi: Y\rightarrow
X$ and the dual bundle of the semi-basic exterior (n+1)+(n+2)-forms on $Y$. Action of geometrical (gauge)
transformations on the constitutive laws $C$ and on the corresponding Poincare-Cartan forms is studied.
Noether Theorem is proved for the symmetry groups of a constitutive law $C$ and the energy-momentum
balance law for a $\nu$-homogenous balance laws is considered.  Entropy principle if formulated for a
general balance systems is formulated and restrictions it put on the constitutive laws are studied. These
considerations are applied to the Rational Extended Thermodynamics (RET) to construct the dual geometrical
picture of RET, present the balance system of RET in an invariant form and to interpret the entropy
principle as the holonomicy of the current component of the constitutive relations.\par

In the second part of this work we will study the partial jet bundles of higher order compatible with the
covariance groups of a balance system (see \cite{MH,YMO,Si}) and extend the scheme presented here to this
situation. Action of the groups of point transformations and the gauge groups on the phase and dual
jet-bundles of a field theory in producing, rearranging and ordering the systems of balance laws
("\emph{balance systems}") of mixed tensorial structure and of different differential order  will be
studied in the framework of the present scheme. More detailed study of the structure of secondary balance
laws of a balance system is the other direction of the future work. Applications to the continuum
mechanics (uniform materials, nonlinear visco-elasticity and the electrodynamics of continua) will be
considered.\par

Another direction of future work would be to extend the constructed scheme to the case of the base
manifolds with the boundary $(X,  \partial X)$.  Even in the case of a homogeneous Thermodynamics the
mathematical (geometrical) description of interaction of a thermodynamical system with the environment
presents a challenge (see, for instance, the works \cite{MD,MB,Mus2}).\par

In the conclusion I would like to express my deep gratitude to Ernst Binz whose interest and discussions
during my short visit to Mannheim in September 2006 were extremely helpful to me and to Professor W.
Muschik for the discussions stimulating my interest to the problems of field thermodynamics and the
entropy principle.

 \vskip1cm

\section{Appendix I. Properties of forms $\eta_{\mu}$.}
Here we collect some properties of the forms $\eta_{\mu}$ that are repeatedly used in the text.\par
 We have
\beq\eta_{\mu}=i_{\partial_{x^\mu}}\eta=(-1)^\mu \sqrt{\vert G\vert}  dx^0 \wedge \ldots \wedge
x^{\mu-1}\wedge dx^{\mu+1}\ldots \wedge dx^n\eeq and $dx^\mu \wedge \eta_{\mu}=\eta$.

The differential $d\eta_{\mu}$ has the form \beq d\eta_{\mu}=(-1)^\mu \sqrt{\vert G\vert}_{,x^\mu}dx^\mu
\wedge dx^0 \wedge \ldots \wedge x^{\mu-1}\wedge dx^{\mu+1}\ldots \wedge dx^n=
(\partial_{x^{\mu}}\lambda_{G})\eta, \eeq where $\lambda_{G}=ln(\sqrt{\vert G\vert})$.\par

  Introduce the (n-1)-forms \[\eta_{\mu
\nu}=i_{\partial_{x^{\nu}}}i_{\partial_{x^{\mu}}}\eta .\]

Then we have \beq \eta_{\mu \nu}=\begin{cases} i_{\partial_{x^{\nu}}}\eta_{\mu} = (-1)^{\mu+\nu}
\sqrt{\vert G\vert}  dx^0 \wedge x^{\nu-1}\wedge dx^{\nu+1} \wedge \ldots \wedge x^{\mu-1}\wedge
dx^{\mu+1}\ldots \wedge dx^n
 ,& \text{if $\nu <\mu$;}\\
i_{\partial_{x^{\nu}}}\eta_{\mu} = (-1)^{\mu+\nu -1} \sqrt{\vert G\vert}  dx^0 \wedge x^{\nu-1}\wedge
dx^{\nu+1} \wedge \ldots \wedge x^{\mu-1}\wedge dx^{\mu+1}\ldots \wedge dx^n,& \text{if $\nu >\mu$.}
\end{cases},
\eeq

 and, in particular, for all $\mu ,\nu$,

\beq \eta_{\mu \nu}=-\eta_{\nu \mu}. \eeq

We also have

\beq
dx^\sigma \wedge \eta_{\mu \nu}=\begin{cases} \eta_{\mu} &\text{if $\sigma =\nu$},\\
 -\eta_{\nu} &\text{if $\sigma =\mu$},\\
0, &\text{otherwise}, \end{cases} \eeq
 for all $\mu,\nu$.  To see this we first check it explicitly for
$\nu <\mu$ and then, for $\nu >\mu$ we use $dx^\sigma \wedge \eta_{\mu \nu}=-dx^\sigma \wedge \eta_{\nu
\mu}$ and use the proved result.\par

For the differentials of these forms we calculate for the case $\mu <\nu$

\begin{multline} d\eta_{\mu \nu}=d[(-1)^{\mu+\nu} \sqrt{\vert G\vert}  dx^0 \wedge x^{\nu-1}\wedge dx^{\nu+1} \wedge
\ldots \wedge x^{\mu-1}\wedge dx^{\mu+1}\ldots \wedge dx^n]=\\ (-1)^{\mu+\nu}\sqrt{\vert
G\vert}[\partial_{x^{\nu}}\lambda_{G})dx^\nu +\partial_{x^{\mu}}\lambda_{G})dx^\mu]\wedge  dx^0 \wedge
x^{\nu-1}\wedge dx^{\nu+1} \wedge \ldots \wedge x^{\mu-1}\wedge dx^{\mu+1}\ldots \wedge dx^n=\\ =(-1)^\mu
\partial_{x^{\nu}}\lambda_{G})\sqrt{\vert G\vert}dx^0   \wedge
\ldots \wedge x^{\mu-1}\wedge dx^{\mu+1}\ldots \wedge dx^n +(-1)^{\nu
+1}\partial_{x^{\mu}}\lambda_{G})\sqrt{\vert G\vert}dx^0   \wedge \ldots \wedge x^{\nu-1}\wedge
dx^{\nu+1}\ldots \wedge dx^n=\\
=((\partial_{x^{\nu}}\lambda_{G})\eta_{\mu}-(\partial_{x^{\mu}}\lambda_{G})\eta_{\nu}).
\end{multline}
and then notice that using (20.5) we get the same result for the case $\nu >\mu$.\par \vskip1cm

\section{Appendix II. Formalism of Rational Extended Thermodynamics (RET).}

Here we describe, in a short form the basic structure of the Rational Extended Thermodynamics developed by
I.Muller and T.Ruggeri, \cite{M,MR}.  For the complete presentation of the formalism of Rational Extended
Thermodynamics we refer to the monograph \cite{MR}, Chapter 3. Here we introduce only necessary material
in the form suited for our purposes.  To be more consistent to the standard notations in the book
\cite{MR} we will use in this section the notations $u^i$ for the basic fields instead of $y^i$.
Constructions of this section are mostly specializations of those of Section 2. \par

\subsection{Space-time base.}

A state of material body will be described by the collection of the time-dependent fields $\{ u^i
,i=1,\ldots ,m\}$ defined in a domain $B\subset E^3$ of the physical euclidian) space $(E,h)$ with the
boundary $\partial B$. We assume that the Pseudo-Riemannian metric $G$ is defined in $X$. An example of
such a metric is the Euclidian metric $g=dt^2+h$ or Lorentz metric. We introduce (global) coordinates
$x^\mu ,\mu =1,2,3$ in $B$ and the time $t=x^0$. Altogether fields $u^i$ are defined in the n-dim physical
space-time $X=\mathbb{R}_{t}\times {\bar B}.$
\par

Denote by $\eta$ the volume n-form $\eta=\sqrt{\vert G\vert}dt\wedge dx^1 \wedge dx^2 \wedge dx^n$
corresponding to the metric $G$.
\par

\subsection{State (configurational) bundle.}
Basic fields of a continuum thermodynamical theory $u^i$ (except of the entropy that will be included
later) take values in the space $U\subset \mathbb{R}^m$ which we will call the {\bf basic state space} of
the system.

Following the framework of a classical field theory (see \cite{BSF,FF}) we organize these fields in the
bundle
\[
\pi_{U}: Y\rightarrow X,\ X=\mathbb{R}_{t}\times {\bar B}, \ Y=X\times U
\]
with the base $X$ being the cylinder $\mathbb{R}\times {\bar B}$ in the Newtonian space-time and the fiber
$U$.
\par

To formulate balance equations in terms of exterior forms we will use the spaces of  (3+4)- exterior forms
in $X=\mathbb{R}^4$ introduced in Section 1. This space has as its basis elements
$\eta_{\mu},\mu=0,1,2,3;\eta$ and is the space of smooth) sections of the bundle of exterior forms of
orders 3 and 4 $\Lambda ^{3+4}(X)= \Lambda^{3}(X)\oplus \Lambda^{4}(X)$ over $X$.\par

Taking the pullback of the bundle $\Lambda^{3+4}(X)\rightarrow X$ to $Y$ (or, what is the same, construct
the fiber product of $\pi_{YX}$ and $\pi_{\Lambda X}$ we get the following commutative diagram

\beq
\begin{CD}
U\times \Lambda^{3+4} @>>> Y\underset{X}{\times} \Lambda^{3+4} @>>> \Lambda^{3+4}(X)\\
@V\pi_{\Lambda U}VV   @V\pi_{\Lambda Y} VV                        @V \pi_{\Lambda X}VV\\
U @>>> Y         @>\pi_{YX} >> X
\end{CD}.
\eeq Left column of this diagram represents a typical fiber of bundle $\pi_{\Omega Y} $ over a point $x\in
X$. Notice also that the sections of the bundle $\pi_{YX} \circ\pi_{\Omega Y} : Y\underset{X}{\times}
\Omega^{3+4} \rightarrow X$ are the "semibasic" (3+4) exterior forms on the space $Y$ of the bundle
$\pi_{YX}$, see \cite{LR}, Sec.4.2.\par

\subsection{Balance Equations.}
Fields $u^i$ are to be determined as solutions of the field equations having the form of {\bf balance
equations} for the currents $F^{\mu}_{i}$, where $F^{0}_{i}=u^i$
\begin{equation}
F^{ \mu}_{i,\mu}=u^{i}_{,t}+F^{\nu}_{i,x^{\nu}}  =\Pi_{i},\ i=1,\ldots, n.
\end{equation}
Here $\Pi_{i}(u,x)$ is called the {\bf production} of the component $u^i$ and $\sum_{\nu=1}^{3}F^{
\nu}_{i}(u,x)\frac{\partial }{\partial x^\nu}$ - the {\bf flow } of the component $u^i$. These quantities
are assumed to be function of the fields $u^i$ and, possibly, of the point $x^\mu \in X.$ Usually in RET
one restricts the attention to the case where there $F^{\mu}_{i} ,\Pi_i$ do not depend explicitly on the
space-time point $x^\mu$.

\begin{remark}
In the rational Extended Thermodynamics one consider a case where balance equations are written \emph{for
all the basic fields in the state space and only for them} and where flows $F^{\mu}_{i}$ and productions
$\Pi_i$ depend on the fields $u^i$ \emph{but not on their gradients or time derivatives}.
\end{remark}

To close system of equations (18.2) for $u^i$ one has to choose the flows and production forms as
functions of $u^i$ - to choose the {\bf constitutive equations} of the body.  Such a choice should be done
for each balance equation.  As we will see below, utilizing of the entropy condition allows to reduce this
process to the choice of {\bf entropy flow} 3-form and to the choice of production 4-forms subject to the
positivity condition.

\subsection{Entropy condition.}

Entropy $h^0(u)$ is assumed to be a function of the basic state variables $u^i$. It satisfies to the
balance law
\begin{equation}
d(h^{\mu}\eta_{\mu})=\Sigma ,
\end{equation}
with the {\bf positive } production 4-form

\beq \Sigma =\sigma (u)\eta, \sigma \geqq 0, \eeq and the flow 3-form $H(u)=\sum_{\nu=1}^{3}h^\nu
\eta_{\nu}$.\par
\begin{remark}
To clarify the geometrical meaning of positivity of an exterior 4-form recall that for each material there
is defined the {\bf mass form} $dM=\rho\eta. $ Using this form we {\bf define a given 4-form $f\eta$ to be
nonnegative (positive) if $f/\rho\geqq 0\ (>0)$}.
\end{remark}

{\bf Entropy principle} requires that any solution of the balance equations (18.2) would also satisfy to
the equation (18.3) and that the production $\sigma$ of entropy (in the system) should be non-negative.

In addition to this a requirement of {\bf convexity}

\beq \frac{\partial^{2}h^0}{\partial u \partial u} \sim \text{negative\ definite} \eeq has to be
fulfilled.\par
\begin{remark}$^*$
The last condition shows that the symmetrical bilinear form \beq g_{ij}(u)=-
\frac{\partial^{2}h^0}{\partial u \partial u}
 \eeq
 can be considered as a \textbf{degenerate Riemannian metric} in the state
 space $U$.  This is the Ruppeiner thermodynamical metric
 (\cite{M}).  It would be interesting to interpret the curvature of
 this metric in the context of RET.
\end{remark}

Requirement of the fulfillment of the entropy balance equation (18.3) for all solutions of balance
equations (18.2) for the fields $u^i$ leads to strong limitations on the form of constitutive equations.
Namely, this condition is equivalent to the following two statements: There exists a functions
$\lambda^{i}(u)$ (Lagrange multipliers) on the space $U$ such that for all values of variables $u^i$
\begin{equation}
\frac{\partial h^\mu }{\partial u^i}=\lambda _{j}\cdot \frac{\partial F^{\mu }_{j}}{\partial
u^i}\Leftrightarrow dh^{\mu}=\lambda_{j}\cdot dF^{\mu}_{j},
\end{equation}
and
\begin{equation}
\Sigma =\lambda^{i}\Pi_{i}\geqq 0.
\end{equation}

\par First of the equation (18.7) defines the Lagrange-Liu
multipliers

\beq \frac{\partial h^0}{\partial u^i}=\lambda^i. \eeq

Differentiating by $u^j$ we get
\[
Hess(h^0 )=\frac{\partial^2 h^0}{\partial u^i \partial u^j}=\frac{\partial \lambda^i}{\partial u^j}
\]
from which it follows that if the entropy density $h^0$ is a strongly convex function of its arguments
$u^i$, then the change of variables $u\rightarrow \lambda $ is {\bf globally invertible}. Thus, we get the
diffeomorphic mapping

\beq \wp : U\rightarrow \Lambda \eeq
 from the state space $U$
onto the space $\Lambda \subset R^n$ of values of variables $\lambda =\{ \lambda^{i} \}.$\par
\subsection{Dual formulation.}
As a result one may present all the quantities as the functions of dual variables $\lambda_{i}$:
\begin{equation}
\tF^{ \mu}_{i}=\tF^{ \mu}_{i}(\lambda ), \tP_{i} \equiv \tP_{i}(\lambda ); \tih^\mu =\tih^\mu (\lambda).
\end{equation}
Combining balance equations (18.2) with this change of variables we rewrite these equations in the form
\begin{equation}
\frac{\partial \tF^{ \mu}_{i}}{\partial \lambda^{j}}\cdot \frac{\partial \lambda^{j}}{\partial
x^{\mu}}=\tP_{i}(\lambda ) \Leftrightarrow \frac{\partial^2 \h^\mu}{\partial \lambda^i
\partial \lambda_j}\frac{\partial \lambda_{j}}{\partial
x^{\mu}}=\tP_{i}(\lambda ),\ i=1,\ldots ,n,
\end{equation}
where the {\bf four-vector potential} (or 3-form)
\begin{equation}
\h^\mu =\lambda^i \cdot \tF^{ \mu}_{i}-\tih^\mu (\lambda )
\end{equation}
was introduced.  In terms of $\hat h$ the relation (18.7) takes the form
\begin{equation}
d\h^\mu (\lambda) =\F^{ \mu}_{i}d\lambda^i,
\end{equation}
summation is assumed by repeating indices.

In terms of 3-forms
\begin{equation}
\h =\lambda^i \cdot \tF^{ \mu}_{i}\eta_{\mu} -\tih^\mu (\lambda )\eta_{\mu}=\lambda_{i}\tF_{i}-\tih.
\end{equation}

From the relation (18.14) it follows that
\begin{equation}
\tF^{\mu}_{i}=\frac{\partial \h^\mu}{\partial \lambda^i} \Rightarrow \tih^{\mu}(\lambda
)=-\h^{\mu}+\lambda^{i}\cdot \frac{\partial \h^\mu}{\partial \lambda^i}.
\end{equation}

As a result, $4n+4$ constitutive functions $\tF^{\mu}_{i}$ and $\tih^{\mu}(\lambda )$ can, in terms of
$\lambda $ variables be derived from the 4 functions $\h^\mu $ - coefficients of 3-form $\h$.

\begin{remark}
As long as we are not dealing with variables $\lambda^i$ as fields in space and time (functions of
$x^\mu$) the presentation of $\h $ as a four-vector potential or as a 3-form in the 4D space-time is pure
formal. We use this representation as the starting point for construction of double bundles of the
geometrical form of RET (see Section 15).
\end{remark}

After presenting currents $\tF^{\mu}_{i}$ in the form (18.16) what is left of the requirements of entropy
principle (provided the condition of convexity of $h^0$ is fulfilled) is the {\bf residual inequality}
\begin{equation}
\Sigma(\lambda )=\lambda^i \Pi_i \geqq 0.
\end{equation}
Two statements containing here determine the entropy production $\Sigma $ in terms of the production
4-forms $\Pi_i $ and {\bf require} positivity of $\Sigma $.\par

Reversing the arguments leading to the statements (18.15) and (18.16) one proves the following basic
result of RET leading to the dual formulation of balance equations (18.2) and the entropy principle (18.3)
\begin{theorem}{\cite{MR}}
The following statements are equivalent under the condition of the convexity of entropy density
$h^{0}(u^i)$ as the function of fields $u^i$:
\begin{enumerate}
\item  Entropy principle is fulfilled for the balance equations
(3.2) and the entropy balance equation (3.3) for given constitutive functions
$F(u),\Pi(u),h(u),\Sigma(u)$.

\item Constitutive fields $F(u),h(u),\Sigma(u)$ are obtained by
the relations (18.12),(18.15), (18.16) from the four-potential $\h(\lambda )$ (formal 3-form) and the
production 4-forms $\Pi_i (\lambda)$ for which the residual inequality
\[
\lambda_i \Pi_i \geqq 0
\]
is fulfilled.
\end{enumerate}
\end{theorem}

\section{Appendix III. Iglesias Differential.}

Differential ${\tilde d}$ is a special case of operators introduced by D. Iglesias and used in \cite{IW}.
\beq
\begin{cases}
 {\tilde d}: \Omega^{k+(k+1)}=\Omega^{k}(X)\oplus
\Omega^{k+1}(X)\rightarrow \Omega^{(k+1)+(k+2)}=\Omega^{k+1}(X)\oplus \Omega^{k+2}(X):\\ {\tilde
d}(\alpha^{k}+\beta^{k+1} )=((-d\alpha+\beta )+d\beta ).
\end{cases} \eeq
\begin{lemma} ${\tilde d}\circ {\tilde d}=0.$
\end{lemma}
\begin{proof} We have
\[{\tilde d} {\tilde d} (\alpha^{k}+\beta^{k+1}) = {\tilde d} ((-d\alpha+\beta )+d\beta )=
[-d(-d\alpha+\beta )+d\beta ]+d(d\beta)=-d\beta +d\beta +0.
\]
\end{proof}
The complex
 \beq
 0\rightarrow \Omega^{1}(X)\oplus
\Omega^{0}(X)\rightarrow \ldots \rightarrow \Omega^{k}(X)\oplus \Omega^{k-1}(X)\rightarrow \ldots
\Omega^{n}(X)\oplus \Omega^{n-1}(X)\rightarrow 0\oplus \Omega^{n}(X)\rightarrow \eeq is generated by de
Rham complex of a manifold $X$ and corresponds to the couples of forms $\alpha^{k}+\beta^{k+1}$.  This
complex can be considered as dual to the complex of chains generated by couples $(C^{k+1},\partial C^{k})$
of submanifolds $C^{k+1}\subset X^n$ of dimension k with the boundary $\partial C^{k}$: Duality is defined
by integration
\[
<\alpha^{k}+\beta^{k+1}, (C^{k+1},\partial C^{k})>=\int_{C}\beta +\int_{\partial C}\alpha .
\]
We have, obviously,
\[
<\tilde d (\alpha^{k}+\beta^{k+1}), (C^{k+1},\partial C^{k})>=0
\]
for all $(C,\partial C )$ iff ${\tilde d}(\alpha^{k}+\beta^{k+1})=0.$

\section{Appendix IV. Reduced horizontal differential.}

here we recall the properties of horizontal differential $d_{H}$ and introduce an augmented horizontal
differential $\hat d$ that is used in Sec.10.

Recall \cite{KV,GMS} that the r-jet bundles $J^{r}(\pi)$ of a bundle $\pi:Y\rightarrow X$ form the inverse
system

\beq X\xleftarrow{\pi} Y\xleftarrow{\pi^{1}_{0}}J^{1}(\pi)\xleftarrow{\pi^{2}_{1}}\xleftarrow \ldots
\xleftarrow{\pi^{r}_{r-1}}J^{r}(\pi)\xleftarrow \ldots \eeq

whose inverse limit $J^{\infty}(\pi)$ is the infinite order jet bundle of the bundle $\pi$.\par Adapted
local coordinates $(x^\mu ,y^i )$ in $Y$ determine the local coordinates $(x^\mu ,y^i ,y^{i}_{\Lambda})$,
where multi-index $\Lambda=(\lambda_{k},\lambda_{k-1},\ldots ,\lambda_{1})$ is a collection of natural
numbers modulo permutations. We denote by
$\partial_{\Lambda}=\partial_{\lambda_{k}}\circ\partial_{\lambda_{k-1}}\circ\ldots \partial_{\lambda_{1}}$
the composition of derivations.\par

Corresponding to the inverse system (20.1) we have the inverse system of projectable vector fields $X_{r}$
on the r-jet bundles $\pi^{r}_{0}:J^{r}(\pi)\rightarrow X.$  \par

Dually, there is the direct system

\beq \Lambda^{*}(X)\xrightarrow{\pi^{*}}
\Lambda^{*}(Y)\xrightarrow{\pi^{1*}_{0}}\Lambda^{*}(J^{1}(\pi))\xrightarrow \ldots \xrightarrow{\pi^{r\
*}_{r-1}}\Lambda^{*}(J^{r}(\pi)))\ldots  \eeq
induced by the pullback of the forms from the lower order jet bundles to the higher order jet bundles.
Limit of this direct system is the exterior $Z$-graded algebra called the bundle
$\mathfrak{D}^{*}_{\infty}=\Lambda^{*}(J^{\infty}(\pi))$ of exterior forms on $J^{\infty}(\pi).$\par

Bundle of algebras $\mathfrak{D}^{*}_{\infty}$ is locally generated by the basic forms $dx^\mu$ and the
contact forms
\[
\theta^{i}_{\Lambda}=dy^{i}_{\Lambda}=y^{i}_{\Lambda+\lambda}dx^\lambda,\ 0\geqq \vert \Lambda \vert.
\]

As a result, the vector subspace $\mathfrak{D}^{s}_{\infty}=\Lambda^{s}(J^{\infty}(\pi))$ of exterior
$s$-forms has the canonical decomposition
\[
\mathfrak{D}^{s}_{\infty}=D^{0,s}_{\infty}\oplus \mathfrak{D}^{1,s-1}_{\infty}\oplus \ldots \oplus
D^{s,0}_{\infty}.
\]
elements of $\mathfrak{D}^{k,s-k}_{\infty}$ are called k-contact forms. Denote by
$h_{k}:\mathfrak{D}^{s}_{\infty}\rightarrow \mathfrak{D}_{\infty}^{k,s-k},\ k\leqq s$ the $k$-contact
projection. Especially important is the horizontal projection $h_{0}:\mathfrak{D}^{s}_{\infty}\rightarrow
\mathfrak{D}^{0,s}_{\infty}$ given by

\beq dx^{\mu}\rightarrow dx^{\mu},\ dy^{i}_{\Lambda}\rightarrow y^{i}_{\lambda +\Lambda}dx^{\lambda}. \eeq
Accordingly, the exterior differential on $\mathfrak{D}^{*}_{\infty}$ is decomposed into the sum

\beq d=d_{h}+d_{v} \eeq

of horizontal differential $d_{H}$ and vertical differential $d_{v}$ so that when \beq \begin{cases}
d:D^{k,s-k}_{\infty}\rightarrow  D^{k+1,s-k}_{\infty}\oplus D^{k,s-k+1}_{\infty},\\
d_{H}:D^{k,s-k}_{\infty}\rightarrow  D^{k,s-k+1}_{\infty},\\
d_{v}:D^{k,s-k}_{\infty}\rightarrow  D^{k+1,s-k}_{\infty}.
\end{cases}
\eeq We have homology properties
\[
d^{2}_{H}=d^{2}_{V}=d_{V}d_{H}+d_{H}d_{V}=0
\]
and the relation
\[
h_{0}\circ d=d_{H}\circ h_{0}.
\]
Introduce the \emph{total derivative} $d_{\mu}$ - lift of partial derivation $\partial_{\mu}$ to the by
the rules to the vector field in $J^{\infty}(\pi)$ in the sense of \cite{KV,O1}:
\[
d_{\mu}f(x,y,z)=\frac{\partial f}{\partial x^\mu }+z^{i}_{\mu}\frac{\partial f}{\partial y^i}+\sum_{i,\{
\mu_{1}\ldots \mu^{k}\mu\}} z^{i}_{\mu_{1}\ldots \mu_{k}\mu}\frac{\partial f}{\partial
z^{i}_{\mu_{1}\ldots \mu_{k}}}.
\]

It acts on the exterior forms by the rules \beq
\begin{cases}
d_{\mu}(\nu \wedge \sigma)=d_{\mu}\nu \wedge \sigma +\nu \wedge d_{\mu}\sigma,\\
d_{\mu}d\sigma =dd_{\mu}\sigma,\ \sigma ,\nu \in \mathfrak{D}^{*}_{\infty}.
\end{cases}
\eeq

 Then the horizontal differential is locally given by expression \beq d_{H}\omega =dx^\mu \wedge
d_{\mu}(\omega),\omega in  \mathfrak{D}^{*}_{\infty}.\eeq

From these properties the following relations follows

\beq
\begin{cases}
d_{H}f=d_{\lambda} fdx^\lambda , f\in \mathfrak{D}^{0}_{\infty},\\
d_{\lambda}(dx^\mu )=0,\ d_H (dx^\mu )=0,\\
d_\lambda (dz^{i}_{\Lambda})=dz^{i}_{\Lambda+\lambda }),\ d_{H}(dz^{i}_{\Lambda} )=dx^\lambda \wedge
dz^{i}_{\Lambda +\lambda },\\
d_{\lambda} (\theta^{i}_{\Lambda} )=\theta^{i}_{\lambda +\Lambda},\ d_{H}(\theta^{i}_{\Lambda}
)=dx^\lambda \wedge \theta^{i}_{\Lambda +\lambda}.
\end{cases}
\eeq \vskip0.5cm

 Directly
from the definition of total derivative the following properties follows
\begin{lemma} Acting on the functions from $C^{\infty}(J^{\infty}(\pi)),$
\begin{enumerate}
\item $[d_{\mu},\partial_{x^\nu} ]=0$,
\item $[d_{\mu},\partial_{y^i} ]=0.$
\end{enumerate}
\end{lemma}

Working with the partial 1-jet bundles $J^{1}_{p}(\pi)$ (see Sec. 4-7) we have to use the reduced version
of the total derivative. We keep the same notation for this derivative silently assuming that when working
on a special kind of partial 1-jet bundle we use the appropriate version of $d_{\mu}.$  Thus, on
$J^{i}_{p}(\pi)$ with the model vector bundle having as its fiber over $y\in Y$ the factor-space of  $
T^{*}(X)\otimes V(\pi)$: for instance for $J^{1}_{K}(\pi)$ the fiber has the form $T^{*}_{K}(X)\otimes
V(\pi)$ of the cotangent bundle $T^{*}(X)$, fiber coincide with $T^{*}(X)\otimes V(\pi)$ in the case of
the full 1-jet bundle and reduces to zero in the RET case.  In local coordinates $(x^\mu ,y^i )$ denote by
$P$ the set of pairs of indices $(\mu ,i)$ such that coordinate $z^{i}_{\mu}$ is defined in
$J^{1}_{p}(\pi)$, or, what is the same, such that $dx^\mu \otimes \partial_{y^i}$ generate the nonzero
element of the fiber of vector model for $Z_{p}$.   Thus, we define,

\beq d_{\mu}f=\partial_{x^\mu}f+\sum_{(\mu ,i)\in P}z^{i}_{\mu}\partial_{y^i}f+\sum_{(\mu ,i)\in
P}z^{i}_{\mu \sigma}\partial_{z^{i}_{\mu}}f.\eeq

It is easy to see that total derivative defined in this way preserves the properties (20.6) and the
properties listed in Lemma 10.\par

In addition to the horizontal differential $d_{H}$ we will be using an "reduced horizontal differential"
$\hat d$. We define operator $\hat d$ by the properties

\beq
\begin{cases}
\hd f =d_{\mu}f dx^\mu,\\
\hd (dx^\nu)= \hd (dy^i )= \hd (dz^{i}_{\Lambda})=0,\\
\hd (\omega_{1}\wedge \omega_{2}=(\hd \omega_{1})\wedge \omega_{2}+\omega_{1}\wedge (\hd \omega_{2}).
\end{cases}
\eeq

In other words we define $\hd$ first on the semi-basic subalgebra of algebra $\mathfrak{D}^{*}_{\infty}$
and then extend the differential $\hat d$ to the whole algebra  by requiring that

\[
{\hat d}dy^i =0,{\hat d}dz^{i}_{\mu_{1}\ldots \mu_{k}}=0.
\]
\begin{lemma}
Operator ${\hat d}: \Lambda^{k}(J^{\infty }(\pi)) \rightarrow \Lambda^{k+1}(J^{\infty}(\pi))$  preserves
the subcomplex $\Lambda^{*}_{Y}(J^{\infty }(\pi))$ of $\pi^{1}_{0}$-semibasic forms (with the generators
$dy^i,dx^\mu$) and maps the subspaces of the forms annulated by $r$ $\pi$-vertical arguments into itself
\[
{\hat d}:\Lambda^{k}_{r}(J^{\infty }(\pi))\rightarrow \Lambda^{k+1}_{r}(J^{\infty }(\pi)).
\]
\end{lemma}
Since $d_{\lambda} d_{\mu}=d_{\mu}d_{\lambda}$ if applied to the functions, it is easy to check that $\hd
\circ \hd =0$, so $\hd$ is the differential operator: $\hd^2=0$.\par

\begin{remark}  Notice that operator $\hd$ does not commute with the usual differential $d$, for instance
$\hd dy^i=0$ but $d \hd y^i =d (z^{i}_{\mu}dx^\mu)\ne 0.$
\end{remark}
\begin{lemma} Let $C\Lambda (J^{\infty}(\pi))$ be the ideal in  $\Lambda (J^{\infty}(\pi))$ of
the contact forms (forms annulating the Cartan distribution), then for any form $\nu \in \Lambda
(J^{\infty}(\pi))$ we have
\[
(\hat d -d )\nu \in C\Lambda (J^{\infty}(\pi)).
\]
\begin{proof} Both operators $d$ and $\hat d$ are derivations of the exterior algebra, therefore it is
sufficient to prove the statement for generators of this algebra $f(x,u,z),dx^\mu ,dy^i ,dz^{i}_{\bar
\mu},\ {\bar \mu}=\mu_{1},\ldots ,\mu_{k}.$  For differentials the result is obvious - both operators
annulate them.  For the functions we have
\[
{\hat d}f=f_{,x^\mu }{\hat d}x^\mu +f_{,y^i}{\hat d}y^i +\sum_{\bar \mu}f_{,z^{i}_{\bar \mu}}{\hat
d}z^{i}_{\bar \mu}=f_{,x^\mu } dx^\mu +f_{,y^i}z^{i}_{\mu}dx^{\mu} +\sum_{\bar \mu}f_{,z^{i}_{\bar
\mu}}z^{i}_{{\bar \mu}\nu}dx^\nu .
\]
Subtracting from this expression the similar (but simpler) expression for $df$ we get
\[
({\hat d}-d)f=f_{,y^i}(z^{i}_{\mu}dx^{\mu}-dy^i ) +\sum_{\bar \mu}f_{,z^{i}_{\bar \mu}}(z^{i}_{{\bar
\mu}\nu}dx^\nu-dz^{i}_{\bar \mu})=-f_{,y^i}\omega^{i} +\sum_{\bar \mu}f_{,z^{i}_{\bar
\mu}}\omega^{i}_{\bar \mu},
\]
that finishes the proof.
\end{proof}
\end{lemma}
\begin{proposition} Let $\phi $ be an automorphism of the bundle $\pi$ and $\phi^\infty $ - its contact
(=flow) prolongation to the $J^{\infty}(\pi)$. Then
\[
{\hat d}\phi^{\infty\ *}\omega \equiv \phi^{\infty\ *}{\hat d}\omega\  \text{mod}\  C\Lambda^{*}
\]
for all $\pi^{1}_{0}$-semibasic forms $\omega$ on $J^{\infty}(\pi).$  Here $C\Lambda^{*}(J^{\infty}(\pi))$
is the ideal in generated by the Cartan forms (forms annulating the Cartan distribution.
\end{proposition}
\begin{proof}
Mapping $\phi^{\infty}_{*}$ of tangent spaces leaves the Cartan distribution invariant, therefore the
pullback $\phi^{\infty \ *}$ of the forms leaves the Contact ideal $C\Lambda^{*}(J^{\infty}(\pi))$
invariant.  Therefore, for all forms $\nu$ by the previous Lemma
\[
\phi^{\infty \ *}({\hat d}-d)\nu \in C\Lambda^{*}(J^{\infty }(\pi )).
\]
On the other hand by the same Lemma $({\hat d}-d)\phi^{\infty \ *}\nu \in C\Lambda^{*}(J^{\infty}(\pi))$
as well.  This last inclusion can be written
\[
{\hat d}\phi^{\infty \ *}\nu -\phi^{\infty \ *}d\nu \in C\Lambda^{*}(J^{\infty}(\pi))
\]
since the pullback commutes with the de Rham differential. Subtracting obtained inclusions we get the
result stated in the Proposition.\par

 Both $\phi^{\infty \ *}$ and ${\hat d}$ are linear and respect the wedge product in the corresponding
sense.  Therefore, one can check the statement for the generators $f,\ dx^\mu , du^i$ only.
\par
For $dx^\mu$ total differential reduces to the usual de Rham differential on $X$ and $\phi^\infty$ acts by
the projection ${\bar \phi}:X\rightarrow X$. Thus, the statement reduces for the usual property of $d$.
\par
For $dy^i$ we have
\[
{\hat d}\phi^{\infty\ *}dy^i={\hat d}d\phi^{i}={\hat d}[\phi^{i}_{,x^\nu }dx^\nu +\phi^{i}_{,y^j }dy^j]=
d_{\mu}\phi^{i}_{,x^\nu }dx^\mu \wedge dx^\nu +d_{\mu}\phi^{i}_{,y^j }dx^\mu \wedge du^j=\]

\[ =
(\partial_{x^\mu }\partial_{x^\nu }\phi^{i}+z^{j}_{\mu}\partial_{y^j} \partial_{x^\nu}\phi^{i})dx^\mu
\wedge dx^\nu +(\phi^{i}_{x^\nu u^j}+z^{k}_{\mu}\phi^{i}_{y^j y^k })dx^\mu \wedge dy^j =
\]

\[ =\partial_{x^\mu
}\partial_{x^\nu }\phi^{i}dx^\mu \wedge dx^\nu +
 (z^{j}_{\mu }dx^\mu -dy^j )\phi^{i}_{y^j x^\nu} \wedge dx^\nu +z^{k}_{x^\mu }dx^\mu \wedge
 \phi^{i}_{u^j y^k }dy^j =
 \]

 \[=\partial_{x^\mu
}\partial_{x^\nu }\phi^{i}dx^\mu \wedge dx^\nu -\omega^j \wedge \phi^{i}_{y^j x^\nu} dx^\nu -(dy^k
-z^{k}_{x^\mu }dx^\mu )\wedge \phi^{i}_{y^j y^k }dy^j +\wedge \phi^{i}_{y^j y^k }dy^k \wedge dy^j=
\]

\[= -\omega^j \wedge \phi^{i}_{y^j x^\nu} dx^\nu-\omega^k \wedge \phi^{i}_{y^j y^k }dy^j =
-\omega^j\wedge (\phi^{i}_{y^j x^\nu} dx^\nu+\phi^{i}_{y^j y^k }dy^k ),
\]
here we canceled two terms with second derivatives of $\phi^i$ due to the antisymmetry of wedge products
of basic forms.  Since $\phi^{\infty\ *}{\hat d}dy^i=0$, statement is proved for $dy^i$.
\end{proof}

\vfill \eject

\end{document}